\documentclass{aa} 

\usepackage{graphicx}
\usepackage{txfonts}
\usepackage{color}
\usepackage[colorlinks=true, allcolors=blue]{hyperref}
\usepackage{multirow}
\usepackage{xspace}
\usepackage{booktabs}
\usepackage{dcolumn}
\newcolumntype{d}[1]{D{.}{.}{#1}}

\usepackage{float}

\usepackage[amsmath,thref,hyperref]{ntheorem}
\usepackage{amsmath}

\usepackage{natbib}
\bibpunct{(}{)}{;}{a}{}{,}

\usepackage{stfloats}

\usepackage{rotating}

\usepackage{ulem}

\newcommand{\uergcm}[1]{erg cm$^{-2}$ s$^{-1}$}
\newcommand{\ergcm}[1]{$\times 10^{#1}$ erg cm$^{-2}$ s$^{-1}$}

\newcommand{\ergs}[1]{$\times 10^{#1}$ erg s$^{-1}$}
\newcommand{\oergs}[1]{$10^{#1}$ erg s$^{-1}$}
\newcommand{\hcm}[1]{$\times 10^{#1}$ cm$^{-2}$}

\newcommand{\kms}{km s$^{-1}$\xspace}
\newcommand{\nh}{\ensuremath{N_\mathrm{H}}}

\newcommand{\Halpha}{H${\alpha}$\xspace}
\newcommand{\Hbeta}{H${\beta}$\xspace}
\newcommand{\HeI}{HeI\,4921\xspace}
\newcommand{\ltsima}{$\buildrel < \over \sim$}
\newcommand{\lsim}{\lower.5ex\hbox{\ltsima}}
\newcommand{\gtsima}{$\buildrel > \over \sim$}
\newcommand{\gsim}{\lower.5ex\hbox{\gtsima}}

\newcommand{\eSASS}{\texttt{eSASS}\xspace}

\newcommand{\ROSAT}{{ROSAT}\xspace}
\newcommand{\nicer}{{NICER}\xspace}
\newcommand{\gaia}{{\it Gaia}\xspace}
\newcommand{\swift}{{\it Swift}\xspace}
\newcommand{\xmm}{{\it XMM-Newton}\xspace}
\newcommand{\nus}{{\it NuSTAR}\xspace}
\newcommand{\cxo}{\hbox{Chandra}\xspace}

\newcommand{\ero}{\mbox{eROSITA}\xspace}
\newcommand{\srg}{{\it SRG}\xspace}
\newcommand{\artxc}{\mbox{ART-XC}\xspace}
\newcommand{\ogle}{{OGLE}\xspace}
\newcommand{\salt}{{SALT}\xspace}

\newcommand{\Poisson}[1]{\mathrm{Poisson}\left(#1\right)}
\newcommand{\unter}[1]{_\mathrm{\tiny{#1}}}

\begin{document} 
        
        \title{A comprehensive catalogue of high-mass X-ray binaries in the Large Magellanic Cloud detected during the first eROSITA all-sky survey}
        
        \author{D. Kaltenbrunner\inst{\ref{mpe}} \and
                C. Maitra\inst{\ref{mpe},\ref{iucaa}} \and
                F. Haberl\inst{\ref{mpe}} \and
                J. Bodensteiner\inst{\ref{eso},\ref{apia}} \and
                D. Bogensberger\inst{\ref{umich},\ref{ups}} \and
                D.\,A.\,H. Buckley\inst{\ref{salt},\ref{uct}}
                M. R. L. Cioni\inst{\ref{liap}} \and
                J. Greiner\inst{\ref{mpe}} \and
                I. Monageng\inst{\ref{salt},\ref{uct}} \and
                A. Udalski\inst{\ref{aowar}} \and
                G. Vasilopoulos\inst{\ref{nkua},\ref{iasa}}\and
                R. Willer\inst{\ref{mpe}}
        } 
        
        \titlerunning{HMXBs in the LMC detected during eRASS1}
        \authorrunning{Kaltenbrunner et al.}

        \institute{
                Max-Planck-Institut f{\"u}r extraterrestrische Physik, Gie{\ss}enbachstra{\ss}e 1, 85748 Garching, Germany\label{mpe}, \email{kald@mpe.mpg.de, cmaitra@mpe.mpg.de}
                \and
                {Inter-University Centre for Astronomy and Astrophysics (IUCAA), Ganeshkhind, Pune 411007, India\label{iucaa}}
                \and
                {ESO $-$ European Southern Observatory, Karl-Schwarzschild-Stra{\ss}e 2, 85748 Garching bei M\"unchen,
                        Germany \label{eso}}
                \and
                Anton Pannekoek Institute for Astronomy, University of Amsterdam, Science Park 904, 1098 XH Amsterdam, the Netherlands\label{apia}
                \and
                Department of Astronomy, The University of Michigan, 1085 South University Avenue, Ann Arbor, Michigan, 48103, USA\label{umich}
                \and
                {Universit\'e Paris-Saclay, Universit\'e Paris Cit\'e, CEA, CNRS, AIM, 91191 Gif-sur-Yvette, France\label{ups}}
                \and
                Southern African Astronomical Observatory, PO Box 9, Observatory Rd, Observatory 7935, South Africa\label{salt}
                \and
                Department of Astronomy, University of Cape Town, Private Bag X3, Rondebosch 7701, South Africa\label{uct}
                \and
                Leibniz Institut für Astrophysik Potsdam, An der Sternwarte 16, D-14482 Potsdam, Germany\label{liap}
                \and
                Astronomical Observatory, University of Warsaw, Warszawa, Poland \label{aowar}
                \and
                {Department of Physics, National and Kapodistrian University of Athens, University Campus Zografos, GR 15784, Athens, Greece\label{nkua}}
                \and
                {Institute of Accelerating Systems \& Applications, University Campus Zografos, Athens, Greece\label{iasa}}
        }
        
        \date{Received ... / Accepted ...}
        \abstract
        {The Magellanic Clouds, the closest star-forming galaxies to the Milky Way, offer an excellent environment to study high-mass X-ray binaries (HMXBs). While the Small Magellanic Cloud (SMC) has been thoroughly investigated with over 120 systems identified, the Large Magellanic Cloud (LMC) has lacked a complete survey due to its large angular size. Most prior studies targeted central or high-star-formation regions. The \srg/\ero all-sky surveys now enable a comprehensive coverage of the LMC, particularly due to its close vicinity to the south ecliptic pole.}
        {This work aims to improve our understanding of the HMXB population in the LMC by building a flux-limited catalogue. This allows us to compare sample properties with those of HMXB populations in other nearby galaxies.}
        {Using detections during the first \ero all-sky survey (eRASS1), we cross-matched X-ray positions with optical and infrared catalogues to identify candidate HMXBs. We assigned flags based on multi-wavelength follow-up observations and archival data, using properties of known LMC HMXBs. These flags defined confidence classes for our candidates.}
        {We detect sources down to X-ray luminosities of a few $10^{34}$ erg s$^{-1}$, resulting in a catalogue of 53 objects, including 28 confirmed HMXBs and 21 new \ero detections. Compared to the SMC, the LMC hosts fewer HMXBs and more systems with supergiant companions. We identify several likely supergiant systems, including a candidate supergiant fast X-ray transient with phase-dependent flares. We also find three Be stars with likely white dwarf companions. Two of the candidate Be/WD binaries show steady luminosities across four \ero scans, unlike the post-nova states seen in the majority of previous Be/WD reports.}
        {Our catalogue is the first to cover the entire LMC since the \ROSAT era, providing a basis for statistical population studies. Using the HMXB population, we estimate the LMC star-formation rate to be $(0.22^{+0.06}_{-0.07})$\,M$_{\odot}$yr$^{-1}$, which is in agreement with results using other tracers.}

        
        \maketitle
        \section{Introduction}
        \label{sec:intro}
        High-mass X-ray binaries (HMXBs) are instrumental in studying the final stages of massive star evolution and their interactions with compact objects such as neutron stars (NSs) or black holes \citep[BHs;][]{2006A&A...455.1165L,2006ARA&A..44...49R}. These systems consist of an early-type (O or B) massive ($\gtrsim8$M$_{\odot}$) star and a compact companion, where the compact object accretes material from the stellar wind or via Roche lobe overflow, leading to diverse X-ray emission, ranging from persistent to highly transient behaviour, and can be among the brightest X-ray sources in the sky \citep{2013A&A...551A...1R}.
        
        The Large Magellanic Cloud (LMC) serves as an excellent environment for investigating HMXBs due to its proximity \citep[$\sim$50 kpc][]{Piertrzynski2019} and low foreground absorption \citep[][]{2009AJ....138.1243H}. Additionally, the LMC exhibits a high specific star formation rate (SFR) similar to that of the Small Magellanic Cloud (SMC) and significantly higher than that of the Milky Way \citep[MW;][]{2004AJ....127.1531H, 2009AJ....138.1243H}, as well as a metallicity approximately half of that of our galaxy, and approximately twice of that of the SMC \citep{2002A&A...396...53R, 1998AJ....115..605L}.
        
        Additionally, the LMC's large angular extent, approximately 10 degrees by 10 degrees on the sky, has limited deep sensitive studies mainly to its central regions \citep{2016A&A...585A.162M} as compared to the SMC \citep{2013A&A...558A...3S}. While \ROSAT covered the entire LMC, the observations were constrained by a relatively low sensitivity and positional precision, and a narrow soft energy range (0.1--2.4\,keV).
        
        The known HMXB population in the LMC is dominated by Be/X-ray binaries (BeXRBs), which are the most common sub-class of HMXBs in metal-poor, actively star-forming environments. These systems typically consist of a NS orbiting a Be-type donor star, with mass transfer occurring episodically through interactions with the circumstellar disc. Their X-ray spectra are characterised by a hard power-law continuum with power-law indices of $\sim$1. Typically, BeXRBs show high variability, seen in Type I outbursts (modulated by the orbital period) and occasional giant Type II outbursts, where the X-ray luminosity can increase by several orders of magnitude \citep[e.g.][]{2020MNRAS.494.5350V,2025MNRAS.536.1357Y}.
        
        The observed HMXB population in the LMC provides valuable insights into how metallicity and star formation history (SFH) influence the formation and evolution of compact object binaries. Compared to the SMC, the fraction of supergiant X-ray binaries (SgXRBs) is higher in the LMC. This can be mainly attributed to differences in the SFH. In the SMC, the HMXB population is caused predominantly by high star formation activity 25$-$40\,Myr ago \citep{2010ApJ...716L.140A}. To date, the SMC is known to host $\sim$130 BeXRBs and only one SgXRB \citep{2016A&A...586A..81H}. The HMXB population in the LMC is associated with a star formation period at an earlier epoch and at a lower HMXB formation efficiency \citep{2016MNRAS.459..528A}. Out of the 59 HMXBs known to date, 8 are SgXRBs, such as LMC X$-$1, LMC X$-$3, and LMC X$-$4, and more recent discoveries such as supergiant fast X-ray transient (SFXT) systems \citep{2018MNRAS.475..220V}. The higher number of HMXBs in the SMC despite its lower stellar mass compared to the LMC \citep[SMC: $\sim3.2\times10^{8}$M$_{\odot}$, LMC: $\sim1.3\times10^{9}$M$_{\odot}$; ][]{2012ApJ...761...42S} can be attributed to the interplay between recent SFH and metallicity \citep{2010ApJ...725.1984L}.
        
        The \ero instrument aboard the {\it Spektrum Roentgen Gamma} (\srg) spacecraft \citep[][]{2021A&A...656A.132S, 2021A&A...647A...1P} has drastically improved the detection and cataloguing of X-ray sources, particularly in the LMC. The first \ero all-sky survey (eRASS1) provides enhanced sensitivity and positional accuracy, which is particularly beneficial in the LMC due to its proximity to the south ecliptic pole (SEP), where all \ero scans overlap. This overlap results in some parts of the LMC, especially those near the SEP, being observed for in total over 50 ks over the first four eROSITA all-sky surveys (eRASS:4), enabling detailed studies of this region \citep[][]{2024A&A...682A..34M}. Compared to \ROSAT, \ero is about 25 times more sensitive in the soft band (0.2--2.3\,keV), while in the hard band (2.3--8.0\,keV) it provides the first imaging survey of the entire LMC \citep{2021A&A...647A...1P}.
        
        In this paper, we present a new catalogue of HMXBs in the LMC based on objects detected during eRASS1. To validate and further characterise these sources, we use properties of known LMC HMXBs. We use photometric data from the Magellanic Clouds Photometric Survey \citep[MCPS;][]{2002AJ....123..855Z, 2004AJ....128.1606Z} and the VISTA Magellanic Cloud Survey \citep[][]{2011A&A...527A.116C}. We assess LMC membership using proper motion measurements from \textit{Gaia} eDR3 \citep{2021A_A...649A...7G}. We study optical variability with light curves from the Optical Gravitational Lensing Experiment \citep[\ogle;][]{2015AcA....65....1U}. We also incorporate optical follow-up spectroscopy with the FLOYDS spectrograph mounted on the 2\,m Las Cumbres Observatory \citep[LCO;][]{2013PASP..125.1031B} telescope at Siding Spring Observatory in Australia, the Robert Stobie Spectrograph (RSS), and the High Resolution Spectrograph (HRS), both on the 9.2\,m Southern African Large Telescope (\salt). Finally, we include X-ray observations from \xmm.
        We also leverage archival data from the VizieR\footnote{\url{https://vizier.cds.unistra.fr/}} database, the ESO Archive Science Portal\footnote{\url{https://archive.eso.org/scienceportal/home}}, and the HILIGT upper limit server\footnote{\url{http://xmmuls.esac.esa.int/upperlimitserver/}} \citep{2022A&C....3800531S}.
        
        For new candidate HMXBs, we apply a system of flags to assess their credibility, taking into account multi-wavelength information and additional follow-up observations. Moreover, X-ray spectra and light curves for these sources are derived from data collected across all four \ero all-sky surveys (eRASS1$-$eRASS4). This approach allows us to make detailed inferences about all the candidates by assigning confidence classes. Following this, we derive a flux-limited catalogue of HMXBs in the entire LMC.
        
        The paper is organised as follows. Section\,\ref{sec:data} focuses on the instruments used for the detailed analysis of sources in our catalogue. In Sect.\,\ref{sec:cat_building} we explain the criteria we used for objects to enter our catalogue. Section\,\ref{sec:analysis} then focuses on how we analysed the multi-wavelength data to obtain the parameters and characteristics of interest. In Sect.\,\ref{sec:cat_results} we discuss the properties of the whole sample, and explain the scheme for the classification of all sources in our catalogue into six confidence classes of (candidate) HMXBs. Additionally, this section explains the flags we used for assigning confidence classes. In Sect.\,\ref{sec:XLF} we discuss the X-ray luminosity function, a fundamental relation that links HMXBs with the SFR of a galaxy. In Sect.\,\ref{sec:discussion} we discuss our results for individual objects, examine optical variability and the classification of optical counterparts, and compare our overall findings with those from previous studies in the SMC. Finally, in Sect.\,\ref{sec:summary} we summarise our findings. Following \citet{Piertrzynski2019}, we assume a distance of $d=49.49$\,kpc to the LMC in this paper.
        
        \section{Observations}
        \label{sec:data}
        \subsection{\ero}
        \label{sec:ero}
        The main instrument we used for our analysis is \ero \citep[][]{2021A&A...647A...1P}, the soft X-ray instrument on board the \srg mission, which was launched in 2019 and surveyed the whole X-ray sky in great circles passing through the ecliptic poles between December 2019 and February 2022 in the energy range of 0.2 to 8\,keV. We investigated sources that were detected during the first \ero all-sky survey \citep[eRASS1,][]{2024A&A...682A..34M} and analysed the combined data products of all \ero scans. Due to its close vicinity to the SEP, sources in the direction of the LMC were monitored for a significantly longer period than the rest of the sky, with average effective exposures (corrected for vignetting) that are higher by between one and two orders of magnitude. During eRASS1, effective exposures (0.2$-$4.5\,keV) in the LMC range from 406\,s to 28218\,s, with a median value of 1064\,s. The effective exposure at the ecliptic equator is $\sim100$\,s \citep{2024A&A...682A..34M}.
        
        \subsection{\xmm}
        \label{sec:xmm}
        \xmm \citep[X-ray Multi-Mirror mission;][]{2001A&A...365L...1J} is an X-ray observatory launched by the European Space Agency in 1999. For our analysis, we utilised data from the three European Photon Imaging Cameras (EPIC), comprising two MOS-CCD cameras \citep{2001A&A...365L..27T} and one pn-CCD camera \citep{2001A&A...365L..18S}. We used data from \xmm
        for more detailed X-ray analysis and to search for possible pulse periods of objects we found during our \ero analysis. \xmm data additionally provides more precise astrometrical positions (median uncertainty of 0.9" for \xmm positions of HMXBs in the LMC performing source detection using the standard \xmm pipeline similar to \citet{2025MNRAS.542..583H}) and a way to validate optical counterparts.
        
        \subsection{\ogle monitoring}
        \label{sec:ogle}
        We used photometric data from the regular monitoring of the LMC by the \ogle project \citep{1992AcA....42..253U}.
        Images were taken at the Las Campanas Observatory in the I and (less frequently) V bands with the 1.3\,m Warsaw telescope starting in 1997 (\ogle-II). During phases \ogle-III (begin 2001) and \ogle-IV (begin 2010), improved CCDs with an increased field of view (FOV) were used. 
        The data were calibrated to the standard I-band
        system in the manner described in \citet{2008AcA....58...69U,2015AcA....65....1U}.
        The \ogle data used in this work are summarised in Table\,\ref{tab:ogle_data}.
        Most of our objects were covered by \ogle-IV for about 14 years, while for those also observed during \ogle-III, light curves $\gtrsim$23 years are available. The typical observing cadence for our sources (median time interval between observations) is two or three days, but can be longer in a few cases (e.g. sources 10 and 28 from Table\,\ref{tab:ogle_data}, with 4 and 7 days, respectively). On the other hand, some selected fields are observed up to eight times per night for selected observing seasons (e.g. sources 36 and 43).
        
        \subsection{LCO/FLOYDS spectroscopy}
        \label{sec:lco}
        To characterise the optical counterpart of objects in our catalogue, we used spectroscopic data from the LCO/FLOYDS spectrograph that was commissioned at Faulkes Telescope South (FTS) at Siding Springs Observatory in 2012. Observations were planned to achieve a signal-to-noise ratio (S/N) of $\approx$100 with a resolving power of 326$-$384 in the 5600--6600\,\AA\ range and  482$-$588 in the 4100--5000\,\AA\ range. To reduce spectra, \texttt{PyRAF} tasks were used as part of the FLOYDS pipeline\footnote{\url{https://lco.global/documentation/data/floyds-pipeline/}} in the manner explained in \citet{2013_FLOYDS_manual}. We analysed the \Halpha and \Hbeta lines that typically appear in emission in Be stars \citep{2003PASP..115.1153P, 2000ASPC..214....1B, 1988PASP..100..770S}. Table\,\ref{tab:spectroscopy} gives details on each observation we used.
        
        \subsection{\salt/RSS and \salt/HRS spectroscopy}
        \label{sec:salt}
        Additional optical spectroscopy was undertaken using the RSS and the HRS on \salt (first light in 2005) under the \salt transient follow-up programme. For the RSS, the PG0900 VPH grating was used, which covers the spectral region 3920--7000\,\AA\ at a resolving power of 632--1129. For the HRS, we used the low resolution mode with a resolving power of $\sim$16000 in the spectral region 3700--8900\,\AA. The \salt pipeline was used to perform primary reductions comprising overscan corrections, bias subtraction, gain correction, and amplifier cross-talk corrections \citep[][]{2010SPIE.7737E..25C}. The remaining steps, which include wavelength calibration, background subtraction, and extracting the 1D spectrum, were executed using IRAF\footnote{\url{https://iraf-community.github.io/}}. Table\,\ref{tab:spectroscopy} lists information on all individual observations taken with \salt.
        
        \subsection{VLT FLAMES/GIRAFFE spectroscopy}
        We made use of archival optical spectra obtained with the Fibre Large Array Multi Element Spectrograph (FLAMES), which started its operations in 2001 at the 8.2\,m Unit Telescope 2 (UT2) of the Very Large Telescope (VLT) at Cerro Paranal, Chile \citep{2002Msngr.110....1P}. The spectra we used were obtained with the GIRAFFE spectrograph, which has a resolving power of 17000 in the spectral region 6299--6691\,\AA, and were processed using the standard data reduction pipeline \citep{2011A&A...530A.108E}.
        
        \begin{table*} 
                \centering
                \caption{Summary of spectroscopic observations done for objects in our catalogue.} 
                \label{tab:spectroscopy} 
                \begin{tabular}{l|lllllll} 
                        \hline\hline\noalign{\smallskip}
                        Target & Observation & V & RA & Dec & Exposure & \Halpha & v$^{a}{\alpha}$\xspace \\
                        ID & Date & mag & \multicolumn{2}{c}{J2016} & s & \AA & km s$^{-1}$\\
                        \noalign{\smallskip}\hline\noalign{\smallskip}
                        1 & 2022-01-11 12:37:41
                        & 15.5 & 04:53:15.1 & $-$69:32:42 & 3200 & $-6.0\pm0.4$ & $160\pm32$ \\
                        3 & 2021-11-16 12:54:44 
                        & 14.2 & 04:57:37.0 & $-$69:27:28 & 1000 & $-8.1\pm0.2$ & $70\pm10$  \\
                        10 & 2021-12-31 10:32:15
                        & 15.0 & 05:16:00.0 & $-$69:16:08 & 1200 & $-6.4\pm0.7$ & $418\pm83$  \\
                        28 & 2023-04-28 09:50:16
                        & 15.0 & 05:50:06.5 & $-$68:14:56 & 1500 & $-17.4\pm0.3$ & $176\pm6$  \\
                        29$^{(b)}$ & 2021-11-02 13:14:15 
                        & 14.9 & 04:43:54.7 & $-$69:29:46 & 1200 & $+5.0\pm0.2$ & $-563\pm16$  \\
                        = & 2023-07-24 18:46:09
                        & = & = & = & 1600 & $-15.0\pm0.3$ & $106\pm6$ \\
                        30 & 2023-11-10 14:51:44 
                        & 15.6 & 04:50:24.5 & $-$69:18:42 & 1700 & $+5.3\pm0.5$ & $152\pm33$ \\
                        32 & 2023-11-12 13:49:52 
                        & 14.9 & 04:52:18.4 & $-$66:32:49 & 1500 & $-20.6\pm0.4$ & $110\pm6$ \\
                        35 & 2023-07-22 18:38:49
                        & 13.5 & 05:02:14.1 & $-$67:46:18 & 1200 & $-28.11\pm0.22$ & $146.8\pm2.0$ \\
                        36 & 2023-11-10 14:05:44 
                        & 14.9 & 05:03:59.7 & $-$70:32:10 & 1500 & $-43.8\pm0.5$ & $-51\pm3$ \\
                        37 & 2023-11-10 13:20:45 
                        & 14.6 & 05:07:06.2 & $-$65:21:47 & 1400 & $-26.8\pm0.4$ & $150\pm4$ \\
                        39$^{(c)}$ & 2023-11-10 16:47:41 
                        & 15.6 & 05:27:26.1 & $-$66:33:08 & 1700 & $+4.9\pm0.7$ & $287\pm723$ \\
                        41$^{(c)}$ & 2021-11-02 13:46:12 
                        & 14.4 & 05:30:49.6 & $-$66:20:11 & 1500 & $+5.0\pm3.3$ & $295\pm3042$ \\
                        = & 2022-11-21 17:03:03
                        & = & = & = & 2300 & $+4.4\pm2.8$ & $612\pm3303$\\
                        43 & 2022-11-18 12:20:46
                        & 14.9 & 05:34:48.9 & $-$69:43:39 & 1200 & $+6.0\pm0.3$ & $219\pm18$ \\
                        = & 2022-01-03 15:32:07
                        & = & = & = & 1500 & $+4.9\pm0.2$ & $140\pm20$\\
                        44$^{(c)}$ & 2023-11-12 15:49:12
                        & 16.2 & 05:40:21.9 & $-$68:56:46 & 2300 & $+3.5\pm1.9$ & $291\pm3718$ \\
                        45 & 2023-04-22 09:39:27
                        & 14.2 & 05:41:37.5 & $-$68:32:33 & 1300 & $-4.8\pm0.4$ & $30\pm52$ \\
                        46 & 2021-11-02 14:23:07
                        & 14.4 & 05:42:41.6 & $-$67:27:55 & 3600 & $-10.7\pm0.2$ & $24\pm8$ \\
                        47$^{(d)}$ & 2021-10-27 14:15:57
                        & 13.5 & 05:44:22.1 & $-$67:27:33 & 900 & $-$ & $-$ \\
                        51 & 2023-11-12 16:49:20
                        & 16.2 & 05:58:50.6 & $-$67:52:27 & 2300 & $-15.9\pm0.6$ & $264\pm20$ \\
                        52 & 2023-11-10 15:47:37
                        & 14.4 & 06:02:13.1 & $-$67:43:06 & 1300 & $+5.2\pm0.3$ & $-102\pm22$ \\
                        \hline\noalign{\smallskip}
                        35 & 2020-09-11 & 13.5 & 05:02:14.1 & $-$67:46:18 & 1200 & $-31.3\pm0.5$ & $9.7\pm2.9$ \\
                        47$^{(e)}$ & 2023-11-05 -- 2023-11-23 & 13.5 & 05:44:22.1 & $-$67:27:33 & 12000 & $-1.630\pm0.012$ & $27.06\pm0.06$ \\
                        \hline\noalign{\smallskip}
                        5$^{(f)}$ & 2008-10-07 -- 2009-01-07 & 15.8 & 05:07:22.2 & -68:47:59 & 16620 & $-55.76\pm0.16$ & $3.9\pm0.6$ \\
                        \noalign{\smallskip}\hline
                \end{tabular}
                \tablefoot{
            The top, middle, and bottom lists refer to observations made with LCO/FLOYDS, \salt, and VLT FLAMES/GIRAFFE, respectively. The target ID refers to the source number in Table\,\ref{tab:MasterTable_known}. RA and Dec refer to J2016 positions of \textit{Gaia} counterparts.
                        \tablefoottext{a}{Radial velocity of \Halpha line with respect to the MW minus 278 km s$^{-1}$ (radial velocity of the LMC).}
                        \tablefoottext{b}{Re-observed due to too low S/N.}
                        \tablefoottext{c}{High uncertainty due to too low S/N.}
                        \tablefoottext{d}{Fit to \Halpha and \Hbeta does not show a significant line due to low S/N. Re-observed with \salt for high-resolution monitoring of \Halpha, \Hbeta, and \HeI to investigate a likely SFXT nature.}
                        \tablefoottext{e}{The source was monitored in ten observations of 1200s exposure each between 5 and 23 November 2023 with \salt/HRS. Values for \Halpha and v are given for the observation with the strongest emission line. See Sect.\,\ref{sec:SFXT} for detailed results.}
                        \tablefoottext{f}{The source was monitored in six observations of 2770\,s exposure each between 7 October 2008 and 7 January 2009 during the programme 082.D-0575 led by PI\,R.\,E.\,Mennickent. Values for \Halpha and v are given for the observation with the strongest emission line. The velocity reported is the average of the two peaks observed. See Sect.\,\ref{sec:SNR_ESO} for detailed results.}
                }
        \end{table*}
        
        \subsection{Broad-band photometry}
        \label{sec:broadband_photo}
        In order to get an additional assessment of whether the spectral energy distribution (SED) is consistent with that of an early-type star in the LMC, we compared all available photometric measurements with synthetic models. For the observations, we used the VizieR photometry tool and obtained all available photometric measurements within a 1" circle around the source position. This approach is supported by the availability of data from survey missions such as 2MASS, \textit{Gaia}, IRAC, and WISE for the majority of our sources. All flux values were converted to common units of erg\,cm$^{-2}$\,s$^{-1}$\,\r{A}$^{-1}$.
        
        \section{Building the catalogue}
        \label{sec:cat_building}
        In this section we explain the catalogues and steps required to create our initial set of HMXBs. In Sect.\,\ref{sec:flags}, these candidates are then rated more precisely by comparing parameters with those of all secure HMXBs in our catalogue and of possible contaminators such as active galactic nuclei (AGNs) or foreground stars.
        
        \subsection{Catalogues used}
        \subsubsection{eRASS1 source catalogue}
        As a basis for our analysis, we used the first \ero All-Sky Survey \citep[eRASS1;][]{2024A&A...682A..34M}. Most HMXBs are spectrally hard sources, which in the LMC can be very useful because they stand out against the soft X-ray emission caused by the hot interstellar medium, which is especially dominant in the centre of the LMC. To utilise this, we used two different eRASS1 catalogues. The first is the publicly released one-band (1B) catalogue for which source detection in the most sensitive band of \ero between 0.2\,keV and 2.3\,keV was applied. As a complementary catalogue to reduce the impact of the hot ISM in the LMC centre and to increase the sensitivity for hard sources, we used a catalogue for which source detection was applied in three energy bands simultaneously (3B catalogue; the bands are 0.2$-$0.6\,keV, 0.6$-$2.3\,keV, and 2.3$-$5.0\,keV). This catalogue was published by \cite{2024A&A...682A..34M}, but they used an additional cut of a minimum detection likelihood in the hardest band \texttt{DET\_LIKE\_3}$\geq$12. This cut proved overly restrictive for our purposes, and we therefore did not apply it.
        
        The positional uncertainties of the 1B catalogue are given in the \texttt{POS\_ERR} column. This column corrects the uncertainties found during source detection (\texttt{RADEC\_ERR}) by analysing distances found during matching the eRASS1 source catalogue with AGN catalogues as is described in Sect.\,6.2 of \citet{2024A&A...682A..34M}. The correlation between \texttt{POS\_ERR} and \texttt{RADEC\_ERR} is given by
        \[
        \texttt{POS\_ERR}=\sqrt{A\cdot\sigma^2+\sigma_0^2},
        \]
        where $\sigma=\texttt{RADEC\_ERR}/\sqrt{2}$, $A$, and $\sigma_0$ are the multiplicative and the systematic correction terms, respectively. To test whether this correction also applies to the higher-exposure LMC region and if it can be applied to the 3B catalogue, we conducted a similar but simplified analysis for \textit{Gaia}-detected AGNs in the LMC region. For both catalogues, we find both correction terms in agreement within uncertainties with those from \citet{2024A&A...682A..34M}, which are $A=1.3$ and $\sigma_0=0.9$. Given our smaller sample compared to the one used by \citet{2024A&A...682A..34M}, we therefore used the \texttt{POS\_ERR} as positional uncertainties of the 1B catalogue and applied the same correction for objects in the 3B catalogue.
        
\begin{figure}
                \centering
                \resizebox{\hsize}{!}{\includegraphics{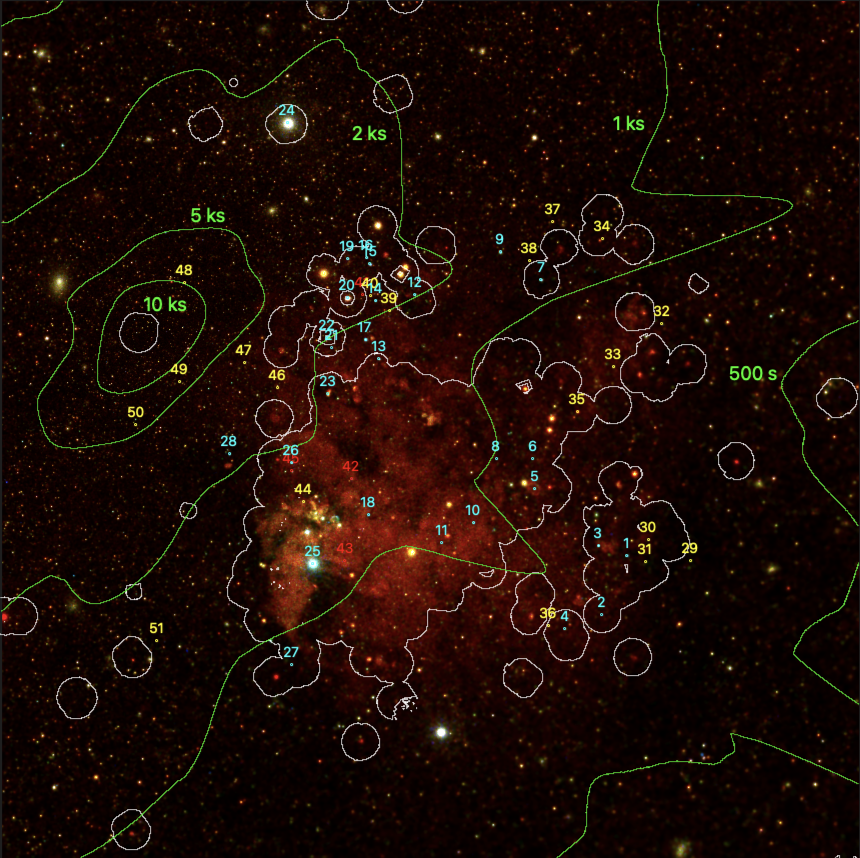}}
                \caption{RGB (r: 0.2$-$1.0\,keV, g: 1.0$-$2.0\,keV, b: 2.0$-$4.5\,keV) of LMC during eRASS\,1. The green contours show the total exposure achieved with eRASS1. The exposure maximum lies at the SEP. Cyan markers represent known HMXBs. Red and yellow markers show candidates, representing previously known objects and those discovered with \ero, respectively. Labels refer to the sequence numbers in Table \ref{tab:MasterTable_known}. The entire region visible in the image, except for the top corners, was investigated during our analysis. White contours indicate the LMC coverage by \xmm. Remarkably, the central LMC has been thoroughly observed by \xmm, while the outskirts remain under-sampled.}
                \label{fig:RGB}
        \end{figure}

        Due to the close vicinity of the LMC to the SEP, the exposure varies highly from the east to the west end (see Fig.\,\ref{fig:RGB}). This has its strongest influence on population analysis, such as completeness (see Sect.\,\ref{sec:flags}) and the extraction of a completeness-corrected luminosity function (see Sect.\,\ref{sec:XLF}).
        
        To reduce the contribution of spurious detections and chance coincidences, we applied three cuts to the eRASS1 catalogue. The first was to select only point-like sources by requiring log-likelihood of the extent probability \texttt{EXT\_LIKE}=0. This cut applied to 7\% and 6\% of detections in the LMC listed in the 1B and 3B catalogues, respectively.
        By selecting only catalogue objects with positional uncertainties within the lower 95 percent of all objects within the LMC, we made sure not to include sources for which a secure identification of the optical and infrared (IR) counterparts could not be achieved. This resulted in a maximum \texttt{POS\_ERR} of 6.92" and 6.96" for the 1B and 3B catalogues, respectively.
        Next, we applied \texttt{DET\_LIKE\_0}$\geq$20, which we found to be most useful in restraining the number of spurious detections caused by the hot ISM. In the 1B catalogue, \texttt{DET\_LIKE\_0} refers to the detection log-likelihood in the 0.2$-$2.3\,keV band; in the 3B catalogue, it refers to the combined detection likelihood in all three bands. For fainter sources, it is typically not possible to securely identify an object as a source, and even less so to constrain the spectral parameters well enough to distinguish an HMXB from contaminating objects. This cut applies to 64\% and 66\% of detections in the LMC listed in the 1B and 3B catalogues, respectively.
        Finally, the remaining spurious objects were identified and sorted out through visual screening using RGB images of eRASS1$-$4 and eRASS:4 covering the sources. This visual screening was conducted during the final step of matching the eRASS1 catalogues with optical and IR catalogues (see Sect.\,\ref{sec:matching}).
        
        \subsubsection{MCPS}
        \label{sec:mcps}
\begin{figure}
                \centering
                \resizebox{\hsize}{!}{\includegraphics{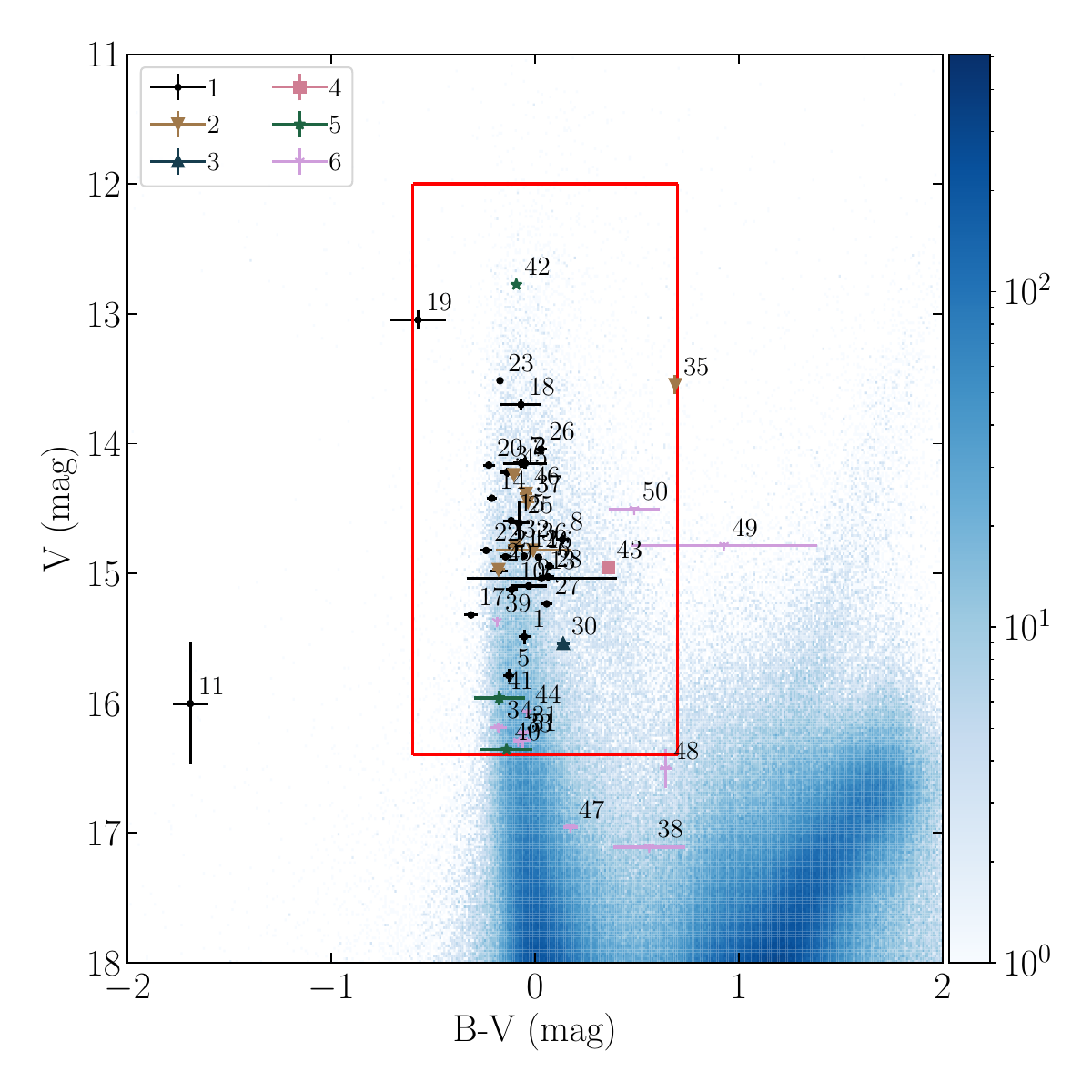}}
                \caption{Selection of MCPS counterparts (red rectangle) compared to the distribution of MCPS entries matched with the \textit{Gaia} proper motion selection for the LMC (colour mesh; found by matching the MCPS catalogue with \textit{Gaia} proper motion selection). The colour scale gives the number of sources in each colour-magnitude bin. The bin sizes are 0.01 mag and 0.02 mag for B$-$V and V, respectively. Data points show the MCPS counterparts of all objects in our catalogue grouped by their confidence classes. Note that \#11 was not used for defining the selection criterion, because the absence of I-band and U-band measurements in the MCPS catalogue indicates a possible measurement error. We include objects \#48 and \#49 in our selection because our criterion allows objects to be considered as long as they fall within the selection region when accounting for uncertainties. Objects \#38 and \#47 are considered candidates because their VMC counterparts meet the VMC selection criteria described in Sect.\,\ref{sec:vmc}.}
                \label{fig:MCPS_selection}
        \end{figure}

        The Magellanic Clouds Photometric Survey is a four-band (U, B, V, and I) survey of the LMC and SMC with a typical astrometric uncertainty of less than 1" \citep[][]{2002AJ....123..855Z, 2004AJ....128.1606Z}. For the LMC, the central $8\times8$ deg$^2$ were covered down to a typical limiting magnitude of V=21\,mag. \cite{2013A&A...558A...3S} have shown that the early-type optical companions in HMXBs can be found using a colour and magnitude selection of secure HMXBs. We used the criteria 12.0\,mag<V<16.4\,mag and $-$0.6\,mag<B$-$V<0.7\,mag, which we got from the distribution of known HMXBs (see Fig.\,\ref{fig:MCPS_selection}). We slightly relaxed the selection criteria by requiring that the selection criteria have to be fulfilled within the uncertainties listed in the MCPS catalogue.
        
        \subsubsection{VMC}
        \label{sec:vmc}
\begin{figure*}
                \centering
                \resizebox{0.495\hsize}{!}{\includegraphics{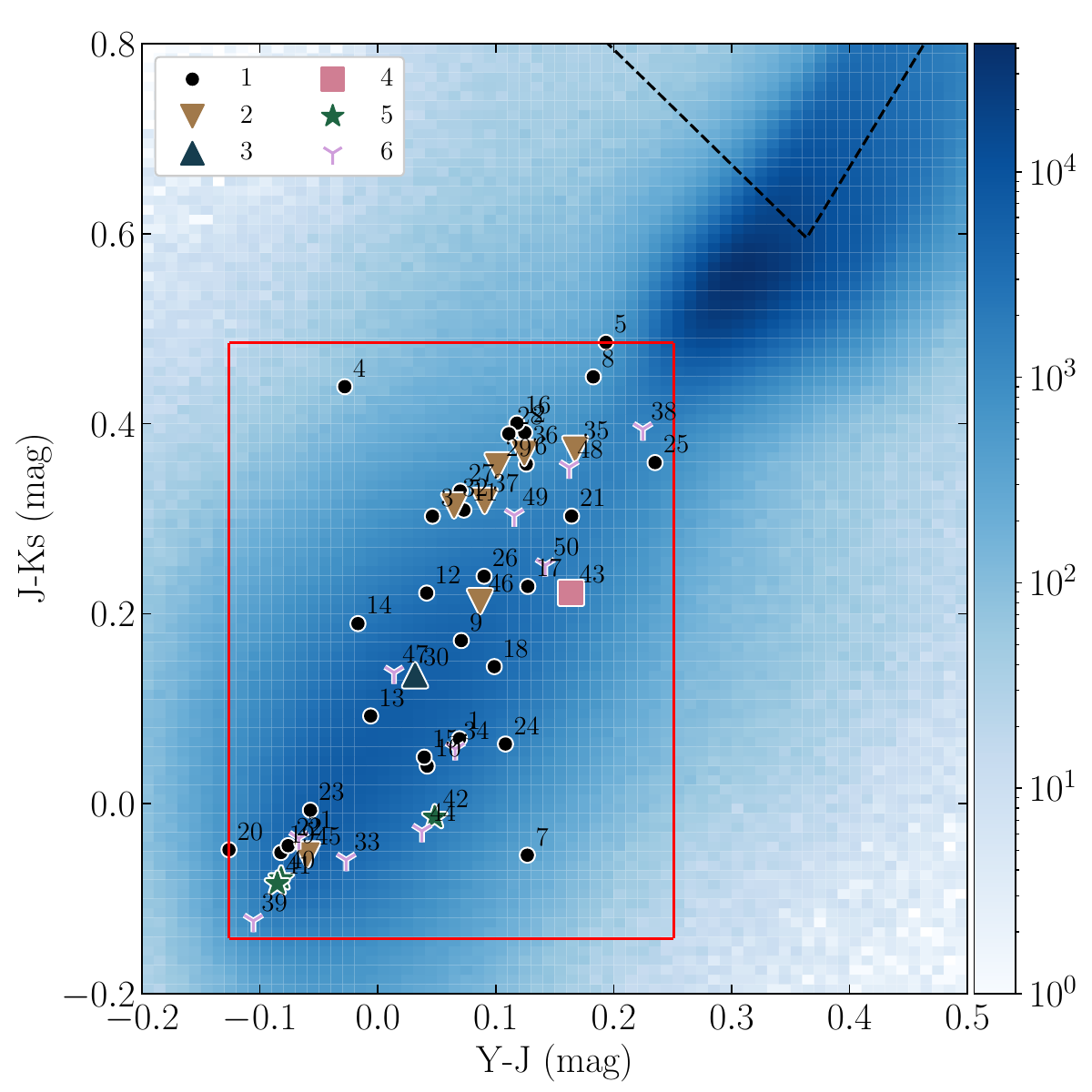}}
                \resizebox{0.495\hsize}{!}{\includegraphics{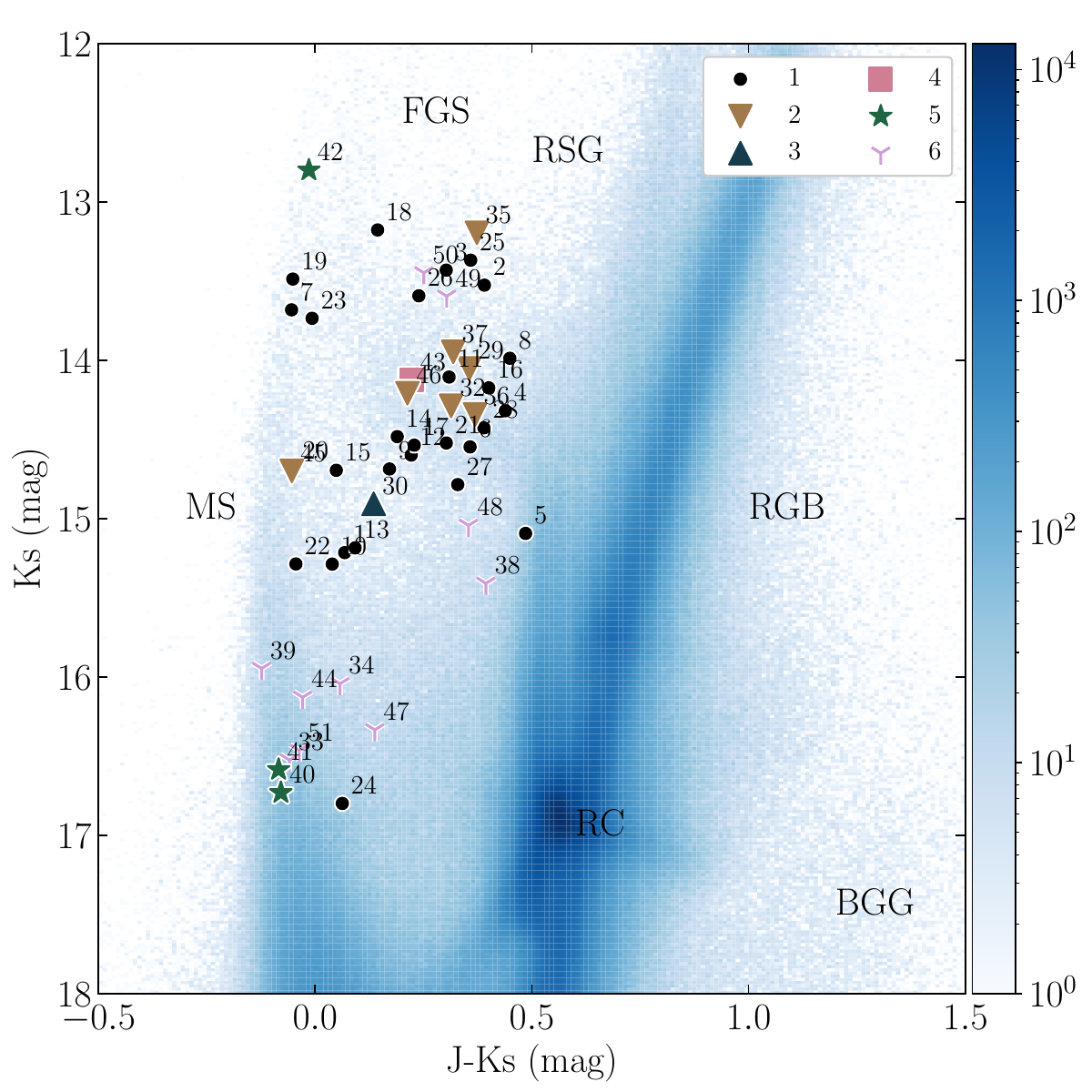}}
                \caption{Left: Distribution of VMC counterparts of all sources in our catalogue grouped by confidence class (see Table \ref{tab:conf_classes}) in the VMC colour-colour plane compared to the typical location of QSOs marked by the dotted line \citep[see][Equ. 1, 2 and 3]{2013A&A...549A..29C} and the distribution of LMC stars (colour mesh; found by matching VMC with \textit{Gaia} proper motion selection). The colour scale gives the number of sources in each colour-colour bin. The bin sizes are 0.01 mag for both axes. Right: Sources in our catalogue in the colour-magnitude plane compared to different stellar evolution stages and features caused by foreground and background objects \citep[see][their Fig. 2 and Sect. 3.1; MS: main-sequence, RGB: red giant branch, RC: red clump, RSG: red supergiants, FGS: foreground Galactic stars, BGG: background galaxies]{2017ApJ...849..149S} and LMC stars (colour mesh). The colour scale gives the number of sources in each colour-magnitude bin. The bin sizes are 0.01 mag and 0.02 mag for Y$-$J and Ks, respectively.}
                \label{fig:VMC_selection}
        \end{figure*}

        The VISTA (VISual and infrared Telescope for Astronomy) near-IR YJK$_S$ survey of the Magellanic Cloud system \citep[VMC;][]{2011A&A...527A.116C} has an astrometric uncertainty of less than 1" for 98.4\% of the sources. By observing stars across multiple wavelengths, VMC aims to analyse stellar populations, map their three-dimensional structure, identify variable stars, and investigate SFH. The survey was designed to achieve S/N=10 at Y=21.1\,mag, J=21.3\,mag, and Ks=20.7\,mag, respectively. For our study we used DR6 of the VMC catalogue \citep{2023yCat.2375....0C}. Similar to the MCPS selection, we used the distribution of known HMXBs to define a selection criterion for the candidates. We required 12.8\,mag<Y<16.6\,mag, $-$0.126\,mag<Y$-$J<0.251\,mag, and $-$0.142\,mag<J$-$Ks<0.485\,mag. As is seen in Fig.\,\ref{fig:VMC_selection}, this selection is well separated from the colour-colour region that quasars typically reside in \citep{2013A&A...549A..29C}.
        
        \subsubsection{\textit{Gaia} EDR3}
        \label{sec:Gaia_pm_selection}
        To verify the LMC membership of the MCPS and VMC counterparts, we used the spatial and proper motion selection criterion of the LMC from \textit{Gaia} EDR3 presented in Sect.\,2 in \citet{2021A_A...649A...7G}. We also used the more precise \textit{Gaia} positions for matching between different optical and IR catalogues.
        
        \subsubsection{Known HMXBs}
        
        We utilised existing literature (\citet{1999A&AS..139..277H, 2000A&AS..143..391S, 2016MNRAS.459..528A, 2013A&A...558A..74V, 2022A&A...662A..22H, 2023A&A...671A..90H, 2019MNRAS.490.5494M, 2021MNRAS.504..326M, 2021A&A...647A...8M,2023A&A...669A..30M, 2018MNRAS.475.3253V}) to identify known HMXBs and their characteristics. By analysing these sources, we established parameter ranges for selecting new HMXB candidates.
        
        \subsubsection{Screening catalogues}
        
        To obtain a reliable HMXB catalogue, it is essential to screen other kinds of securely identified X-ray point sources. The majority of sources possibly contaminating our catalogue are AGNs behind the LMC that match an LMC early-type star by chance coincidence. The accretion of matter onto supermassive BHs can produce similar X-ray spectra as HMXBs, but they can more easily be discriminated using data from other wavelengths, especially using mid-IR and far-IR data. Other possible contaminators of our catalogue are galaxies behind the LMC that appear as point sources for \ero, cataclysmic variable stars, or foreground MW stars. For screening, we used the catalogues listed in Table\,\ref{tab:screening_cats}. We used only subsets defined by the rules listed in the table to avoid falsely excluding HMXBs listed in those catalogues. Entries of the catalogues that fulfil the criteria in the 'Flags' column are treated as candidates. Each source in our catalogue that matches one of those candidates is flagged and considered an HMXB candidate of lower reliability.
        
        \begin{table} 
                \centering
                \caption{Catalogues used for matching (top) and screening of foreground and background sources (bottom).} 
                \label{tab:screening_cats} 
                \begin{tabular}{l|ll} 
                        \hline\hline\noalign{\smallskip}
                        Catalogue & Screening & Flags\\
                        \noalign{\smallskip}\hline\noalign{\smallskip}
                        ZHT04 & \multicolumn{2}{l}{see Sect.\,\ref{sec:mcps}} \\
                        C03 & \multicolumn{2}{l}{see Sect.\,\ref{sec:vmc}} \\
                        GLC21 & \multicolumn{2}{l}{see Sect.\,\ref{sec:Gaia_pm_selection}} \\
                        \noalign{\smallskip}\hline\noalign{\smallskip}
                        HFM00$^{(a)}$ & $\lvert \texttt{pmRA}\rvert$>0,$\lvert \texttt{pmDE}\rvert$>0 & -- \\
                        F23 & $\texttt{Z}\geq0$ & None \\
                        SGA19 & None & -- \\
                        KOK13 & None & -- \\
                        KK09 & $\texttt{z}\geq0$ & $\texttt{Type}=\textrm{QSO-*}$ \\
                        SDD15 & $\texttt{z}>0$ & None \\
                        ASN18 & -- & R90 \\
                        MAC12 & -- & None \\
                        BFA19 & -- & $\texttt{pqso}>0.75$ \\
                        KPT12 & -- & None \\
                        PvL21$^{(b)}$ & -- & None \\
                        KZA21$^{(c)}$ & -- & None \\
                        \noalign{\smallskip}\hline
                \end{tabular}
                \tablefoot{
            For screening catalogues, the columns ‘Screening’ and ‘Flags’ indicate which catalogue entries we used for screening and flagging, respectively. Catalogues that have ‘-’ listed in one of the two selection columns were not used for the corresponding task. If ‘None’ is listed, the entire catalogue is used. Screening catalogues without footnote annotations are catalogues of background AGNs.
            \tablefoottext{a}{Tycho catalogue of bright foreground stars.}
                        \tablefoottext{b}{Australian Square Kilometre Array Pathfinder (ASKAP) 888 MHz radio continuum survey of the LMC.}
                        \tablefoottext{c}{Heraklion Extragalactic Catalogue (HECATE) of galaxies..}

            \textbf{References:}
            ZHT04: \citet{2004AJ....128.1606Z},
            C03: \citet{2023yCat.2375....0C},
            GLC21: \citet{2021A_A...649A...7G},
            HFM00: \citet{2000yCat.1259....0H},
            F23: \citet{2023yCat.7294....0F},
            SGA19: \citet{2019yCat..36240145S},
            KOK13: \citet{2013ApJ...775...92K},
            KK09: \citet{2009ApJ...701..508K},
            SDD15: \citet{2015ApJS..221...12S},
            ASN18: \citet{2018ApJS..234...23A},
            MAC12: \citet{2012MNRAS.426.3271M},
            BFA19: \citet{2019yCat.7285....0B},
            KPT12: \citet{2012ApJ...747..107K},
            PvL21: \citet{2021yCat..75063540P},
            KZA21: \citet{2021MNRAS.506.1896K}
        }
        \end{table}
        
        \subsection{Catalogue matching}
        \label{sec:matching}
\begin{figure*}
                \centering
                \resizebox{\hsize}{!}{\includegraphics{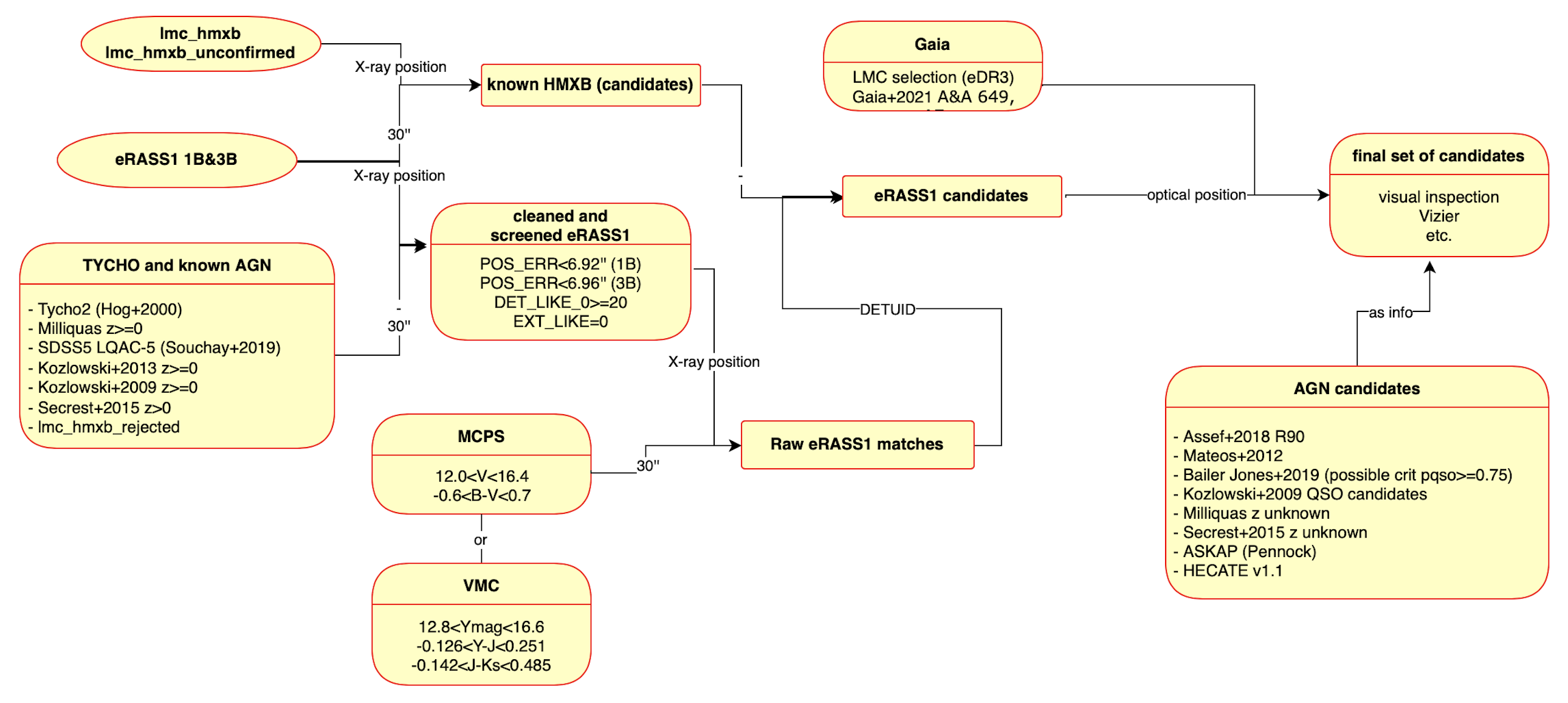}}
                \caption{Flowchart summarising the selection process for new candidates included in the catalogue. A detailed description can be found in Sect.\,\ref{sec:matching}.}
                \label{fig:selection_flowchart}
        \end{figure*}

        A schematic of the matching process to arrive at the final list of HMXB candidates is shown in Fig.\,\ref{fig:selection_flowchart}. The fundamental catalogues of HMXB candidates were derived from matching the eRASS1 1B catalogue with optical or IR counterparts from the MCPS and VMC catalogues, respectively. Complementary to this, we utilised the eRASS1 3B catalogue to search for very faint, hard sources that would be missed when using the 1B catalogue alone. In total, this results in four lists that we combined into a final catalogue of HMXB candidates in the last step. The detailed steps are the following:
        \begin{itemize}
                \item \textbf{Known HMXB (candidates) in eRASS1:} We matched the full \ero catalogue with the X-ray positions of our list of previously known HMXB (candidates) with a maximum separation of 30" to minimise the risk of chance-coincidences. Additionally, we implied a maximum separation of less than or equal to 3$\sigma$. For a Rayleigh distribution, this is equivalent to $Separation\leq3.439\cdot\sqrt{\mathtt{POS\_ERR}^2+\sigma_{X-ray}^2}$, where $\sigma_{X-ray}$ is the X-ray positional error of previous observations. If \ero improved the positional uncertainty of a candidate and the new position no longer matched that of an early-type star from our MCPS and VMC selections, we rejected the candidate as a possible HMXB.
                \item \textbf{eRASS1 LMC catalogue:} Contains all eRASS1 sources with distances <10 degrees to the centre of the LMC at RA=05h 23m 34.00s and Dec=$-$69d 45m 22.0s (1B: 50148 objects, 3B: 50914 objects).
                \item \textbf{Cleaned eRASS1 LMC catalogue:} From the eRASS1 LMC catalogue, we selected only the objects with \texttt{DET\_LIKE\_0}$\geq20$ and \texttt{EXT\_LIKE}=0 (1B: 15770 objects, 3B: 15229 objects).
                \item \textbf{Screened and cleaned eRASS1 LMC catalogue:} To clean the eRASS1 catalogue from known foreground and background X-ray sources, we removed from the eRASS1 catalogue matches with the screening catalogues found in Table\,\ref{tab:screening_cats} within a search radius of 30" and a maximum separation of less than or equal to 3$\sigma=3.439\cdot\sqrt{\mathtt{POS\_ERR}^2+\sigma_{screen}^2}$ with the positional error listed in the respective screening catalogue $\sigma_{screen}$ (1B: 14992 objects, 3B: 14458 objects).
                \item \textbf{Raw eRASS1 matches:} The screened and cleaned eRASS1 catalogue was then matched with our selection of the MCPS or VMC catalogue, respectively, within 30" and 3$\sigma=3.439\cdot\sqrt{\mathtt{POS\_ERR}^2+\left(1"\right)^2}$, where we used 1" as an upper limit for the expected positional uncertainty of the MCPS and VMC catalogues (VMC: 1B: 167 objects, 3B: 160 objects; MCPS: 1B: 124 objects, 3B: 123 objects).
                \item \textbf{Final set of \ero candidates:} As a final step, we matched the optical/IR positions of the raw eRASS1 matches with the LMC selection of \textit{Gaia} eDR3 objects within 1" to secure LMC membership of the counterpart (VMC: 1B: 74 objects, 3B: 74 objects; MCPS: 1B: 65 objects, 3B: 63 objects; VMC and MCPS: 1B: 48 objects, 3B: 51 objects; VMC or MCPS: 1B: 88 objects, 3B: 85 objects). Additionally, we visually inspected the \ero RGB images for eRASS1, 2, 3, 4 and :4/5 to reject spurious objects, which typically appear as random fluctuations in the background count rates. We also rejected clear foreground or background objects based on their VizieR and Simbad\footnote{\url{https://simbad.cds.unistra.fr/simbad/}} matches. In total, we excluded 43 matched objects from the 1B catalogue and 5 from the 3B catalogue.
                
                The final set of eROSITA candidates was matched with the lists of AGN candidates indicated in Table\,\ref{tab:screening_cats}. Matches and corresponding flags were added to the catalogue as information. Table \ref{tab:MasterTable_known} reports the results for known and new HMXBs.
        \end{itemize}
        
        \subsection{Chance coincidence}
        To assess contamination from chance-coincidence misidentifications, we estimated the number of objects that would appear in our final set of \ero candidates when using a simulated eRASS1 catalogue, following a method similar to \citet{2019A&A...622A..29M}. This fake catalogue was generated by shifting and rotating the entire original eRASS1 catalogue as a whole. Specifically, we started with the screened and cleaned eRASS1 LMC catalogue, then applied a uniform translation in RA and Dec by a random value between 180" and 540", followed by a single rotation around the LMC centre using a random angle within the same range. We then applied the same selection procedure used for the real eRASS1 dataset to obtain the final set of \ero candidates. For VMC matches, the contamination rates are $69 \pm 9\%$ and $63 \pm 8\%$ for the 1B and 3B catalogues, respectively. For MCPS matches, the contamination rates are $51 \pm 9\%$ and $47 \pm 8\%$. Among objects with counterparts in both the MCPS and VMC selections, we find contamination rates of $54 \pm 11\%$ and $45 \pm 8\%$ for the 1B and 3B catalogues, respectively. For objects with a counterpart in at least one of the two selections, the expected contamination rates due to chance coincidences are $67 \pm 8\%$ and $62 \pm 8\%$. The relatively high contamination rate is consistent with the fact that nearly half of the initial matches were rejected as spurious during visual inspection, indicating that the majority of contaminants were effectively identified and removed.
        
        \section{Data reduction and analysis}
        \label{sec:analysis}
        
        \subsection{Methodology}
        \label{sec:methodology}
        To distinguish between the region we are discussing and the origin of photons, we shall use the terms ‘on’ and ‘off’ to refer to the regions with and without the source, respectively. The terms ‘source’ and ‘background’ are used for the photon origin only and do not correspond to regions. As an example, we expect all photons in the off-region to be background photons. Photons in the on-region are the sum of source and background photons.
        
        For the extraction of source products from \ero data, we used the \ero Standard Analysis Software System \citep[\eSASS version {\tt eSASSusers\_211214;}][]{2022A&A...661A...1B}. For the extraction of events in the on- and off-regions, we used the \eSASS task \texttt{srctool}.
        For light curves, we used the combined data of all telescope modules (TMs 1$-$7) and typically used a cut on fractional exposure of 0.15 (\texttt{FRACEXP}>0.15) to avoid highly vignetted data (except for very bright objects such as LMC X$-$1, LMC X$-$3, and LMC X$-$4). For spectra, we used the combined data of cameras with an on-chip optical block filter (TMs 1$-$4 and 6). TMs 5 and 7 suffer from light leak \citep[][]{2021A&A...647A...1P} and no reliable energy calibration is available as of yet. As an on-region for the large majority of sources, we used circles of approximately 50", depending on source brightness, to optimise the S/N. The exceptions are LMC X$-$1, LMC X$-$3, and LMC X$-$4, where the high source flux leads to photon pile-up in the central region. To counter this, we used annuli for these three objects, which lowered the overall fractional exposure by approximately a factor of 1000, such that we had to adjust the fractional exposure cut for the light curves by this factor. Where possible, we used two circles of the same size placed on both sides of the \ero scan direction as an off-region. This provided a homogeneous background fraction during the whole observation. For some objects, it was not possible to apply this method due to strong changes in the background caused by the hot ISM.
        
        There are two fundamental schemes to group observations into time periods for \ero. The first is to group the data by eRASSs, which are defined by specific dates seen in Table\,\ref{tab:eRASS_times}. The second way is to group it into epochs of continuous observation. For the majority of the sky, this is the same thing, but due to the fact that the starting line of the eRASSs lies in the LMC, there are objects that are observed at the start and end of each eRASS. For those, it makes most sense to divide the data not by eRASS but by the gaps between observations. In the following, these two period schemes are referred to as ‘eRASS’ and ‘epoch’, respectively.
        
        \begin{table} 
                \centering
                \caption{eRASS start and stop times.} 
                \label{tab:eRASS_times} 
                \begin{tabular}{l|ll} 
                        \hline\hline\noalign{\smallskip}
                        eRASS & Start & Stop\\
                        & (UTC) & (UTC) \\
                        \noalign{\smallskip}\hline\noalign{\smallskip}
                        pre & 2019-12-11 10:50 & 2019-12-11 21:30\\
                        1 & 2019-12-11 21:30 & 2020-06-11 11:00 \\
                        2 & 2020-06-11 11:00 & 2020-12-15 12:30 \\
                        3 & 2020-12-15 12:30 & 2021-06-16 16:00 \\
                        4 & 2021-06-16 16:00 & 2021-12-19 17:30 \\
                        5 & 2021-12-19 17:30 & 2022-02-26 00:00\\
                        \noalign{\smallskip}\hline
                \end{tabular}
        \tablefoot{Note that the ‘pre’ eRASS identifies test scans that were taken before the official start of eRASS1, and eRASS5 is the unfinished final all-sky survey that covers approximately half of the LMC.}
        \end{table}
        
        \subsection{\ero timing analysis: Searching for variability}
        \label{sec:time-analysis}
        
        \subsubsection{Creation of light curves}
        \label{sec:LCs}
        For our most fundamental approach to analysing the time-dependent variability of our sources, we extracted light curves. Due to the observation strategy of \ero to create a meaningful light curve, it is essential to take into account the changing fractional exposure at all times. In addition, the presence of a large number of faint sources in our sample, combined with high background count rates resulting from the hot interstellar medium and the high source density in the LMC, led us to employ a Bayesian approach to model background and source count rates in each time bin.
        
        The natural time bin of \ero is one full rotation of the satellite (four hours), also called an eROday. However, due to the scans moving over the source, the time for which a source is visible during one eROday is variable (up to 40\,s). Due to the close vicinity to the SEP, the total exposure during one eRASS can add up to more than 10\,ks in the NE part of the LMC and is approximately 2\,ks in the central regions. To not integrate over too long times when the source is at the edges of the FOV, we chose to use 1\,s bins for the initial extraction with srctool and rebin the light curve thereafter. Before rebinning, we applied a cut in fractional exposure of 0.15 (with a few exceptions noted for individual objects) to eliminate very noisy data points at the beginning and end of scans, which occur due to high off-axis angles. For rebinning, we used two different approaches. The first one was to rebin into eROdays, for which we used the temporal offset between two primary bins, defining the start of a new eROday when the offset to the previous bin exceeded 3600\,s. The second rebinning pattern we used was motivated by the low count rates of many of our objects. To obtain statistically meaningful count rates, we rebinned our light curves by ensuring a minimum number of photon counts in the on-region while allowing the bin width to vary freely. By default, we set this minimum to 10. We implied that such a bin cannot extend over the end of a period. If the last bin of a period did not have a sufficient number of photon counts, it was merged with the previous bin. The only exception to merging was when the bin was the first of its period, in which case the period was treated as a single time bin, even if it did not meet the minimum photon count threshold. In addition, we did not allow a time bin to end during an eROday. This was motivated by the large time gap between eROdays relative to their length, leading to count rates that are expected to vary more between eROdays than within a single scan.
        
        Rebinning the light curves by a minimum number of photon counts can wash out flares for faint objects, which is why we create both types of light curves -- binned by scans and binned by counts -- for all of our objects.
        
        Once the light curves were rebinned, we fitted for count rates in each new time bin using the \texttt{CmdStanPy} interface\footnote{\url{https://mc-stan.org/cmdstanpy/}} to \texttt{Stan} \citep{stan2024} for using Bayesian inference. For this, we assume a constant count rate in each new bin, respectively, and assume that counts in the on- and off-region are drawn as Poisson variables as follows:
        \begin{equation}
                \label{equ:LC_poisson}
                \begin{split}
                        n\unter{on,i} &\sim \Poisson{\left(src+bkg\right)\cdot\mathrm{d}t\unter{i}\cdot f\unter{E,i}}\\
                        n\unter{off,i} &\sim \Poisson{\frac{bkg}{r\unter{i}}\cdot\mathrm{d}t\unter{i}\cdot f\unter{E,i}},
                \end{split}
        \end{equation}
        where $n\unter{on,i}$ and $n\unter{off,i}$ are the number of photon counts in the on and off-region, respectively, $src$ and $bkg$ are the count rates of the source and background in the on-region which are fit for, $\mathrm{d}t\unter{i}$ is the time interval of an initial time bin, $f\unter{E,i}$ are the fractional exposures and $r\unter{i}$ is the factor by which the background counts should be scaled in order to estimate the number of background counts within the on-region. The subscript $i$ refers to the initial time bins that are included in the new bin. We note that the criterion mentioned for rebinning to a minimum of 10 counts can be written as $\sum_{i} n\unter{on,i}\geq10$. We used a log-uniform prior for the source and background count rates. As uncertainties, we show the 1\,$\sigma$ percentiles of the source and background count rate posterior distributions in each time bin.

        \subsubsection{Variability}
        \label{sec:MAV}
        As a measure to quantify the variability of our sources and compare them to the spectrally similar AGN, we used the ratio, $var$, of the maximum and the minimum measured count rates as
        \begin{equation}
                \label{equ:MAV}
                var = \frac{src\unter{max}}{src\unter{min}},
        \end{equation}
        where $src$ and $\sigma$ are the source count rates and corresponding uncertainties at the bins where $src-\sigma$ has its maximum and where $src+\sigma$ has its minimum (subscripts `max' and `min', respectively). We used these definitions of maximum and minimum values to avoid being dominated by high-uncertainty values. We calculated this value for light curves extracted as explained in Sect.\,\ref{sec:LCs} and \ref{sec:long_var}.
        
        \subsubsection{Bayesian blocks}
        
        Bayesian blocks \citep[][]{1998ApJ...504..405S, 2013ApJ...764..167S, 2013arXiv1304.2818S} is an algorithm to find change points in binned data or a list of photon arrival times. Bayesian blocks decides whether to put a change point or not based on Bayesian model comparison. The most fundamental application of Bayesian blocks is to use it to analyse count light curves or a list of arrival times without any predefined binning. However, as is mentioned in Sect.\,\ref{sec:ero}, due to the scanning procedure of \ero, the fractional exposure of a source changes during an observation. This means that one cannot directly compare the number of counts in different time bins; however, it is necessary to extract the corresponding count rates. Another difficulty for the standard Bayesian blocks algorithm is that, especially for fainter sources, the background is not negligible. A modification of the Bayesian blocks algorithm to incorporate those two aspects would be desirable, but this is beyond the scope of this work. Instead, we used the Gaussian Bayesian blocks implementation by \texttt{astropy} \citep{astropy:2013, astropy:2018, astropy:2022}, which allows for the application of Bayesian blocks on a sequence of (non-integer) measured data points with Gaussian errors. For this, we used the extracted light curve as explained in Sect.\,\ref{sec:LCs}. Due to the method we used for extracting our light curves, we typically do not find Gaussian or symmetric errors. As an estimator for the Bayesian blocks algorithm, we used the maximum of the upper and lower value at each data point and cap the value by the count rate in the given bin, such that $src-\sigma\geq0$. We then applied the \texttt{astropy} Bayesian blocks algorithm (\texttt{bayesian\_blocks}) with the \texttt{fitness} parameter set to `measures' and the false alarm probability parameter, $p_{0}$, set to 0.05. We want to note that $p_{0}$ does not exactly correspond to a probability due to several reasons:
        \begin{itemize}
                \item $p_{0}$ enters the Bayesian blocks algorithm by modifying the prior for the number of changing points as a function of the number of initial bins $N$. \cite{2013ApJ...764..167S} did extensive simulations for binned event data and this way determined the prior as a function of $N$ and $p_{0}$ empirically as 
                \begin{equation}\label{equ:ncp_prior}
                        ncp\_prior=4-\log(73.53p_{0}N^{-0.478}).
                \end{equation}
                Note the correction done by \cite{2013arXiv1304.2818S}. This prior was developed for event data only, and it does not match the relation \cite{2013ApJ...764..167S} find for point measures done for $p_{0}=0.05$.
                \item In the case of event data in the \texttt{astropy} package, there is a note that $p_{0}$ does not seem to accurately represent the false alarm probability. While the same functional form is used for $ncp\_prior$ for all three cases, there is no such comment for the case of point measures.
                \item As was mentioned earlier, our data does not exhibit Gaussian errors, and we are forced to apply an estimation so we can use the Bayesian blocks algorithm.
        \end{itemize}
        For these reasons, we refrain from stating a false alarm probability of 5\%. Instead, $p_{0}$ should be understood only as a parameter in our analysis. However, since we did not use Bayesian blocks to measure the variability of our sources but only as a tool to look for outbursts or flares that could otherwise be missed, this was sufficient for our needs, and we could tune the $p_{0}$ parameters using a select number of objects for which we knew which behaviour to expect.
        
        \subsubsection{Hardness ratio light curves}
        
        Drastic variability in flux in HMXB arise from variations in accretion rate and geometry, often causing spectral changes \citep{2013A&A...551A...1R}. For example, in addition to Compton up-scattering at the accretion column, at higher accretion rates, photons can be produced through blackbody-like emission in a newly formed accretion disc. Meanwhile, cold, dense material surrounding the compact object make the spectrum appear harder through absorption.
        
        To study these effects, we analysed spectral changes using hardness ratio (HR) light curves. We extracted 1\,s binned light curves in three energy ranges: a reference band (0.2$-$5.0\,keV; same as the band for spectral fitting) used for fractional exposure cuts, a soft band (0.2$-$2.0\,keV), and a hard band (2.0$-$5.0\,keV). The bands were chosen to highlight absorption effects in the soft band while ensuring a sufficient number of photons in the hard band for sources with expected power-law spectra. For better statistics, we then rebinned the data such that each new bin fulfilled two criteria:
        \begin{itemize}
                \item The number of counts in the full band had to be at least 10.
                \item The number of counts in each of the sub-bands had to be at least 1 (for higher time resolution) or 10 (for better statistics).
        \end{itemize}
        We then applied Eq. \ref{equ:LC_poisson} for the soft and hard bands simultaneously to fit for the count rates and calculate the HR as
        \begin{equation}
                \label{equ:HR_LC}
                HR=\frac{src\unter{hard}-src\unter{soft}}{src\unter{hard}+src\unter{soft}}.
        \end{equation}
        As uncertainties, we again used the 1\,$\sigma$ percentiles of the posterior distribution.
        
        \subsection{Spectral fitting of the \ero data}
        \label{sec:spec-analysis}
        
        \subsubsection{BXA}
        \label{sec:bxa}
        
        For the spectral analysis of our sources, we used Bayesian X-ray analysis \citep[\texttt{BXA};][]{2021JOSS....6.3045B}. \texttt{BXA} allows one to use X-ray models from \texttt{Xspec} \citep[][]{1996ASPC..101...17A} together with nested sampling from \texttt{UltraNest} \citep[][]{2021JOSS....6.3001B} to explore the entire model parameter space. This provides a computationally efficient tool for an unsupervised search of the best estimate parameters, utilising the Bayesian theorem.
        
        \subsubsection{Models and priors used}
        \label{sec:ero_models}
        
        Most of our sources are well described by an absorbed power law (\texttt{powerlaw} in \texttt{Xspec}), a black body (\texttt{bbodyrad}), or a combination of the two. The power-law model is commonly used for Comptonised thermal emission in HMXBs. The blackbody emission typically originates in the accretion disc or appears on the polar cap or the surface of supersoft sources (SSSs), such as binary systems of Be stars with white dwarfs (Be/WD). For absorption, we used a \texttt{tbabs} component to account for MW foreground absorption. We used the weighted average value from \cite{1990ARA&A..28..215D} as a reference and scaled it up by a factor of 1.25 to account for the minimum contribution to absorption by molecular gas as suggested in \cite{2013MNRAS.431..394W}. If it improved the fit, we added a \texttt{tbvarabs} model for local absorption with LMC metallicity \citep[elemental abundances fixes at 0.49;][]{2002A&A...396...53R, 1998AJ....115..605L} if needed or a \texttt{tbpcf} component to account for partially covered sources. For all absorption models, we used abundances from \citet{2000ApJ...542..914W}.
        
        For the background spectrum, we used a principal component analysis (PCA) model provided by \texttt{BXA} and fit the source and background spectra simultaneously. To determine the spectral shape of the PCA component, we first fit the off-region spectrum alone. We then kept the shape frozen and fitted the normalisation simultaneously to the off-region and, together with the source model, to the on-region, while tying the normalisations in the two regions to one another using the \texttt{BACKSCAL} value as a factor. \texttt{BACKSCAL} is the keyword calculated by \texttt{srctool}, which links the sizes of the on- and off-regions with one another.
        
        The final important component for our models was the priors we used for the fit parameters. These are described in Table\,\ref{tab:priors}.
        
        \begin{table} 
                \centering
                \caption{Model priors for the spectral fits.} 
                \label{tab:priors} 
                \begin{tabular}{llll} 
                        \hline\hline\noalign{\smallskip}
                        Parameter & Spectral & Prior & Range\\
                        & Model$^{a}$ &  & \\
                        \noalign{\smallskip}\hline\noalign{\smallskip}
                        N$_{\mathrm{H}}$ (cm$^{-2}$) & \texttt{tbvarabs} & Jeffreys & 10$^{17}-10^{24}$ \\
                        N$_{\mathrm{H}}$ (cm$^{-2}$) & \texttt{tbpcf} & Jeffreys & 10$^{17}-10^{24}$ \\
                        pcf & \texttt{tbpcf} & uniform & $0-1$ \\
                        $\Gamma$ & \texttt{powerlaw} & uniform & $-2-4$ \\
                        norm (keV$^{-1}$cm$^{-2}$s$^{-1}$) & \texttt{powerlaw} & Jeffreys & $10^{-8}-10^{2}$ \\
                        kT (keV) & \texttt{bbodyrad} & uniform & $10^{-3}-10$ \\
                        norm$^{b}$ & \texttt{bbodyrad} & Jeffreys & $10^{-3}-10^{7}$ \\
                        T$_{in}$ (keV) & \texttt{diskbb} & uniform & $10^{-3}-0.3$ \\
                        norm$^{b}$ & \texttt{diskbb} & Jeffreys & $10^{-6}-10^{7}$ \\
                        \noalign{\smallskip}\hline
                \end{tabular} 
                \tablefoot{
                        \tablefoottext{a}{Model names from \texttt{Xspec}.}
                        \tablefoottext{b}{$R^2_{km}/D^2_{10}$, where $R_{km}$ is the source radius in km and $D_{10}$ is the distance to the source in units of 10\,kpc.}
                }
        \end{table}
        
        \subsection{Investigating the variability of the optical counterpart through \ogle}
        \label{sec:ogle-analysis}
        We used \ogle I-band light curves to investigate the long-term variability of our sources and to search for orbital periods of the (candidate) HMXB systems. In Table\,\ref{tab:ogle_data} we present \ogle information about the optical counterparts. While we focus on new systems or systems without published \ogle data (\ogle-IDs are listed in Table\,\ref{tab:ogle_data}), we provide references for already published \ogle data. In three cases, stars are placed near CCD gaps, which can lead to fewer measurements (indicated by the letter `D' in the \ogle-IDs). For systems which show high variability in their I-band light curves ($>$0.3 mag), we used V-band data to investigate colour-magnitude diagrams. V-band data can be identified by the letter `v' in their \ogle-IDs).
        
        Our Fig.\,\ref{fig:ogle_IVlc} and Fig.\,\ref{fig:ogle_Ilc} present \ogle light curves of the systems investigated in this work. I- and V-band light curves are only presented for highly variable cases ($>$0.3 mag in I). For the latter, we created colour (V$-$I) magnitude (I) diagrams, as is described, for example, in \citet{2022A&A...662A..22H} and shown in Fig.\,\ref{fig:ogle_IVlc}.
        
        To search for periodic variations in the \ogle I-band light curves, we used the Lomb-Scargle (LS) periodogram analysis \citep{1976Ap&SS..39..447L,1982ApJ...263..835S}, implemented in the \texttt{astropy} package of \texttt{Python}\footnote{\url{https://docs.astropy.org/en/stable/timeseries/lombscargle.html}}. As applied to new LMC HMXBs in the past \citep[e.g.][]{2022A&A...662A..22H,2023A&A...671A..90H}, we first removed long-term trends from the light curves by subtracting a smoothed version of the light curve. To avoid false positive signals and remove true signals, we used two different methods of smoothing: applying 1) a Savitzky–Golay filter with different window lengths \citep{1964AnaCh..36.1627S} and 2) a spline fit \citep[rspline from \texttt{wotan};][]{2019AJ....158..143H} with a window of 200 and a break tolerance of 500 \citep[see][for an analysis of BeXRBs in the SMC]{2025A&A...694A..43T}. Candidate periods, which could indicate the orbital period of the binary system, are only accepted when found by both methods. In addition, we created LS periodograms from the light-curve window functions. Strong peaks are only found at 1\,d, $\sim$0.5\,year, $\sim$1\,year, and beyond several years.
        
        We produce LS periodograms for the three period ranges 2$-$20\,days, 20$-$200\,days, and 200\,days to one third of the total observing time for the original and both sets of detrended light curves. 
        The split into three period ranges allows for individual scalings, providing better visualisation of peaks with different heights.
        When significant peaks were found near 2\,days, we also looked at shorter periods to check for aliasing effects with the sampling period of 1\,day and/or short periods, which most likely are caused by non-radial pulsations (NRPs) of the Be star \citep[e.g.][]{2003A&A...411..229R}. 
        
\begin{figure*}
                \centering
                \resizebox{0.33\hsize}{!}{\includegraphics{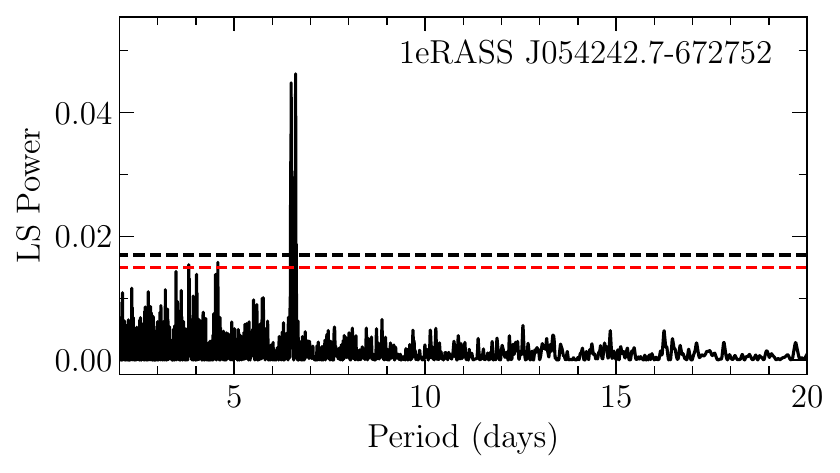}}
                \resizebox{0.33\hsize}{!}{\includegraphics{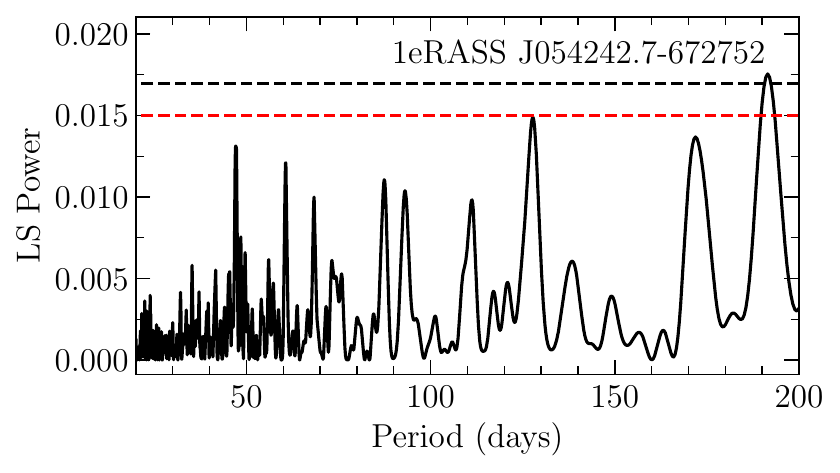}}
                \resizebox{0.33\hsize}{!}{\includegraphics{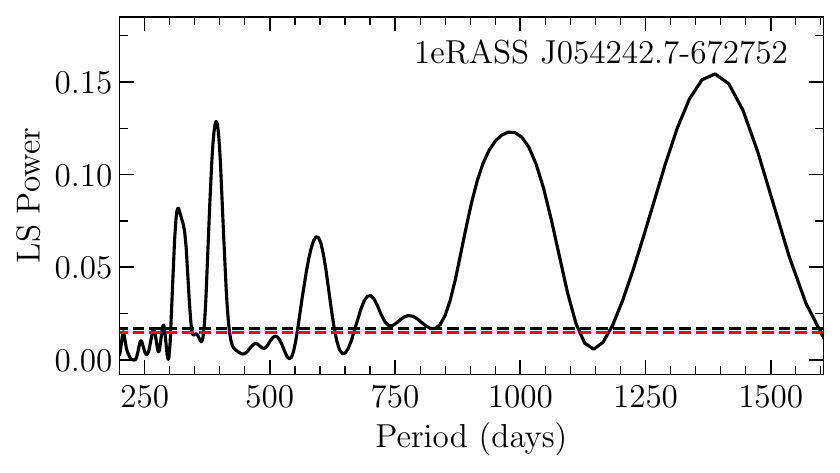}}
                \resizebox{0.33\hsize}{!}{\includegraphics{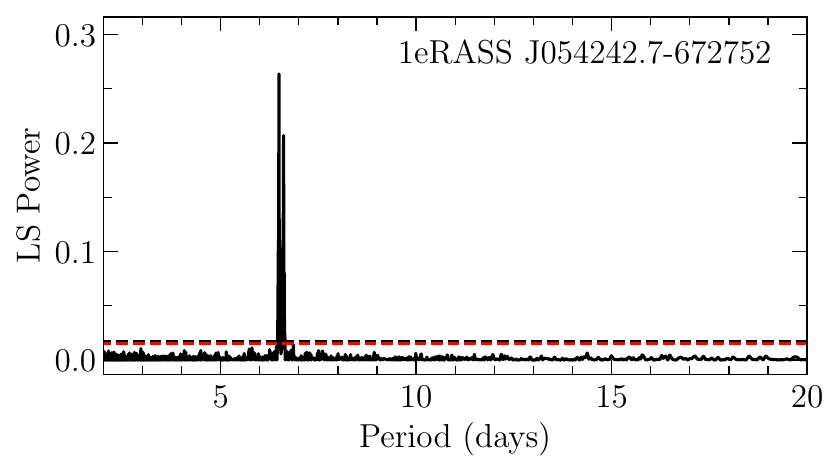}}
                \resizebox{0.33\hsize}{!}{\includegraphics{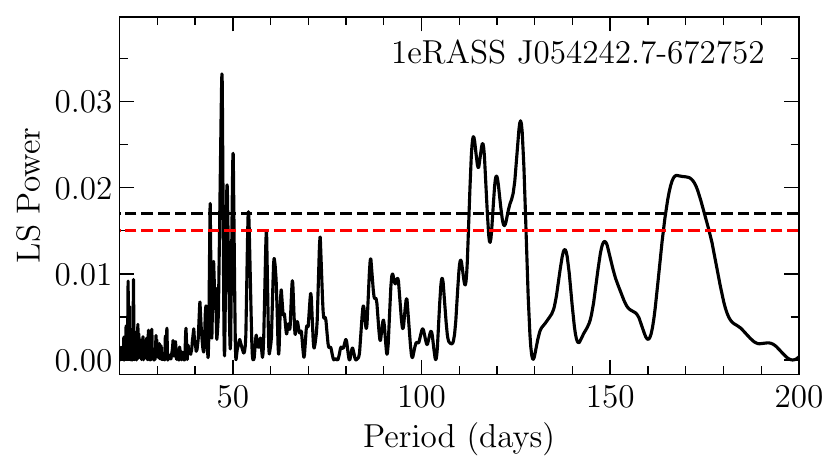}}
                \resizebox{0.33\hsize}{!}{\includegraphics{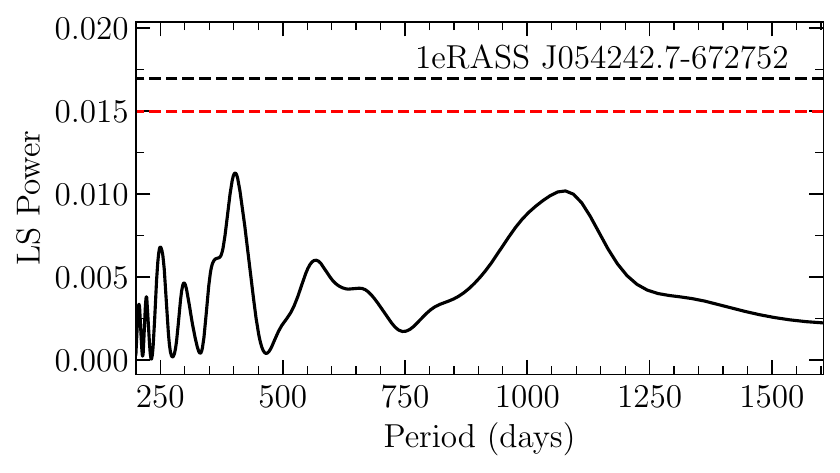}}
                \resizebox{0.33\hsize}{!}{\includegraphics{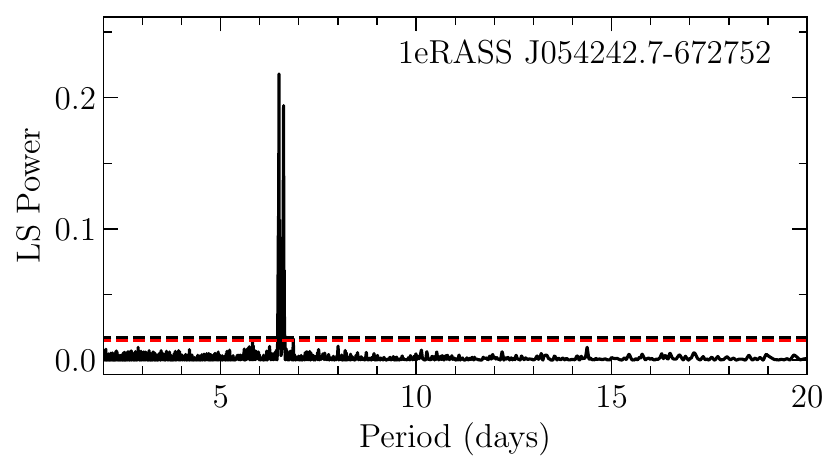}}
                \resizebox{0.33\hsize}{!}{\includegraphics{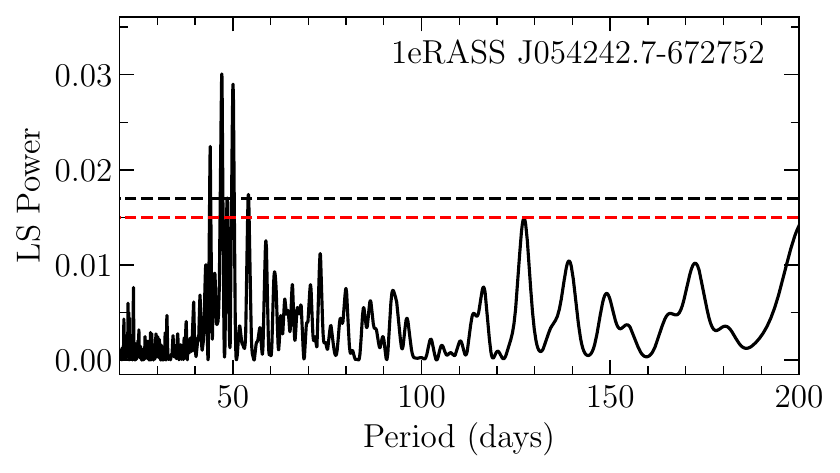}}
                \resizebox{0.33\hsize}{!}{\includegraphics{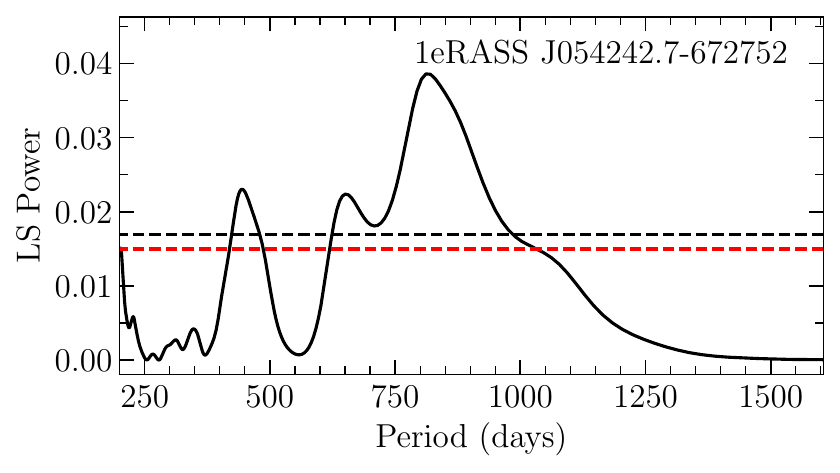}}
                \resizebox{0.33\hsize}{!}{\includegraphics{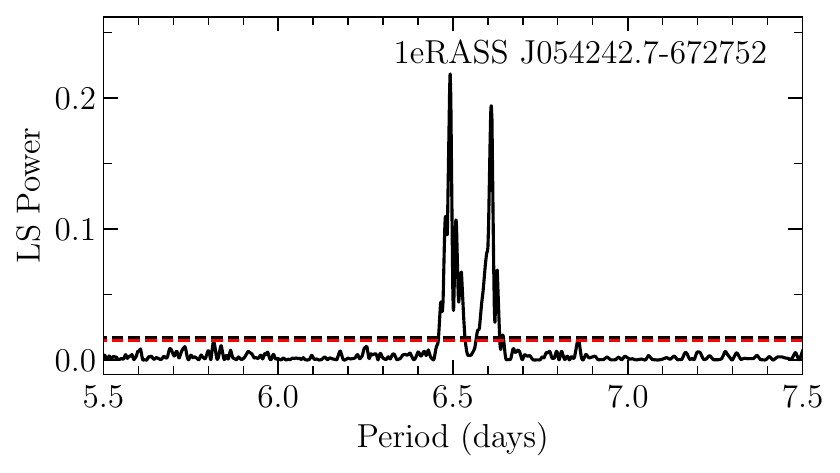}}
                \resizebox{0.33\hsize}{!}{\includegraphics{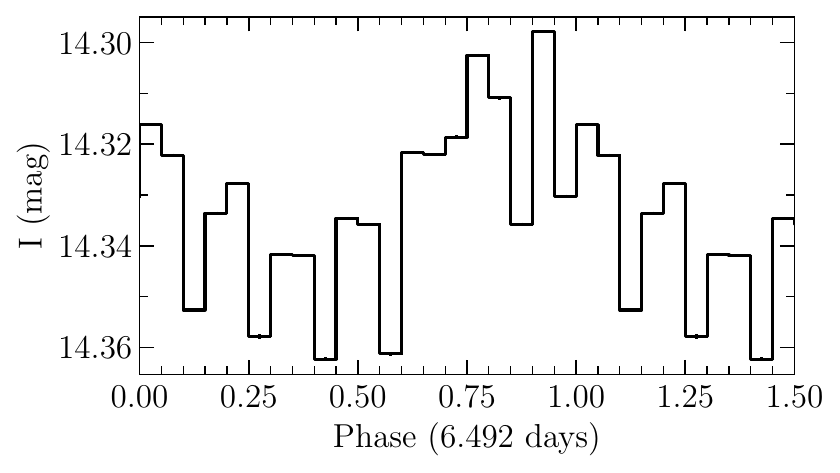}}
                \resizebox{0.33\hsize}{!}{\includegraphics{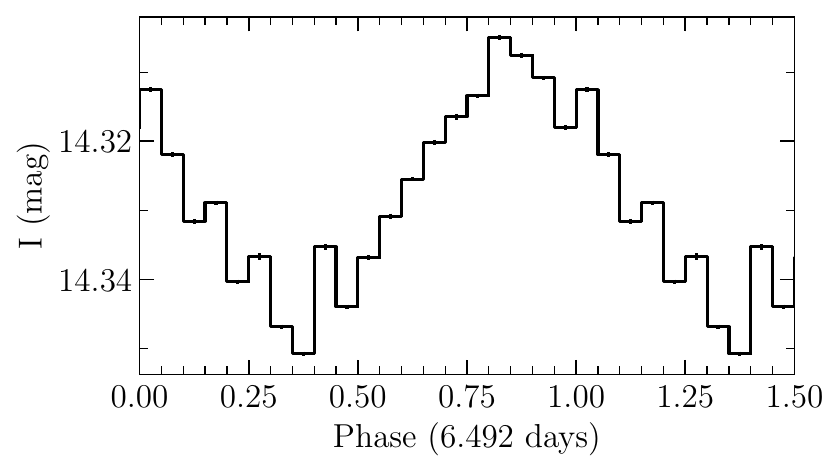}}
                \caption{LS analysis of the \ogle I-band light curve of 1eRASS\,J054242.7$-$672752 (see source 46 in Table\,\ref{tab:MasterTable_known} and Fig.\,\ref{fig:ogle_IVlc}). The top three rows show the LS periodogram split into three period ranges (for better visualisation): 2$-$20\,d, 20$-$200\,d, and 200\,d to $\sim$1600\,d, which is one-third of the monitoring period. The first row pertains to the original, the second row to the spline fit, and the third to the Savitzky–Golay filtered light curves (window 101). The bottom row shows a zoom of the LS periodogram and the light curves folded with the period with the highest power at 6.49\,d (middle panel for original and right panel for spline-detrended light curves). The dashed red and black lines mark the 95\% and 99\% confidence levels.}
                \label{fig:example:ogle_ls}
        \end{figure*}

        An example for the period analysis is shown in Fig.\,\ref{fig:example:ogle_ls} for 1eRASS\,J054242.7$-$672752. While various peaks appear in the periodograms at periods longer than 20\,s, a highly significant pair of sharp peaks is consistently found around 6.5\,s across the different detrending methods. We newly detect periods from eleven systems, seven of which are new HMXB candidates discovered during eRASS1. The periods are listed in Table\,\ref{tab:ogle_data}, marked with `(TW)', and we provide additional information on individual systems in the following.
        
        \subsection{Spectroscopic follow-up of optical counterparts}
        \label{sec:halpha-analysis}
        To securely identify the optical counterpart of our sources as Be stars, we used flux-calibrated spectra obtained using SALT, LCO/FLOYDS, and VLT FLAMES/GIRAFFE. For all three instruments, we analysed the Balmer series \Halpha and \Hbeta lines; for SALT, we additionally fitted the \HeI line. Depending on the spectrum, we fitted a single- or double-peaked Gaussian or a Lorentzian line profile to the locally normalised spectrum. From the position of the line, we extracted the radial velocity with respect to the MW to verify LMC membership \citep[radial velocity of LMC with respect to the Sun: 278$\pm$2\,km\,s$^{-1}$,][]{1987A&A...171...33R}. For a double-peaked line, we used the average velocity of the two lines as a reference. We determined all parameter uncertainties through Monte Carlo simulations, assuming the deviations in the line-less part of the fitted spectra are stochastic in nature. The \Halpha emission line profile allows us to constrain the disc inclination towards the line of sight \citep{1988A&A...189..147H} and the equivalent width can be related to the size of the decretion disc \citep{2006ApJ...651L..53G}.
        
        \subsection{SED fitting}
        \label{sec:SED-fitting}
        As an additional tool to evaluate the credibility of candidates, we tested whether their archival broadband SEDs are consistent with those of a Be star. We used publicly available data from VizieR as explained in Sect.\,\ref{sec:broadband_photo}. The fitting procedure was adapted from the one described in \citet{Bodensteiner2023}, where more details can be found.
        
        The synthetic SEDs were taken from the TLUSTY OSTAR2002 and BSTAR2006 grid \citep{Lanz2003, Lanz2007} assuming LMC metallicity. Here, we selected a constant $log\,g$ of 4 (given that the SED is not directly sensitive to the surface gravity) and varied the effective temperature over all available models (that is, between 15 000 and 50 000 K). The TLUSTY models provide Eddington flux at the stellar surface, which we converted to the observed flux by scaling for the LMC distance of $d=$49.59\,kpc \citep{Piertrzynski2019}, the stellar radius, and interstellar extinction. We varied the radius from 2 to 80 solar radii, typical of OB MS stars and supergiants (SGs). For the extinction, we used the extinction map from \citet{Skowron2021} to obtain an overall reddening value E(V$-$I) at the position of the star, which we converted to E(B$-$V) following $E(V-I) = 1.237 E(B-V)$ as indicated by the authors. To account for local variations in reddening, we treated the extinction as a free parameter, allowing it to vary between half and twice the value indicated in the extinction map. We further assume the extinction model from \citet{Gordon2023} and a constant R$_V=3.5$. 
        The IR excess is a well-known signature of the Be disc \citep[e.g.][]{2006A&A...456.1027D}, while it can also result from X-ray irradiation during major outbursts \citep{2025A&A...698A..26V}. To reduce potential contamination from the Be disc, we limited our analysis to photometric data at wavelengths below 10 microns, where its contribution is comparatively weaker.
        
        For each model, we finally obtained the observed flux by convolving it with the corresponding filter transmission curves \citep{Rodrigo2012, Rodrigo2020}. To find the best-fit model, we computed an overall $\chi^2$ for all observation-model pairs and located the minimum. We then assessed the quality of fit to determine whether the observations can be well reproduced by an early-type star located at the LMC distance. 
        
        \section{Catalogue results}
        \label{sec:cat_results}
        Our final catalogue comprises 53 objects that meet our selection criteria (with relaxed criteria for previously known objects) and are not classified as spurious, foreground, or background sources. Among these, we detect 28 of the 59 HMXBs known prior to our study. The large number of non-detected known HMXBs can be attributed to their high intrinsic variability and the fact that they fall below \ero's detection threshold. Additionally, we identify 25 candidate HMXBs, 21 of which were detected with \ero for the first time. One known HMXB and two candidate HMXBs were added based on their 3B detection; they would not have been included in our final catalogue had we relied solely on the eRASS1 1B catalogue (see Table \ref{tab:MasterTable_known} for details).
        
        Of the 53 objects in our catalogue, 38 meet all the criteria required for new candidates. Given a total of 88 objects that satisfy the selection criteria for either the MCPS or VMC catalogue and an expected contamination rate of $67 \pm 8\%$ due to chance coincidences, we anticipate approximately 11 misidentified objects in our catalogue. Later in this section, we define a classification scheme into confidence classes based on optical and X-ray properties for all objects in our catalogue. With this scheme, 11 objects fall into our lowest confidence class, suggesting that the remaining catalogue has a high level of purity.
        
        \begin{table*}
        \centering
                \caption{Known and candidate HMXB detected during eRASS1.}
                \label{tab:MasterTable_known}
                \begin{tabular}{l|lllll}
                        \hline\hline\noalign{\smallskip}
                        \# & X-ray Name & RA & Dec & RADEC ERR & Conf. Class and Flags \\
                        \noalign{\smallskip}\hline\noalign{\smallskip}
                        1 & XMMU\,J045315.1$-$693242 & 04 53 15.0 & $-$69 32 40.0 & 2.1 & 1, xrb, po, oo, xs, oi:, em, ix \\
                        2$^{(a)}$ & Swift\,J045558.9$-$702001 & 04 55 58.4 & $-$70 20 1.0 & 5.1 & 1, xrb, xs:, oi, em, ix \\
                        3 & XMMU\,J045736.9$-$692727 & 04 57 37.6 & $-$69 27 25.0 & 2.0 & 1, ps, xs, oi, em, ix \\
                        4$^{(a)}$ & RX\,J0501.6$-$7034 & 05 01 24.5 & $-$70 33 32.0 & 5.8 & 1, ps, po, oo, xvl, xvs, xs, oi, em \\
                        5 & XMMU\,J050722.1$-$684758 & 05 07 22.3 & $-$68 47 58.0 & 4.2 & 1, ps, po, oo, xvl, xs, oi:, em, ix \\
                        6$^{(b)}$ & XMMU\,J050755.3$-$682506 & 05 07 56.4 & $-$68 25 7.0 & 5.3 & 1, ps, po:, oo:, xvl, xvs, xs, oi:, ix \\
                        7 & eRASSU\,J050810.4$-$660653 & 05 08 10.1 & $-$66 06 54.0 & 1.5 & 1, ps, xvl, xs, oi, em \\
                        8$^{(a)}$ & 3XMM\,J051259.8$-$682640 & 05 12 59.8 & $-$68 26 36.0 & 4.8 & 1, ps, po, oo, xs:, oi, em, ix \\
                        9 & Swift\,J0513.4$-$6547 & 05 13 28.5 & $-$65 47 16.0 & 1.4 & 1, ps, px, po, os, xvl, xvs, xs, oi, em, ix \\
                        10 & RX\,J0516.0$-$6916 & 05 15 59.5 & $-$69 16 7.0 & 2.8 & 1, xrb, po, oo, xvl, xvs, oi:, em \\
                        11$^{(a)}$ & RX\,J0520.5$-$6932 & 05 20 30.3 & $-$69 31 53.0 & 4.8 & 1, ps, po, oo, xvl, xvs, xs, oi:, em, ix \\
                        12 & RX\,J0524.2$-$6620 & 05 24 11.7 & $-$66 20 52.0 & 1.5 & 1, ps, xs, oi, em \\
                        13 & 4XMM\,J052858.4$-$670946 & 05 28 58.7 & $-$67 09 46.0 & 1.7 & 1, xrb, po:, oo:, xs, oi, em \\
                        14 & eRASSU\,J052914.9$-$662446 & 05 29 14.3 & $-$66 24 46.0 & 1.6 & 1, ps, po, oo, xvl, xs, oi, em \\
                        15 & RX\,J0529.8$-$6556 & 05 29 48.1 & $-$65 56 41.0 & 1.3 & 1, ps, xvl, xvs, oi, em \\
                        16 & 4XMM\,J053011.3$-$655123 & 05 30 11.5 & $-$65 51 26.0 & 1.7 & 1, ps:, po, oo, xvl, xs, oi, em, sx, ix \\
                        17 & Swift\,J053041.9$-$665426 & 05 30 42.2 & $-$66 54 31.0 & 1.3 & 1, ps, xvl, xvs, xs, oi, em, ix \\
                        18 & XMMU\,J053108.3$-$690923 & 05 31 8.2 & $-$69 09 25.0 & 3.6 & 1, ps, xvl, xvs, xs, oi:, em \\
                        19 & RX\,J0532.5$-$6551 & 05 32 32.9 & $-$65 51 41.0 & 1.7 & 1, po, oo, xvl, xvs, xs, oi: \\
                        20 & LMC\,X$-$4 & 05 32 49.5 & $-$66 22 12.0 & 3.0 & 1, ps, px, po, os, xvl, xvs, oi \\
                        21$^{(a)}$ & RX\,J0535.0$-$6700 & 05 35 7.2 & $-$67 00 15.0 & 4.6 & 1, xrb, po, oo, xvl, xs:, oi, em, ix \\
                        22 & 1A\,0535$-$66 & 05 35 41.3 & $-$66 51 56.0 & 2.5 & 1, ps, po, oo, xvl, xvs, oi:, em \\
                        23 & CXOU\,J053600.0$-$673507 & 05 35 60.0 & $-$67 35 8.0 & 2.6 & 1, xrb, px, po, oxo, xvl, xs, oi \\
                        24 & LMC X$-$3 & 05 38 56.1 & $-$64 04 58.0 & 6.6 & 1, xrb, xvl, xvs, oi:, em \\
                        25 & LMC X$-$1 & 05 39 37.9 & $-$69 44 32.0 & 6.5 & 1, xrb, xvl, oi:, ix \\
                        26 & XMMU\,J054134.7$-$682550 & 05 41 34.4 & $-$68 25 50.0 & 1.6 & 1, ps, po, oo, xvl, xs, oi, em, ix \\
                        27$^{(a)}$ & RX\,J0544.1$-$7100 & 05 44 4.8 & $-$71 00 50.0 & 5.1 & 1, ps, xvl, xs, oi:, em, ix \\
                        28 & Swift\,J0549.7$-$6812 & 05 50 6.7 & $-$68 14 55.0 & 1.7 & 1, ps, po, oo, xvs, xs, oi, em, sx, ix \\
                        \noalign{\smallskip}\hline\noalign{\smallskip}
                        29 & 1eRASS\,J044354.7$-$692949 & 04 43 54.7 & $-$69 29 49.0 & 4.1 & 2, xs, oi:, em, sx \\
                        30 & 1eRASS\,J045024.0$-$691839 & 04 50 24.0 & $-$69 18 40.0 & 4.8 & 3, xs, oi:, sx \\
                        31 & 1eRASS\,J045028.2$-$693558 & 04 50 28.2 & $-$69 35 58.0 & 4.0 & 6, po, oo, xs, oi, qc, nv \\
                        32 & 1eRASS\,J045218.7$-$663250 & 04 52 18.7 & $-$66 32 51.0 & 3.4 & 2, xvs, xs, oi, em, sx, ix \\
                        33 & 1eRASS\,J045759.6$-$670935 & 04 57 59.6 & $-$67 09 35.0 & 7.1 & 6, oi:, sx: \\
                        34 & 1eRASS\,J050053.3$-$653209 & 05 00 53.3 & $-$65 32 9.0 & 3.7 & 6, po, oo, oi, qc, nv \\
                        35 & eRASSU\,J050213.8$-$674620 & 05 02 14.6 & $-$67 46 16.0 & 1.6 & 2, xvl, xvs, wd, oi, em, sx, ix \\
                        36 & 1eRASS\,J050359.5$-$703206 & 05 03 59.5 & $-$70 32 7.0 & 3.3 & 2, xs, oi:, em, sx, ix \\
                        37 & 1eRASS\,J050705.9$-$652149 & 05 07 5.9 & $-$65 21 49.0 & 2.9 & 2, po, oo, xvl, wd, oi, em, sx, ix \\
                        38 & 1eRASS\,J050945.8$-$655237 & 05 09 45.8 & $-$65 52 37.0 & 3.0 & 6, xvl, oi, ix, qc, nm, ns \\
                        39 & 1eRASS\,J052727.0$-$663303 & 05 27 27.0 & $-$66 33 4.0 & 4.1 & 6, oi \\
                        40 & 1eRASS\,J052948.1$-$662058 & 05 29 48.1 & $-$66 20 58.0 & 3.6 & 5, xvl, oi \\
                        41 & 4XMM\,J053049.6$-$662010 & 05 30 49.7 & $-$66 20 12.0 & 2.4 & 5, xvl, oi:, sx: \\
                        42 & XMMU\,J053320.8$-$684122 & 05 33 21.4 & $-$68 41 22.0 & 3.2 & 5, xvl, xvs, xs, oi \\
                        43$^{(c)}$ & 4XMM\,J053449.0$-$694338 & 05 34 49.3 & $-$69 43 43.0 & 9.3 & 4, po, oo, xvl, oi:, sx \\
                        44 & 1eRASS\,J054022.9$-$685644 & 05 40 22.9 & $-$68 56 45.0 & 3.2 & 6, oi \\
                        45 & RX\,J0541.6$-$6832 & 05 41 37.7 & $-$68 32 32.0 & 2.1 & 2, po, oo, xvl, xs, oi, em \\
                        46 & 1eRASS\,J054242.7$-$672752 & 05 42 42.7 & $-$67 27 53.0 & 4.1 & 2, po, oo, wd, oi:, em, sx, ix \\
                        47$^{(d)}$ & 1eRASS\,J054422.3$-$672729 & 05 44 22.3 & $-$67 27 29.0 & 4.2 & 4, po, oo, xvs, oi, em \\
                        48 & 1eRASS\,J054647.4$-$670608 & 05 46 47.4 & $-$67 06 8.0 & 3.7 & 6, xs, oi, sx:, qc, nm \\
                        49 & 1eRASS\,J055318.4$-$655953 & 05 53 18.4 & $-$65 59 53.0 & 2.3 & 6, oi:, ns \\
                        50 & 1eRASS\,J055536.1$-$671444 & 05 55 36.1 & $-$67 14 45.0 & 3.1 & 6, xvl, xs, oi, nv, ns \\
                        51$^{(d)}$ & 1eRASS\,J055849.9$-$675220 & 05 58 49.9 & $-$67 52 21.0 & 3.5 & 4, po, oo, oi, em, sx \\
                        52 & 1eRASS\,J060212.5$-$674305 & 06 02 12.5 & $-$67 43 6.0 & 2.4 & 6, po:, oo:, xs, oi, ns \\
                        53 & 1eRASS\,J060425.8$-$702920 & 06 04 25.8 & $-$70 29 20.0 & 4.3 & 6, oi \\
                        \noalign{\smallskip}\hline\noalign{\smallskip}
                        \noalign{\smallskip}\hline\noalign{\smallskip}
                \end{tabular}
                \tablefoot{
            Known (top) and candidate (bottom) HMXB detected during eRASS1. The full version of the catalogue is available at the CDS and regular updates will be posted at \url{https://projects.mpe.mpg.de/heg/lmc_eROSITA/index.html}
                        \tablefoottext{a}{eRASS1 counterpart with DET\_LIKE<20.}
                        \tablefoottext{b}{eRASS1 counterpart belongs to 3B catalogue.}
                        \tablefoottext{c}{eRASS1 counterpart shows EXT\_LIKE>0 but \xmm counterpart does not show extent.}
                        \tablefoottext{d}{eRASS1 1B counterpart with DET\_LIKE<20, 3B counterpart DET\_LIKE>20.}
        }
    \end{table*}

        \subsection{Gaia colours}
        
\begin{figure}
                \centering
                \resizebox{\hsize}{!}{\includegraphics{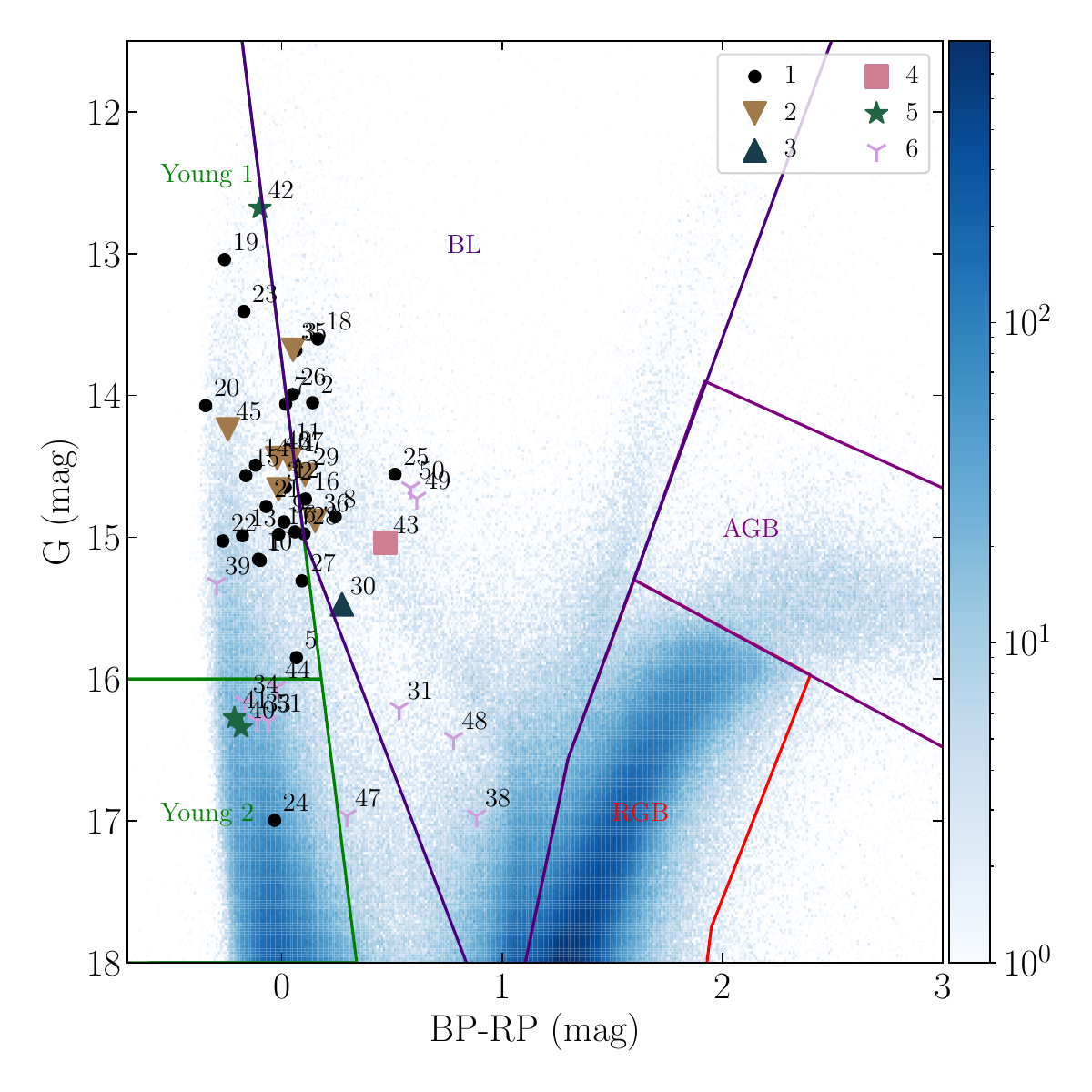}}
                \caption{\textit{Gaia} colour-magnitude diagram for objects in our catalogue labelled by confidence classes compared to the distribution of LMC stars (colour mesh) selected by proper motions and with overlaid contours for different stellar types (proper motion selection and stellar evolutionary phases from \citet{2021A_A...649A...7G} evolutionary phases in Sect. 2.3.1 and figure 2; Young 1: very young main sequence (age < 50 Myr), Young 2: young main sequence (50 Myr < age < 400 Myr), BL: blue loop, RGB: red giant branch, AGB: asymptotic giant branch). The colour scale of the colour mesh gives the number of sources in each colour-magnitude bin. The bin sizes are 0.01 mag and 0.02 mag for BP$-$RP and G, respectively. Note that the majority of objects from confidence classes 1$-$5 lie in or close to the area of very young main-sequence stars, which is to be expected for Be stars. Several objects of confidence class 6 deviate strongly from this, which might indicate misidentifications.}
                \label{fig:Gaia_HR}
        \end{figure}

        Figure\,\ref{fig:Gaia_HR} shows the distribution of the \textit{Gaia} counterparts of objects in the catalogue compared to the entire proper motion selected sample of \textit{Gaia} sources in the LMC as described in \citet{2021A_A...649A...7G}. Coloured lines separate regions in the diagram, populated by different stellar types, as shown in Fig. 2 of that work. The vast majority of objects from confidence classes 1$-$5 lie in or close to the area of very young main-sequence stars, which is to be expected for Be stars and SGs. Several objects of the confidence class 6 deviate strongly from this, which might indicate misidentifications.
        
        \subsection{IR colours}
\begin{figure}
                \centering
                \resizebox{\hsize}{!}{\includegraphics{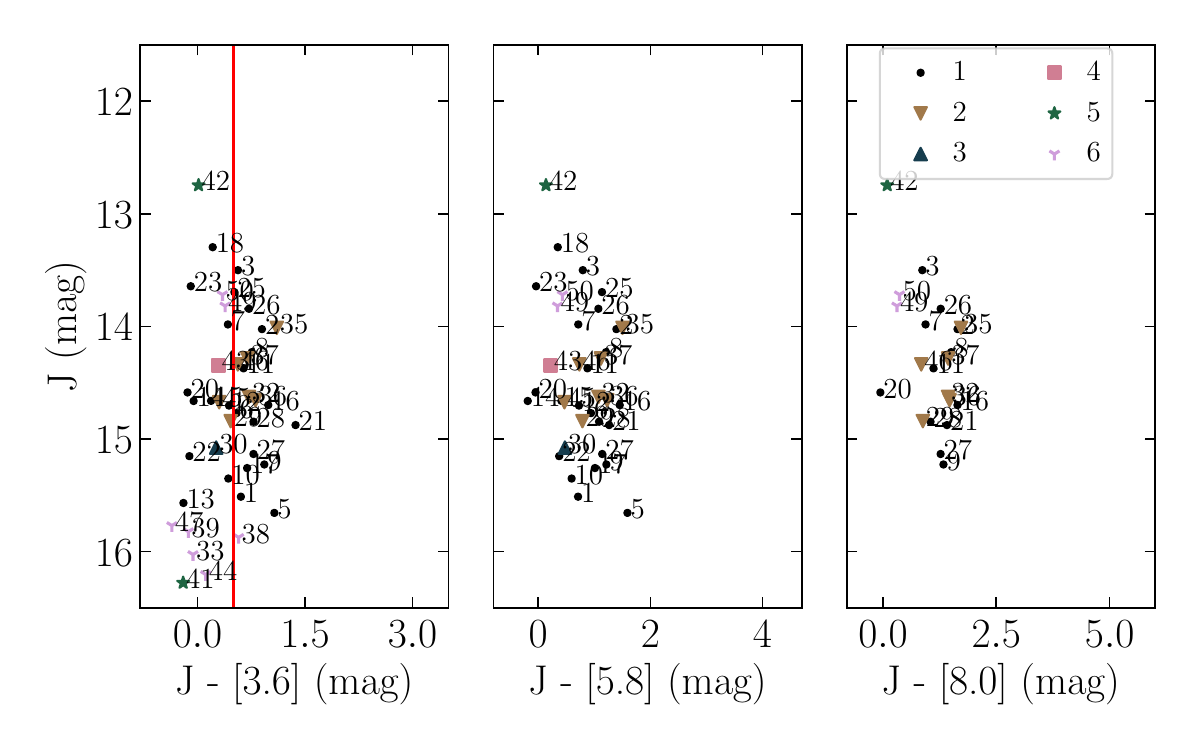}}
                \caption{2MASS J magnitudes over SAGE IR colours for objects in our catalogue labelled by confidence classes. \#42 stands out due to its high luminosity, caused by the SG nature of its optical companion. The notable clustering of confidence class 6 objects at the bottom left in the left figure, which accounts for their faint nature without hints of any IR excess, can be seen as an indication of misidentification of those objects. Objects missing in one or several of the plots are caused by a lack of entries in the corresponding SAGE bands.}
                \label{fig:IR_excess}
        \end{figure}

        Similar to the SED after photometric fitting (see Sect.\,\ref{sec:SED-fitting}), the IR excess caused by the Be disc can also be observed via IR colours. \citet{2010AJ....140..416B} used Spitzer IRAC fluxes at 3.6, 4.5, 5.8 and 8.0 $\mu$m relative to J magnitudes and found that Be stars typically exhibit high values in the IR colour J-[3.6]. They further define a photometric Be star classification for objects with J-[3.6]>0.5. Figure \ref{fig:IR_excess} shows IR colours for the objects in our catalogue. Remarkably, a large portion of confidence class 6 objects fall in the bottom left corner of the J over J-[3.6] plot. This corresponds to fainter and bluer objects, hinting at possible misidentifications.
        
        \subsection{Optical colour-colour distribution}
\begin{figure*}
                \centering
                \resizebox{0.495\hsize}{!}{\includegraphics{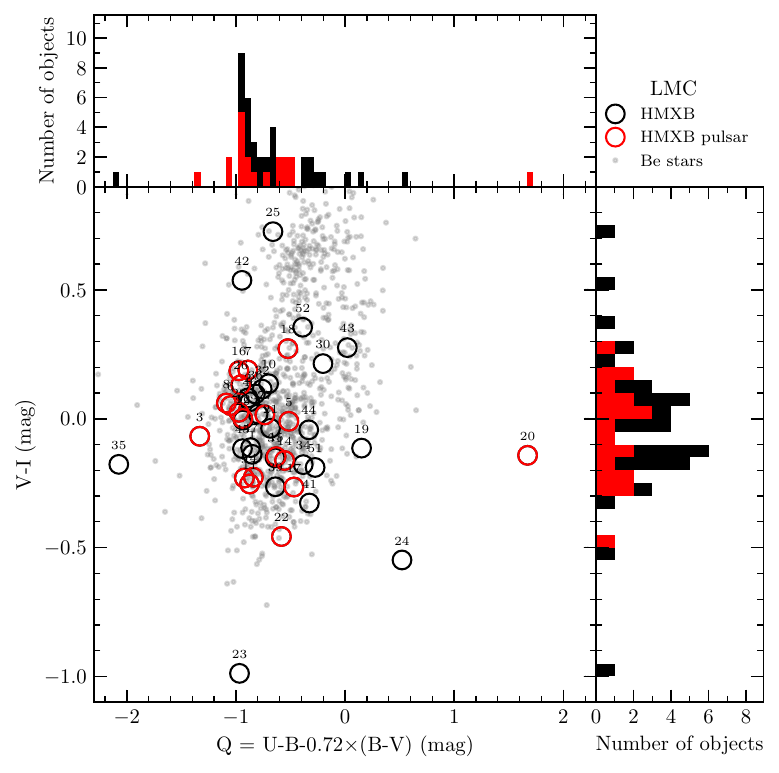}}
                \resizebox{0.495\hsize}{!}{\includegraphics{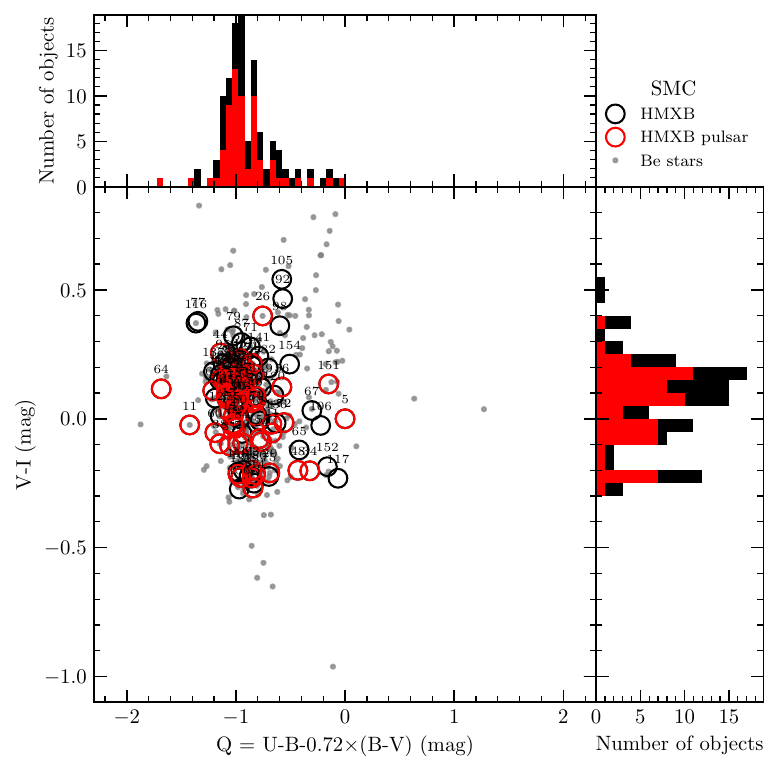}}
                \caption{V$-$I over Q plotted for the optical companion stars in (candidate) HMXBs (black circles) and HMXBs with X-ray pulsar (red circles) as detected during eRASS1 in the LMC (left, labelled by source number from Table\,\ref{tab:MasterTable_known}). Be stars (V<18 mag) from the list of \citet{2005MNRAS.361.1055S} are marked as faint grey dots. For comparison, the same is plotted for the SMC (right, source numbers from \citet{2016A&A...586A..81H} and Be stars from \citet{2002A&A...393..887M}). The larger spread in the LMC compared to the SMC is evident in all systems.}
                \label{fig:V-I_Q}
        \end{figure*}

        To test the similarity of Be stars among each other, we used the plots shown in Fig. \ref{fig:V-I_Q}, which display the colour V$-$I over the reddening-free Q parameter \citep[$\textrm{Q}=\textrm{U}-\textrm{B}-0.72\times(\textrm{B}-\textrm{V})$; see][]{1998BaltA...7..605S, 2023Galax..11...31A}. The striking difference between the LMC and SMC populations is evident in Be stars within HMXB systems, as well as in the entire Be population. The significantly higher spread observed for the LMC population suggests a greater variety in the physical properties of Be stars therein.
        
        \subsection{X-ray variability}
        \label{sec:var}
        A defining aspect of HMXB is their variability in X-ray brightness, which can be attributed to changes in accretion. This variability manifests itself both in the short and long terms.
        \subsubsection{Short-term}
        \label{sec:short_var}
\begin{figure}
                \centering
                \resizebox{\hsize}{!}{\includegraphics{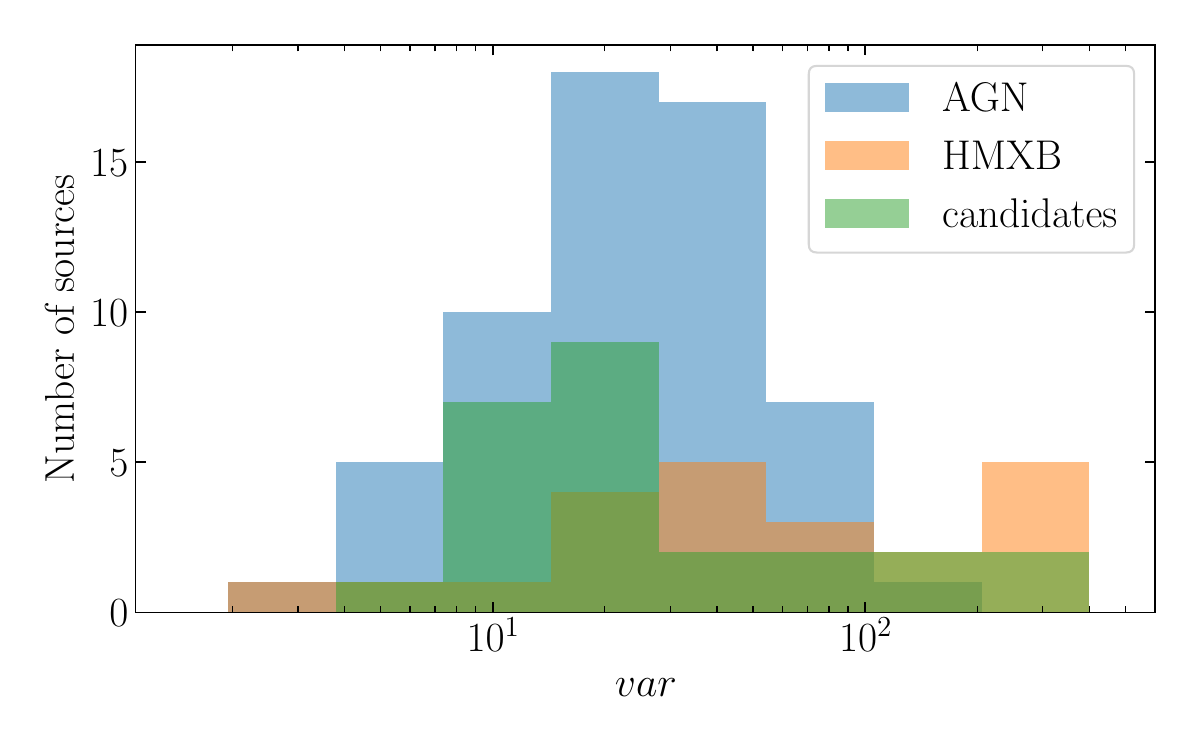}}
                \caption{Short-term variability of AGN, HMXBs, and HMXB candidates represented by their $var$. For readability, the plot is capped at a maximum variability of 400. Six additional HMXBs exceed this $var$ threshold, while no AGN or HMXB candidates show larger values.}
                \label{fig:short_var}
        \end{figure}

        We define short-term variability as one observable over the span of the two years during which objects were observed by \ero. Figure \ref{fig:short_var} shows the distribution of $var$s (see Sect.\,\ref{sec:MAV}) we find for 0.2$-$5.0\,keV light curves using time bins with a minimum of 10 net counts per time bin. We compare the distributions for secure HMXB (class I, see Table\,\ref{tab:conf_classes}) with candidates of different confidence levels and the same set of AGN as described in Sect.\,\ref{sec:eRO_HR}. We find that the distributions of variability in HMXBs and AGN show large similarities, with the exception that very high values of variability are predominantly observed in HMXBs. We therefore give objects with a short-term $var$ of >100 a positive flag, indicating a high chance of being an HMXB.
        
        \subsubsection{Long-term}
        \label{sec:long_var}
        To study long-term variability, we used data available on the HILIGT upper limit server for \xmm (slew and pointed), \ROSAT (pointed and survey) and \swift and included data points for the average flux in each eRASS. Lacking knowledge of spectral changes of objects during observations included in the upper limit server, we used an absorbed power-law spectral model with a power-law index of 1 and \nh\ of 3\hcm{20} for the flux normalisation, also applying this to \ero data points for comparability. If the lowest flux point in the long-term light curve is an upper limit, the resulting $var$ is a lower limit. Similar to short-term variability, we find that large values of variability are predominantly observed in HMXBs. For long-term variability, this corresponds to values greater than 30, which matches those of earlier works \citep{2016A&A...586A..81H}. The individual long-term LCs with $var$ values for each source are shown in Fig.\,\ref{fig:UL_lightcurves}.
        
        \subsection{X-ray hardness ratios}
        \label{sec:eRO_HR}
        As a qualitative measure of the spectral differences between HMXBs and AGNs $-$ our main expected contributors to misidentifications caused by chance-coincidence matches $-$ X-ray HRs can be used. For this purpose, we used four energy bands (1: 0.2$-$0.5\,keV, 2: 0.5$-$1.0\,keV, 3: 1.0$-$2.0\,keV, 4: 2.0$-$5.0\,keV) and defined the HR as
        \[
        HR_{i}=\frac{r_{i+1}-r_{i}}{r_{i+1}+r_{i}}
        \]
        and uncertainties as
        \[
        HR\_err_{i} = 2\frac{\sqrt{\left({r_{i}*r\_err_{i+1}}\right)^2+\left({r_{i+1}*r\_err_{i}}\right)^2}}{\left(r_{i}+r_{i+1}\right)^2},
        \]
        where $r_{i}$ is the \ero count rate of energy band $i$ and $r\_err_{i}$ is the corresponding uncertainty. We used the values given in the forced photometry catalogue \citep{2024A&A...682A..34M}.
        
\begin{figure}
                \centering
                \resizebox{\hsize}{!}{\includegraphics{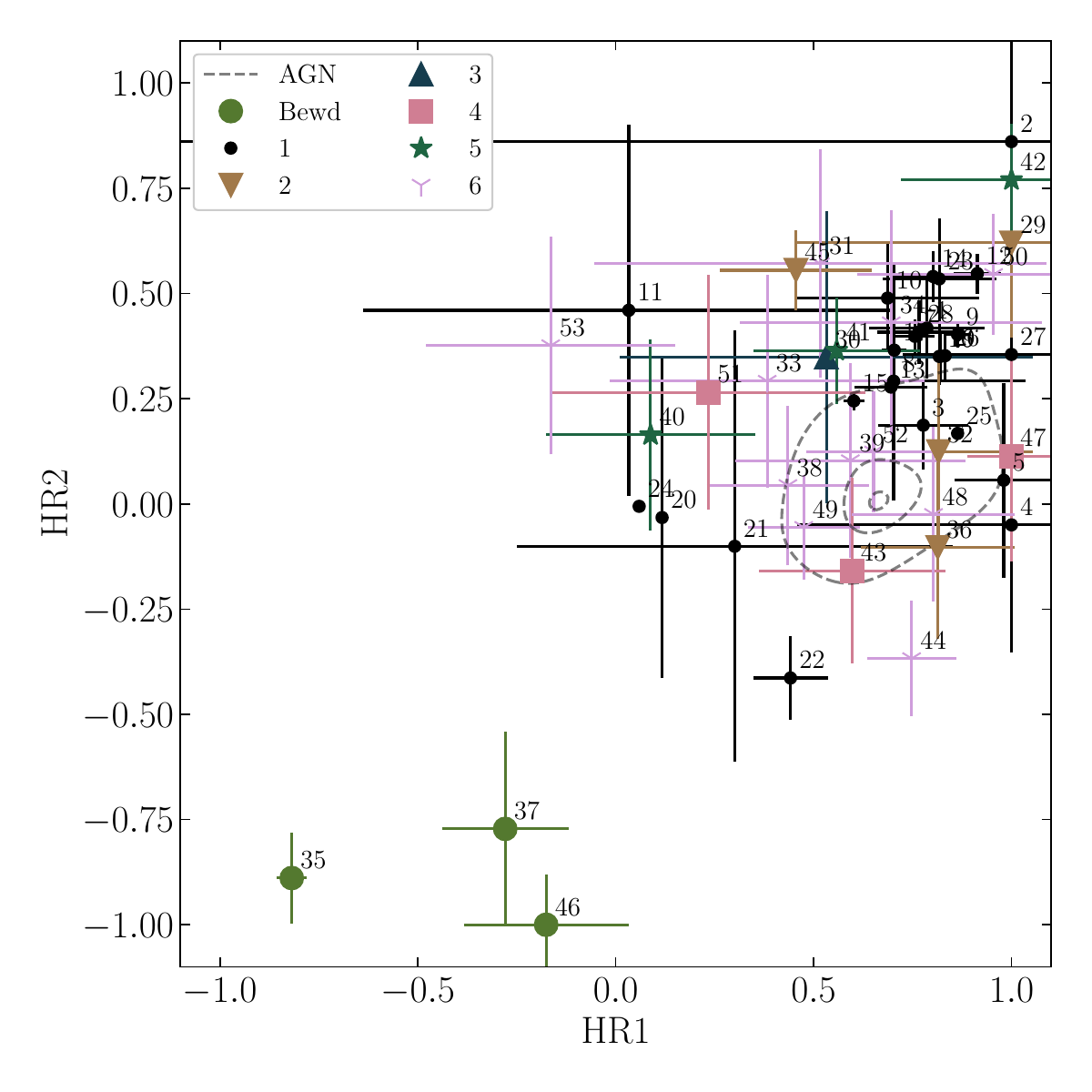}}
                \caption{Hardness ratio diagram of \ero detections of HMXBs and HMXB candidates compared to contours showing where the eRASS counterparts of AGN in \citet{2013ApJ...775...92K} with maximum separations of 2" lie, for which 1, 2 and 3 $\sigma$ contours of the entire sample are plotted. The large error bars of fainter sources make a spectral identification through HR alone impossible. We identify objects \#35, \#37, and \#46 as Be/WD candidates with spectra fit best by absorbed blackbody spectra (see Sect. \ref{sec:bewd}).}
                \label{fig:eRO_HR}
        \end{figure}

        Figure\,\ref{fig:eRO_HR} shows the distribution of objects in our catalogue compared to \ero counterparts of known AGN. To obtain a highly clean AGN sample, we matched objects in \citet{2013ApJ...775...92K} with known redshifts and of classes QSO-Aa or QSO-Ba with the eRASS\,1 catalogue in the LMC region and maximum separations of 2", minimising chance-coincidence matches. This results in a list of 61 AGN, which is comparable to the number of objects in our catalogue. One can clearly see two distributions, but the large overlap between the two distributions and the large uncertainties prevent a clear discrimination between the two groups.
        
        Three objects that clearly stand out are located at the bottom left of the HR diagrams, which correspond to the softest sources observed. Those three objects are the Be/WD candidates, well explained by soft blackbody emitters with no detection above 2\,keV (see Sect. \ref{sec:bewd}).
        
        \subsection{Filtering of HMXB candidates}
        \label{sec:flags}
        
        Candidate systems presented in this catalogue were found by matching the positions of early-type LMC stars with positions of detected X-ray sources. Despite cleaning the X-ray catalogue for known fore- and background sources (see Table\,\ref{tab:screening_cats}), the high source density in the LMC paired with the size of typical X-ray error circles in the eRASS1 catalogue bears the risk of a large number of chance-coincidence matches. This would lead to a high fraction of mis-identifications without adequate filtering.
        
        To assess the credibility of candidates, we used a combination of various properties in multiple wavelengths. We then assigned positive and negative flags and divide objects into different classes of confidence, applying an approach similar to \citet{2016A&A...586A..81H}. Table\,\ref{tab:flags} lists all flags we used to rate candidates. Table\,\ref{tab:conf_classes} assigns confidence classes according to the presence of these flags and shows the number of objects in each class.
        \begin{table} 
                \centering
                \caption{Identification flags.} 
                \label{tab:flags} 
                \begin{tabular}{ll} 
                        \hline\hline\noalign{\smallskip}
                        Flag & Description\\
                        \noalign{\smallskip}\hline\noalign{\smallskip}
                        ps & X-ray variability indicating spin period \\
                        xrb & other secure HMXB identification\\
                        & (e.g. BH binary) \\
                        po & optical variability indicating orbital period \\
                        os & orbital solution exists \\
                        ox & assuming X-ray period as orbital period \\
                        oo & assuming optical period as orbital period \\
                        oxo & periods from X-ray and optical are consistent \\
                        em & Balmer (\Halpha) emission detected \\
                        sx & IR excess found from SED fit \\
                        ix & IR excess found in IR colours \\
                        xvs & X-ray variability from \ero larger than 100$^{(a)}$ \\
                        xvl & long term variability larger than 30$^{(a)}$ \\
                        xs & typical X-ray spectrum\\
                        & (well described by power-law (PL) with index <1.3) \\
                        xs: & X-ray spectrum well fit with PL index \\
                        & frozen to 1$^{(b)}$ \\
                        wd & X-ray spectrum typical for Be/WD XRB \\
                        oi & good optical match \\
                        \noalign{\smallskip}\hline\noalign{\smallskip}
                        qc & match with a catalogue contradicting\\
                        & HMXB nature \\
                        nv & no counterpart in VMC selection found \\
                        nm & no counterpart in MCPS selection found \\
                        ns & SED fit contradicting OB classification \\
                        \noalign{\smallskip}\hline
                \end{tabular} 
                \tablefoot{
            Flags indicating either typical behaviour of a HMXB (top part) or of contamination to the catalogue caused by chance-coincidence (bottom part). If not otherwise mentioned, flags with a colon indicate an uncertain property or a weaker version of the original flag.
                        \tablefoottext{a}{Using $var$ as described in Sect.\,\ref{sec:MAV} as a measure. The analysis of short- and long-term variability in comparison to AGNs is described in Sect.\,\ref{sec:var}.}
                        \tablefoottext{b}{For faint sources, a high local absorption can lead to similar spectra as harder PL indices. A large amount of absorbing circumstellar matter is often observed in HMXBs.}
                }
        \end{table}
        
        \begin{table} 
                \centering
                \caption{Confidence classes.} 
                \label{tab:conf_classes} 
                \begin{tabular}{llr} 
                        \hline\hline\noalign{\smallskip}
                        Class & Flags & \# of sources\\
                        \noalign{\smallskip}\hline\noalign{\smallskip}
                        I & ps $\cup$ ps: $\cup$ xrb & 28 \\
                        II & (xs $\cup$ wd) $\cap$ em & 7 \\
                        III & (xs $\cup$ wd) $\cap$ (sx $\cup$ ix) & 1 \\
                        IV & po & 3 \\
                        V & xvl $\cup$ xvs & 3 \\
                        VI & qc $\cup$ nv $\cup$ nm $\cup$ ns $\cup$ oi & 11 \\
                        \noalign{\smallskip}\hline
                \end{tabular}
        \tablefoot{Confidence classes using the flags of Table\,\ref{tab:flags} as criteria. Class VI includes all objects that either have at least one negative flag or do not fall into any other category due to positive flags.}
        \end{table}
        
        \subsection{Rejected HMXB candidates}
        The improved sensitivity and positional accuracy of \ero allow us to re-evaluate HMXB candidates from older surveys, such as those detected by \ROSAT. In particular, we can now test whether the X-ray source positions align with early-type stars, which are the optical counterparts expected for HMXBs. If an X-ray source detected by \ero does not match the position of an early-type star, we can confidently reject it as an HMXB candidate, assuming that the two X-ray detections correspond to the same source. In our study, we identified five HMXB candidates, originally reported by \citet{1999A&AS..139..277H} and listed in the catalogue of HMXBs in the Magellanic Clouds of \citet{2005A&A...442.1135L}, which can now be ruled out with high confidence as HMXB candidates. This demonstrates the capability of \ero to refine and update earlier catalogues by improved positional uncertainties. In Table\,\ref{tab:rejectHMXB}, a summary of the five X-ray sources can be found.

        \begin{table*} 
                \centering
                \caption{Rejected HMXB candidates.} 
                \label{tab:rejectHMXB} 
                \begin{tabular}{llllll} 
                        \hline\hline\noalign{\smallskip}
                        \ROSAT      & \ROSAT    & \ero            & \ero      & dist.     & [HP99] \\
                        name        & perr      & name            & perr      & ROS-eRO   &        \\
                        RX\,J       & (\arcsec) & 1eRASS\,J       & (\arcsec) & (\arcsec) &        \\
                        \noalign{\smallskip}\hline\noalign{\smallskip}
                        0527.1$-$7005 &  11.9   & 052706.8$-$700459 &  1.74   &  5.7    &  1078 \\
                        0530.7$-$6606 &  11.4   & 053048.6$-$660600 &  3.22   &  16.3   &   247 \\
                        0532.3$-$7107 &  24.3   & 053218.4$-$710746 &   1.4   &  25.1   &  1238 \\
                        0535.8$-$6530 &  13.0   & 053554.5$-$653038 &   5.0   &  9.4    &   131 \\
                        0543.9$-$6539 &   8.6   & 054358.3$-$653952 &   1.2   &  1.9    &   148 \\
                        \noalign{\smallskip}\hline
                \end{tabular} 
                \tablefoot{
                        HP99: \citet{1999A&AS..139..277H}
                }
        \end{table*}
        
        \section{The X-ray luminosity function}
        \label{sec:XLF}
        To study the luminosity distribution of HMXBs in the LMC and to compare the population of that in the MW, SMC, and other galaxies in the local Universe with a large variety in underlying conditions, we extracted the X-ray luminosity function (XLF) of HMXBs. As was shown by \citet{2003ChJAS...3..257G}, there is a close correlation between the recent SFR and the HMXB XLF. Close to the SEP, \ero allows us to probe the LMC HMXB XLF down to luminosities of $\approx10^{34}$\,erg s$^{-1}$ at a completeness of 10\% (see Fig.\,\ref{fig:senscurve}). However, due to \ero's scanning scheme, the sensitivity is highly dependent on the position within the LMC (see Figs.\,\ref{fig:RGB} and \ref{fig:senscurve}).
        In order to correct our XLF for completeness, we binned our sources by luminosity and \ero sky tile \citep[see Sect. 4.1 in][]{2024A&A...682A..34M}. We modelled the number of sources in each luminosity bin by assuming a homogeneous source distribution throughout the LMC, that the observed number of objects is a Poisson variable, and that the detection probability equals the average sensitivity in the luminosity bin.
        
        The luminosities we obtained by scaling the eRASS1 count rates assuming an absorbed power-law model with N$_H=6$\hcm{20} and $\Gamma=$1. For comparison with other works in the literature, we also calculated the expected flux in the 2$-$12\,keV range and created the corresponding XLF. Note that this should be viewed as an extrapolation and interpreted with caution. We only include objects from confidence classes 1$-$5 to minimise the impact of possible source contamination. Additionally, we exclude Swift\,J045558.9$-$702001, RX\,J0501.6$-$7034, 3XMM\,J051259.8$-$682640, RX\,J0520.5$-$6932, RX\,J0535.0$-$6700, RX\,J0544.1$-$7100, 4XMM\,J053449.0$-$694338, 1eRASS\,J054422.3$-$672729, and 1eRASS\,J055849.9$-$675220 from the analysis, because they would not have been included in the catalogue using the eRASS1 1B catalogue alone (see Table \ref{tab:MasterTable_known} for details). Furthermore, eRASSU\,J050213.8$-$674620, 1eRASS\,J050705.9$-$652149 and 1eRASS\,J054242.7$-$672752 were excluded due to their likely Be/WD nature, which is not typically included in the study of HMXB populations. We extracted a second luminosity function using only HMXBs of high confidence (confidence classes 1 and 2).
        
\begin{figure*}
                \centering
                \resizebox{0.495\hsize}{!}{\includegraphics{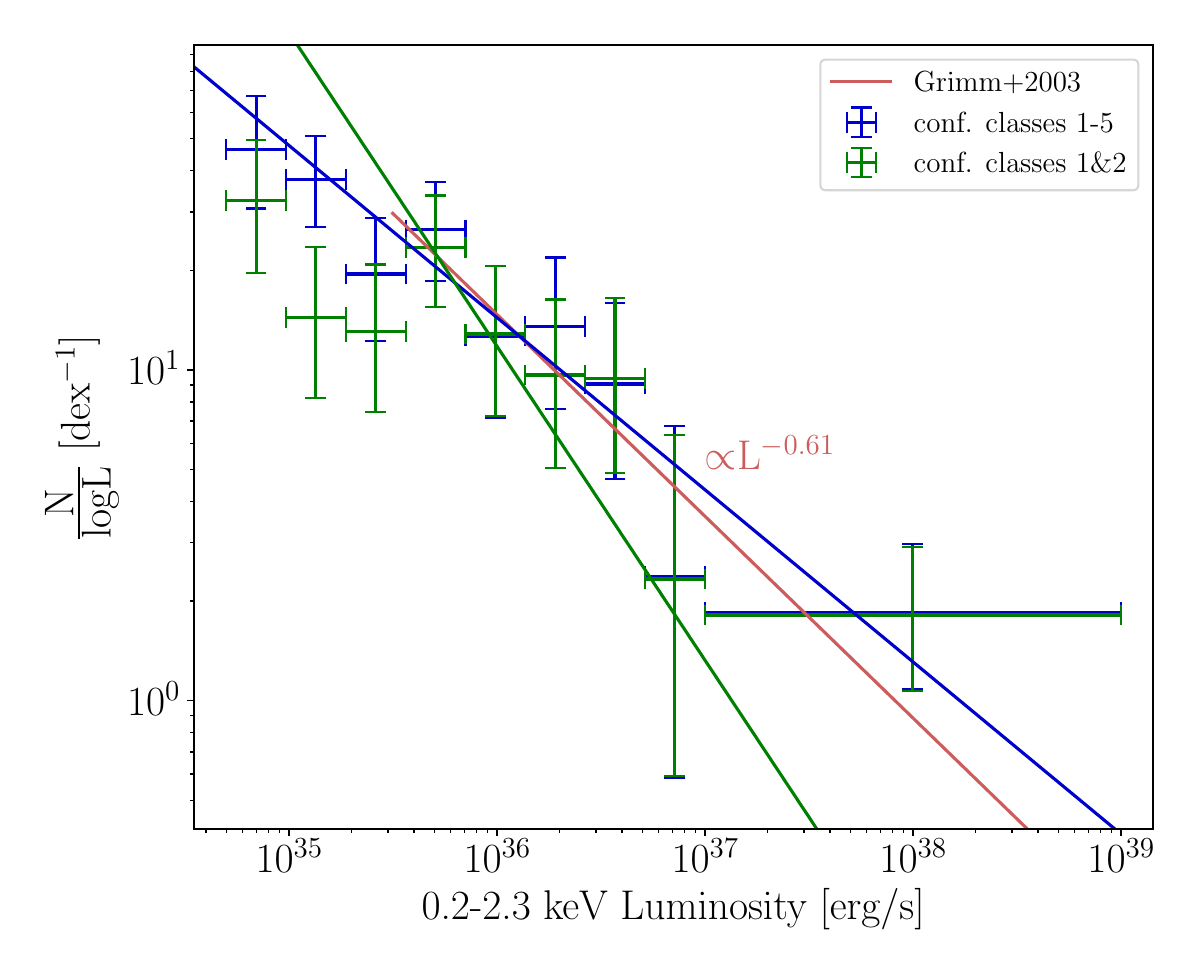}}
                \resizebox{0.495\hsize}{!}{\includegraphics{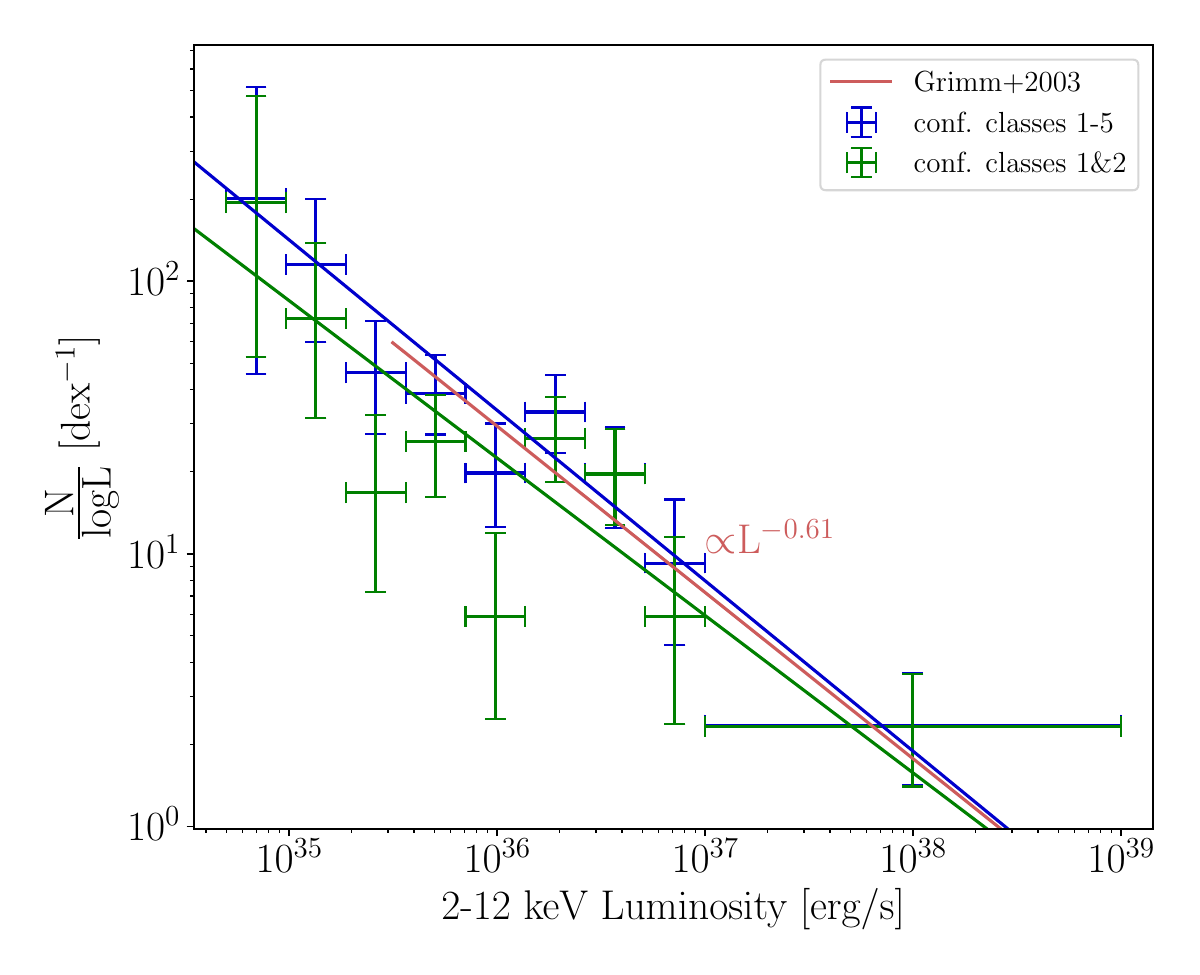}}
                \caption{X-ray luminosity functions of HMXB detected during eRASS1. Green and blue points mark the expected number of sources per dex. A Poisson distribution of measured numbers is assumed. The blue and green lines mark the best-fit power-law correlation for the luminosity functions of confidence classes 1$-$5 and 1\&2, respectively. The red line marks the universal power-law correlation found by \citet{2003ChJAS...3..257G}. Left: HMXB XLF in the detection band of the eRASS1 catalogue scaling catalogue count rates assuming an absorbed power-law model with N$_{H}=6$\hcm{20} and $\Gamma=1$. Right: Extrapolation to $2-12$\,keV range.}
                \label{fig:LumFunc}
        \end{figure*}

\begin{figure}
                \centering
                \resizebox{\hsize}{!}{\includegraphics{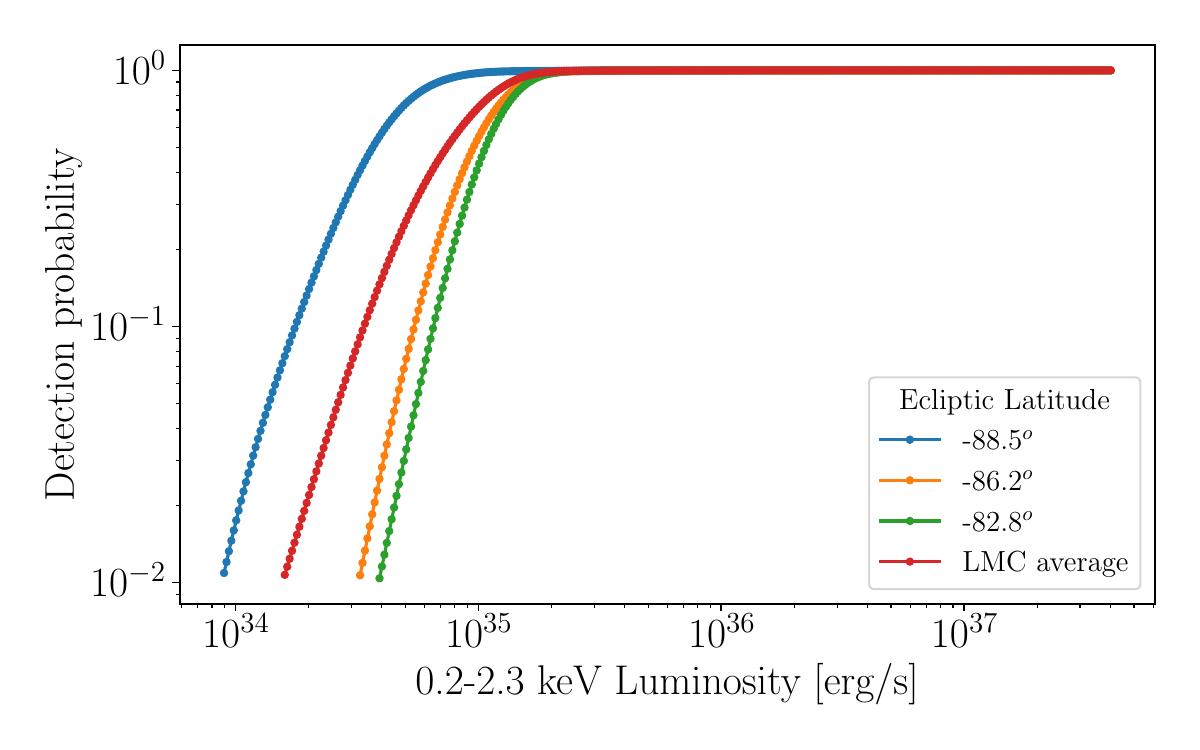}}
                \caption{Sensitivity function for eRASS\,1 detection assuming a spectra defined by an absorbed power-law with power-law index 1 for three positions within the LMC (sky tile closest to SEP, sky tile in the centre of the LMC and sky tile most distant from the SEP) as well as the LMC average sensitivity.}
                \label{fig:senscurve}
        \end{figure}

        Figure\,\ref{fig:LumFunc} shows the results of the XLF fit. Assuming a power-law distribution of luminosities (0.2$-$12\,keV), we followed \citet{1970ApJ...162..405C} to fit an unbinned power-law function and find power-law indices of $0.34^{+0.14}_{-2.94}$ and $0.55^{+0.53}_{-0.11}$, using confidence classes 1\&2 and 1$-$5, respectively. Those results are similar to what \citet{2003ChJAS...3..257G} find as a universal value for nearby galaxies (note that the power-law indices we find are lower by 1, due to the change in the y axis). The small difference might be caused by including objects down to luminosities approximately one order of magnitude lower than \citet{2003ChJAS...3..257G} and a possible change in power-law index at lower luminosities, as was already suggested by \citet{2005MNRAS.362..879S}.
        
        \section{Discussion}
        \label{sec:discussion}
        \subsection{X-ray luminosity function}
        
        To test the correlation between the luminosity function and the LMC SFR, we needed the luminosity function normalisation, which we obtained from an additional fit. For this purpose, we fitted a straight line in the log-log plane of figure \ref{fig:LumFunc} to the binned luminosity function, freezing the slope of the line to the power-law indices we obtain with the unbinned fit. Using this fit, we could then estimate the expected number of objects with luminosities >2\ergs{38}, which is the value used by \citet{2003ChJAS...3..257G} to relate SFR and the HMXB population of galaxies, using the Antennae Galaxies as a reference. By then scaling this number by a factor of $SRF_{Antennae}/N(L>2$\ergs{38}$)_{Antennae}$ we get an estimate for the SFR of the LMC given its HMXB population, which we calculate as $(0.21^{+0.07}_{-0.05})$\,M$_{\odot}$yr$^{-1}$ and $(0.22^{+0.06}_{-0.07})$\,M$_{\odot}$yr$^{-1}$ for confidence classes 1\&2 and 1$-$5, respectively. Both values match the 0.2,M$_{\odot}$yr$^{-1}$ found by \citet{2009AJ....138.1243H} using photometric MCPS data. Estimating the SFR from the HMXB luminosity function is inherently limited by the stochastic variability of the instantaneous HMXB sample size and the fact that the HMXB population reflects the SFR from several tens of millions of years in the past. This is further supported by \citet{2003ChJAS...3..257G}, who found that the luminosity functions of the galaxies in their sample vary by a factor of $\sim$2 when scaled with the SFR. Given these limitations, this analysis serves as an initial step towards linking HMXBs to the LMC SFH, a connection that we shall explore in detail in our next work.
        
        \subsection{Subpopulations of HMXBs}
        \label{sec:individuals}
        In this section, we discuss our analysis of (groups of) objects outstanding in their multi-wavelength properties and due to follow-up observations we conducted. \ero timing and spectral analysis for all objects are based on Sects. \ref{sec:time-analysis} and \ref{sec:spec-analysis}, respectively; optical spectra use the methods described in \ref{sec:halpha-analysis}.
        \subsubsection{BeXRBs detected in outburst during eRASS1}
        \textbf{RX\,J0529.8$-$6556} is a BeXRB pulsar that was first discovered in ROSAT data \citep{1997A&A...318..490H} with a pulse period of 69.5\,s and at a flux of 3.8\ergcm{-12} (0.1$-$2.4\,keV). During a deep \xmm observation \citep{2003A&A...406..471H} the source was found at a flux of 2.2\ergcm{-13} (0.2$-$10\,keV). \ero detected RX\,J0529.8$-$6556 in outburst \citep{2020ATel13828....1H} at a flux of 2.6\ergcm{-11} (0.3$-$8\,keV) and \citet{2021MNRAS.503.6187T} conducted targeted \nicer observations to characterise the outburst. Additionally, they studied the optical behaviour using \salt spectroscopy and the 10-year \ogle V- and I-band light curves. The \ero spectrum is fit best with an absorbed combination of a power-law and a blackbody (\texttt{tbabs*(pow+bbodyrad)} in \texttt{Xspec}). No additional absorption component to account for absorption in the LMC is necessary (note that we always use a component to account for MW foreground absorption as described in Sect.\,\ref{sec:spec-analysis}). For the power law, we find an index of 0.98$\pm$0.04. The blackbody component shows a temperature of 73$^{+26}_{-21}$\,eV and a normalisation corresponding to an emission region of 107$^{+199}_{-57}$\,km at LMC distance. The power-law component accounts for 99.1\,\% of the total flux of (24.3$\pm$0.6)\ergcm{-13} between 0.2 and 5\,keV. During the outburst (end of eRASS1, beginning of eRASS2) we find a flux of (12.35$\pm$0.16)\ergcm{-12} (0.2$-$5\,keV), corresponding to an absorption-corrected luminosity of (40.7$\pm$0.5)\ergs{35}. During all other eRASSs, the source is fainter by a factor of $\sim$100.
        
        \textbf{eRASSU\,J050810.4$-$660653} was discovered in a bright state during eRASS1 \citep{2020ATel13609....1H} and with pointed \xmm observations \citet{2021ATel15133....1H, 2023A&A...671A..90H} confirmed it as an HMXB pulsar with a pulse period of 40.6\,s. \citet{2022MNRAS.514.4018S} then later observed eRASSU\,J050810.4$-$660653 during an episode of enhance X-ray activity with \srg/\artxc, \swift and \nus, where an orbital period of $\sim$38\,d was found. We find the source in a bright state during all eRASSs. In the eRASS1 catalogue, the source is listed under the name 1eRASS\,J050810.1$-$660653. The merged \ero spectrum is best fit with an absorbed power-law with power-law index 0.87$^{+0.06}_{-0.05}$ and at a flux of (3.61$\pm$0.11)\ergcm{-12} (0.2$-$5\,keV). The spectrum is best fit with an additional absorption component accounting for gas in the LMC, with an absorption column of $N_{H,LMC}=$(9.9$^{+2.2}_{-1.8}$)\hcm{20}. Assuming standard LMC distance the luminosity corrected for absorption is (11.8$\pm$0.3)\ergs{35}. During eRASS1 we find a flux of (15.4$\pm$0.7)\ergcm{-13} (0.2$-$5\,keV), corresponding to an absorption-corrected luminosity of (50.3$\pm$2.3)\ergs{34}. Across all eRASSs, we find a $var$ of $\sim$34.
        
        \textbf{eRASSU\,J052914.9$-$662446} was discovered in a bright state during eRASS1 and confirmed as a BeXRB pulsar with a spin period of 1412\,s and an orbital period of $\sim$151\,days \citep{2020ATel13610....1M, 2020ATel13650....1M}. \citet{2023A&A...669A..30M} studied the broadband spectral and timing behaviour using \ero, \swift and \nus data in X-rays and \ogle and \salt data in optical. In the eRASS1 catalogue, the source is listed under the name 1eRASS\,J052914.3$-$662445. Across all eRASSs, we find variability of a factor $\leq$4. The merged \ero is fit best by an absorbed power-law spectrum without an additional absorption component for the LMC. We find a power-law index of 0.30$^{+0.08}_{-0.07}$ and a flux of (6.2$\pm$0.4)\ergcm{-13} (0.2$-$5\,keV), which corresponds to an absorption-corrected luminosity of (19.0$^{+1.1}_{-1.0}$)\ergs{34}.
        
        \textbf{Swift\,J053041.9$-$665426} was discovered during an outburst with a luminosity of 9.7\ergs{36} in November 2011 during a \swift observation of the LMC nova 2009B \citep{2011ATel.3747....1S}. The re-detection with higher positional accuracy during \swift follow-up observations and spectral properties of the optical counterpart led to the classification as a BeXRB candidate \citep{2011ATel.3753....1S}. \citet{2011ATel.3751....1C} reported variability of the optical counterpart. \citet{2013A&A...558A..74V} then confirmed Swift\,J053041.9$-$665426 as a BeXRB, detecting X-ray pulsations at $\sim$28.78\,s and a luminosity of 5.53\ergs{35}. The source was detected in outburst at the end of eRASS1, eRASS2, and eRASS3. In each successive eRASS, the outburst appeared at progressively later scan positions, consistent with an orbital period of $\sim$195\,d. This is in line with orbital variability \citep{2013A&A...558A..74V}. The merged \ero spectrum can be best fit by an absorbed power-law spectrum without an additional absorption component for the LMC. We find a power-law index of 0.63$\pm$0.05 and a flux of (11.6$\pm$0.5)\ergcm{-13} (0.2$-$0.5\,keV), which corresponds to an absorption-corrected luminosity of (35.8$\pm$1.4)\ergs{34}. During the outbursts, the source displays a luminosity of (13.6$\pm$0.4)\ergs{35}, in quiescence, the source is fainter by a factor of $\sim$100.
        
\begin{figure}
                \centering
                \resizebox{\hsize}{!}{\includegraphics{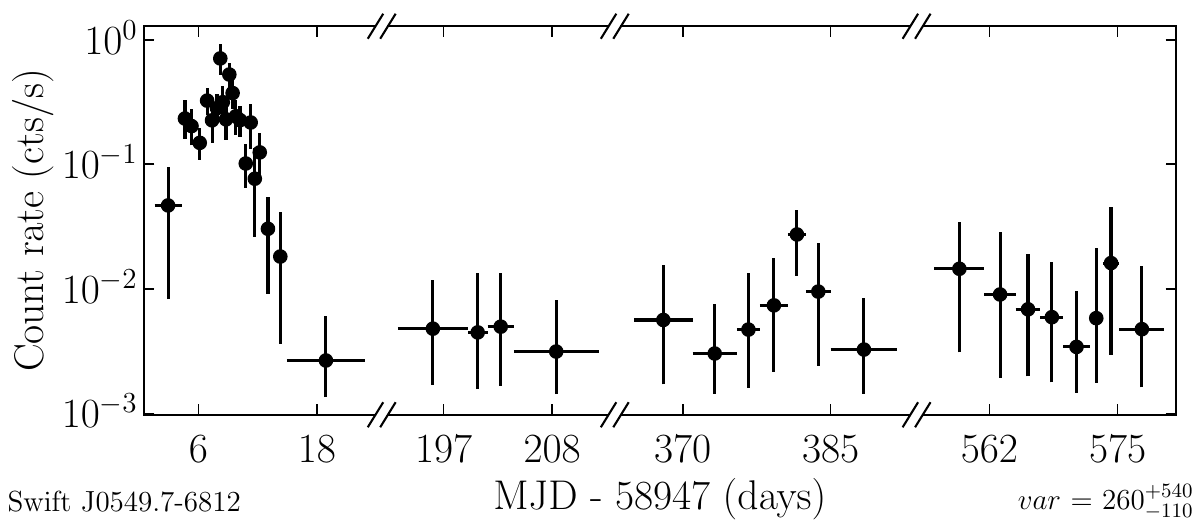}}
                \caption{\ero light curve for the BeXRB Swift\,J0549.7$-$6812. During the outburst in eRASS1, the flux rises to a maximum of $\sim$3.5\ergs{35}. For readability, white spaces between the individual eRASSs are removed. The light curve is rebinned to have a minimum number of 10 net source counts per time bin as described in Sect. \ref{sec:LCs}. The variability $var$ (found in the bottom right) is calculated as described in Sect.\,\ref{sec:MAV}.}
                \label{fig:LC_J0549}
        \end{figure}

        \textbf{Swift\,J0549.7$-$6812} is a BeXRB system that was discovered serendipitously in 2013 with the hard X-ray transient monitor of the \swift Burst Alert Telescope \citep[BAT,][]{2013ATel.5286....1K}. \citet{2013ATel.5293....1K} found the source to be consistent with the position of the faint \ROSAT\, object 1RXS\,J055007.0$-$681451, which was detected at a luminosity of 6.4\ergs{34} (assuming a typical spectral power-law index of 1). \citet{2013ATel.5309....1K} later conducted targeted \swift\, observations that revealed X-ray pulsations with a 6.2\,s period, and associated the X-ray source with an early-type star, identifying it as an HMXB. They reported that during a deep \cxo\, observation (OBSID 3850; 67.5 ks on 2 October 2003), no X-ray source consistent with Swift\,J0549.7$-$6812 was detected, suggesting the source must have been fainter by a factor of at least $\sim$100 compared to the \ROSAT\, observations. \citet{2023MNRAS.524.3263C} reported that the 2013 \swift\, observations coincided with an outburst, during which the source showed a peak luminosity of 8.1\ergs{37} (0.3$-$10\,keV), and that long-term \swift/BAT observations over 10\,yr have not shown any additional X-ray outbursts. They also conducted \salt/RSS observations of the optical counterpart, which revealed \Halpha and \Hbeta in emission, confirming the BeXRB nature of the source. Swift\,J0549.7$-$6812 was detected in an outburst in eRASS1 at an average flux of (3.9$\pm$0.4)\ergcm{-13} (0.2$-$5\,keV), which corresponds to an absorption-corrected luminosity of (1.19$\pm$0.11)\ergs{35}. At the peak of the outburst during eRASS1, the source reached a luminosity of $\sim$7.0\ergs{35} with the event lasting for $\sim$17 days. These characteristics are consistent with a type I outburst. The entire \ero\, light curve can be seen in Fig.\,\ref{fig:LC_J0549}. During all other eRASSs, the source is fainter by a factor of at least $\sim$10 and lies below the background flux. The merged spectrum can be best fit by an absorbed power law without an additional absorption component for the LMC. The best-fit power-law index is 0.56$\pm$0.20.
        
        \textbf{eRASSU\,J050213.8-674620} was first detected during an outburst in eRASS1. As a Be/WD candidate, the source is discussed in more detail in Sect.\,\ref{sec:bewd}.
        
        \subsubsection{A population of SgXRBs}
        \label{sec:SgXRB}
        SgXRBs consist of a NS accreting from an OB SG, primarily through stellar winds or Roche-lobe overflow. They typically show persistent or moderately variable X-ray emission \citep[$10^{35}$–$10^{38}$ erg s$^{-1}$;][]{2017SSRv..212...59M} and short orbital periods of the order of a few days \citep{2015A&ARv..23....2W}. Many SgXRBs display pulsations due to the spin of a magnetised NS, as well as eclipses and orbital modulation driven by interactions with the dense stellar wind. Some also exhibit superorbital variability.
        
        \textbf{RX\,J0532.5$-$6551} was first detected by \citet{1995A&A...303L..49H}. Based on the optical classification of the proposed counterpart, they classified it as an SgXRB. During the \ROSAT observations between 1990 and 1994, they observed a change in flux by a factor of $\sim$50 with an average X-ray luminosity of a few \oergs{34} and a maximum luminosity of 1.9\ergs{35} (0.1$-$2.4\,keV). Later, \citet{2002A&A...385..517N} and \citet{2001PASP..113.1130J} independently confirmed the SgXRB classification, and \citet{2002A&A...385..517N} further confirmed the system to be the first wind-fed SgXRB system known in the LMC. The merged \ero spectrum can best be fit by an absorbed power-law without an additional component to account for local absorption. We find a power-law index of 0.58$\pm$0.05 and a flux of (8.5$\pm$0.3)\ergcm{-13} (0.2$-$5\,keV), corresponding to an absorption-corrected luminosity of (2.62$^{+0.09}_{-0.10})$\ergs{35}.
        
        \textbf{LMC\,X$-$1} was the first extra-galactic HMXB to be detected \citep{1969ApJ...155L.143M}. It consists of an O-type SG, and a BH with a mass of 10.9\,M$_\odot$ \citep{2009ApJ...697..573O}. LMC\,X$-$1 is characterised by soft X-ray spectrum at a relatively constant luminosity of $\sim$2\ergs{38}. We fit the merged \ero spectrum with an absorbed disc-blackbody spectrum with an additional absorption component to account for local absorption (\texttt{tbabs}*\texttt{tbvarabs}*\texttt{diskbb} in \texttt{Xspec}). The best-fit parameters are N$\unter{H}$=(4.91$^{+0.18}_{-0.17}$)\hcm{21} and T$\unter{in}$=(800$^{+15}_{-14}$)\,eV at a flux of (5.22$\pm$0.07)\ergcm{-10} (0.2$-$5\,keV). This corresponds to an absorption-corrected luminosity of (2.42$\pm$0.03)\ergs{38}. Similar to what was observed in the past, LMC\,X$-$1 showed low variability across all eRASS, with a factor of $\sim$4.
        
        \textbf{LMC\,X$-$3} is another binary system of an SG filling its Roche lobe and a BH with a mass of 6.98\,M$_\odot$ \citep{2001A&A...365L.273S,2014ApJ...794..154O}. We fitted the merged \ero spectrum with an absorbed combination of a power-law and a disc-blackbody model (\texttt{tbabs}*(\texttt{pow}+\texttt{diskbb}) in \texttt{Xspec}). We find a power-law index of (1.82$^{+0.11}_{-0.08}$) and a temperature of $T\unter{eff}$=(0.79$\pm$0.03)\,keV at an average flux of (7.09$\pm$0.07)\ergcm{-10}, which corresponds to an absorption-corrected luminosity of (24.94$\pm$0.24)\ergs{37}.
        
        \textbf{LMC\,X$-$4} is an eclipsing high-mass binary pulsar. An orbital solution of the binary shows a mass of 18\,M$_\odot$ for the O-type SG and 1.57\,M$_\odot$ for the NS \citep[][and references therein]{2015A&A...577A.130F}. The merged \ero spectrum is best fit by an absorbed combination of a blackbody, a hard and a soft power-law (\texttt{tbabs}*(\texttt{bbodyrad}+\texttt{pow}+\texttt{pow}) in \texttt{Xspec}) without an additional local absorption component. For the blackbody component we find $T\unter{eff}$=(43$^{+4}_{-3}$)\,eV. For the soft and hard power-law components, we find $\alpha=2.17\pm0.05$ and $\alpha=-$1.7$^{+0.4}_{-0.5}$, respectively. The average flux during all eRASSs is (3.06$\pm0.06$)\ergcm{-11} (0.2$-$5\,keV), which corresponds to an absorption-corrected luminosity of (1.83$^{+0.10}_{-0.09}$)\ergs{37}.
        
        \textbf{XMMU\,J053108.3$-$690923}, \textbf{XMMU\,J053320.8$-$684122} and \textbf{1eRASS\,J054422.3$-$672729} are (candidate) systems of the sub-class of SFXTs and are discussed in Sect.\,\ref{sec:SFXT}.
        
        \subsubsection{Supergiant fast X-ray transients}
        \label{sec:SFXT}
        Supergiant fast X-ray transients are a subtype of SgXRBs characterised by a faint nature during the majority of the time, interrupted by sporadic strong flares that only last for a few hours and during which the flux typically rises by a factor of a hundred or more. Often, those flares appear grouped during periods of higher activity and are accompanied by spectral changes. The exact mechanics leading to this behaviour are still a matter of debate. Plausible explanations include clumpiness of the stellar wind \citep{2005A&A...441L...1I}, magnetic or centrifugal gates due to the NS rotation and strong magnetic fields \citep{2008ApJ...683.1031B}, as well as the onset of a long-standing settling accretion regime \citep{2012MNRAS.420..216S}.
        
        In eRASS1, we detect both SFXTs in the LMC known from the literature. Additionally, find one new object that displays a typical SFXT behaviour, 1eRASS\,J054422.3$-$672729, showing an increase in flux by a factor of $\sim$100 and $\sim$50 during two flares, respectively. We propose that this should be seen as a lower limit for the variability of this system, because it is likely that the system's brightest state was not observed due to \ero's scanning scheme. Typical SFXT dynamical ranges can be up to $\sim10^3-10^5$ between quiescence and outburst \citep{2015JHEAp...7..126R}.
        
        \textbf{XMMU\,J053320.8$-$684122} was first detected by \citet{2012ATel.3993....1S} during a \swift survey and a follow-up observation in 2012, at fluxes of 5.4\ergcm{-13} and 8.1\ergcm{-13}, respectively (Swift\,J053321.3$-$684121). It was later re-detected during observations for the \xmm large survey of the LMC. \citet{2018MNRAS.475..220V} then conducted a detailed study of X-ray data from \xmm and \swift combined with optical spectra obtained with the Fibre-fed Extended Range Optical Spectrograph \citep[FEROS;][]{1999Msngr..95....8K} and photometric data from the Gamma-Ray Burst Optical and Near-Infrared Detector \citep[GROND;][]{2008PASP..120..405G}. They found a mean flux of 8.4\ergcm{-13} (0.3$-$10\,keV) with a flux increase by a factor of $\sim$13 towards the end of the \xmm observation. Given the optical data indicating an SG companion, it is proposed that XMMU\,J053320.8$-$684122 belongs to the class of SFXTs. The \ero light curve shows a $var$ of $\sim$120, and our Bayesian blocks analysis reveals that it stays in a quiescent state most of the time to brighten up by a factor of $\sim$10 for several days during eRASS1, 2 and 4. The spectrum is fit best with an absorbed power-law at a mean flux of (3.5$^{+0.5}_{-0.4}$)\ergcm{-13} (0.2$-$5\,keV) with a power-law index of 0.6$\pm$0.4 and with a high local absorption of (1.1$^{+0.6}_{-0.5}$)\hcm{22}. This corresponds to an absorption-corrected luminosity of 13.3$^{+1.7}_{-1.4}$\ergs{34}.
        
        \textbf{XMMU\,J053108.3$-$690923} was first reported as an SgXRB by \citet{2018MNRAS.475..220V} by using optical spectroscopy for the SG classification of the optical companion. Using \xmm they detected X-ray variability at a period of $\sim$2013\,s, which is very high for typical pulsars in HMXBs. \citet{2021A&A...647A...8M} later verified the high spin period of $\sim$2020\,s with \ero and \xmm observations. During two of the \ero pointed observations, they also observed strong flares, during which the flux increased by a factor of $\sim$15 and $\sim$30, respectively, leading them to classify the system as an SFXT. Across all the eRASSs, we do not detect any significant changes in the luminosity of XMMU\,J053108.3$-$690923. The merged \ero spectrum is fit best by an absorbed power-law with an additional component for LMC absorption. The best-fit parameters are an absorption of $N_{H,LMC}$=(3.3$^{+2.0}_{-1.5})$\hcm{22}, a power-law index of 0.9$^{+0.8}_{-0.7}$ and a flux of (2.2$^{+0.5}_{-0.4}$)\ergcm{-13} (0.2$-$5\,keV), which corresponds to an absorption-corrected luminosity of (1.1$^{+0.8}_{-0.3})$\ergs{35}.
        
        We find \textbf{1eRASS\,J054422.3$-$672729} as a new HMXB candidate in the course of this study. The optical properties of its companion match those of an SG in terms of colour and brightness. Analysis of \ogle I-band data revealed a highly significant 5.25\,d period, which likely is the orbital period of the system. The source is seen in a state of increased flux twice, once during eRASS2 and once during eRASS4, where the flux increased by a factor of $\sim$100 and $\sim$50, respectively, and which each last for $\sim$1\,d. We propose 1eRASS\,J054422.3$-$672729 as a new SFXT. The merged \ero spectrum can be best fit with an absorbed power law, with an additional component to account for LMC absorption. The best-fit parameters are $N_{H,LMC}$=(0.8$^{+0.4}_{-0.3}$)\hcm{22} for the absorption, 2.1$\pm$0.6 for the power-law index and (5.8$^{+1.5}_{-1.2}$)\ergcm{-14} (0.2$-$5\,keV) for the flux, which corresponds to and absorption-corrected luminosity of (4.2$^{+3.3}_{-1.0}$)\ergs{34}. We conducted a series of ten 1200\,s exposure \salt observations between  2023-11-05 and 2023-11-23 to study the optical counterpart of 1eRASS\,J054422.3$-$672729. We specifically studied the equivalent widths and kinematics of the Lyman \Halpha and \Hbeta lines, as well as those of the \HeI lines. The line profiles are best fit by a wide absorption line with a narrow emission line for \Halpha and \Hbeta, and a single absorption line for \HeI. See figure \ref{fig:SFXT_SALT} for the normalised spectra with best-fit functions at the \Halpha line for all observations. Table~\ref{tab:SFXT_SALT} lists the best-fit parameters for the \Hbeta and \HeI lines. We tested whether the velocity shift or the change in equivalent width could be linked to the orbital period derived from \ogle, but found no indication of such a link for any of the three lines. In the case of the \HeI line, this may be due to low S/N.
        
\begin{figure}
                \centering
                \resizebox{\hsize}{!}{\includegraphics{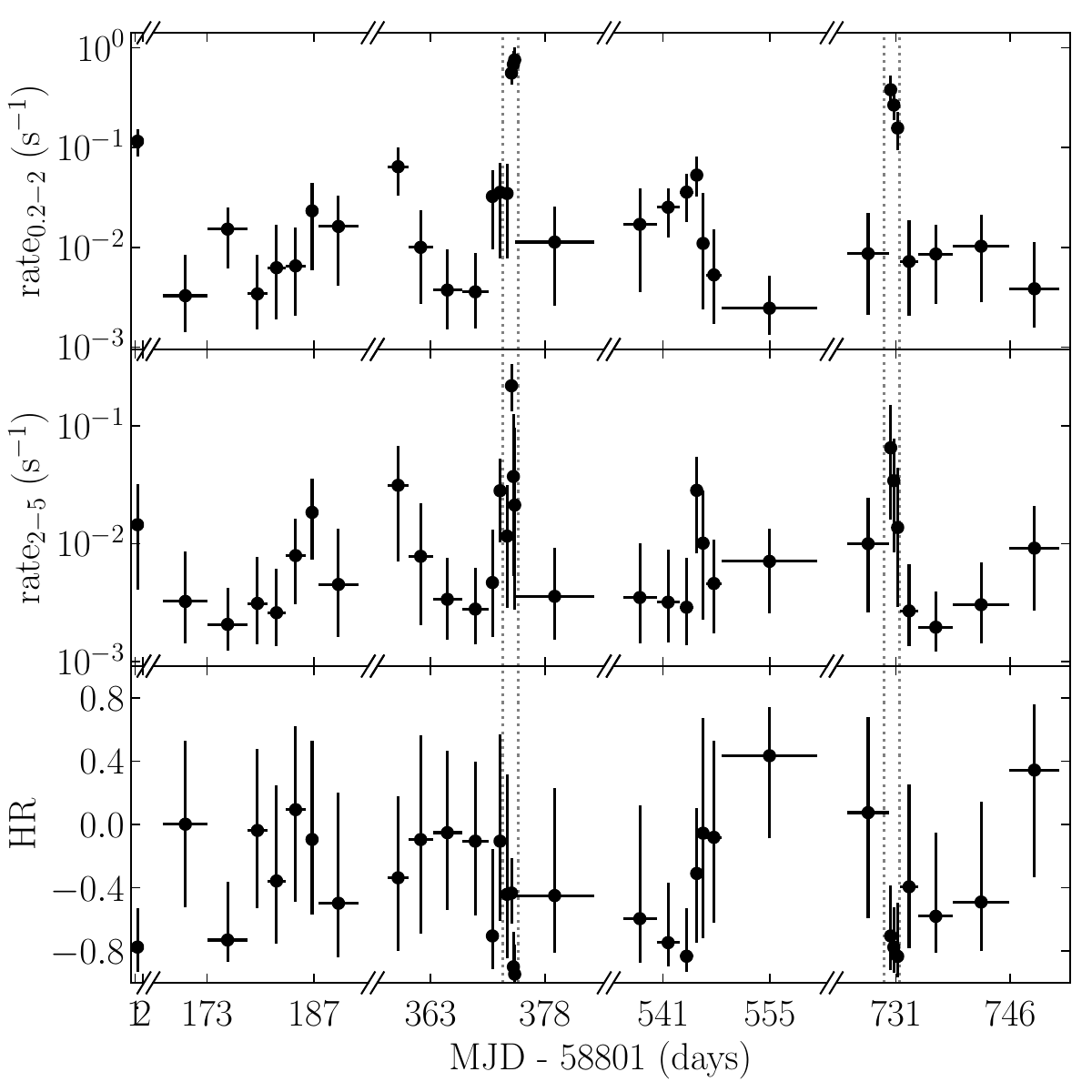}}
                \caption{\ero light curve of 1eRASS\,J054422.3$-$672729 in the energy bands 0.2$-$2.0\,keV (soft; top), 2.0$-$5.0\,keV (hard; middle) and the HR defined from those two bands (bottom). The light curve is binned such that there is a minimum of 1 net count in each band and a minimum of 10 net counts in the sum of the two bands per time bin. The flares at $\sim$373 and $\sim$730 days are marked with dotted grey lines and can be best seen in the soft band. The HR light curve indicates spectral softening at those times. Note that time gaps between eRASSs are cut out for readability.}
                \label{fig:SFXT_HR}
        \end{figure}

\begin{figure}
                \centering
                \resizebox{\hsize}{!}{\includegraphics{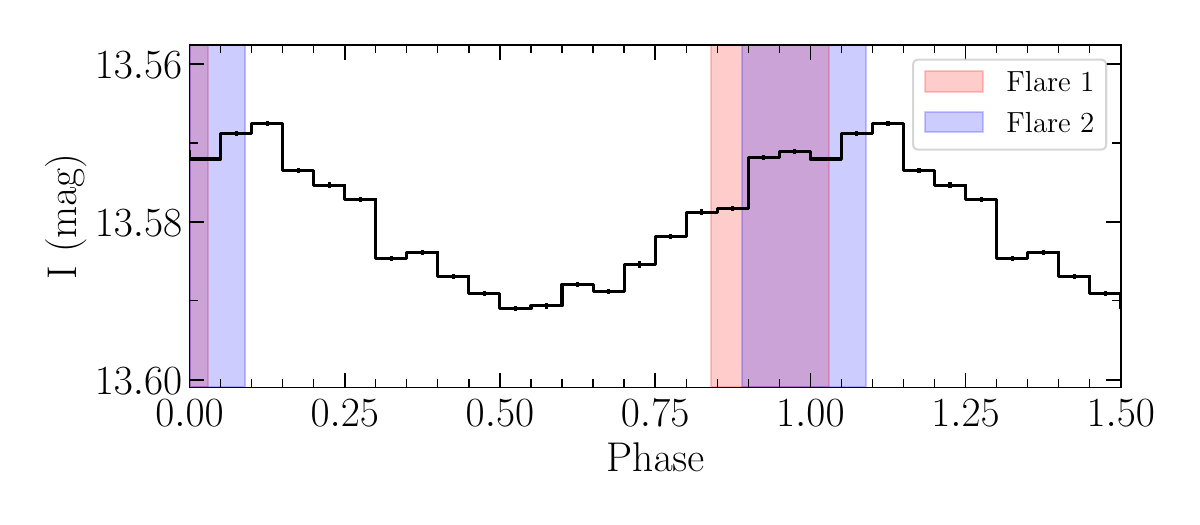}}
                \caption{\ogle phase-folded light curve of 1eRASS\,J054422.3$-$672729 with timing of the flares marked. Flares 1 and 2 refer to the flares at ~373 and ~730 days in Fig.\,\ref{fig:SFXT_HR}, respectively.}
                \label{fig:SFXT_phase}
        \end{figure}

\begin{figure}
                \centering
                \resizebox{\hsize}{!}{\includegraphics{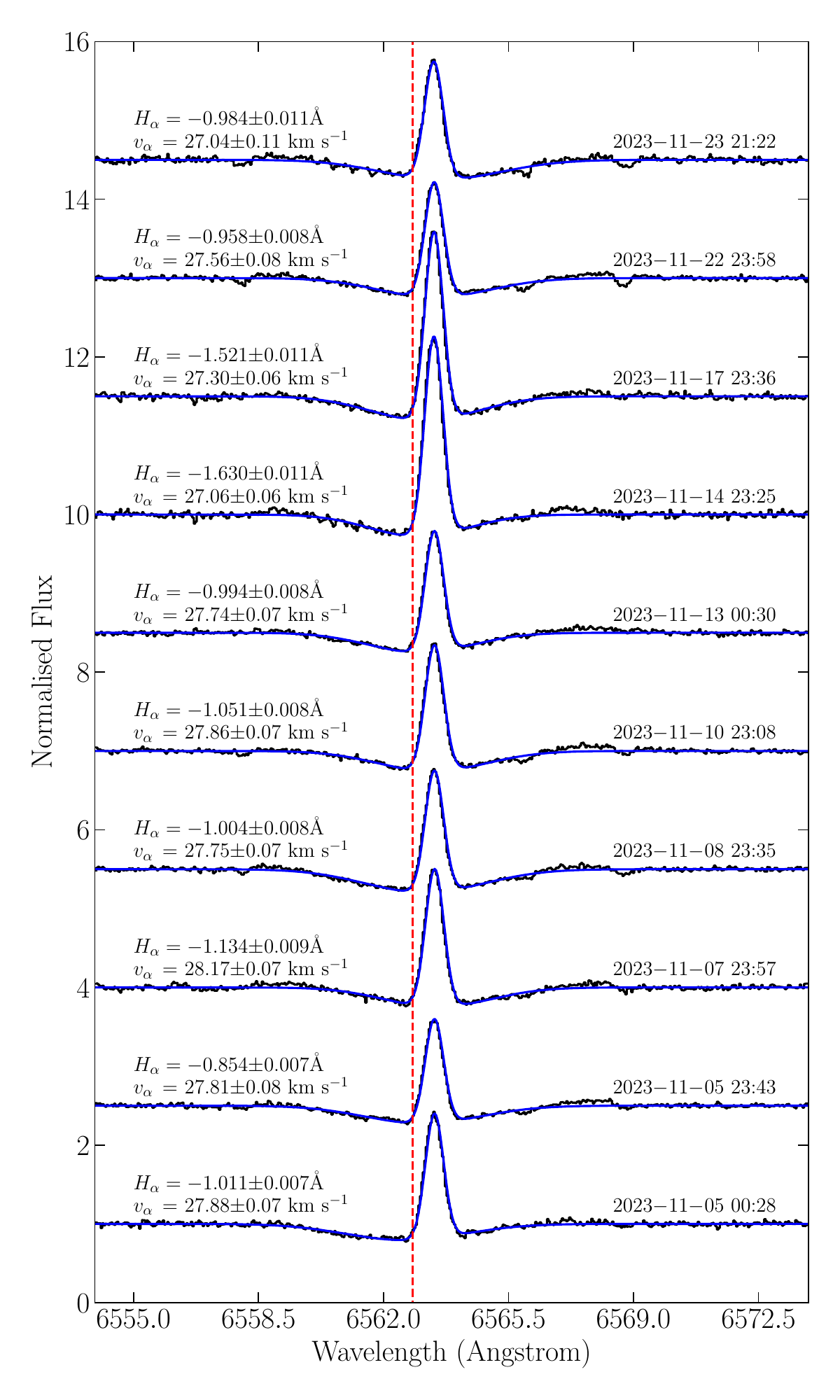}}
                \caption{\Halpha profiles of \salt observations of the optical counterpart of 1eRASS\,J054422.3$-$672729 between 5 and 23 November 2023 shifted to the rest-frame of the LMC. The black line shows the data, the blue line shows the best-fit spectrum using a wide absorption line and a narrow emission line feature, and the dashed red line indicates the rest-frame wavelength of \Halpha. Between individual observations lies a shift of 1.5 for the normalised flux for readability. The equivalent widths and radial velocities are given for the narrow absorption line.}
                \label{fig:SFXT_SALT}
        \end{figure}

        The light curve of 1eRASS\,J054422.3$-$672729 can be seen in Fig.\,\ref{fig:SFXT_HR} in two energy bands and the HR of those two bands as defined in Eq. \ref{equ:HR_LC}. Two flares at ~373 days and ~730 days are clearly visible in the soft band and are less pronounced in the hard band. The HR indicates a spectral softening of the spectrum during the flares. It is possible that this spectral softening, as opposed to the often observed hardening, is caused by low statistics during quiescence. Alternatively, the observed behaviour could resemble that of the objects in \citet{2022A&A...664A..99F}, which showed no spectral hardening. This may indicate alternative mechanisms to gas clump accretion, or that the clumps are not located along the LOS. Also, the observed dynamic ranges for all three systems observed by \ero should be understood as a lower limit, as \ero is likely to have missed the main flare due to its scanning scheme. An interesting feature we found is that both flares seem to happen at the same orbital phase, using the orbital period we found by analysing \ogle data (see Sect. \ref{sec:ogle-results}). This can be seen in Fig. \ref{fig:SFXT_phase}. In general, X-ray flares appear more concentrated around periastron passage in many SFXT systems \citep[e.g.][]{2006A&A...450L...9S, 2015A&A...583L...4G}, which could be what is seen in 1eRASS\,J054422.3$-$672729 and agrees with the scenario of accretion from the clumpy wind of an SG. Nevertheless, flares also appear during other orbital phases \citep[e.g.][]{2012MNRAS.422.2661S, 2013MNRAS.434.2182G, 2023A&A...671A.150S}, preventing any definitive interpretation in the absence of an orbital solution.
        
        \begin{table} 
                \centering
                \caption{Best-fit results for \Hbeta and \HeI lines for \salt spectra of 1eRASS\,J054422.3$-$672729.} 
                \label{tab:SFXT_SALT} 
                \begin{tabular}{lllll} 
                        \hline\hline\noalign{\smallskip}
                        Obs.      & \Hbeta    & $v_{\beta}$            & \HeI      & $v_{He}$    \\
                        \#        & $\times10^{-2}$\,\AA       &  km s$^{-1}$       & $\times10^{-2}$\,\AA       & km s$^{-1}$ \\
                        \noalign{\smallskip}\hline\noalign{\smallskip}
                        1  & $-11.6\pm0.6$ & $28.2\pm0.6$ & $70.2\pm1.3$ & $20.3\pm2.2$ \\
                        2  & $-7.8\pm1.3$  & $27\pm4$     & $73.0\pm1.1$ & $26.8\pm1.3$ \\
                        3  & $-12.0\pm0.7$ & $26.8\pm0.4$ & $71.7\pm1.4$ & $35.4\pm1.9$ \\
                        4  & $-10.7\pm0.6$ & $26.8\pm0.4$ & $77.3\pm1.1$ & $30.5\pm1.3$ \\
                        5  & $-10.2\pm0.6$ & $27.2\pm0.6$ & $74.5\pm1.2$ & $50.1\pm1.4$ \\
                        6  & $-10.1\pm0.5$ & $26.1\pm0.5$ & $75.9\pm1.3$ & $44.2\pm1.4$ \\
                        7  & $-15.0\pm0.8$ & $28.0\pm0.5$ & $74.4\pm1.8$ & $29.3\pm2.1$ \\
                        8  & $-16.1\pm1.0$ & $26.8\pm0.5$ & $70.3\pm1.6$ & $21.7\pm2.0$ \\
                        9  & $-9.1\pm0.6$  & $27.2\pm0.6$ & $75.0\pm1.0$ & $40.0\pm1.5$ \\
                        10 & $-9.1\pm1.9$  & $26.8\pm1.2$ & $69.2\pm1.7$ & $32.9\pm2.3$ \\
                        \noalign{\smallskip}\hline
                \end{tabular} 
                \tablefoot{Obs. \#: Number of observation by date, 1: 2023-11-05 (00:28), 2: 2023-11-05 (23:43), 3: 2023-11-07, 4: 2023-11-08, 5: 2023-11-10, 6: 2023-11-13, 7: 2023-11-14, 8: 2023-11-17, 9: 2023-11-22, 10: 2023-11-23. \Hbeta: The \Hbeta line profile is best fit by the sum of a wide absorption line and a narrow emission line. The value given is the equivalent width of the emission line. $v_{\beta}$: Radial velocity of \Hbeta emission line relative to the LMC. \HeI: Equivalent width of \HeI absorption line. $v_{He}$: Radial velocity of \HeI absorption line relative to the LMC.
                }
        \end{table}
        
        \subsubsection{Be/WD binaries}
        \label{sec:bewd}
        The vast majority of HMXBs found to date consist of a Be star and an NS. This contradicts the expectations of binary evolution models, which predict that there should be a factor of seven more systems with a WD as a compact object than the number of Be/NS binaries \citep{1991A&A...241..419P, 2001A&A...367..848R}. However, to date, only a few such systems are known \citep{2006A&A...458..285K,2012A&A...537A..76S,2020MNRAS.497L..50C,2021MNRAS.508..781K, 2025ApJ...980L..36M}.
        
        Among the candidates newly detected with \ero, we find three objects with spectral properties that cannot be explained by the model of a Be/NS system with a hard X-ray spectrum. All three show soft blackbody spectra with close to no or no emission above 2\,keV, which is typical for SSSs \citep{2022A&A...657A..26M}. This points towards their nature as Be systems containing a WD or BH. Be/BHs are elusive objects that are mostly expected to be X-ray faint \citep{2021A&A...652A.138S}. Recently, BPS models have predicted a fraction of Be/BH systems in the luminosity range of \oergs{31-35} with a non-thermal spectrum \citep{2024A&A...690A.256S}.
        
        \textbf{eRASSU\,J050213.8$-$674620} was first detected during eRASS1 \citep{2020ATel13789....1H} and as an optical companion AL\,55 \citep{2013A&A...555A.141H}, a star with typical features of an early type star, was proposed. In the eRASS1 catalogue, the source is listed under the name 1eRASS J050214.6-674616. We conducted follow-up LCO/FLOYDS observations for a spectral classification of the optical counterpart and found a strong \Halpha emission line of ($-28.11\pm0.22$)\,\AA, which confirms its Be nature. The line is at a radial velocity of ($146.8\pm2.0$)\,\kms relative to the LMC rest frame. The \ero spectrum is best described by an absorbed blackbody spectrum without an additional absorption component for the LMC. The best-fit temperature is  (45.8$^{+2.3}_{-2.1}$)\,eV. eRASSU\,J050213.8$-$674620 shows a flux of (6.4$\pm0.5$)\ergcm{-13} (0.2$-$2.0\,keV) during eRASS1. During all other eRASSs, the flux lies below the background flux, which requires a change of at least two orders of magnitude. Assuming LMC distance the flux during eRASS1 corresponds to an emission radius of (2800$^{+700}_{-500}$)\,km and an absorption-corrected luminosity of ($1.52\pm0.12$)\ergs{36}. This suggests emission from the entire surface of a WD.
        
        For the optical counterpart of \textbf{1eRASS\,J050705.9$-$652149} we find strong \Halpha emission of ($-26.8\pm0.4$)\,\AA\ at a radial velocity of ($150\pm4$)\,\kms relative to the LMC. It shows a higher flux during eRASS1, but in contrast to eRASSU\,J050213.8$-$674620, the flux goes down only by a factor of $\sim$10. We find a best-fit blackbody temperature of (85$^{+20}_{-14}$)\,eV. During eRASS1, we find a flux of (3.0$\pm$0.7)\ergcm{-14}, which at LMC distance corresponds to an absorption-corrected luminosity of (2.1$\pm$0.5)\ergs{34} and an emission radius of ($63^{+54}_{-28}$)\,km. Averaging over all eRASSs we find a flux of (9.1$\pm2.6$)\ergcm{-15}, corresponding to an absorption-corrected flux of (6.2$^{+2.7}_{-2.1}$)\ergs{33} or an emission radius of ($35^{+30}_{-16}$)\,km, possibly originating from a hot spot on the surface of a WD.
        
        The LCO/FLOYDS spectrum of the optical companion of \textbf{1eRASS\,J054242.7$-$672752} shows \Halpha at ($-$10.7$\pm$0.2)\,\AA\ at a radial velocity of (24$\pm$8)\,\kms, confirming its Be nature. The \ero data show a constant flux across all eRASSs, which is best fit with an absorbed blackbody spectrum without an additional LMC absorption component. We find a best-fit blackbody temperature of $88\pm6$\,eV and a flux of (2.69$^{+0.30}_{-0.28}$)\ergcm{-14} (0.2$-$2.0\,keV). Assuming LMC distance, the flux corresponds to an emission radius of (65$^{+15}_{-12}$)\,km or an absorption-corrected luminosity of (2.4$^{+0.4}_{-0.3}$)\ergs{34}. Similar to 1eRASS\,J050705.9$-$652149, this can be explained by emission from a hot spot on the surface of a WD.
        
        To evaluate whether the \ero spectra of the three sources can be alternately explained by BH binaries, we tested two standard models. First, we applied an absorbed power-law model, allowing for local absorption and power-law indices in the range 0.0 to 4.0, using a flat prior (\texttt{tbabs*tbvarabs*pow} in \texttt{Xspec}). This type of model is typically associated with the low-hard state observed in BH binaries. For eRASSU\,J050213.8$-$674620, this model can be clearly rejected. In contrast, for 1eRASS\,J050705.9$-$652149 and 1eRASS\,J054242.7$-$672752, the low source flux relative to the background prevents a definitive rejection. However, in both cases, the posterior distribution for the power-law index peaks at the upper boundary of the prior, indicating a potentially even higher value. Since power-law indices observed in BH systems generally lie below $\sim$3, this model is also disfavoured for the two fainter sources.
        
        The second model we tested was an absorbed multi-blackbody model (\texttt{tbabs*tbvarabs*diskbb} in \texttt{Xspec}), which is typically observed only in the high-soft state during periods of high accretion. The parameters were constrained within the expected range for Schwarzschild BHs with masses between $10/3~M_\odot$ and $30~M_\odot$, assuming a face-on accretion disc. For the inner disc radius, we adopted the innermost stable circular orbit (ISCO):
        \[
        R_{\mathrm{ISCO}} = \frac{6GM}{c^2},
        \]
        where $M$ is the BH mass, $G$ the gravitational constant, and $c$ the speed of light.
        
        The temperature at the inner disc radius was taken as the maximum temperature in the distribution described by \citet{1973A&A....24..337S}
        \[
        T_{\mathrm{max}} = \frac{6\sqrt{6}}{7 \cdot 7^{3/4}} \left(\frac{3}{8\pi \cdot 6^3}\right)^{1/4} \left(\frac{\dot{M} c^6}{G^2 M^2 \sigma}\right)^{1/4} \approx 0.075 \left(\frac{\dot{M} c^6}{G^2 M^2 \sigma}\right)^{1/4},
        \]
        at $r = \frac{49}{36} R_{\mathrm{ISCO}}$, where $\dot{M}$ is the mass accretion rate and $\sigma$ is the Stefan–Boltzmann constant.
        
\begin{figure}
                \centering
                \resizebox{\hsize}{!}{\includegraphics{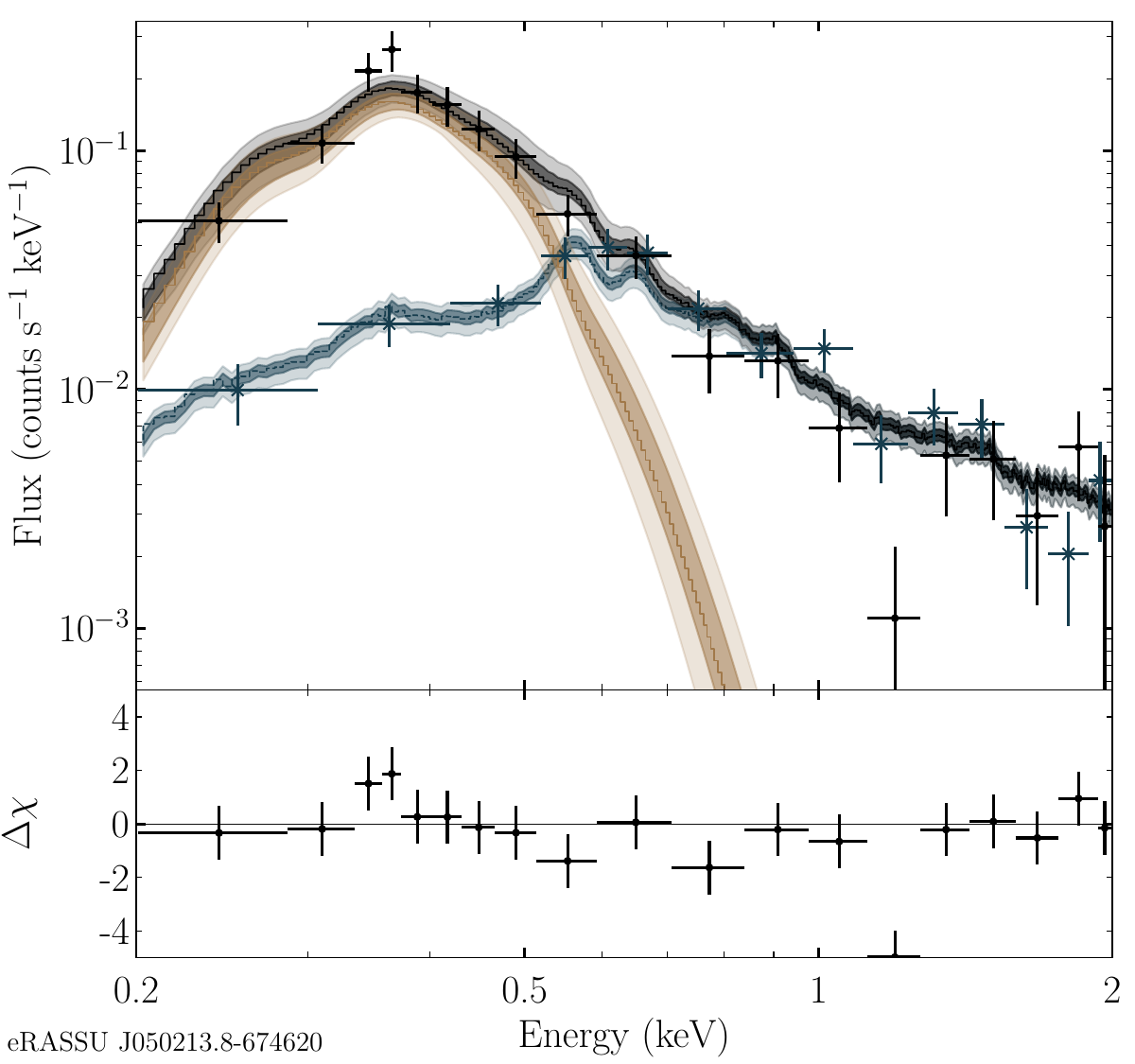}}
                \caption{\ero spectra of the Be/WD candidate eRASSU\,J050213.8$-$674620 merged over all eRASS. Black data points represent the flux in the on-region, while blue data points represent the flux in the off-region, scaled by the size difference between the on- and off-regions. The dark yellow curve represents the best-fit model for the source (absorbed black body), and the shaded areas surrounding it indicate the 1 and 3 sigma regions for the model. The blue curve and shaded area show the best fit for the PCA background model. The black curve and shaded area are the sum of the source and background models.}
                \label{fig:BeWD_spectrum1}
        \end{figure}

\begin{figure}
                \centering
                \resizebox{\hsize}{!}{\includegraphics{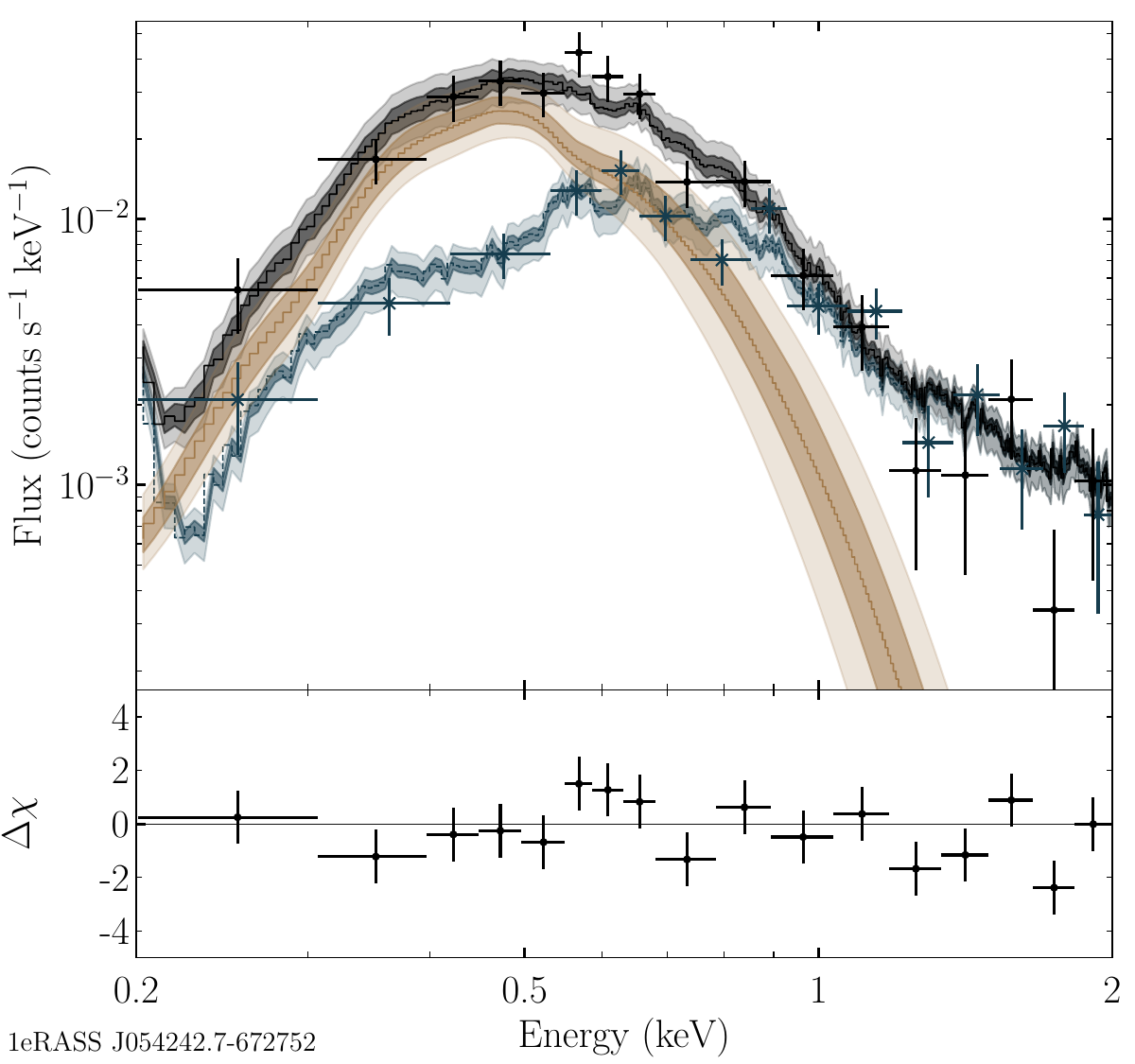}}
                \caption{Same as Fig.\,\ref{fig:BeWD_spectrum1} but for 1eRASS\,J054242.7$-$672752 without local absorption.}
                \label{fig:BeWD_spectrum2}
        \end{figure}

\begin{figure}
                \centering
                \resizebox{\hsize}{!}{\includegraphics{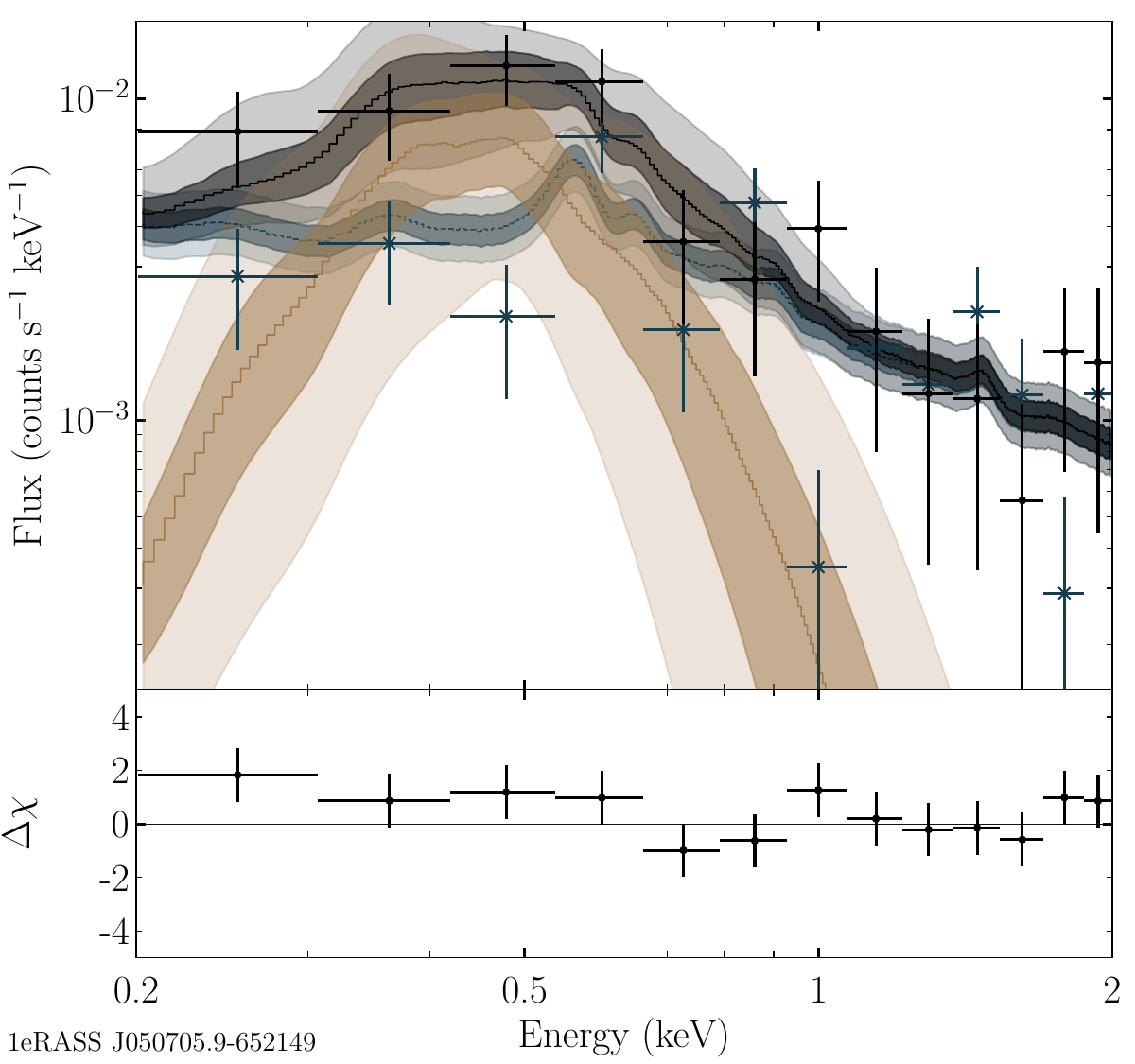}}
                \caption{Same as Fig.\,\ref{fig:BeWD_spectrum1} but for 1eRASS\,J050705.9$-$652149 without local absorption.}
                \label{fig:BeWD_spectrum3}
        \end{figure}

        As a lower limit for the accretion rate, we adopted $\dot{M} \geq 10^{-2} \dot{M}_{\mathrm{Edd}} \approx 10^{-10} \frac{M}{M_\odot} \frac{M_\odot}{\mathrm{yr}}$, where $\dot{M}_{\mathrm{Edd}}$ is the Eddington mass accretion rate, following \citet{1997ApJ...489..865E}, who find that accretion discs are not formed below this threshold. This model results in a poor fit and can be robustly rejected for all three sources.
        We therefore propose all three objects as new Be/WD candidates. Figures \ref{fig:BeWD_spectrum1}, \ref{fig:BeWD_spectrum2}, and \ref{fig:BeWD_spectrum3} show the merged \ero spectra with best-fit absorbed black body models for the three sources. 
        
        Be/WDs have been relatively elusive, with only a few such systems identified, and these are usually observed during X-ray outbursts, attributed to a thermonuclear runaway. eRASSU\,J050213.8$-$674620 falls into this pattern. It is detected only during eRASS1, where the entire WD surface appears to be emitting X-rays at a temperature of $\sim$50\,eV. 
        In contrast, it appears that 1eRASS\,J054242.7$-$672752 and 1eRASS\,J050705.9$-$652149 are in a persistent emission state. Their spectral fits agree with temperatures of $\sim$100\,eV and luminosities corresponding to emission regions of $\sim$50\,km. Given those higher blackbody temperatures and much smaller emission regions, this suggests emission from hot spots on the WD surface. \citet{2012A&A...537A..76S} observed the SSS XMMU\,J010147.5$-$715550 in the SMC in a similar state.
        
        \subsubsection{Persistent BeXRBs}
        \label{sec:persistent}
        
        High-mass X-ray binaries, predominantly systems with a Be star counterpart (BeXRBs), are characterised by eccentric orbits, resulting in a transient nature. During a large portion of the orbit, the systems stay in an X-ray quiescent state until they undergo a major outburst during periastron passage.
        
        Persistent HMXBs are so far scarcely studied sub-class of HMXBs that are characterised by relatively low luminosities of $\lesssim 10^{34}$ erg s$^{-1}$ and a lack of major outbursts. \citet{2002ApJ...574..364P} proposed that this behaviour is caused by long orbital periods and low eccentricities. Due to their wide orbits, no tidal circularisation could have occurred after the supernova explosion that created the NS. Therefore, the low-eccentricity orbit must have formed in a supernova explosion without a significant kick. It was found that NSs in such systems exhibit longer spin periods of the order of a few hundred to a thousand seconds \citep{1999MNRAS.306..100R, 2007A&A...474..137L, 2009A&A...505..947L, 2012A&A...539A..82L}. This can be explained by quasi-spherical accretion or very high magnetic fields of the NS, leading to gating mechanisms. In the subsonic regime, accretion occurs where the plasma remains hot until it reaches the magnetosphere, forming a hot, quasi-spherical shell \citep{2012MNRAS.420..216S}.
        
\begin{figure}
                \centering
                \resizebox{\hsize}{!}{\includegraphics{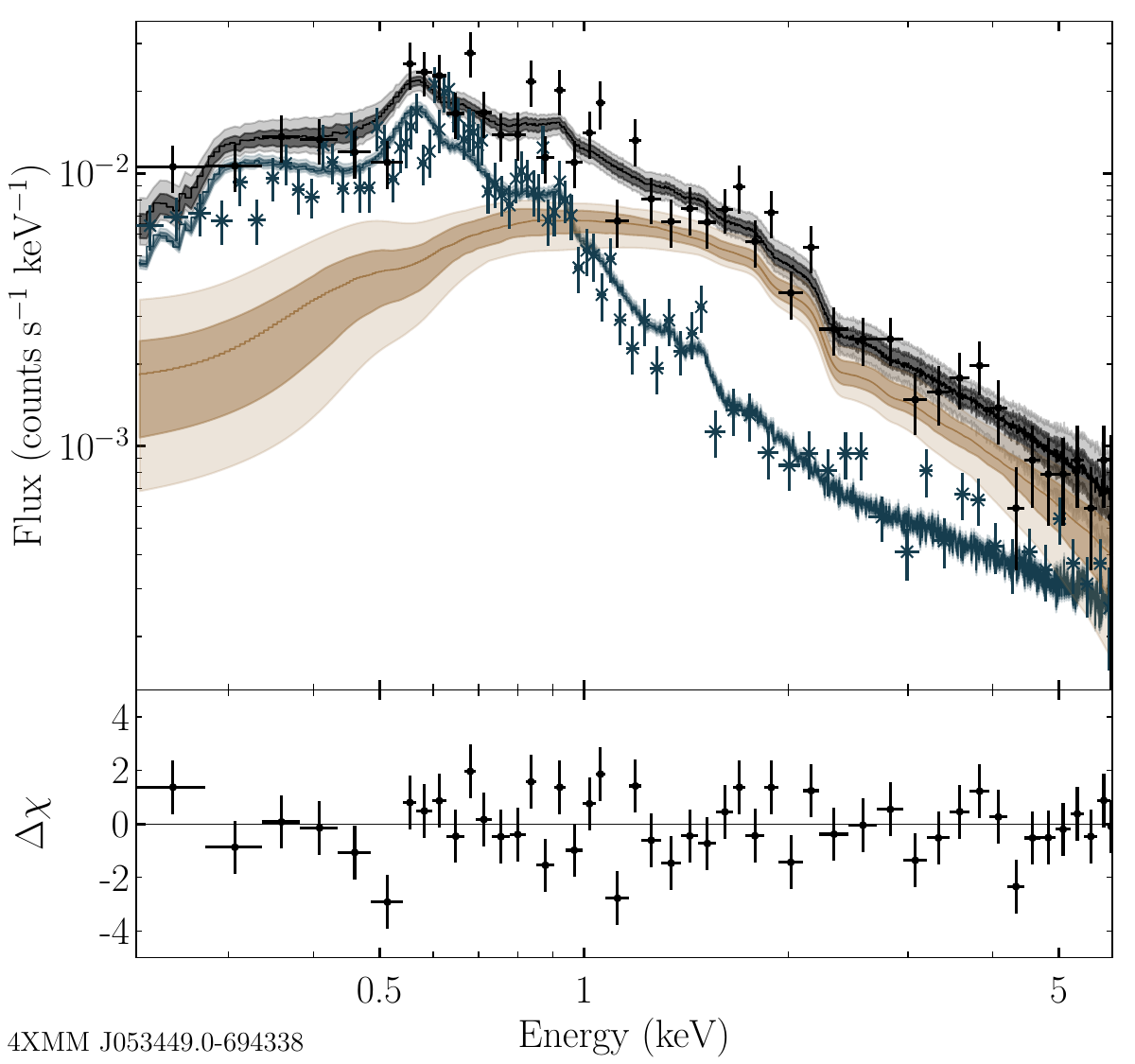}}
                \caption{EPIC-pn spectrum of 4XMM\,J053449.0$-$694338 with colours as in Fig. \ref{fig:BeWD_spectrum1} visually rebinned for better readability.}
                \label{fig:4XMMJ053449.0-694338_xmm}
        \end{figure}

        \textbf{4XMM\,J053449.0$-$694338} is a new HMXB candidate we identified in this work. The source shows long-term variability (see Sect.\,\ref{sec:long_var}) of a factor $\sim$10 over a period of 30 years. The merged \ero spectrum is fit best with an absorbed power-law without an additional absorption component for the LMC, a power-law index of 1.6$\pm$0.4 and a flux of (5.0$^{+1.5}_{-1.4}$)\ergcm{-14}. This corresponds to an absorption-corrected luminosity of (1.9$\pm$0.5)\ergs{34}. To study the characteristics of 4XMM\,J053449.0$-$694338 in more detail, we conducted \xmm follow-up observations (obs.-ID 0943900201; obs. time: 2024-07-18 22:47 -- 2024-07-19 16:34). The exposure time after background-flare screening for EPIC-pn is 40520\,s.
        The spectral analysis was conducted on events with \texttt{PATTERN} 1--4 using the conservative event filtering with \texttt{FLAG}=0.
        The response files were computed with the \texttt{XMMSAS} (version \texttt{xmmsas\_22.1.0-a8f2c2afa-20250304}\footnote{\url{https://www.cosmos.esa.int/web/xmm-newton/sas-release-notes-2210/}}) tasks \texttt{arfgen} and \texttt{rmfgen}.
        To optimise the S/N, we used data up to 6.0\,keV for our fit. For the spectral shape of the background, we used a PCA model provided by \texttt{BXA} similar to our analysis of \ero data (see Sect. \ref{sec:ero_models}). This PCA model is fit-able only down to detector channel 41, which confines our spectral fit to energies larger than $\sim$0.22\,keV. Our best-fit model for the source is an absorbed combination of a power-law and a blackbody with an additional absorption component to account for gas in the LMC (\texttt{tbabs*tbvarabs*(pow+bbodyrad)} in \texttt{Xspec}). The best-fit parameters are a power-law index of 1.50$^{+0.44}_{-0.23}$, an effective temperature of 0.77$^{+0.21}_{-0.22}$\,keV and a local absorption of (7.5$^{+14.1}_{-7.0}$)\hcm{20}. For the flux we find (5.2$\pm$0.4)\ergcm{-14} (0.2--6\,keV), which corresponds to a luminosity corrected for absorption of (1.85$^{+0.28}_{-0.18}$)\ergs{34}. The blackbody normalisation corresponds to an emission region of (340$^{+160}_{-190}$)\,m and the blackbody component accounts for $\sim$30\% of the total flux. The EPIC-pn spectrum with the best-fit model is shown in Fig. \ref{fig:4XMMJ053449.0-694338_xmm}. Correcting the event arrival times to the Solar System barycentre, we created an EPIC-pn light curve and conducted a search for periodic variability using an LS periodogram in the range 10--1000\,s. We do not find any significant periodicity in the data.
        
        Similarly, \textbf{4XMM\,J053049.6$-$662010} is another new HMXB candidate identified in this work. The source exhibits long-term variability of a factor of $\leq10$ over a 30-year period. The merged \ero spectrum is fit best with an absorbed power-law without an additional absorption component for the LMC, a power-law index of $1.43\pm0.17$ and a flux of $(9.0^{+1.1}_{-1.0})$\ergcm{-14}. This corresponds to an absorption-corrected luminosity of $(3.1\pm0.3)$\ergs{34}.
        
        The LCO/FLOYDS spectra of the optical companions of both 4XMM\,J053449.0$-$694338 and 4XMM\,J053049.6$-$662010 are characteristic of main sequence O or B stars but show \Halpha in absorption, while their results for the SED fits indicate IR excess (see Sect.\,\ref{sec:SED-fitting}). This can indicate the loss or truncation of the Be disc due to binary interactions, which is in agreement with the low X-ray luminosity scenario despite the short orbital period we identified with \ogle for 4XMM\,J053449.0$-$694338 (see Sect.\,\ref{sec:ogle-results}).
        
        \subsubsection{A gamma-ray binary embedded in a supernova remnant}
        While hundreds of interacting binary systems containing a high-mass star and a compact object show emission in X-rays \citep{2006A&A...455.1165L, 2016MNRAS.459..528A, 2016A&A...586A..81H}, very few systems are detectable in gamma-rays. For the creation of gamma-ray emission, non-thermal processes are necessary, such as particle acceleration in the shock of colliding winds of a fast-rotating pulsar and its high-mass companion \citep{2006A&A...456..801D} or high-velocity jets in BH-binaries \citep{1998Natur.392..673M}. It is expected that HMXBs hosting NSs undergo a short gamma-ray emitting phase shortly after formation, leading to an expected number of approximately 30 such systems in the MW \citep{1989A&A...226...88M}. Finding fewer gamma-ray binaries might suggest that NSs are born with relatively slow rotation.
        
        \textbf{CXOU\,J053600.0$-$673507} was proposed as an HMXB candidate inside the pulsar wind nebula DEM L241 \citep{2012ApJ...759..123S,2006A&A...450..585B}. \citet{2016ApJ...829..105C} later identified it as the first extragalactic gamma-ray binary with its gamma-ray counterpart called LMC\,P3. They consistently find orbital modulation at a period of 10.3 days in radio, X-rays, and gamma-rays.
        
\begin{figure}
                \centering
                \resizebox{\hsize}{!}{\includegraphics{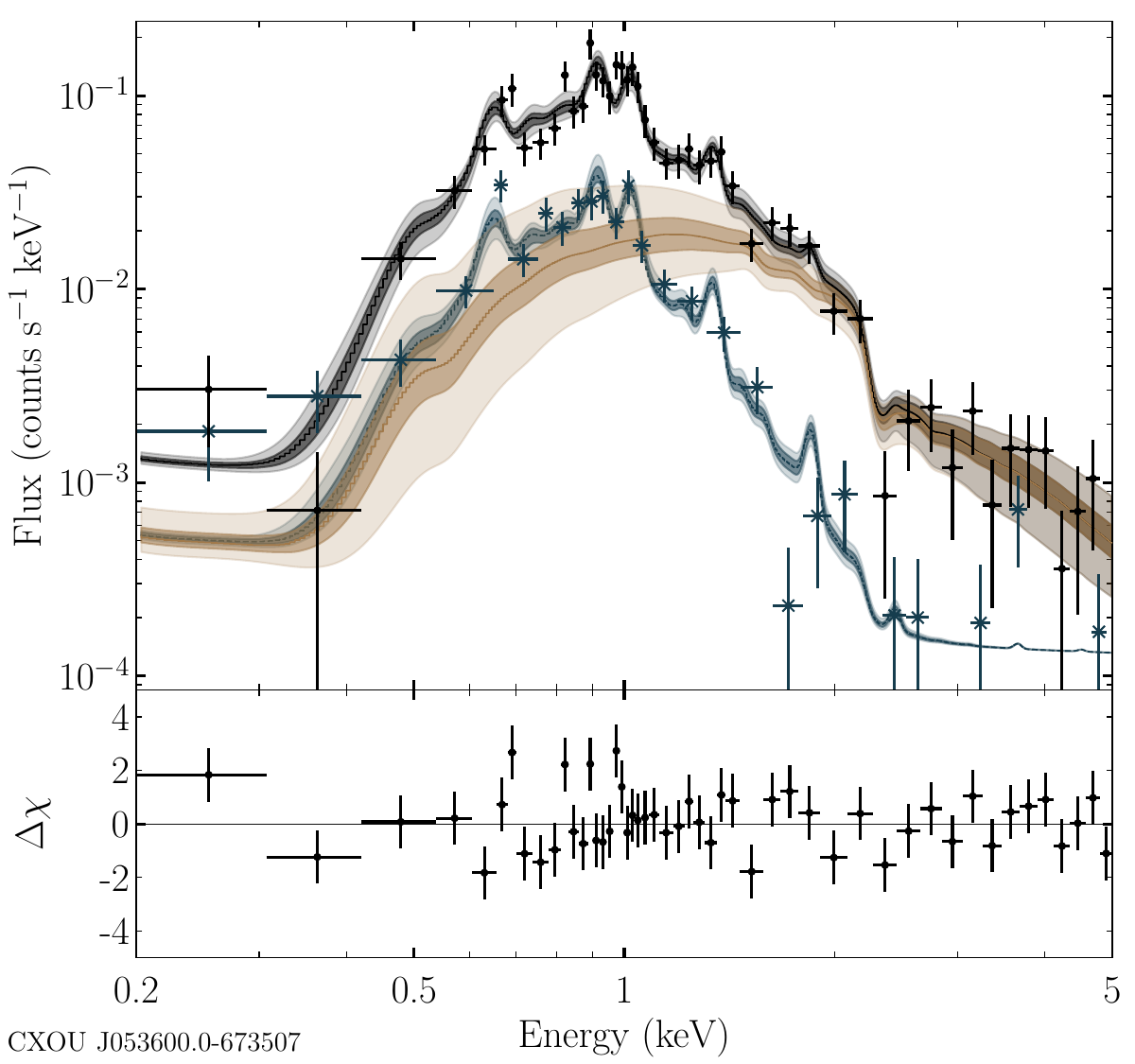}}
                \caption{\ero spectra of the Gamma-ray binary CXOU\,J053600.0$-$673507 merged over all eRASS like Fig. \ref{fig:BeWD_spectrum1}. The dark yellow curve shows the best-fit model for the source (absorbed power law). The blue curve and shaded area show the best fit for a sum of particle background and a \texttt{vapec} model for the SNR contribution in the on-region. The background spectrum is dominated by the SNR emission up to $\sim$2\,keV, where the particle background takes over. The black curve and shaded area are the sum of the source and background models.}
                \label{fig:CXOUJ053600.0-673507}
        \end{figure}

        \ero detects CXOU\,J053600.0$-$673507 as a point source surrounded by the extended emission of DEM L241. To correctly fit the X-ray spectrum of the HMXB, we need to apply a more sophisticated spectral fit to the background spectrum than the PCA model described in Sect.\,\ref{sec:spec-analysis}. For the off-region, we chose a circle well inside the South head of the supernova remnant (SNR) \citep[see][]{2012ApJ...759..123S}. We then follow \citet{2023A&A...676A...3Y} (Appendix A) to estimate the particle background (PB) contribution in the on- and the off-regions and fit for the sum of PB and the SNR in the off-region and the sum of SNR, PB, and HMXB contributions in the on-region. We account for the positionally non-constant background contribution by the SNR by allowing the normalisations to change between on- and off-regions while tying the spectral shapes. Figure \ref{fig:CXOUJ053600.0-673507} shows the best-fit models and Table \ref{tab:CXOUJ053600.0-673507} gives the best-fit parameters. We detect the source at an average flux of ($4.5^{+0.4}_{-0.3}$)\ergcm{-13} (0.2$-$5\,keV), which corresponds to an absorption-corrected luminosity of (10.9$^{+1.2}_{-1.1}$)\ergs{34}.
        
        \begin{table} 
                \centering
                \caption{Best-fit parameters for CXOU\,J053600.0$-$673507.} 
                \label{tab:CXOUJ053600.0-673507} 
                \begin{tabular}{lll} 
                        \hline\hline\noalign{\smallskip}
                        Parameter & Model & Value\\
                        \noalign{\smallskip}\hline\noalign{\smallskip}
                        N$_{\mathrm{H}}$ (cm$^{-2}$) & \texttt{tbvarabs} & (3.5$\pm$0.4)$\times$10$^{21}$\\
                        $\Gamma$ & \texttt{powerlaw} & 1.23$\pm0.29$ \\
                        T$_{plasma}$ (eV) & \texttt{vapec} & 405$^{+14}_{-13}$ \\
                        O & \texttt{vapec} & 0.43$^{+0.08}_{-0.07}$ \\
                        Ne & \texttt{vapec} & 0.83$\pm$0.08 \\
                        Fe & \texttt{vapec} & 0.070$^{+0.014}_{-0.013}$ \\
                        norm$_{on}$ & \texttt{vapec} & $(8.1^{+1.2}_{-1.0})\times10^{-4}$ \\
                        norm$_{off}$ & \texttt{vapec} & $(4.9^{+0.6}_{-0.5})\times10^{-4}$ \\
                        
                        \noalign{\smallskip}\hline
                \end{tabular}
        \tablefoot{Best-fit parameters for CXOU\,J053600.0$-$673507 using the \texttt{Xspec} model \texttt{tbabs*tbvarabs*(pow+vapec)} in the on-region and \texttt{tbabs*tbvarabs*vapec} in the off-region. \texttt{tbabs} and \texttt{tbvarabs} account for the MW foreground absorption and the local absorption, respectively, as described in Sect.\,\ref{sec:spec-analysis}. The PB contribution is added separately, using only the RMF, not the ARF. The normalisation of the PB is frozen to a value scaled by the \texttt{BACKSCAL} parameter of the respective spectral files and the one used for modelling the PB in \citet{2023A&A...676A...3Y}. All element abundances not specified for the \texttt{vapec} model did not improve the fit and were set to the LMC value of 0.49. The subscripts $on$ and $off$ indicate model parameters in the on- and off-regions, respectively.}
        \end{table}
        
        \subsection{Other BeXRBs}
        \label{sec:SNR_ESO}
        
\begin{figure}
                \centering
                \resizebox{\hsize}{!}{\includegraphics{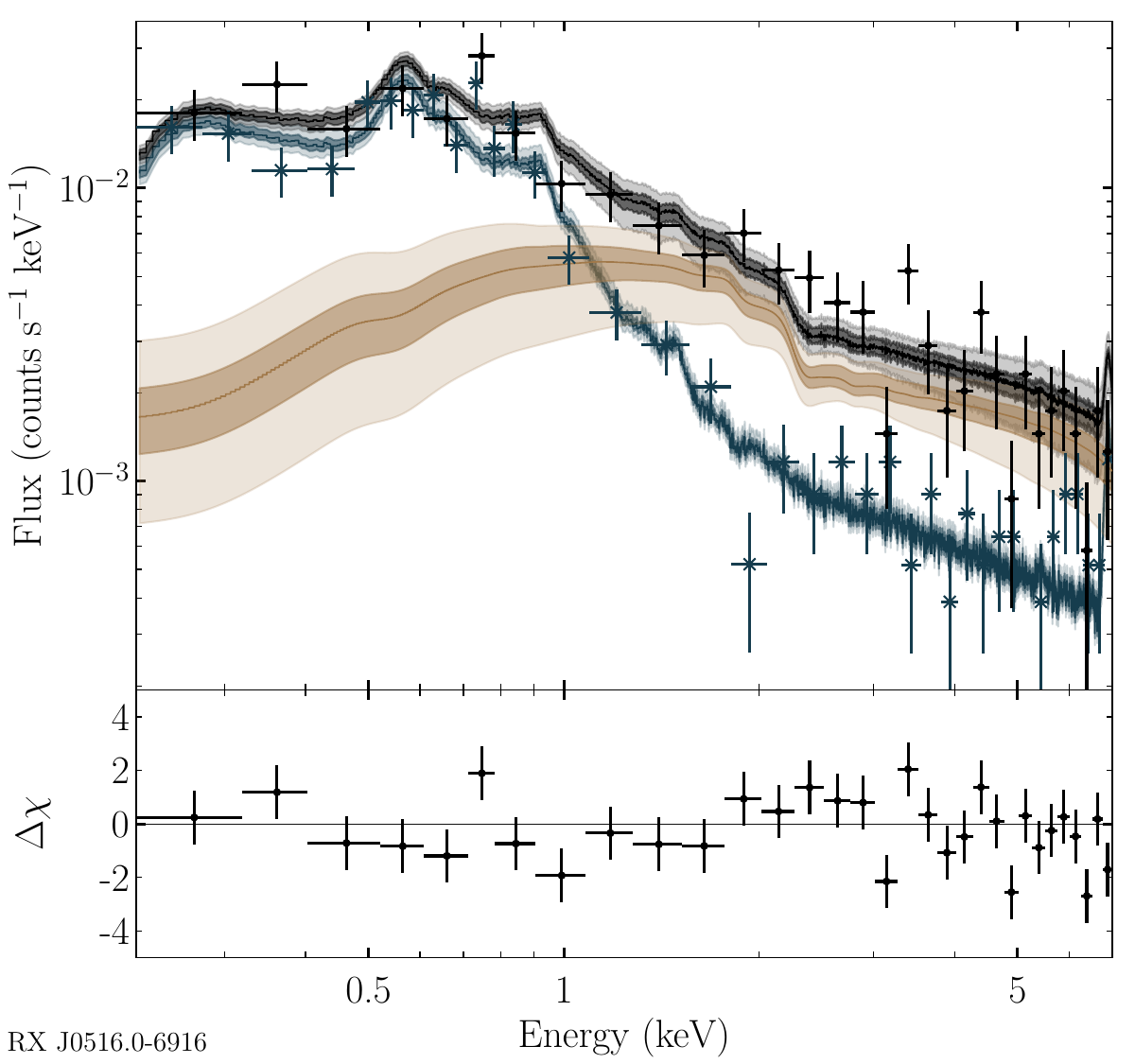}}
                \caption{Same as Fig. \ref{fig:4XMMJ053449.0-694338_xmm} but for RX\,J0516.0$-$6916.}
                \label{fig:RXJ0516.0-6916_xmm}
        \end{figure}

        Due to its stable luminosity and long-term behaviour, we conducted \xmm\, follow-up observations for \textbf{RX\,J0516.0$-$6916} to test its possible membership as a class of persistent BeXRB. Here we present our spectral and timing analysis conducted with \ero\, and \xmm\, observations. \citet{1997PASP..109...21C} first suggested RX\,J0516.0$-$6916 as a BeXRB by confirming the optical counterpart of the X-ray source as an early type star, which is claimed to be a Be star. Later works \citep{2002A&A...385..517N, 2005A&AT...24..151R} questioned the identification of the optical counterpart as a Be star due to the lack of display of any characteristic Be behaviour. \citet{2012A&A...539A.114R} found the source to exhibit \Halpha\, emission, which suggests the correct classification as a BeXRB with the Be star switching between periods with and without a decretion disc. During our LCO/FLOYDS observation, the source again shows \Halpha\, emission with an equivalent width of ($-6.4\pm0.7$)\,\AA\ at a radial velocity relative to the LMC of ($418\pm83$)\,\kms, further confirming the Be classification. Across all eRASSs, we find RX\,J0516.0$-$6916 to vary by a factor of $\leq10$. The merged \ero\, spectrum is best fit by an absorbed power-law with an additional component to account for absorption in the LMC. We find a local absorption of $N_{H,LMC}$=(2.1$^{+2.2}_{-1.7}$)\hcm{21}, a power-law index of $1.4\pm0.4$ and a flux of $(2.7^{+0.5}_{-0.4})$\ergcm{-13} for the merged \ero\, spectrum. This corresponds to an absorption-corrected luminosity of $(11.1^{+1.7}_{-1.1})$\ergs{34}. The \xmm\, observations for RX\,J0516.0$-$6916 were conducted on 2023-11-13 11:28 -- 20:10 (obs.-ID 0923160101). The exposure time after background-flare screening for pn is 13780\,s. The spectral and timing analysis was performed similarly to that of 4XMM\,J053449.0$-$694338 (see Sect.\,\ref{sec:persistent}), with fitting restricted to the range of 0.22 to 7.0\,keV for higher S/N and due to the PCA model used for the spectral shape of the background. Our best-fit model is an absorbed power law, without an additional component to account for LMC absorption, with a power-law index of 0.66$\pm0.14$. For the flux we find (9.4$^{+1.0}_{-0.9}$)\ergcm{-14} (0.2--0.7\,keV), corresponding to an absorption-corrected luminosity of (2.87$^{+0.28}_{-0.26}$)\ergs{34}. The EPIC-pn spectrum with the best-fit model is shown in Fig. \ref{fig:RXJ0516.0-6916_xmm}. Correcting the event arrival times to the Solar System barycentre, we created an EPIC-pn light curve and conducted a search for periodic variability using an LS periodogram in the range 10--1000\,s. We do not find any significant indication of a period present in the data.
        
        \textbf{XMMU\,J050722.1$-$684758} was discovered as a new BeXRB by \citet{2021MNRAS.504..326M}, probably associated with the SNR MCSNR\,J0507$-$6847. The authors confirm the NS nature of the compact object by detecting X-ray pulsations with a periodicity of 570\,s and find an absorption-corrected X-ray luminosity of $8.5$\ergs{34} (0.2$-$12\,keV).
        
\begin{figure}
                \centering
                \resizebox{\hsize}{!}{\includegraphics{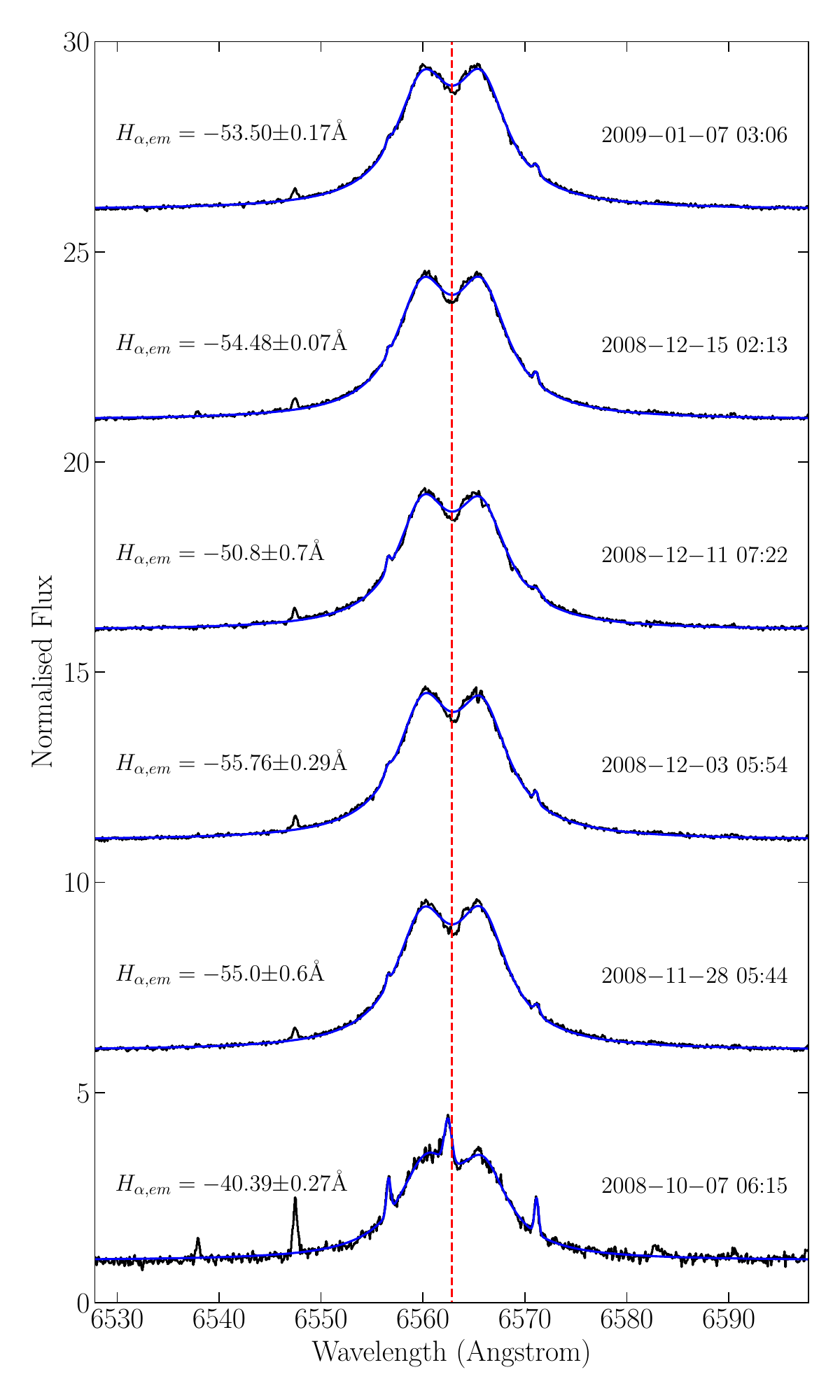}}
                \caption{\Halpha profiles of VLT FLAMES/GIRAFFE observations of the optical counterpart of XMMU\,J050722.1$-$684758 between 2008-10-07 and 2009-01-07 shifted to the rest-frame of the LMC. The black line shows the data, the blue line shows the best-fit model using a double-peaked Lorentzian line profile, and the dashed red line indicates the rest-frame wavelength of \Halpha. Between individual observations, there is a shift of 5.0 in the normalised flux for readability.}
                \label{fig:SNR_ESO_Halpha}
        \end{figure}

\begin{figure}
                \centering
                \resizebox{\hsize}{!}{\includegraphics{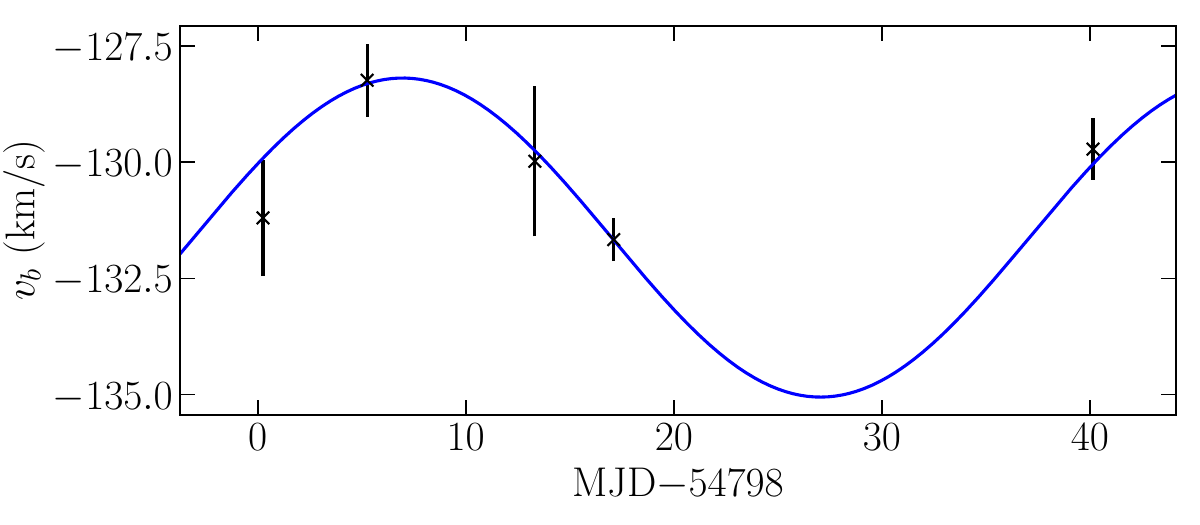}}
                \caption{Radial velocity change of the blue component in the double-peaked \Halpha lines observed from XMMU\,J050722.1$-$684758 between 2008-11-28 and 2009-01-07 compared to a sinusoidal fit with a period frozen at the orbital period found with \ogle.}
                \label{fig:SNR_ESO_v_t}
        \end{figure}

        The merged \ero spectrum can best be fit by an absorbed power-law without an additional component to account for local absorption. We find a power-law index of 1.2$\pm$0.6 and a flux of (2.9$^{+1.6}_{-1.0}$)\ergcm{-14} (0.2$-$5\,keV), corresponding to an absorption-corrected luminosity of (1.0$^{+0.4}_{-0.3}$)\ergs{34}. The optical counterpart of XMMU\,J050722.1$-$684758 was observed with the VLT FLAMES/GIRAFFE spectrograph during six observations between 2008-10-07 and 2009-01-07 (PI:\,R.\,E.\,Mennickent, programme ID 082.D-0575 in the wavelength range  6299$-$6691\,\AA\ (red) at spectral resolutions between 10.2 and 39.1. During our analysis, we found strong double-peaked \Halpha emission in all six observations, which can be best fit with two Lorentzian line profiles. The red and blue lines lie at radial velocities of approximately $+100$\,\kms and $-100$\,\kms with respect to the LMC. During the first observation on 2008-10-07, the source was fainter by a factor of $\sim$5 compared to the other five observations. The variability is consistent with a sinusoidal function with a period of 40.16\,d (from \ogle analysis, see Sect. \ref{sec:ogle-results}), excluding the first of the six observations. No such behaviour could be observed for the red line. The results of our spectral analysis can be found in Fig.\,\ref{fig:SNR_ESO_Halpha} and Fig.\,\ref{fig:SNR_ESO_v_t}.
        
        \subsection{Variability and classification of the optical counterparts}
        \label{sec:ogle-results}
        
        We analysed the \ogle I-band light curves of 28 (candidate) HMXB systems in the LMC in order to find periodicities that are likely related to the binary orbit of the systems. We report eleven new periods (Table\,\ref{tab:ogle_data}), seven from the new candidates discovered in the eROSITA eRASS1 data, supporting their identification as HMXBs. Four periods were identified from previously known systems. From one of them (Swift\,J0549.7$-$6812 with period 45.7\,d) also the spin period of 6.2\,s is known \citep{2013ATel.5309....1K}. From the eleven new periods, six are found between 45\,d and 170\,d, as typical for orbital periods of BeXRBs. Orbital periods of HMXBs shorter than 10\,d are usually associated with wind-fed SgXRBs.
        
        Note that this section is self-contained and independent of Sect.\,\ref{sec:individuals}; objects introduced there may also appear here. The analysed \ogle I-band light curves can be found in Fig.\,\ref{fig:ogle_Ilc}.
        
        \subsubsection{Systems with periods shorter than 10\,d}
        
\begin{figure*}
                \centering
                \resizebox{0.33\hsize}{!}{\includegraphics{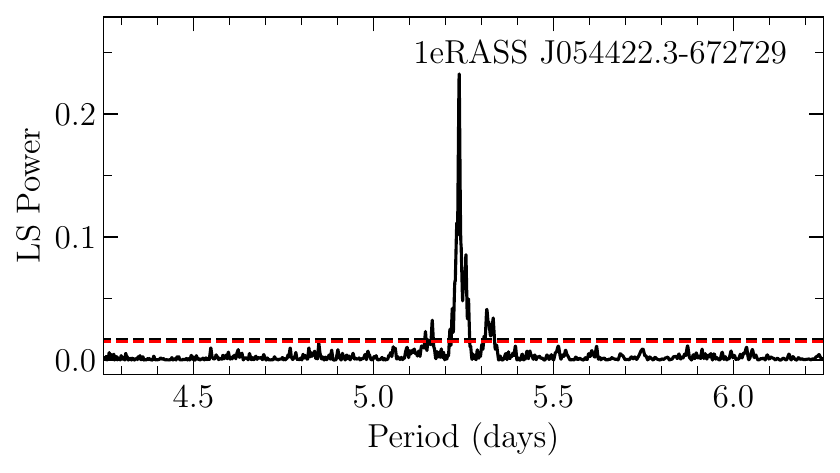}}
                \resizebox{0.33\hsize}{!}{\includegraphics{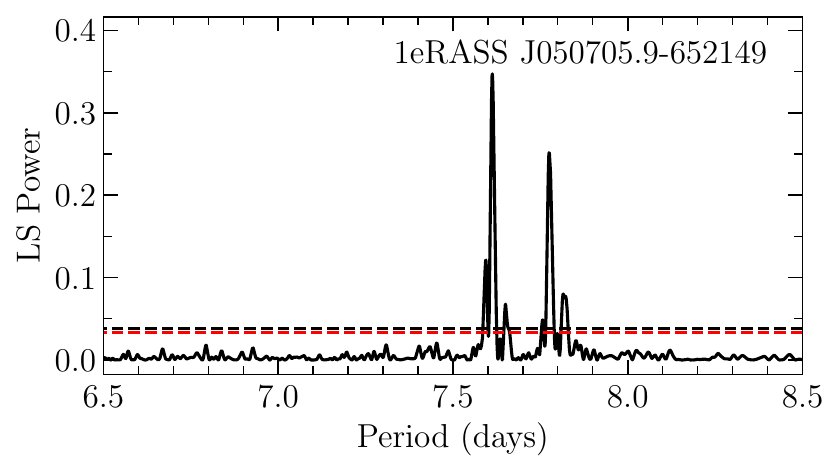}}\\
                \resizebox{0.33\hsize}{!}{\includegraphics{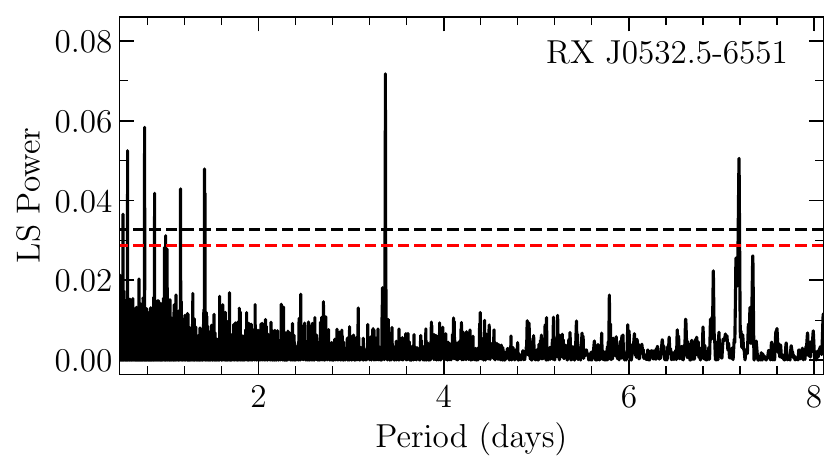}}
                \resizebox{0.33\hsize}{!}{\includegraphics{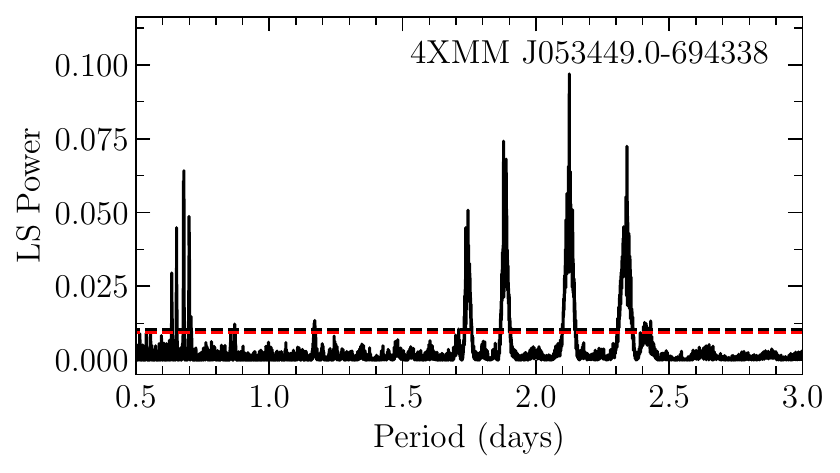}}
                \caption{LS periodograms with highly significant peaks at periods below 10\,d.
                }
                \label{fig:ogle_ls_short}
        \end{figure*}

        In this section and the following, we discuss our findings for the analysis described in Sect.\,\ref{sec:ogle-analysis}. Similar to 1eRASS\,J054242.7$-$672752, we find four additional systems with highly significant signals at periods below 10\,d. The LS-periodograms around the periods are collected in Fig.\,\ref{fig:ogle_ls_short}.
        
        \textbf{1eRASS\,J054242.7$-$672752}, used as example in Fig.\,\ref{fig:example:ogle_ls}, shows a pair of peaks in the periodograms at 6.49\,d and 6.61\,d. The periods are consistent with the 1\,yr alias (the difference in frequency space is 1\,year$^{-1}$) of each other. We suggest the peak with the highest significance at 6.49\,d as the likely orbital period of the system, but cannot rule out the period of 6.61\,d.
        
        \textbf{1eRASS\,J050705.9$-$652149} is another such case with the most significant peak at 7.61\,d together with the 1\,yr alias at 7.77\,d. Similarly to the previous case, We propose 7.61\,d as the orbital period of the HMXB system, but cannot rule out the period of 7.77\,d.
        
        Both 1eRASS\,J050705.9$-$652149 and 1eRASS\,J054242.7$-$672752 are candidate Be/WD systems. Likely orbital periods for similar systems are known with 21.5\,d \citep[Swift\,J004427.3$-$734801,][]{2020ATel13709....1H,2020MNRAS.497L..50C} and 17.4\,d \citep[Swift\,J011511.0$-$725611,][]{2021ATel14341....1K,2021MNRAS.508..781K} from two Be stars associated with supersoft X-ray sources in the SMC. Orbital periods at the short end of the period distribution seen from BeXRBs suggest tighter orbits for systems with a white dwarf as a compact object with respect to classical BeXRBs with a NS.
        
        The periodograms of \textbf{1eRASS\,J054422.3$-$672729} are characterised by a highly significant peak at 5.25\,d with weaker peaks on both sides, which are consistent with 1\,yr aliases. This, together with the relatively bright optical counterpart (V = 13.52\,mag) and the X-ray temporal behaviour, identifies this source as an SFXT candidate with a SG in a 5.25\,d orbit. While commonly SFXT systems show larger orbital periods than Roche-lobe overflow SgXRBs, there are known SFXT systems with orbital periods of the order of a few days \citep[e.g.][]{2009MNRAS.397L..11J}.
        
        The highest peaks in the periodograms of \textbf{RX\,J0532.5$-$6551} are found at 3.37\,d (P1) and 7.19\,d (P2). Smaller peaks around P2 can be related to 1\,yr aliases. One-day aliases of P1 and P2 also appear at periods around 1\,d. We cannot find a relation between P1 and P2, suggesting two independent periods in the system: 7.19\,d, which could be the orbital period, and 1.42\,d, which could be the period of NRPs.
        
        \textbf{4XMM\,J053449.0$-$694338} shows a series of four peaks at 1.737\,d (P1), 1.879\,d (P2), 2.125\,d (P3), and 2.341\,d (P4), with their corresponding series of 1\,d aliases between 0.6\,d and 0.7\,d. Also, P1 and P4 are 1\,d aliases, as well as P2 and P3. P4 could be the alias period of 1\,d and 3$\times$P3. It remains unclear if the highest-power peak indicates the fundamental period of 2.125\,d, which could be the orbital period of the system, or if it is itself an alias of 0.68\,d, suggesting NRPs of a Be star. In addition, the Floyds spectrum showed \Halpha in absorption (Table\,\ref{tab:spectroscopy}) and the U$-$B colour is unusually red (fainter in U than in B), adding a question mark behind the identification of the optical counterpart as O/Be star \citep{2016A&A...586A..81H}. On the other hand, the optical counterpart is too faint (V $\sim$15 mag) for an SG at LMC distance, and the identification as HMXB needs further confirmation.
        
        \subsubsection{Systems with longer periods}
        
\begin{figure*}
                \centering
                \resizebox{0.33\hsize}{!}{\includegraphics{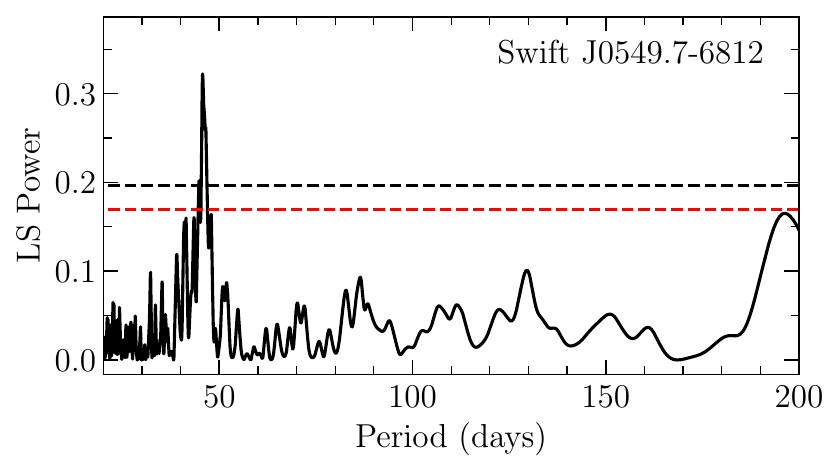}}
                \resizebox{0.33\hsize}{!}{\includegraphics{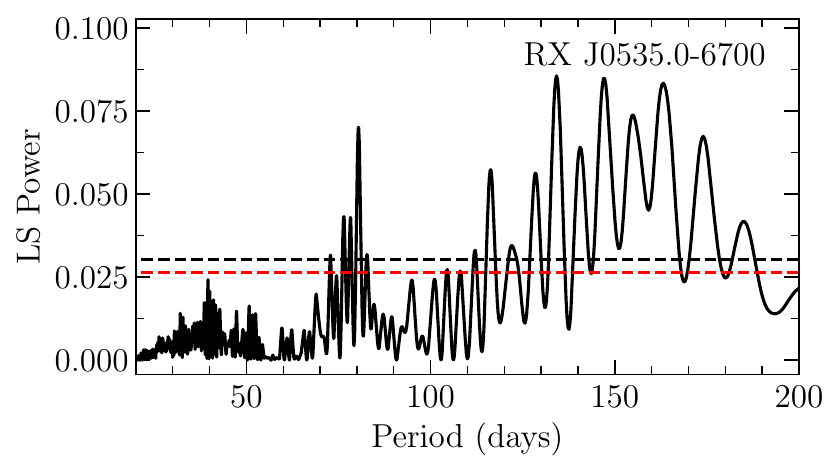}}
                \resizebox{0.33\hsize}{!}{\includegraphics{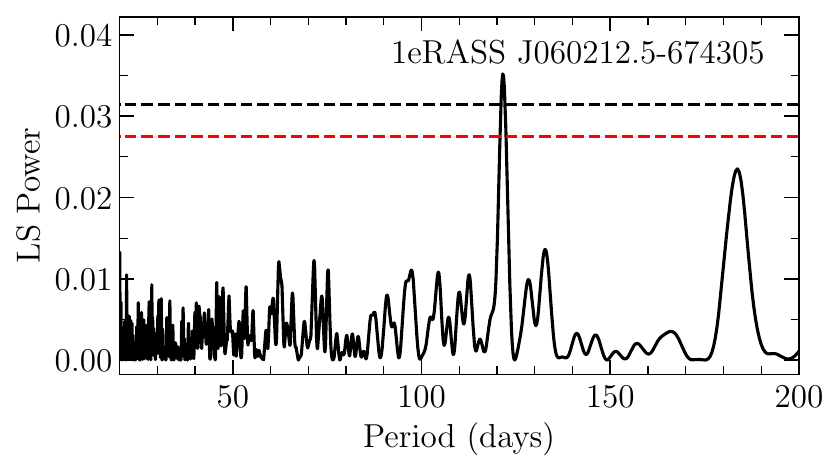}}
                \resizebox{0.33\hsize}{!}{\includegraphics{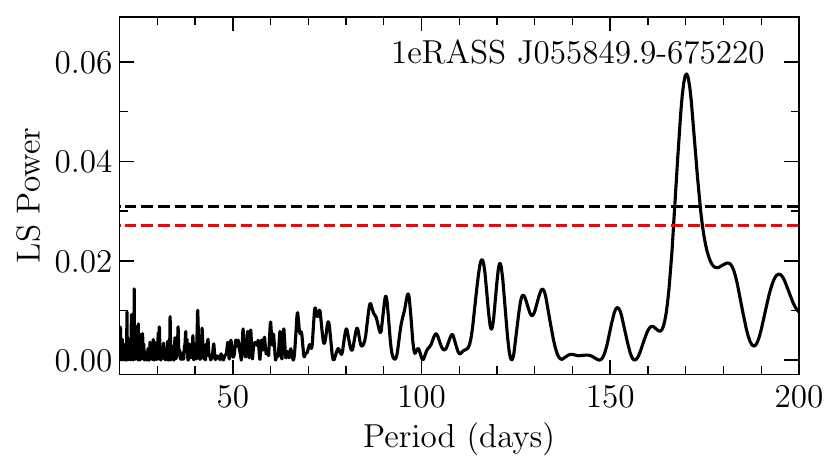}}
                \resizebox{0.33\hsize}{!}{\includegraphics{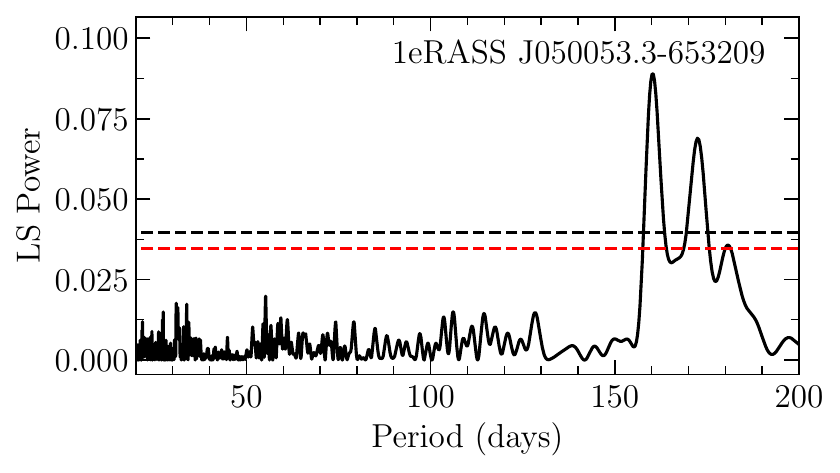}}
                \resizebox{0.33\hsize}{!}{\includegraphics{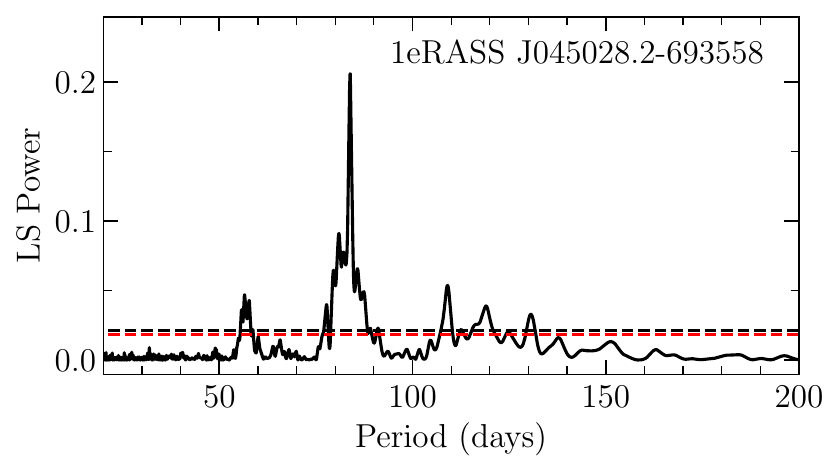}}
                \caption{LS periodograms with peaks at longer periods.}
                \label{fig:ogle_ls_long}
        \end{figure*}

        Our timing analysis of the \ogle light curves revealed six systems with periods ranging from 45.7\,d to 170.3\,d. Figure\,\ref{fig:ogle_ls_long} presents the corresponding LS-periodograms, highlighting peaks at these longer periods. (Fig.\,\ref{fig:ogle_ls_long}).
        
        The I-band light curve of \textbf{Swift\,J0549.7$-$6812} is only sparsely populated with measurements. Nevertheless, a clear peak at 45.7\,d is found, with a light curve folded at that period with sinusoidal shape and an amplitude (maximum minus minimum) of $\sim$0.05\,mag. We suggest 45.7\,d as the orbital period of an HMXB.
        
        \citet{2002A&A...385..517N} confirmed \textbf{RX\,J0535.0$-$6700} as a BeXRB in the LMC. 
        Therefore, the period of 241\,d reported by \citet{1988MNRAS.232...53R} from a UK Schmidt telescope photographic survey was suggested as the likely orbital period of the binary by \citet{1999A&A...344..521H}. 
        In our LS periodograms, we detect a relatively sharp peak at 80.5\,d, consistent with one-third of the previously reported value, which we cannot confirm in our periodograms. There is also a signal at $\sim$161\,d, which is difficult to filter out of many nearby peaks. Inspection of the light curve between HJD\,2458000 and HJD\,2459000 reveals a series of strong outbursts, which are 80.5\,d apart. However, every other outburst is much stronger than the one before, suggesting that the true underlying period is 161\,d. Therefore, we suggest 161\,d as the revised orbital period of RX\,J0535.0$-$6700. The 241\,d period reported earlier might be explained by the much sparser sampling of the UK Schmidt survey compared to \ogle.
        
        The \ogle-IV light curve of \textbf{1eRASS\,J060212.5$-$674305} is characterised by a slight decrease in brightness over the years. Superimposed are short flares, which increase in their maximum brightness over time. The LS periodograms reveal a peak at $\sim$122\,d, which is also present when removing the flares from the light curve (using only data with I$>$14.04\,mag). Phase folding the light curve data shows that the flares occur at irregular phases. We suggest an orbital period of $\sim$122\,d with the optical flaring activity unrelated to the orbit.
        
        The I-band light curve of \textbf{1eRASS\,J055849.9$-$675220} shows a scatter of $\pm$0.02 mag without any significant long-term trend. The LS analysis of the light curve without detrending reveals only one significant peak across the full investigated period range at 170\,d, which likely indicates the orbital period. The folded light curve is consistent with a sinusoidal profile.
        
        \textbf{1eRASS\,J050053.3$-$653209} shows a somewhat larger scatter of $\pm$0.04 mag on top of small 0.02 mag long-term variations. The most significant peak in the periodograms is found at 160\,d, most likely the orbital period. A one-year alias is seen at 285\,d. Two additional peaks are found at $\sim$173\,d and $\sim$180\,d. While the latter is probably associated with the 0.5\,year period of the window function, 173\,d could be an alias of $\sim$2000\,d, related to the length of the light curve. The folded light curve has a triangular profile. However, no significant signal for harmonics are seen in the periodogram.
        
        \textbf{1eRASS\,J045028.2$-$693558} was covered by \ogle-III and \ogle-IV. In addition to a $\pm$0.05\,mag scatter, short flares and sharp dips are seen. 
        A highly significant peak is observed in the LS periodograms at 83.89\,d. The folded profile is characterised by a broad bump about 0.8 wide in phase and a minimum 0.06\,mag fainter than the maximum.
        
        \subsection{Comparison with SMC population}
        
        Compared to the MW, the MCs, particularly the SMC, host a disproportionately large number of HMXBs relative to their stellar mass, likely due to recent episodes of enhanced star formation \citep{2002A&A...385..517N, 2010ApJ...716L.140A, 2016MNRAS.459..528A} and metallicity differences between the galaxies \citep{2010ApJ...725.1984L, 2015A&A...579A..44D}. To date the SMC is known to host $\sim$130 BeXRBs and only one SgXRB \citep{2014MNRAS.438.2005M, 2016A&A...586A..81H, 2017MNRAS.470.4354V, 2020ATel13823....1K,2023ATel15886....1M}. In comparison, in the LMC 59, HMXBs are known, of which 8 are SG systems. The higher abundance of SG systems compared to the SMC can be explained by differences in when recent star formation peaked in the two galaxies \citep{2010ApJ...716L.140A,2016MNRAS.459..528A}. The different populations of optical counterparts are also apparent in the V$-$I over Q plane (see Fig. \ref{fig:V-I_Q}), where counterparts in the LMC display a significantly larger spread than those in the SMC.
        
        Compared to the findings of \citet{2013A&A...558A...3S} for the SMC, our X-ray luminosity function is less steep at higher luminosities and more closely resembles those reported by \citet{2003ChJAS...3..257G}. Additionally, we do not observe a significant change in the power-law index below $\sim$4\ergs{35}, as reported by \citet{2013A&A...558A...3S}. However, this discrepancy may be due to statistical limitations and needs to be confirmed through follow-up observations that probe lower luminosities.
        
        \section{Summary}
        \label{sec:summary}
        
        We compiled a comprehensive catalogue of 53 HMXBs in the LMC detected during the first \ero all-sky survey (eRASS1), which is complete down to $\sim$\oergs{35}. By cross-matching the eRASS1 sources with MCPS, VMC, and a proper-motion selected subsample of the \textit{Gaia} eDR3 catalogue, we identified and characterised the HMXB population in the LMC.
        
        Out of 60 previously known HMXBs in the LMC, 28 were detected during eRASS1. Additionally, 9 out of 23 previously suggested HMXB candidates were detected, out of which we rejected 5 as HMXBs. We identified 21 new HMXB candidates based on their multi-wavelength properties.
        
        Our analysis utilised the multi-wavelength properties of known HMXBs to define classification flags based on both archival data and follow-up observations, which we used to define six confidence classes. In X-rays, we assessed the spectral properties, the long- and short-term variability, and pulsations. Optical and IR characteristics included H$\alpha$ emission, IR excess searches from SED fitting, IR colour investigations, and long-term optical variability identification using decades of \ogle data.
        
        The population of HMXBs in the LMC differs notably from that in the SMC, with a higher abundance of systems distinct from the stereotypical Be/NS XRBs, including seven SgXRBs and three (candidate) Be/WD systems.
        Among the SgXRBs, we identified a new candidate SFXT with phase-dependent flares. Two of the candidate Be/WDs appear in a low-luminosity state, which can be explained by stable accretion on the hotspot of a magnetised WD.
        These findings highlight the distinct evolutionary pathways and environmental factors influencing HMXB populations in the two Magellanic Clouds.
        
        We constructed the sensitivity-corrected X-ray luminosity function for systems of high confidence classes. A power-law fit to the luminosity function yields an index of $0.55^{+0.53}_{-0.11}$, in agreement with a previous \cxo-based study that examined the XLFs in several nearby galaxies. Using the X-ray luminosity function, we find that the LMC's SFR can be estimated with a value of $(0.22^{+0.06}_{-0.07})$\,M$_{\odot}$yr$^{-1}$, which is in agreement with results from other tracers.

        \section{Data availability}
        The full version of our catalogue is available in electronic form at the CDS via anonymous ftp to cdsarc.u-strasbg.fr (130.79.128.5) or via \url{http://cdsweb.u-strasbg.fr/cgi-bin/qcat?J/A+A/}.
        
        \begin{acknowledgements}
                We thank the referee for useful comments and suggestions. 
                
                This work is based on data from \ero, the soft X-ray instrument aboard SRG, a joint Russian-German science mission supported by the Russian Space Agency (Roskosmos), in the interests of the Russian Academy of Sciences represented by its Space Research Institute (IKI), and the Deutsches Zentrum für Luft- und Raumfahrt (DLR). The \srg spacecraft was built by Lavochkin Association (NPOL) and its subcontractors, and is operated by NPOL with support from the Max Planck Institute for Extraterrestrial Physics (MPE).
                
                The development and construction of the \ero X-ray instrument was led by MPE, with contributions from the Dr. Karl Remeis Observatory Bamberg \& ECAP (FAU Erlangen-Nürnberg), the University of Hamburg Observatory, the Leibniz Institute for Astrophysics Potsdam (AIP), and the Institute for Astronomy and Astrophysics of the University of Tübingen, with the support of DLR and the Max Planck Society. The Argelander Institute for Astronomy of the University of Bonn and the Ludwig Maximilians Universität Munich also participated in the science preparation for eROSITA.
                
                The eROSITA data shown here were processed using the eSASS software system developed by the German eROSITA consortium.
                
                This research has made use of the VizieR catalogue access tool, CDS, Strasbourg, France \citep{10.26093/cds/vizier}. The original description of the VizieR service was published in \citet{vizier2000}.
                
                This research has made use of the Spanish Virtual Observatory (https://svo.cab.inta-csic.es) project funded by MCIN/AEI/10.13039/501100011033/ through grant PID2020-112949GB-I00.
                
                The LCO observations have been made possible by the support of the Deutsche Forschungsgemeinschaft (DFG, German Research Foundation) under Germany’s Excellence Strategy-EXC-2094-390783311.
                
                Based on observations collected at the European Southern Observatory under ESO programme(s) 082.D-0575(A) and/or data obtained from the ESO Science Archive Facility with DOI(s) under https://doi.org/10.18727/archive/27
                
                RW is funded by the Deutsche Forschungsgemeinschaft (DFG, German Research Foundation) – 446281683.
                
                GV acknowledges support from the Hellenic Foundation for Research and Innovation (H.F.R.I.) through the project ASTRAPE (Project ID 7802).
                
                AU acknowledges support from the “Copernicus 2024 Award” of the Polish FNP and German DFG agencies.
        \end{acknowledgements}
        
        \bibliographystyle{aa}
        \bibliography{references}
        
        \appendix
        
        \section{Supplementary X-ray material}
        
        \begin{table}[!htbp]
                \centering
                \caption{Alternative names and literature references for discovery of the X-ray sources with confidence class 1 in our HMXB catalogue.} 
                \label{tab:Xray_names} 
                \begin{tabular}{lll} 
                        \hline\hline\noalign{\smallskip}
                        \# & X-ray Names & Ref.$^{(a)}$ \\
                        \noalign{\smallskip}\hline\noalign{\smallskip}
                        1 &  & HMV22 \\
                        2 &  & VSH13 \\
                        3 &  & HMV22 \\
                        4 & CAL\,9, 2E\,0501$-$7038, 1E\,0501.8$-$7036 & SCH96,  \\
                        &  & SCF94 \\
                        5 &  & MHM21 \\
                        6 & LXP169 & MHS13 \\
                        7 &  & HMC20 \\
                        8 &  & HIR17 \\
                        9 & LXP27.2 & KCE09,  \\
                        &  & CFB15 \\
                        10 &  & CSM97 \\
                        11 & LXP8.04 & SCF94,  \\
                        &  & CNB01 \\
                        12 &  & KdBB01,  \\
                        &  & HP99 \\
                        13 &  & vJBM \\
                        14 &  & MHC20,  \\
                        &  & MHK20 \\
                        15 & XMMU\,J052947.4$-$655639, LXP69.2,  & HDP97,  \\
                        & RX\,J0529.7$-$6556 & HDP03 \\
                        16 & XMMU\,J053011.2$-$655122, LXP272,  & HDP03 \\
                        & RX\,J0530.1$-$6551 & \\
                        17 & LXP28.8 & VMH13 \\
                        18 &  & VMH18 \\
                        19 & XMMU\,J053232.4$-$655139,  & HPD95 \\
                        & 1RXS\,J053224.1$-$655112 & \\
                        20 & 2A\,0532$-$664, LXP13.5, CAL\,49,  & JR84 \\
                        & RX\,J0532.8$-$6622,  & \\
                        & 1RXS\,J053246.1$-$662203, RASS\,232,  & \\
                        & 4U\,0532$-$66 & \\
                        21 &  & NC02 \\
                        22 & 1A\,0535$-$668, RX\,J0535.6$-$6651,  & CBD83 \\
                        & LXP0.07, CAL\,G, 1A\,0538-66, & \\
                        & 1RXS\,J053539.0$-$665158 & \\
                        23 &  & BUN06,  \\
                        &  & CCC16 \\
                        24 & 1H\,0538$-$641, CAL\,70, 4U\,0538$-$64,  & CCH83 \\
                        & XMMU\,J053856.7$-$640503,  & \\
                        & 1RXS\,J053855.6$-$640457,  & \\
                        & 1RXS\,J053855.5$-$640457,  & \\
                        & RX\,J0538.9$-$6405, 3A\,0539$-$641  & \\
                        25 & 3A\,0540$-$697, 1RXS\,J053938.8$-$694515 & HCC83,  \\
                        &  & WM84 \\
                        26 & LXP60.8 & SG05,  \\
                        &  & ITM09 \\
                        27 & 1SAX\,J0544.1$-$7100, LXP96.1,  & CNB01 \\
                        & AX\,J0548$-$704, AX\,J0544.1$-$7100,  & \\
                        & 1WGA\,J0544.1$-$7100 & \\
                        28 & LXP6.2, 1RXS\,J055007.0$-$681451 & KBB13, \\
                        & & KGH13\\
                        \noalign{\smallskip}\hline
                \end{tabular} 
                \tablefoot{
                        \tablefoottext{a}{Paper of discovery and/or identification as an HMXB (candidate).}
                        
                        \textbf{References:} See Table \ref{tab:Xray_names_cand}.
                }
        \end{table}
        
        \begin{table}[!htbp]
                \centering
                \caption{Same as Table \ref{tab:Xray_names} for confidence classes 2$-$6.} 
                \label{tab:Xray_names_cand} 
                \begin{tabular}{lm{6cm}m{12mm}} 
                        \hline\hline\noalign{\smallskip}
                        \# & X-ray Names & Ref.$^{(a)}$ \\
                        \noalign{\smallskip}\hline\noalign{\smallskip}
                        29 &  & TW \\
                        30 &  & TW \\
                        31 &  & TW \\
                        32 &  & TW \\
                        33 &  & TW \\
                        34 &  & TW \\
                        35 &  & HMG20 \\
                        36 &  & TW \\
                        37 &  & TW \\
                        38 &  & TW \\
                        39 &  & TW \\
                        40 &  & TW \\
                        41 &  & TW \\
                        42 & Swift\,J053321.3$-$684121 & SHP12, VMH18 \\
                        43 &  & TW \\
                        44 &  & TW \\
                        45 & RX\,J0541.5$-$6833 & SHP00 \\
                        46 &  & TW \\
                        47 &  & TW \\
                        48 &  & TW \\
                        49 &  & TW \\
                        50 &  & TW \\
                        51 &  & TW \\
                        52 &  & TW \\
                        53 &  & TW \\
                        \noalign{\smallskip}\hline
                \end{tabular} 
                \tablefoot{
                        \tablefoottext{a}{Paper of discovery and/or identification as an HMXB (candidate).}
                        
                        \textbf{References:}
                        HMV22: \citet{2022A&A...662A..22H},
                        VSH13: \citet{2013ATel.5540....1V},
                        SCF94: \citet{1994PASP..106..843S},
                        SCH96: \citet{1996PASP..108..668S},
                        MHM21: \citet{2021MNRAS.504..326M},
                        MHS13: \citet{2013A&A...554A...1M},
                        HMC20: \citet{2020ATel13609....1H},
                        HIR17: \citet{2017A&A...598A..69H},
                        KCE09: \citet{2009ATel.2011....1K},
                        CFB15: \citet{2015MNRAS.447.1630C},
                        CSM97: \citet{1997PASP..109...21C},
                        KdBB01: \citet{2001A&A...371..816K},
                        CNB01: \citet{2001MNRAS.324..623C},
                        HP99: \citet{1999A&AS..139..277H},
                        vJBM18: \citet{2018MNRAS.475.3253V},
                        MHC20: \citet{2020ATel13610....1M},
                        MHK20: \citet{2020ATel13650....1M},
                        HDP97: \citet{1997A&A...318..490H},
                        HDP03: \citet{2003A&A...406..471H},
                        VMH13: \citet{2013A&A...558A..74V},
                        VMH18: \citet{2018MNRAS.475..220V},
                        HPD95: \citet{1995A&A...303L..49H},
                        JR84: \citet{1984ARA&A..22..537J},
                        NC02: \citet{2002A&A...385..517N},
                        CBD83: \citet{1983MNRAS.202..657C},
                        BUN06: \citet{2006A&A...450..585B},
                        CCC16: \citet{2016ApJ...829..105C},
                        CCH83: \citet{1983ApJ...272..118C},
                        HCC83: \citet{1983ApJ...275L..43H},
                        WM84: \citet{1984ApJ...281..354W},
                        SG05: \citet{2005A&A...431..597S},
                        ITM09: \citet{2009MNRAS.395.1662I},
                        KBB13: \citet{2013ATel.5286....1K},
                        KGH13: \citet{2013ATel.5309....1K},
                        HMG20: \citet{2020ATel13789....1H},
                        SHP12: \citet{2012ATel.3993....1S},
                        SHP00: \citet{2000A&AS..143..391S}.
                }
        \end{table}
        
        The merged \ero spectra, \ero light curves and long-term light curves for all systems investigated in this work can be found in Fig.\,\ref{fig:eRO_spectra}, Fig.\,\ref{fig:eRO_LCs} and Fig.\,\ref{fig:UL_lightcurves}, respectively.
        For all sources, Tables\,\ref{tab:eRASSexp_1} and \ref{tab:eRASSexp_2} list the effective eRASS:5 exposure times and luminosities derived from fits to the merged eRASS:5 spectra. Note that the effective exposures of LMC\,X$-$1 (\#25), LMC\,X$-$3 (\#24), and LMC\,X$-$4 (\#20) are significantly lower because their extraction regions were restricted to mitigate photon pile-up (see Sect.\,\ref{sec:methodology}). Tables\,\ref{tab:Xray_names} and \ref{tab:Xray_names_cand} list alternative names of the X-ray sources in our catalogue, and their discovery and/or identification papers as (candidate) HMXBs. Table\,\ref{tab:MasterTable_known_DETUID} lists the DETUIDs under which sources in our catalogue can be found in the eRASS1 catalogue.

    \begin{table}
        \centering
                \caption{eRASS1 DETUIDs.}
                \label{tab:MasterTable_known_DETUID}
                \begin{tabular}{l|l}
                        \hline\hline\noalign{\smallskip}
                        \# & DETUID \\
                        \noalign{\smallskip}\hline\noalign{\smallskip}
                        1 & em01\_074159\_020\_ML00045\_009\_c010 \\
                        2$^{(a)}$ & em01\_074159\_020\_ML01344\_009\_c010 \\
                        3 & em01\_074159\_020\_ML00034\_009\_c010 \\
                        4$^{(a)}$ & em01\_077162\_020\_ML00586\_009\_c010 \\
                        5 & em01\_074159\_020\_ML00301\_009\_c010 \\
                        6$^{(b)}$ & -- \\
                        7 & em01\_080156\_020\_ML00031\_012\_c010 \\
                        8$^{(a)}$ & em01\_074159\_020\_ML00432\_009\_c010 \\
                        9 & em01\_080156\_020\_ML00024\_012\_c010 \\
                        10 & em01\_082159\_020\_ML00210\_010\_c010 \\
                        11$^{(a)}$ & em01\_082159\_020\_ML01885\_010\_c010 \\
                        12 & em01\_080156\_020\_ML00039\_012\_c010 \\
                        13 & em01\_080156\_020\_ML00059\_012\_c010 \\
                        14 & em01\_080156\_020\_ML00050\_012\_c010 \\
                        15 & em01\_080156\_020\_ML00013\_012\_c010 \\
                        16 & em01\_080156\_020\_ML00062\_012\_c010 \\
                        17 & em01\_080156\_020\_ML00015\_012\_c010 \\
                        18 & em01\_082159\_020\_ML00704\_010\_c010 \\
                        19 & em01\_087156\_020\_ML00127\_011\_c010 \\
                        20 & em01\_087156\_020\_ML00002\_011\_c010 \\
                        21$^{(a)}$ & em01\_087156\_020\_ML03969\_011\_c010 \\
                        22 & em01\_087156\_020\_ML00369\_011\_c010 \\
                        23 & em01\_082159\_020\_ML00017\_010\_c010 \\
                        24 & em01\_085153\_020\_ML00001\_002\_c010 \\
                        25 & em01\_082159\_020\_ML00001\_010\_c010 \\
                        26 & em01\_082159\_020\_ML00044\_010\_c010 \\
                        27$^{(a)}$ & em01\_086162\_020\_ML00663\_011\_c010 \\
                        28 & em01\_090159\_020\_ML00084\_008\_c010 \\
                \noalign{\smallskip}\hline\noalign{\smallskip}
                        29 & em01\_074159\_020\_ML00481\_009\_c010 \\
                        30 & em01\_074159\_020\_ML00402\_009\_c010 \\
                        31 & em01\_074159\_020\_ML00577\_009\_c010 \\
                        32 & em01\_073156\_020\_ML00370\_003\_c010 \\
                        33 & em01\_073156\_020\_ML00642\_003\_c010 \\
                        34 & em01\_073156\_020\_ML00665\_003\_c010 \\
                        35 & em01\_074159\_020\_ML00017\_009\_c010 \\
                        36 & em01\_077162\_020\_ML00158\_009\_c010 \\
                        37 & em01\_080156\_020\_ML00358\_012\_c010 \\
                        38 & em01\_080156\_020\_ML00375\_012\_c010 \\
                        39 & em01\_080156\_020\_ML00972\_012\_c010 \\
                        40 & em01\_080156\_020\_ML00602\_012\_c010 \\
                        41 & em01\_080156\_020\_ML00271\_012\_c010 \\
                        42 & em01\_082159\_020\_ML00375\_010\_c010 \\
                        43$^{(c)}$ & em01\_082159\_020\_ML00593\_010\_c010 \\
                        44 & em01\_082159\_020\_ML00119\_010\_c010 \\
                        45 & em01\_082159\_020\_ML00102\_010\_c010 \\
                        46 & em01\_087156\_020\_ML02678\_011\_c010 \\
                        47$^{(d)}$ & em01\_087156\_020\_ML02478\_011\_c010 \\
                        48 & em01\_087156\_020\_ML01942\_011\_c010 \\
                        49 & em01\_087156\_020\_ML00738\_011\_c010 \\
                        50 & em01\_087156\_020\_ML01184\_011\_c010 \\
                        51$^{(d)}$ & em01\_090159\_020\_ML01821\_008\_c010 \\
                        52 & em01\_090159\_020\_ML00615\_008\_c010 \\
                        53 & em01\_090159\_020\_ML01287\_008\_c010 \\
                        \noalign{\smallskip}\hline\noalign{\smallskip}
                        \noalign{\smallskip}\hline\noalign{\smallskip}
                \end{tabular}
                \tablefoot{
                        \tablefoottext{a}{eRASS1 counterpart with DET\_LIKE<20.}
                        \tablefoottext{b}{eRASS1 counterpart belongs to the 3B catalogue, and therefore no eRASS1 DETUID.}
                        \tablefoottext{c}{eRASS1 counterpart shows EXT\_LIKE>0 but \xmm counterpart does not show extent.}
                        \tablefoottext{d}{eRASS1 1B counterpart with DET\_LIKE<20, 3B counterpart DET\_LIKE>20. 1B DETUID noted for reference despite selection from 3B catalogue.}
                }
        \end{table}
        
        \clearpage

        \clearpage
        \thispagestyle{empty}
        \renewcommand{\arraystretch}{1.3}
        \begin{table}[h]
        \centering
        \caption{\ero effective eRASS:5 exposures (corrected for vignetting) and best-fit luminosities for sources with confidence class 1. The energy band used for spectral fitting and determining the effective exposure used was 0.2$-$5\,keV, except for objects with the entry 'soft' in the column 'E band', for which 0.2$-$2\,keV were used.} 
        \label{tab:eRASSexp_1} 
        \begin{tabular}{llll} 
        \hline\hline\noalign{\smallskip}
        \# & exp\_eff & L & E band \\
         & s & erg s$^{-1}$ &   \\
        \noalign{\smallskip}\hline\noalign{\smallskip}
        1 & 1993.5 & (6.8$^{+1.4}_{-1.1})\times10^{34}$ & \\
        2 & 2162.8 & (6$^{+5}_{-3})\times10^{34}$ & \\
        3 & 2113.2 & (13.2$^{+1.6}_{-1.4})\times10^{34}$ & \\
        4 & 2421.8 & (24.8$^{+1.4}_{-1.4})\times10^{34}$ & \\
        5 & 2112.1 & (10$^{+4}_{-3})\times10^{33}$ & \\
        6 & 2847.4 & (6.6$^{+1.8}_{-1.1})\times10^{34}$ & \\
        7 & 4459.5 & (11.79$^{+0.27}_{-0.27})\times10^{35}$ & \\
        8 & 2966.7 & (5.9$^{+1.2}_{-1.1})\times10^{34}$ & \\
        9 & 5271.1 & (6.12$^{+0.17}_{-0.16})\times10^{35}$ & \\
        10 & 2971.6 & (11.1$^{+1.7}_{-1.1})\times10^{34}$ & \\
        11 & 3089.7 & (19$^{+5}_{-4})\times10^{33}$ & \\
        12 & 6393.4 & (3.07$^{+0.12}_{-0.11})\times10^{35}$ & \\
        13 & 4683.5 & (13.9$^{+0.9}_{-0.8})\times10^{34}$ & \\
        14 & 7359.6 & (19.0$^{+1.1}_{-1.0})\times10^{34}$ & \\
        15 & 9035.0 & (8.02$^{+0.19}_{-0.19})\times10^{35}$ & \\
        16 & 8344.4 & (11.8$^{+0.8}_{-0.6})\times10^{34}$ & \\
        17 & 5916.1 & (3.58$^{+0.14}_{-0.14})\times10^{35}$ & \\
        18 & 3721.2 & (11.4$^{+7.5}_{-2.9})\times10^{34}$ & \\
        19 & 9107.7 & (26.2$^{+0.9}_{-1.0})\times10^{34}$ & \\
        20 & 427.2 & (18.3$^{+1.0}_{-0.9})\times10^{36}$ & \\
        21 & 6189.2 & (4.6$^{+1.0}_{-0.9})\times10^{34}$ & \\
        22 & 6934.0 & (13.3$^{+3.0}_{-2.2})\times10^{34}$ & \\
        23 & 3861.0 & (10.9$^{+1.2}_{-1.1})\times10^{34}$ & \\
        24 & 79.3 & (24.94$^{+0.24}_{-0.24})\times10^{37}$ & \\
        25 & 61.5 & (24.24$^{+0.30}_{-0.28})\times10^{37}$ & \\
        26 & 5617.5 & (3.31$^{+0.14}_{-0.13})\times10^{35}$ & \\
        27 & 3443.4 & (4.1$^{+0.9}_{-0.7})\times10^{34}$ & \\
        28 & 7602.6 & (3.4$^{+0.6}_{-0.5})\times10^{34}$ & \\

        \noalign{\smallskip}\hline
        \end{tabular} 
        \end{table}
        \renewcommand{\arraystretch}{1.0}
        
        \renewcommand{\arraystretch}{1.3}
        \begin{table}[h]
        \centering
        \caption{Same as Table \ref{tab:eRASSexp_1} for confidence classes 2$-$6.} 
        \label{tab:eRASSexp_2} 
        \begin{tabular}{llll} 
        \hline\hline\noalign{\smallskip}
        \# & exp\_eff & L & E band \\
         & s & erg s$^{-1}$ &   \\
        \noalign{\smallskip}\hline\noalign{\smallskip}
        29 & 2067.3 & (8.1$^{+1.5}_{-1.3})\times10^{34}$ & \\
        30 & 1623.9 & (17$^{+7}_{-5})\times10^{33}$ & \\
        31 & 1921.3 & (29$^{+7}_{-6})\times10^{33}$ & \\
        32 & 2585.2 & (7.9$^{+1.2}_{-1.1})\times10^{34}$ & \\
        33 & 2391.8 & (16$^{+4}_{-3})\times10^{33}$ & \\
        34 & 3851.2 & (19$^{+4}_{-3})\times10^{33}$ & \\
        35 & 2665.2 & (3.5$^{+0.5}_{-0.4})\times10^{35}$ & soft\\
        36 & 2214.4 & (3.8$^{+0.8}_{-0.6})\times10^{34}$ & \\
        37 & 4643.1 & (6.2$^{+2.7}_{-2.1})\times10^{33}$ & soft\\
        38 & 4476.3 & (23.2$^{+3.5}_{-2.9})\times10^{33}$ & \\
        39 & 6179.6 & (10.1$^{+2.3}_{-2.0})\times10^{33}$ & \\
        40 & 6831.4 & (9.9$^{+1.8}_{-1.8})\times10^{33}$ & \\
        41 & 7246.7 & (3.12$^{+0.30}_{-0.28})\times10^{34}$ & \\
        42 & 4419.8 & (13.3$^{+1.7}_{-1.4})\times10^{34}$ & \\
        43 & 3506.6 & (19$^{+5}_{-4})\times10^{33}$ & \\
        44 & 4277.6 & (6$^{+7}_{-3})\times10^{34}$ & \\
        45 & 3685.1 & (4.5$^{+0.8}_{-0.7})\times10^{34}$ & \\
        46 & 7362.1 & (24$^{+4}_{-3})\times10^{33}$ & soft\\
        47 & 7937.9 & (4.2$^{+3.3}_{-1.0})\times10^{34}$ & \\
        48 & 9535.8 & (10.7$^{+3.1}_{-2.4})\times10^{33}$ & \\
        49 & 29555.8 & (9.1$^{+0.8}_{-0.8})\times10^{33}$ & \\
        50 & 19615.8 & (15.7$^{+2.7}_{-2.4})\times10^{33}$ & \\
        51 & 12356.1 & (3.5$^{+1.0}_{-0.8})\times10^{33}$ & \\
        52 & 17731.8 & (3.6$^{+1.8}_{-1.2})\times10^{33}$ & \\
        53 & 5215.6 & (10.7$^{+2.5}_{-2.1})\times10^{33}$ & \\

        \noalign{\smallskip}\hline
        \end{tabular} 
        \end{table}
        \renewcommand{\arraystretch}{1.0}
        
        \begin{figure*}
            \centering
            \resizebox{0.33\hsize}{!}{\includegraphics{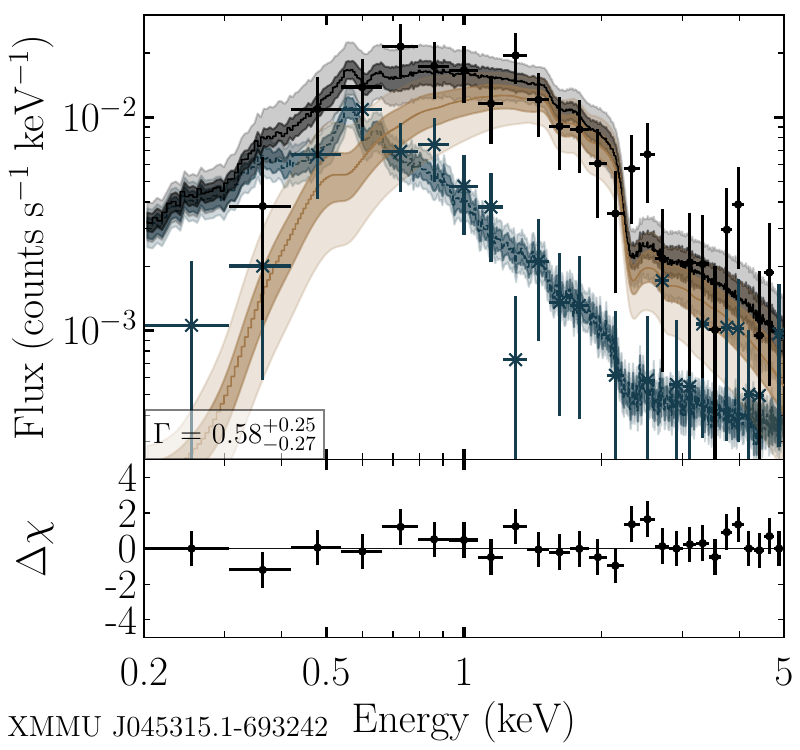}}
            \resizebox{0.33\hsize}{!}{\includegraphics{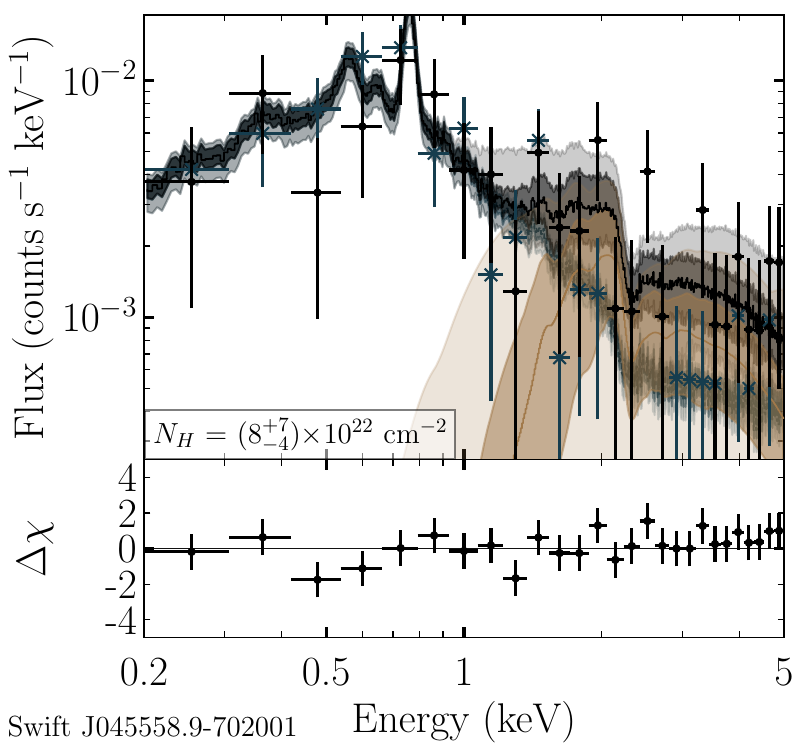}}
            \resizebox{0.33\hsize}{!}{\includegraphics{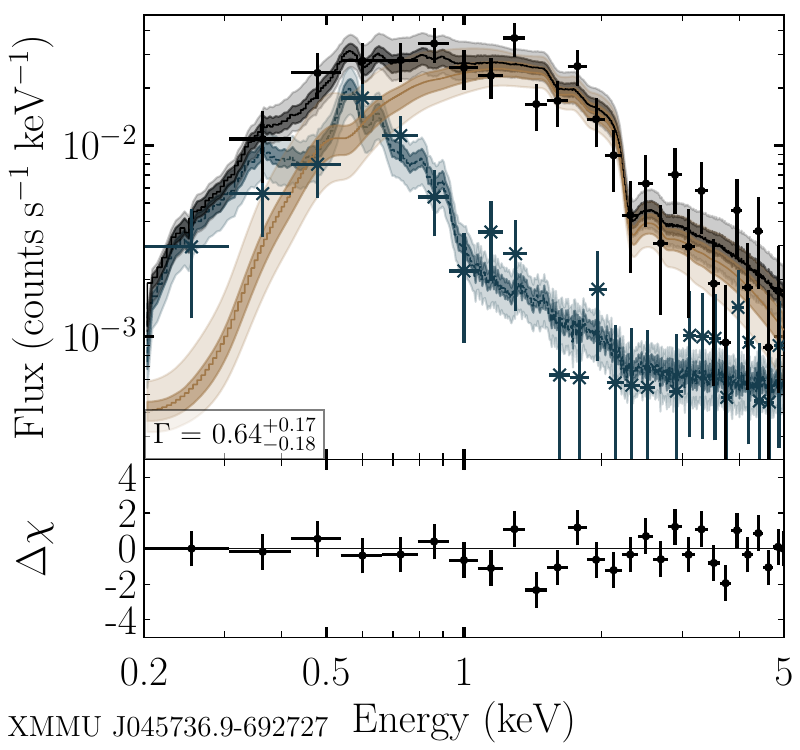}}
            \resizebox{0.33\hsize}{!}{\includegraphics{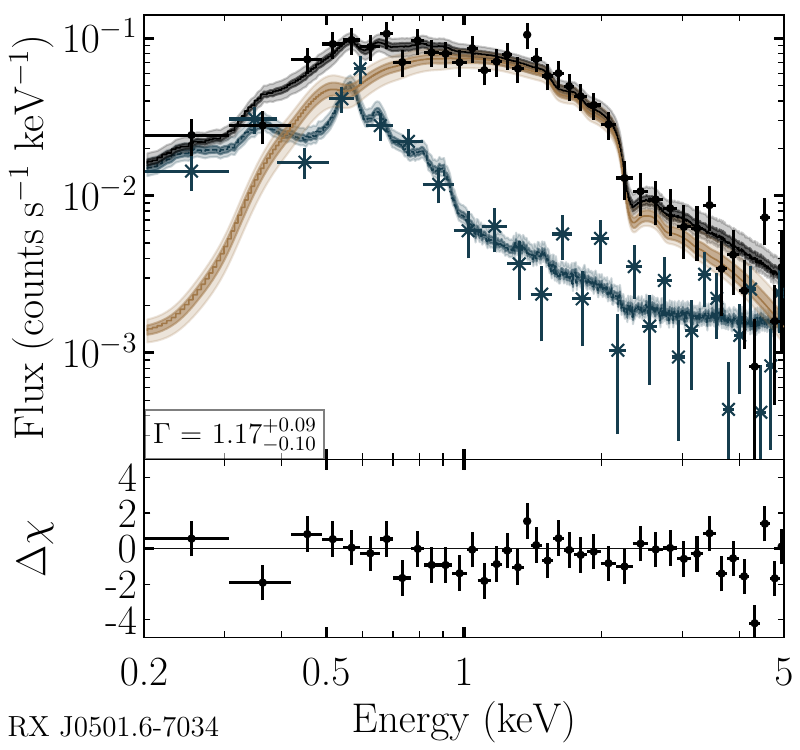}}
            \resizebox{0.33\hsize}{!}{\includegraphics{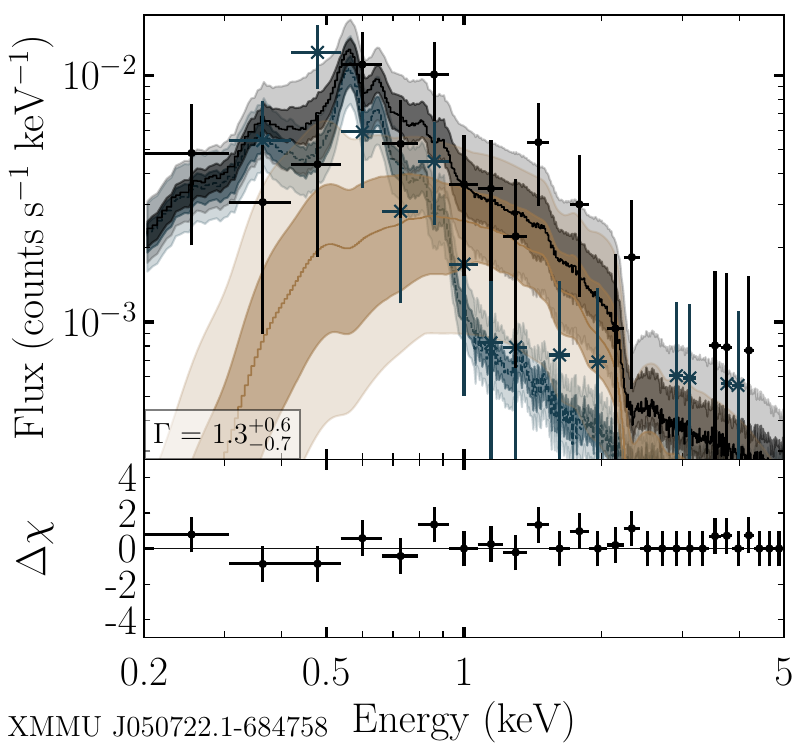}}
            \resizebox{0.33\hsize}{!}{\includegraphics{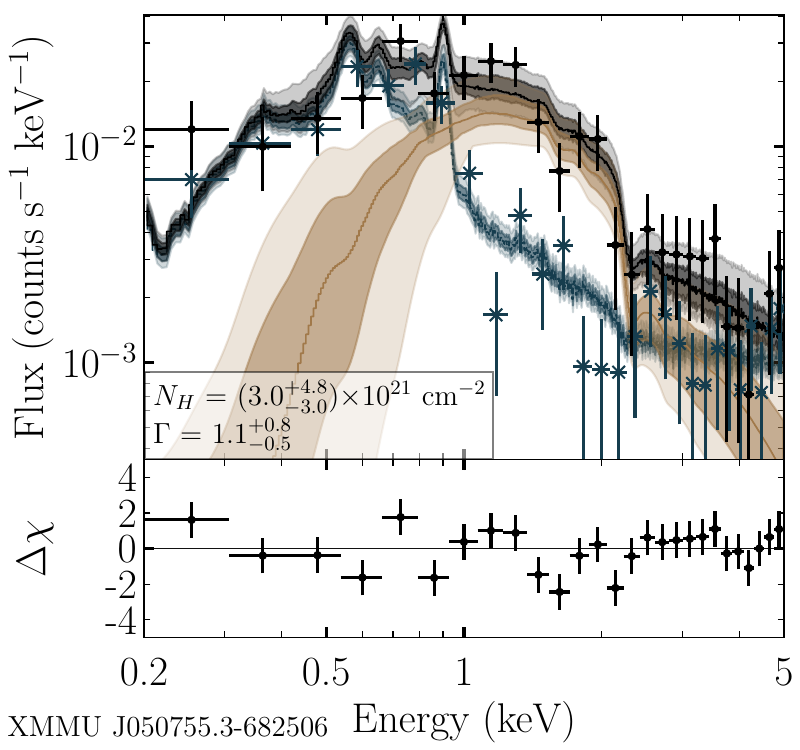}}
            \resizebox{0.33\hsize}{!}{\includegraphics{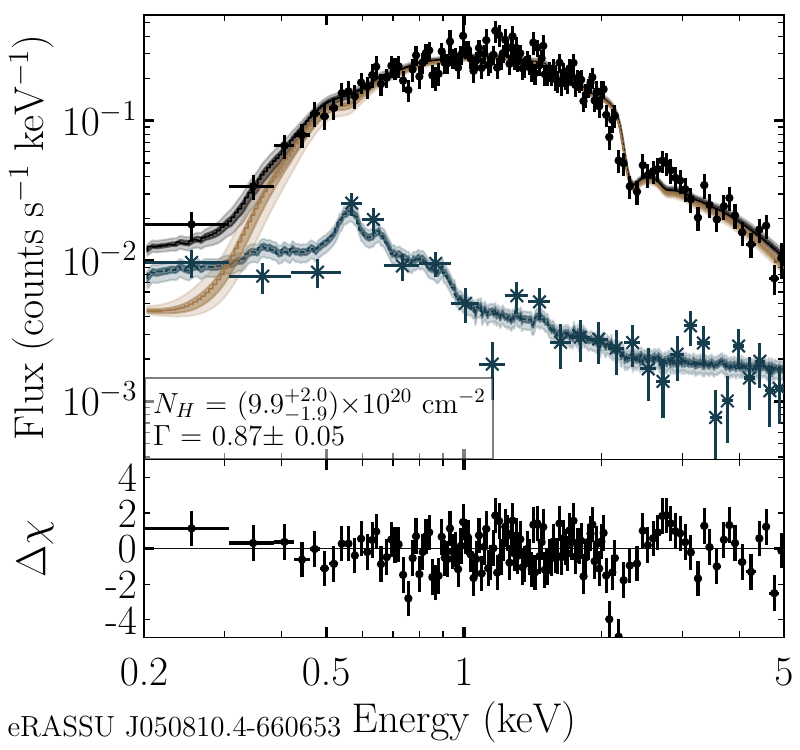}}
            \resizebox{0.33\hsize}{!}{\includegraphics{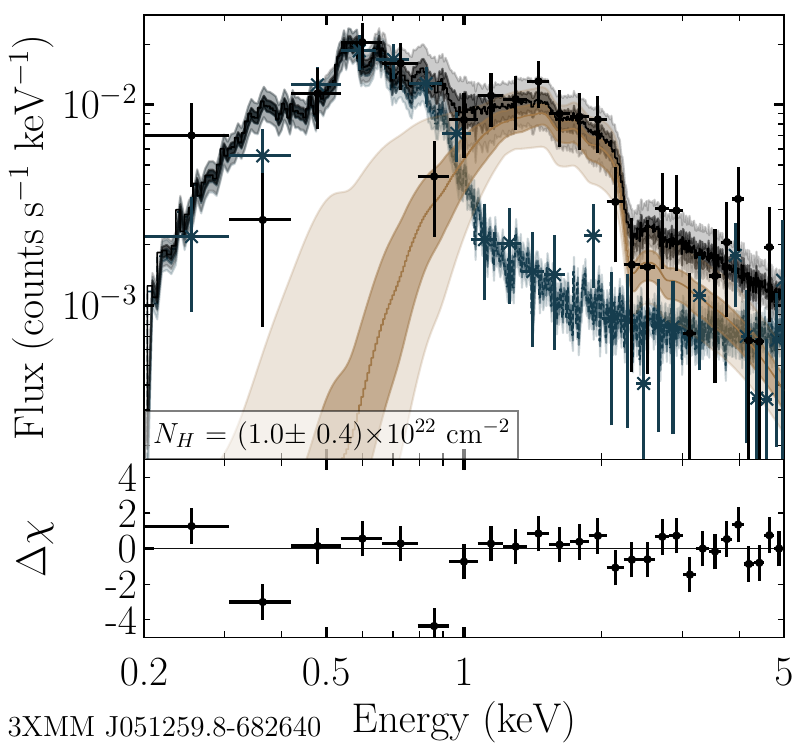}}
            \resizebox{0.33\hsize}{!}{\includegraphics{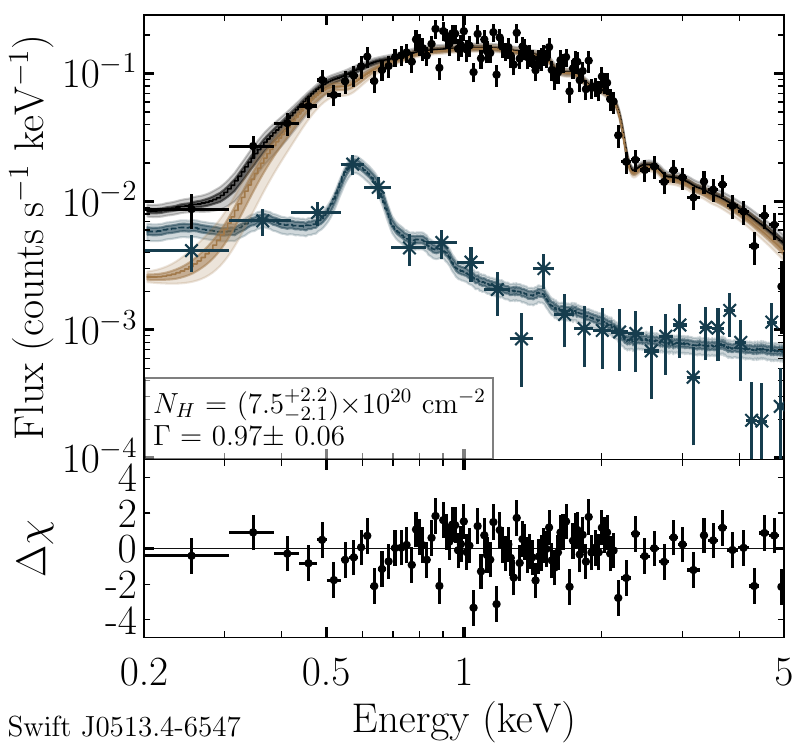}}
            \resizebox{0.33\hsize}{!}{\includegraphics{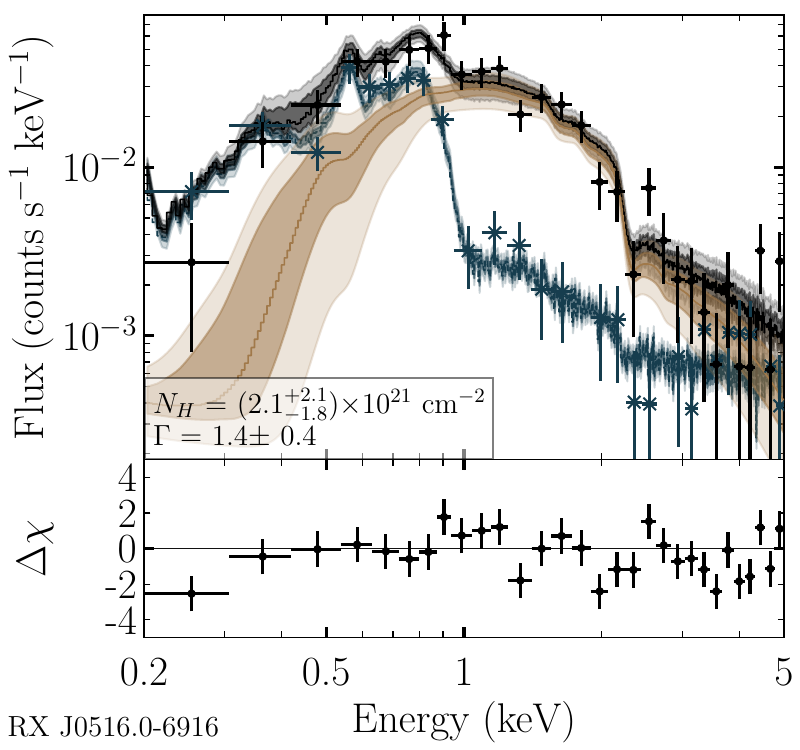}}
            \resizebox{0.33\hsize}{!}{\includegraphics{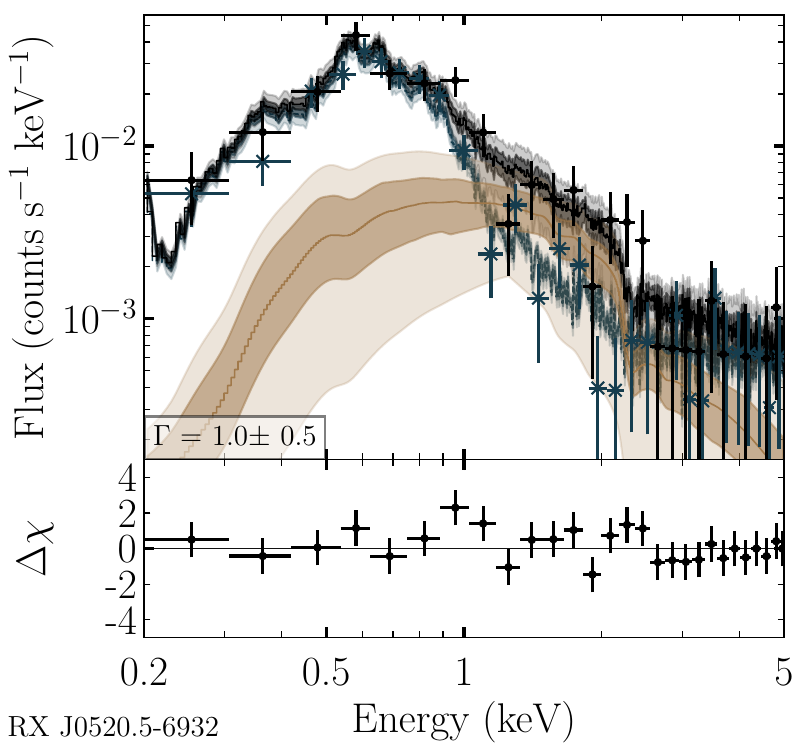}}
            \resizebox{0.33\hsize}{!}{\includegraphics{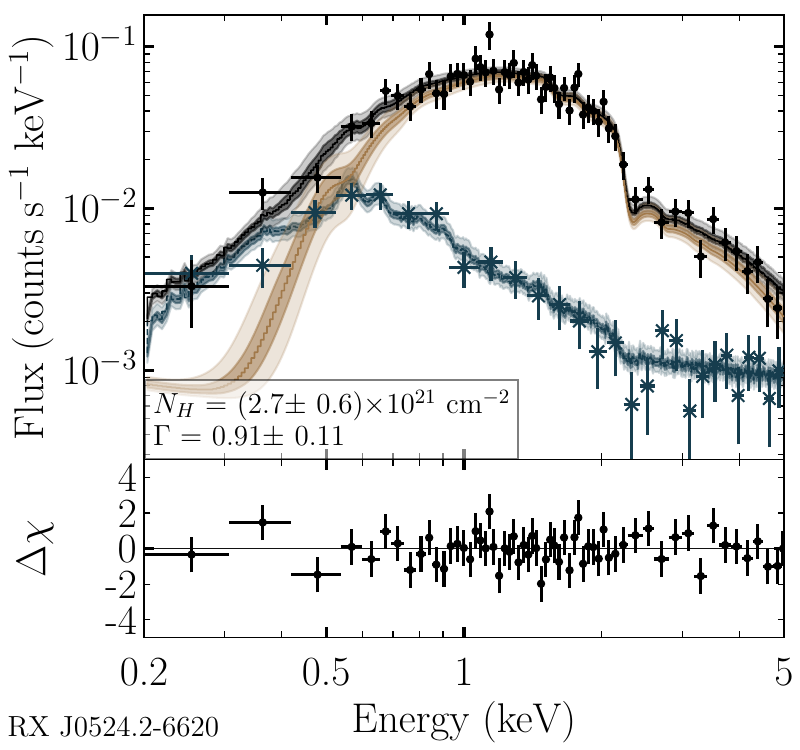}}
            \caption{Merged \ero spectra with best fit models for \#1 to \#12. Black and blue data points are fluxes in the on- and off-region, respectively. Blue and yellow lines show the best-fit models for the background and source, the black line shows the sum of both. Shaded areas display the 1$\sigma$ (dark) and 3$\sigma$ (light) confidence for the models.}
            \label{fig:eRO_spectra}
        \end{figure*}
        \addtocounter{figure}{-1}\begin{figure*}
            \centering
            \resizebox{0.33\hsize}{!}{\includegraphics{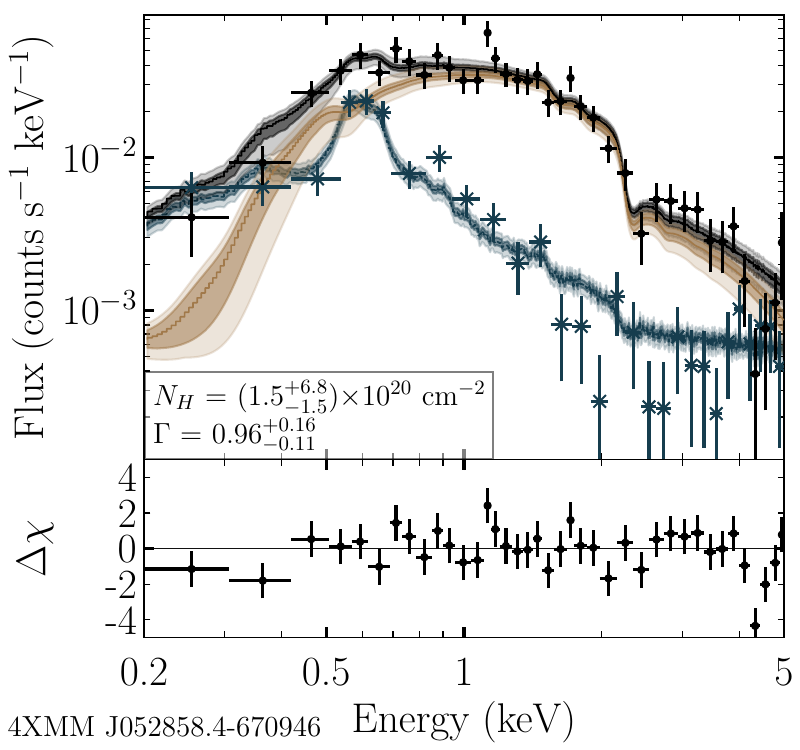}}
            \resizebox{0.33\hsize}{!}{\includegraphics{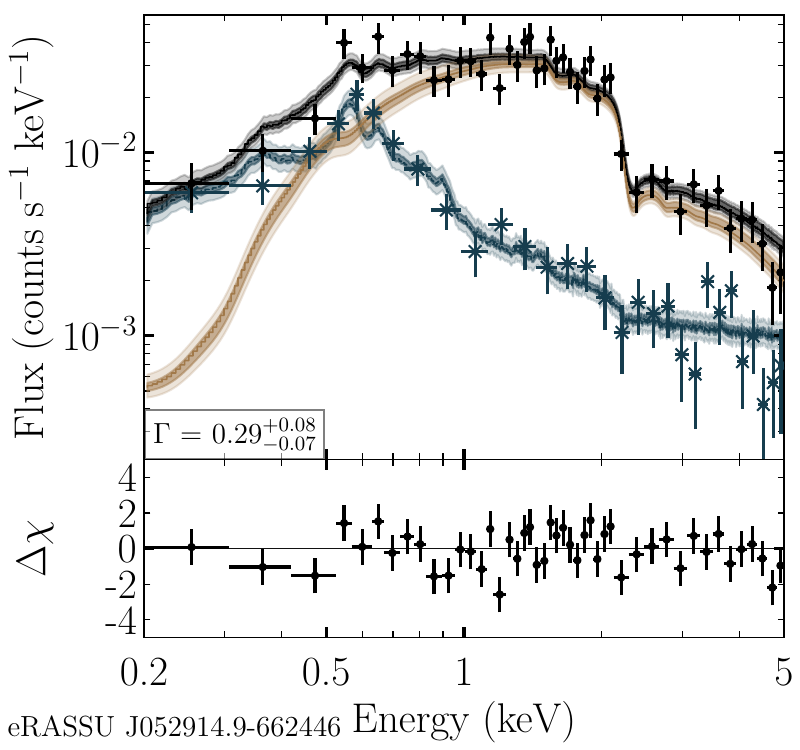}}
            \resizebox{0.33\hsize}{!}{\includegraphics{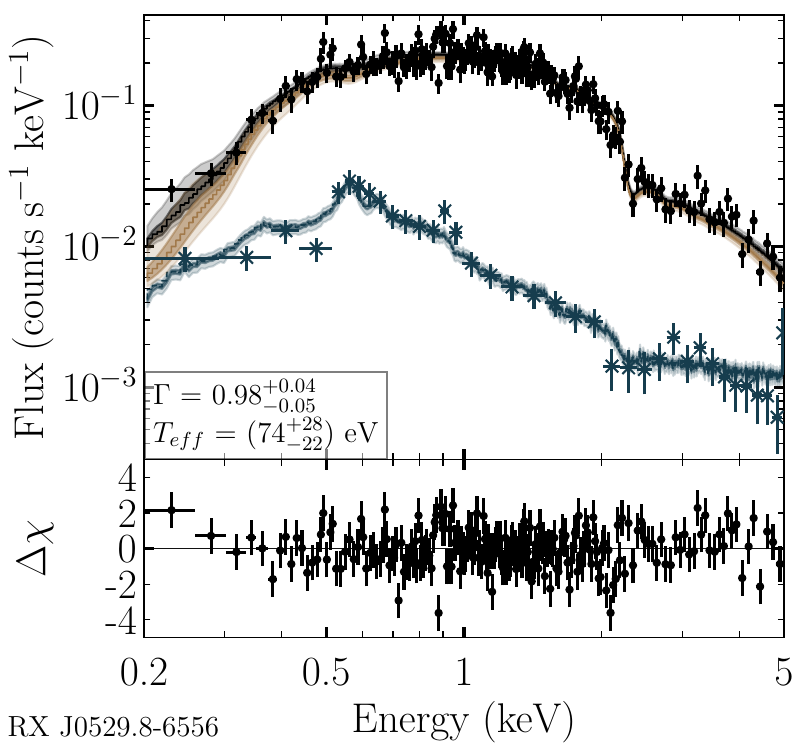}}
            \resizebox{0.33\hsize}{!}{\includegraphics{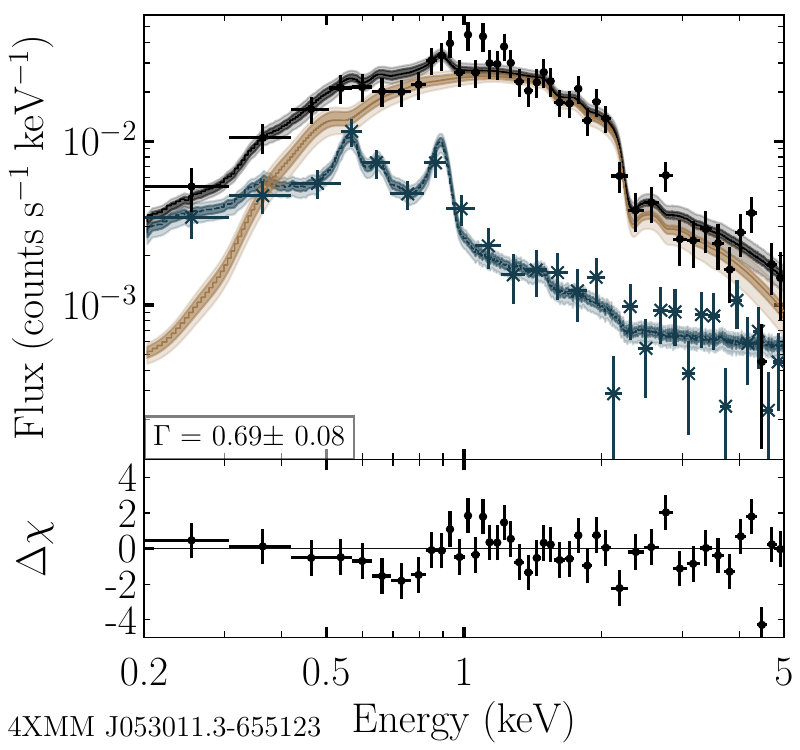}}
            \resizebox{0.33\hsize}{!}{\includegraphics{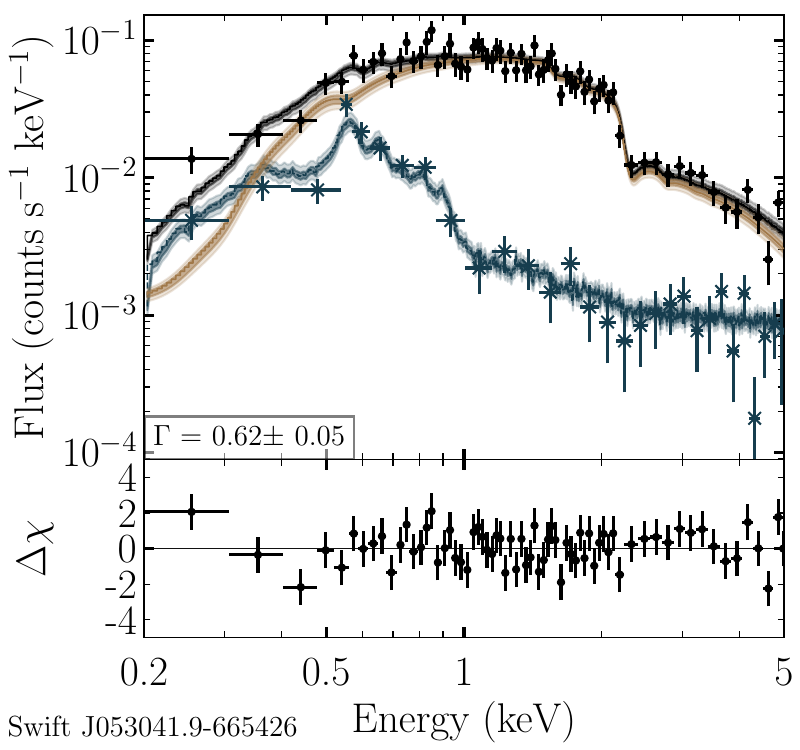}}
            \resizebox{0.33\hsize}{!}{\includegraphics{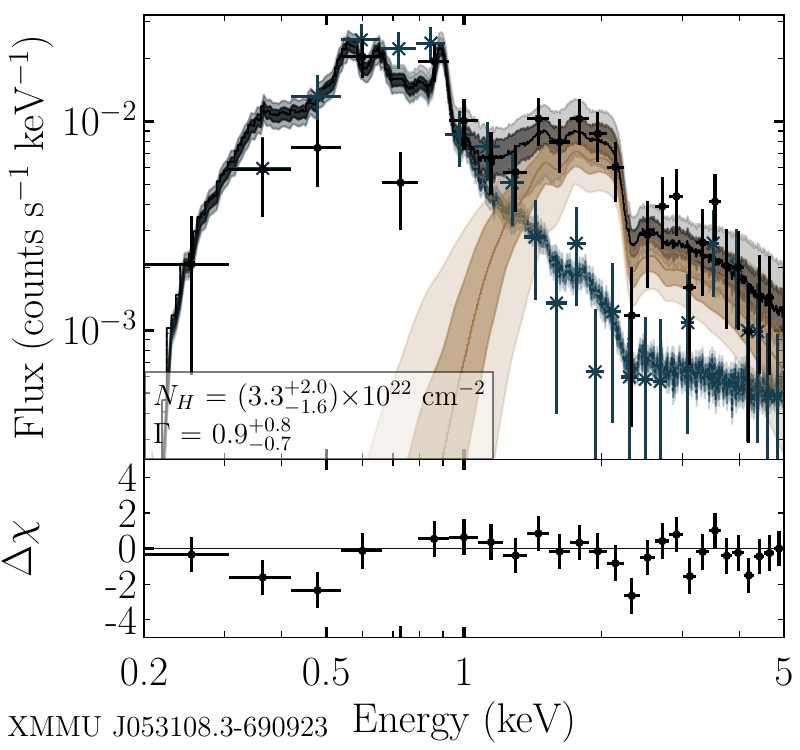}}
            \resizebox{0.33\hsize}{!}{\includegraphics{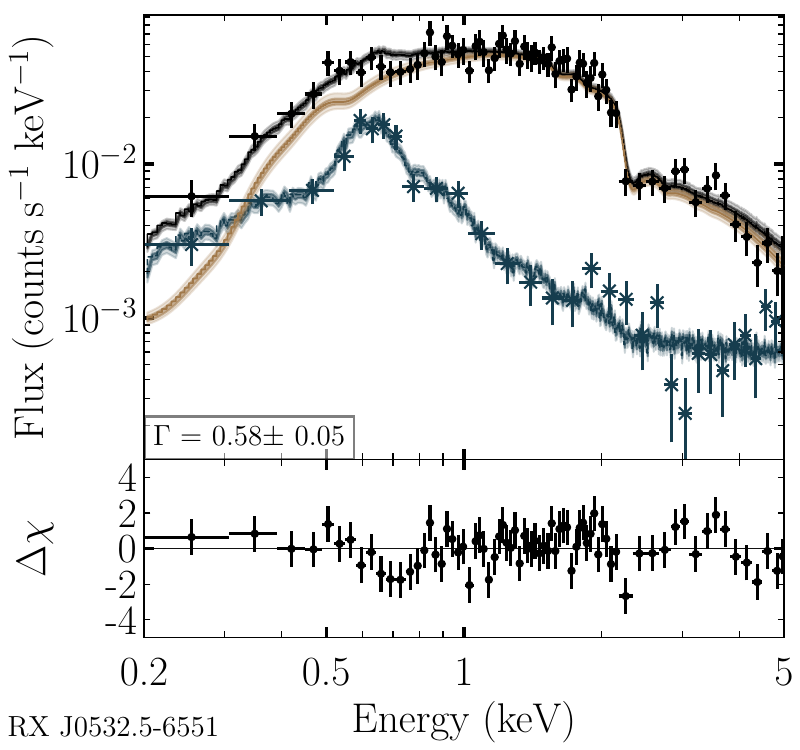}}
            \resizebox{0.33\hsize}{!}{\includegraphics{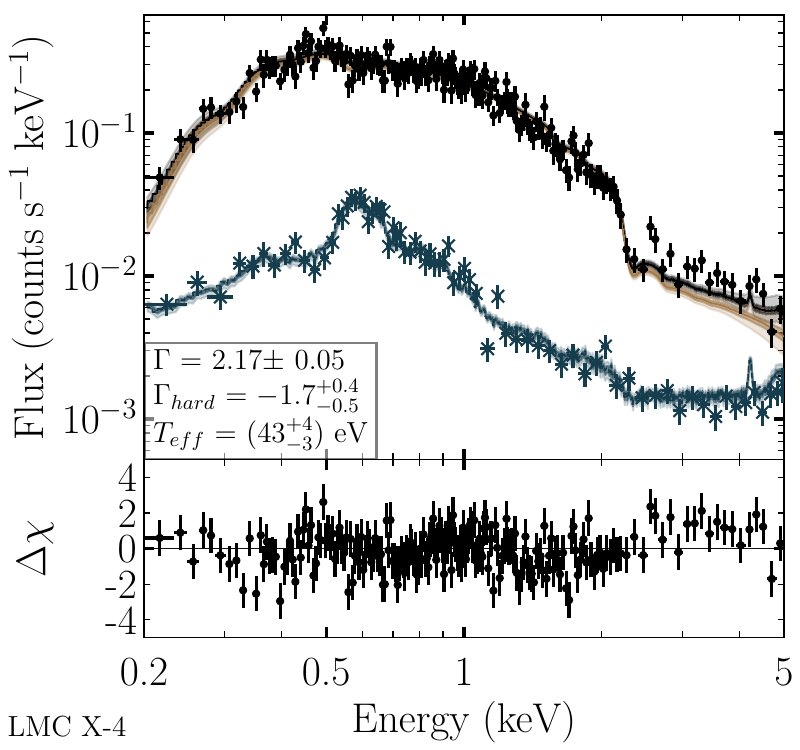}}
            \resizebox{0.33\hsize}{!}{\includegraphics{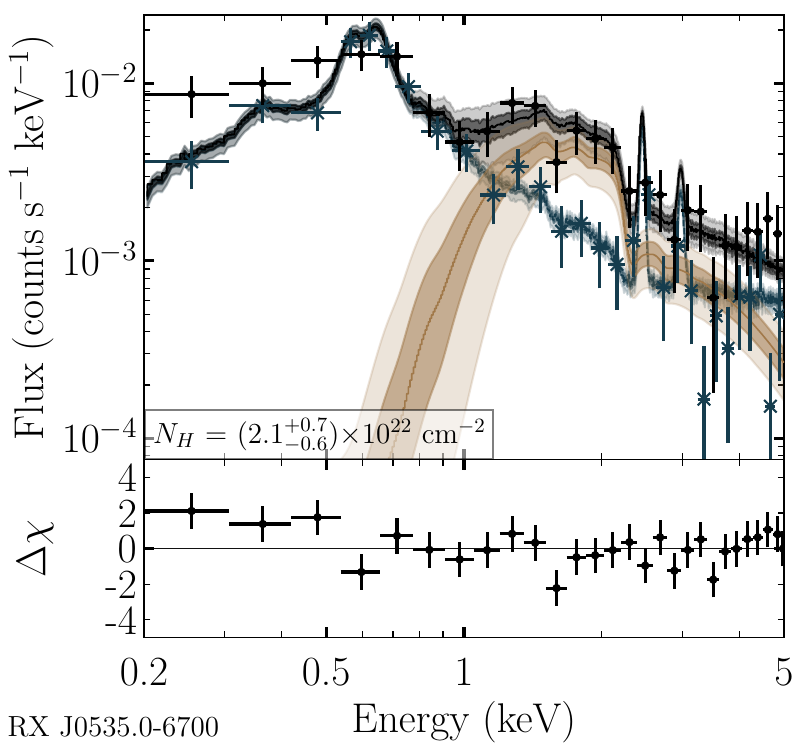}}
            \resizebox{0.33\hsize}{!}{\includegraphics{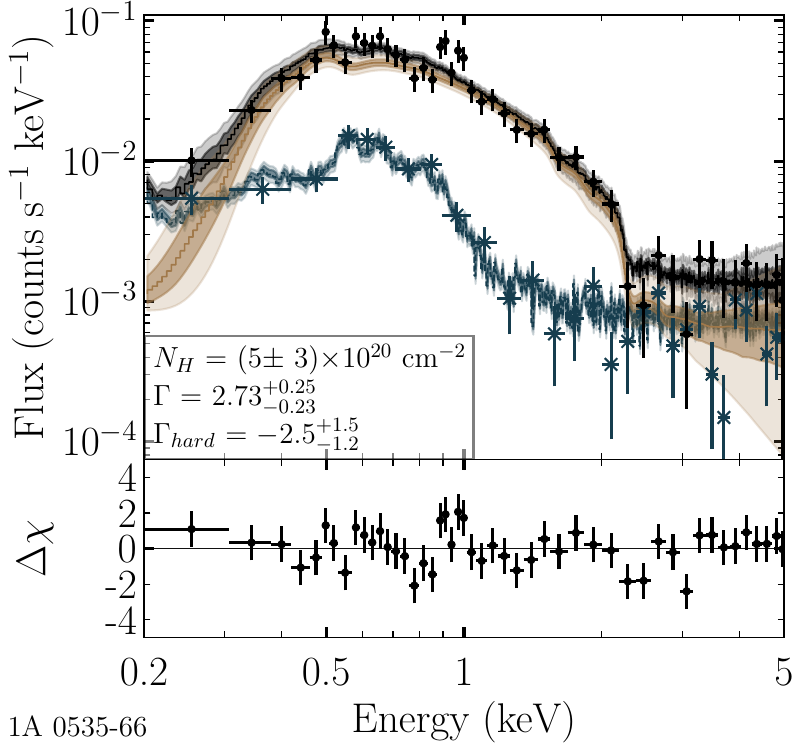}}
            \resizebox{0.33\hsize}{!}{\includegraphics{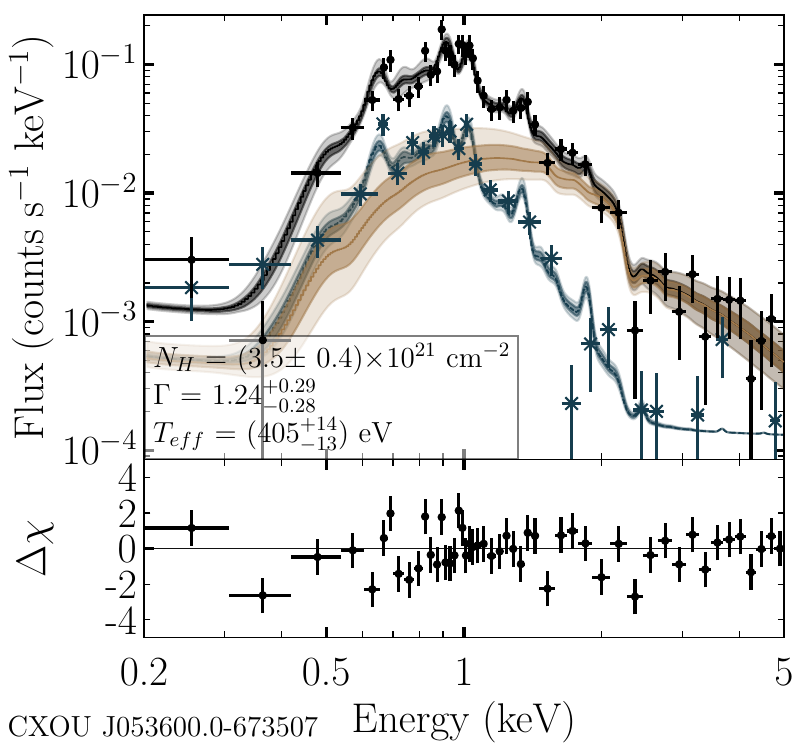}}
            \resizebox{0.33\hsize}{!}{\includegraphics{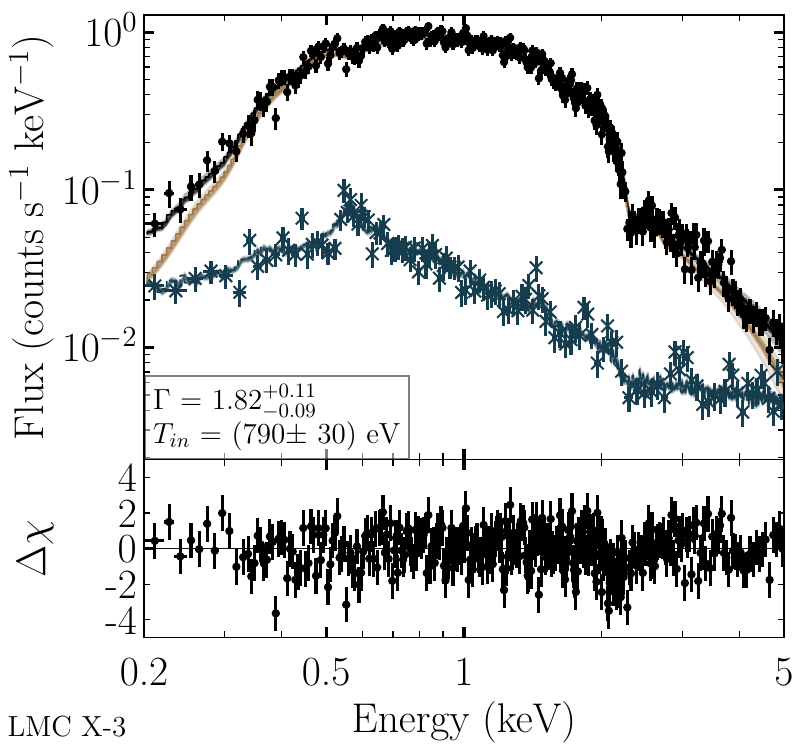}}
            \caption{continued for \#13 to \#24}
        \end{figure*}
        \addtocounter{figure}{-1}\begin{figure*}
            \centering
            \resizebox{0.33\hsize}{!}{\includegraphics{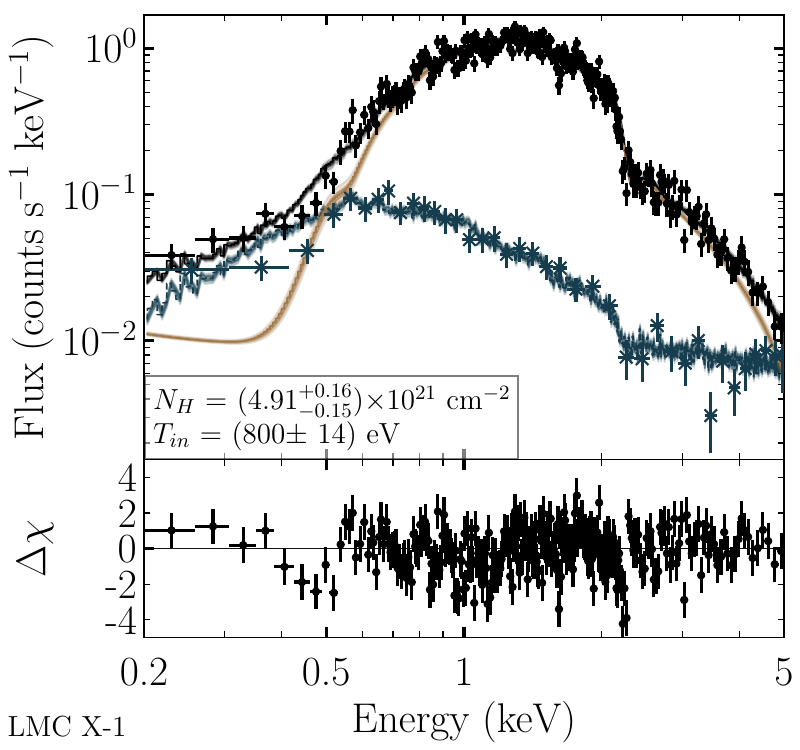}}
            \resizebox{0.33\hsize}{!}{\includegraphics{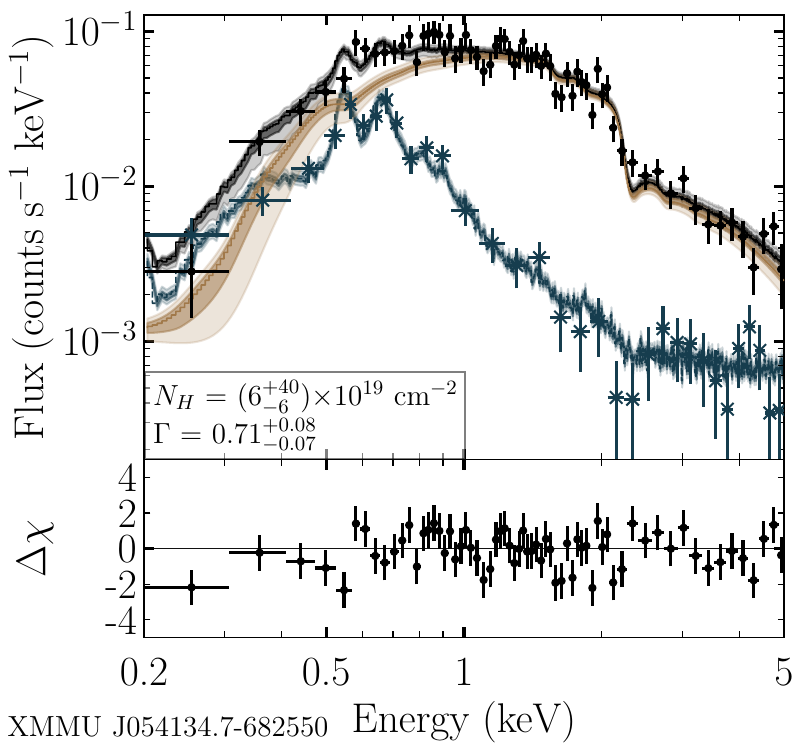}}
            \resizebox{0.33\hsize}{!}{\includegraphics{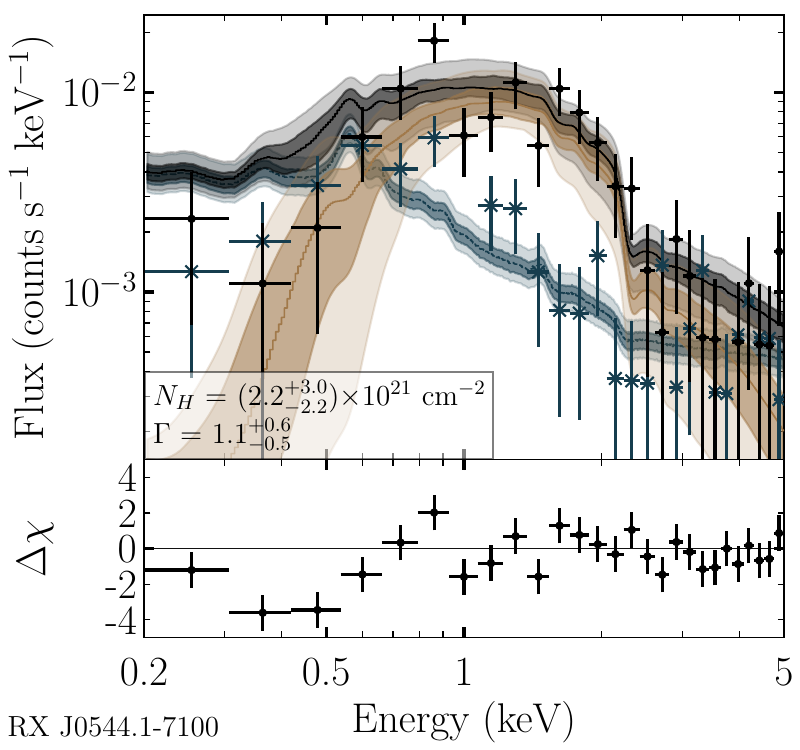}}
            \resizebox{0.33\hsize}{!}{\includegraphics{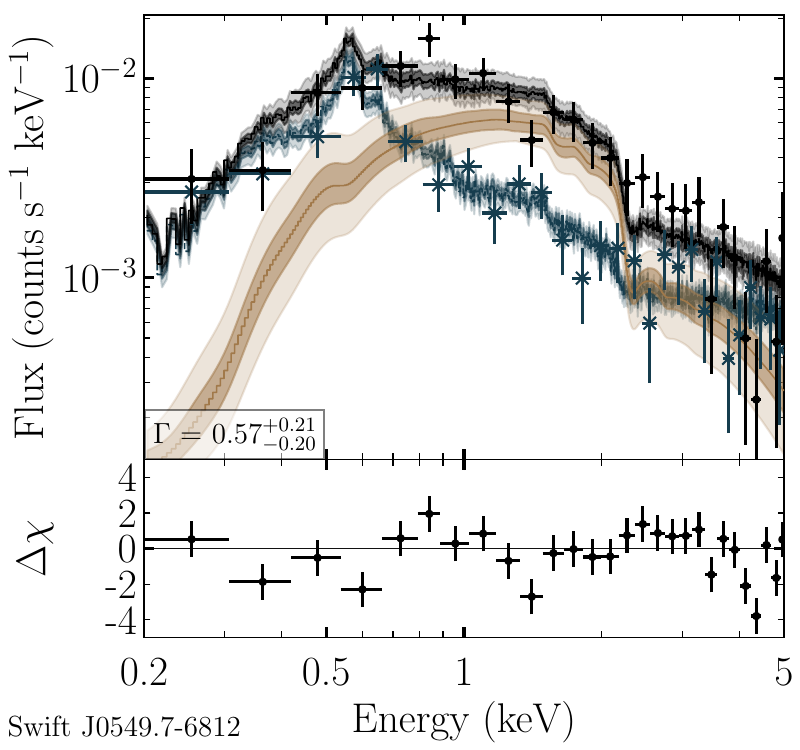}}
            \resizebox{0.33\hsize}{!}{\includegraphics{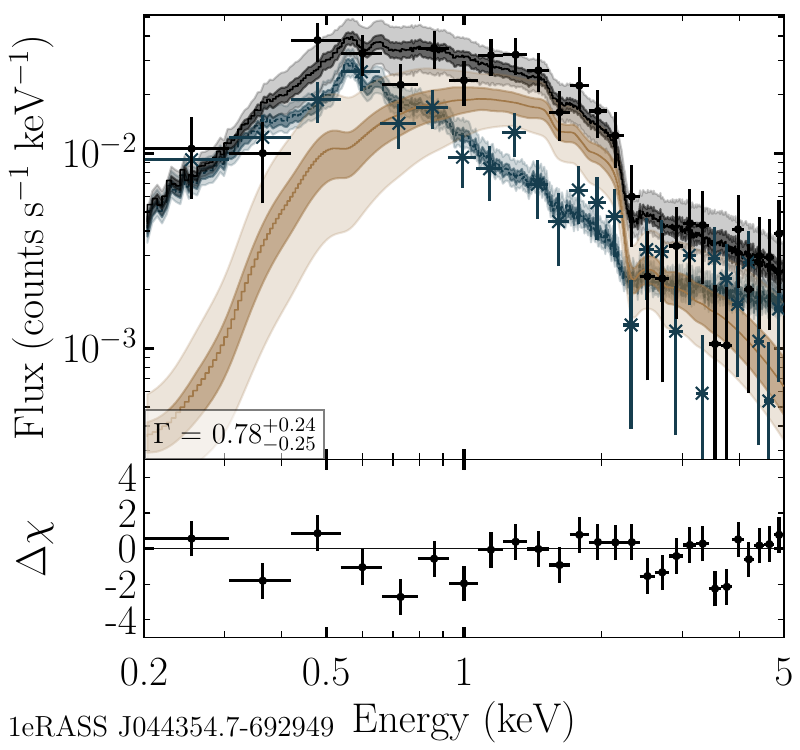}}
            \resizebox{0.33\hsize}{!}{\includegraphics{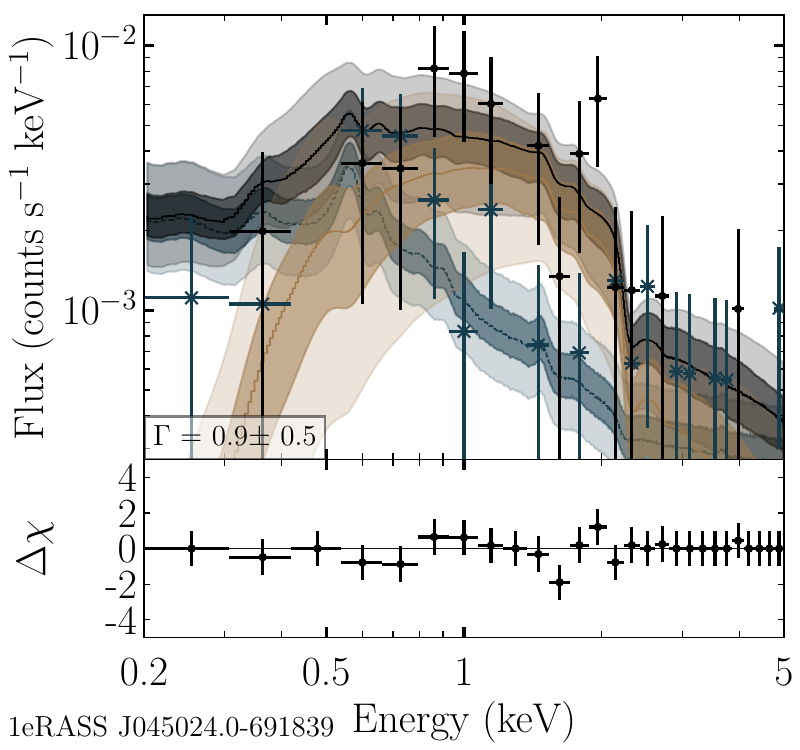}}
            \resizebox{0.33\hsize}{!}{\includegraphics{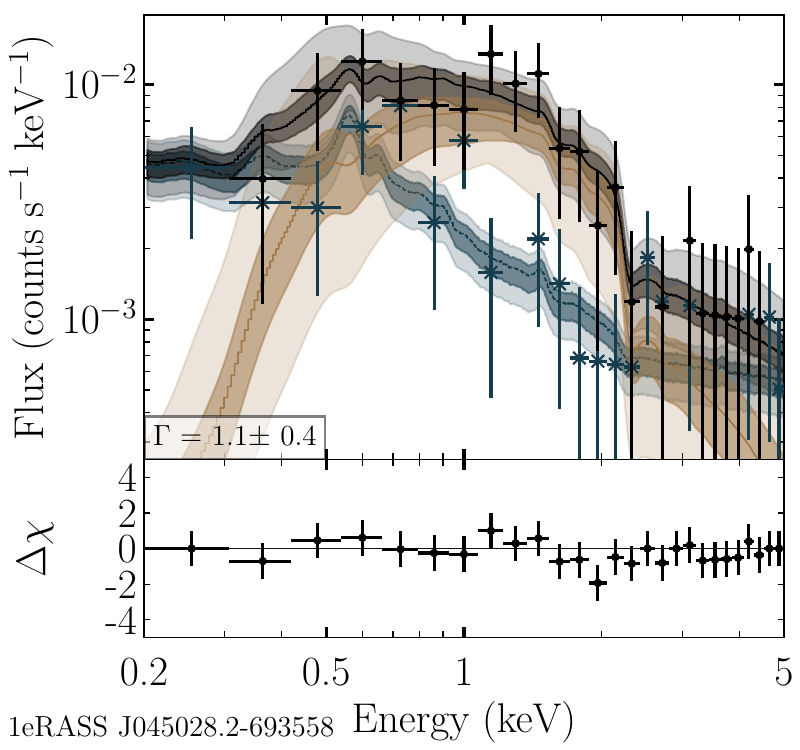}}
            \resizebox{0.33\hsize}{!}{\includegraphics{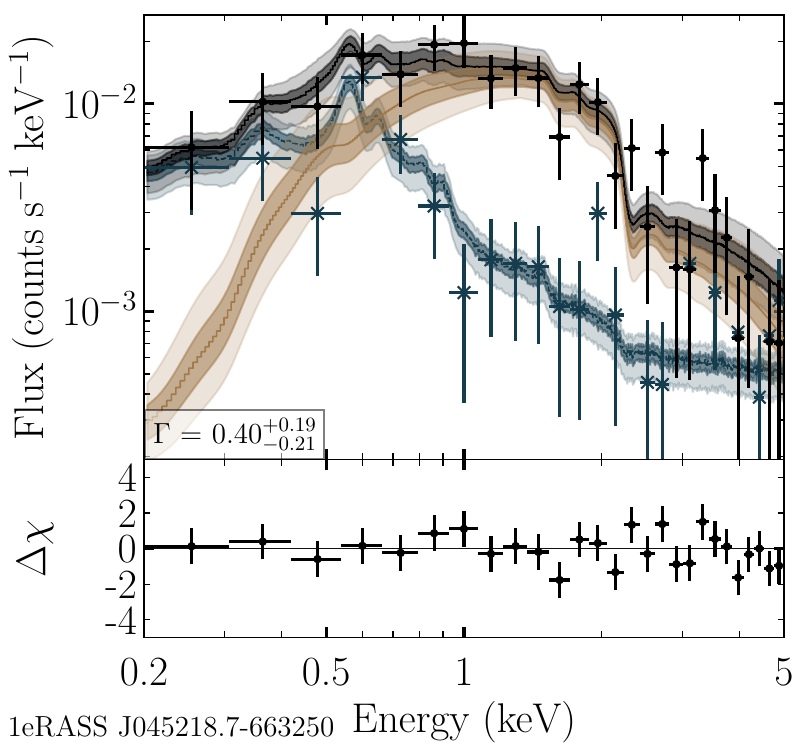}}
            \resizebox{0.33\hsize}{!}{\includegraphics{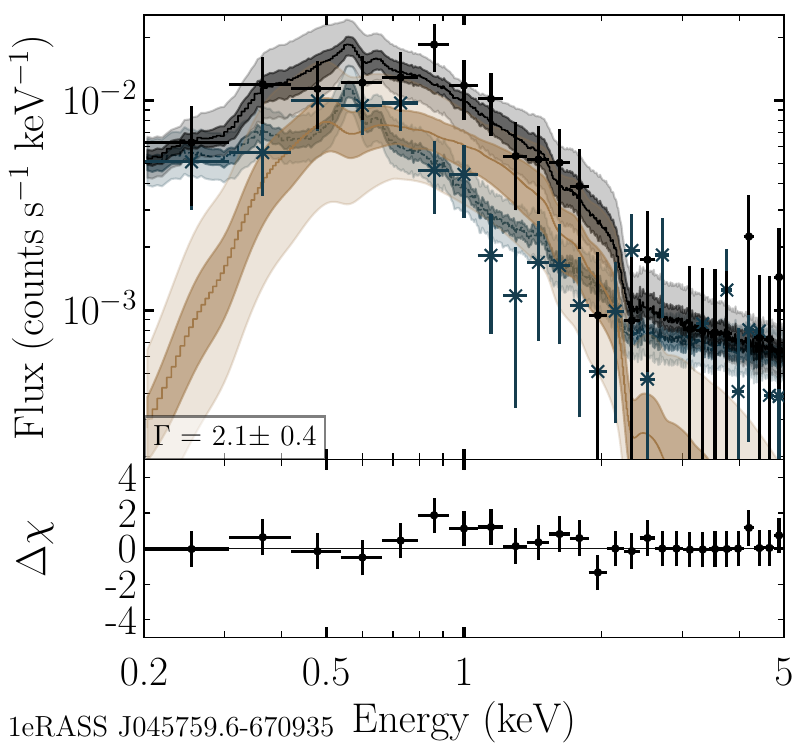}}
            \resizebox{0.33\hsize}{!}{\includegraphics{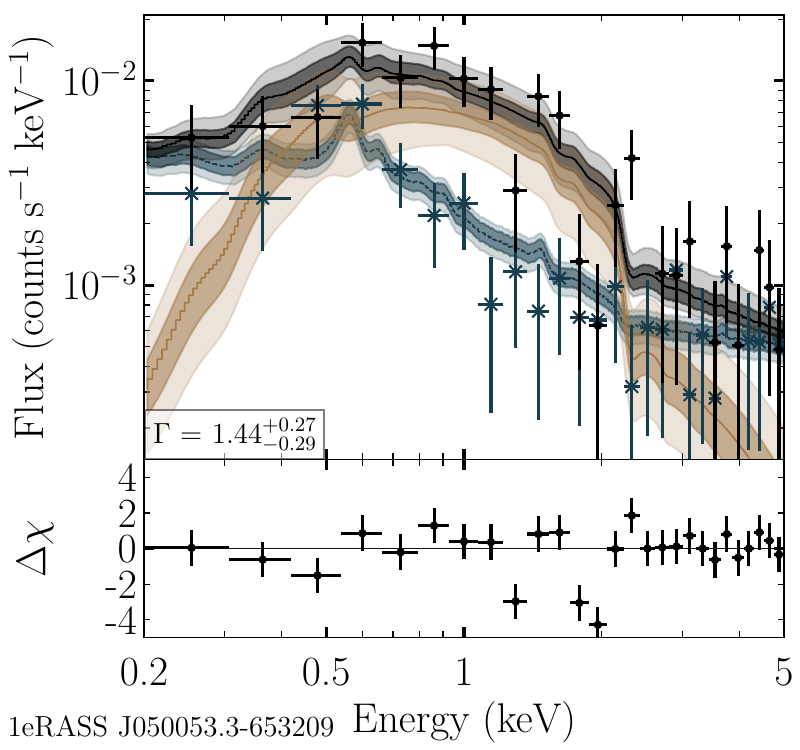}}
            \resizebox{0.33\hsize}{!}{\includegraphics{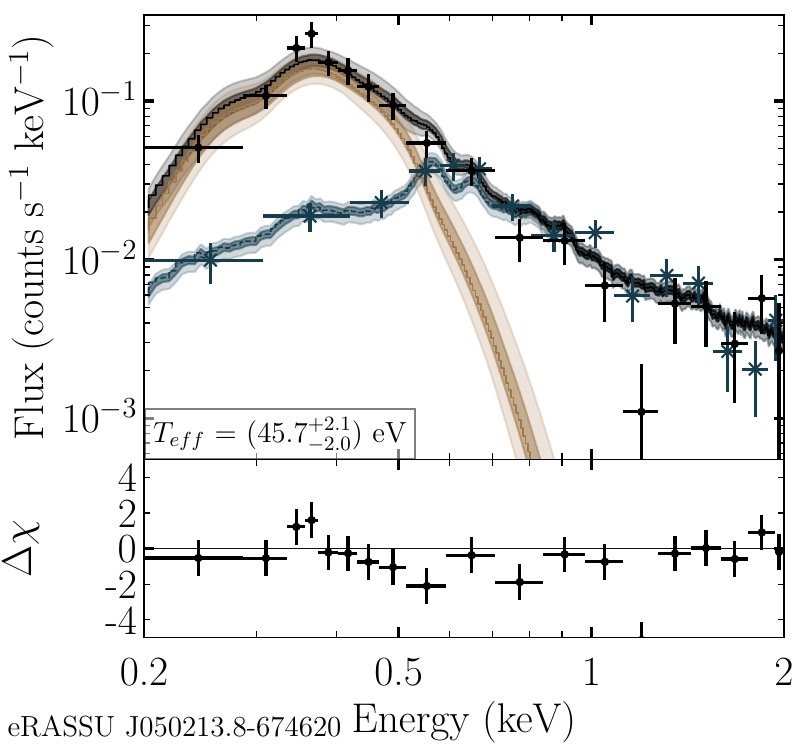}}
            \resizebox{0.33\hsize}{!}{\includegraphics{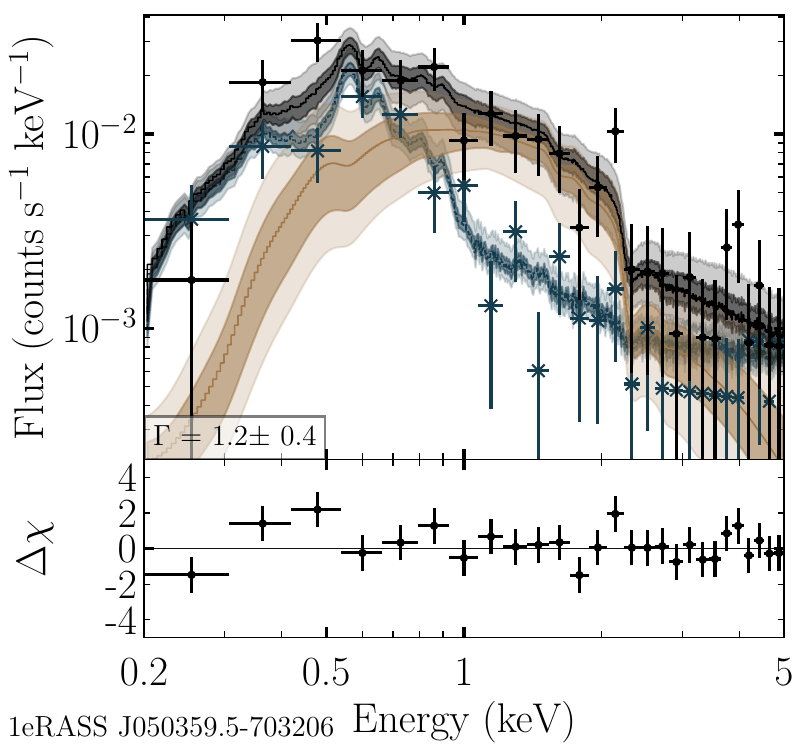}}
            \caption{continued for \#25 to \#36}
        \end{figure*}
        \addtocounter{figure}{-1}\begin{figure*}
            \centering
            \resizebox{0.33\hsize}{!}{\includegraphics{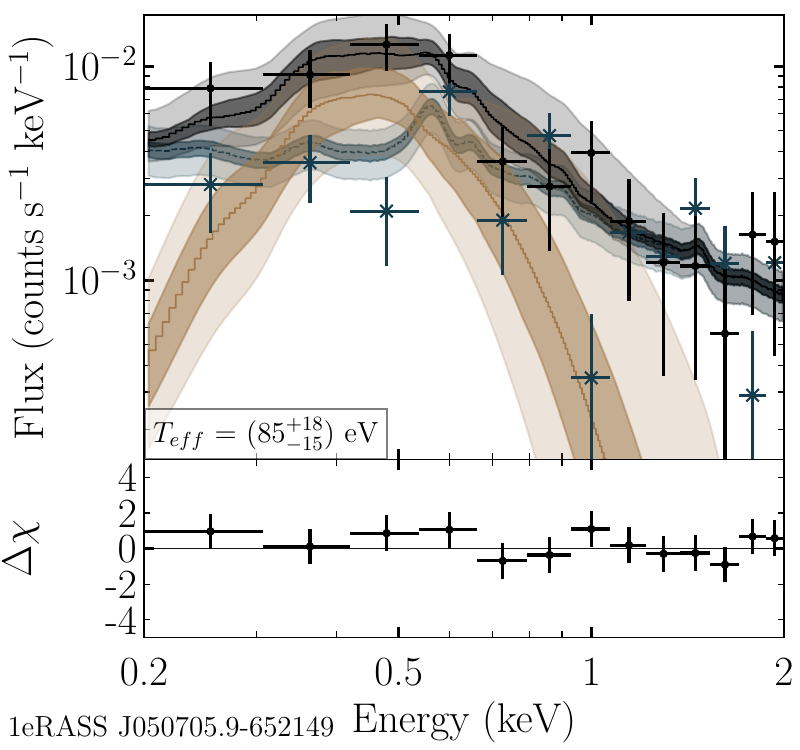}}
            \resizebox{0.33\hsize}{!}{\includegraphics{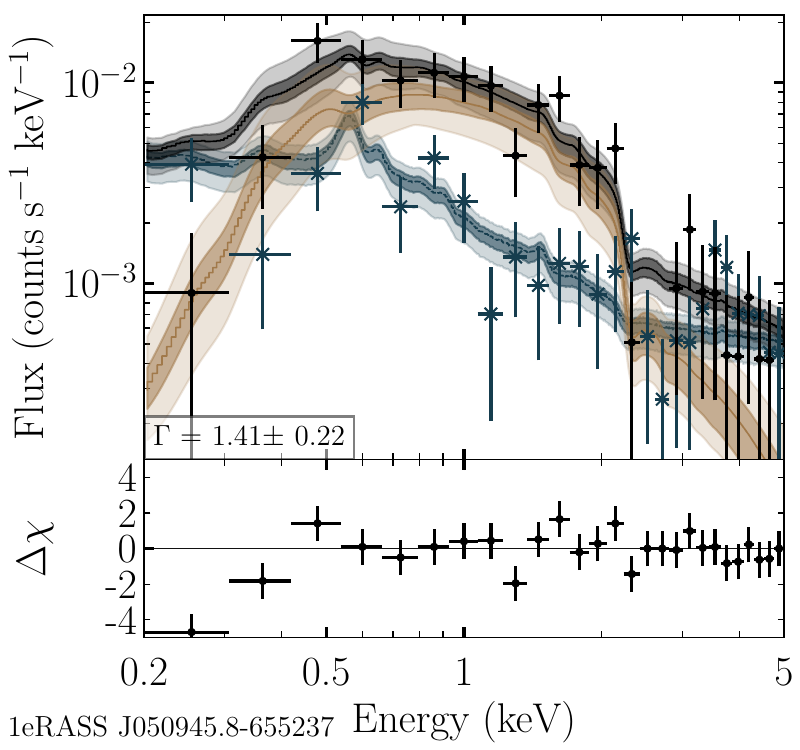}}
            \resizebox{0.33\hsize}{!}{\includegraphics{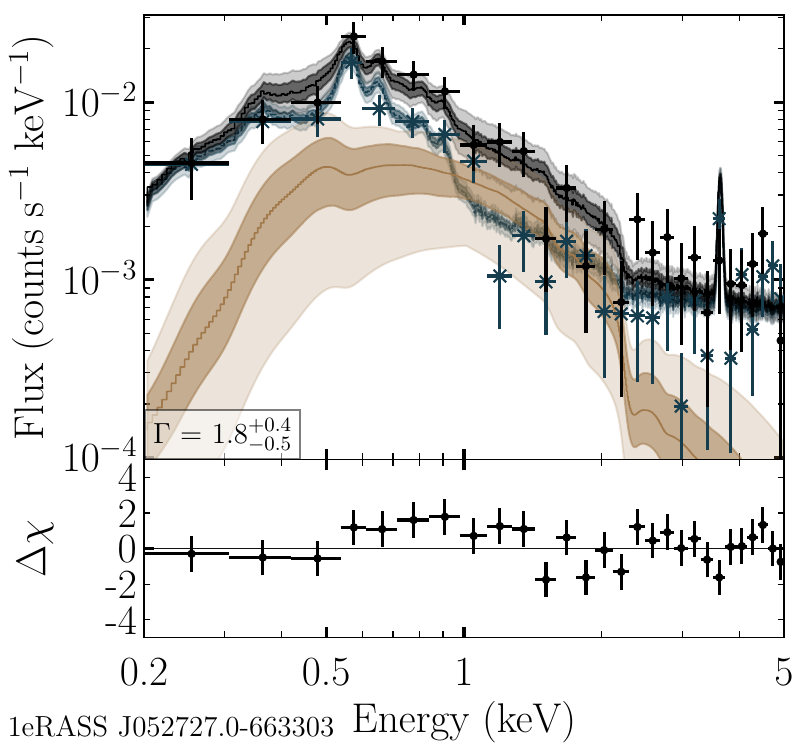}}
            \resizebox{0.33\hsize}{!}{\includegraphics{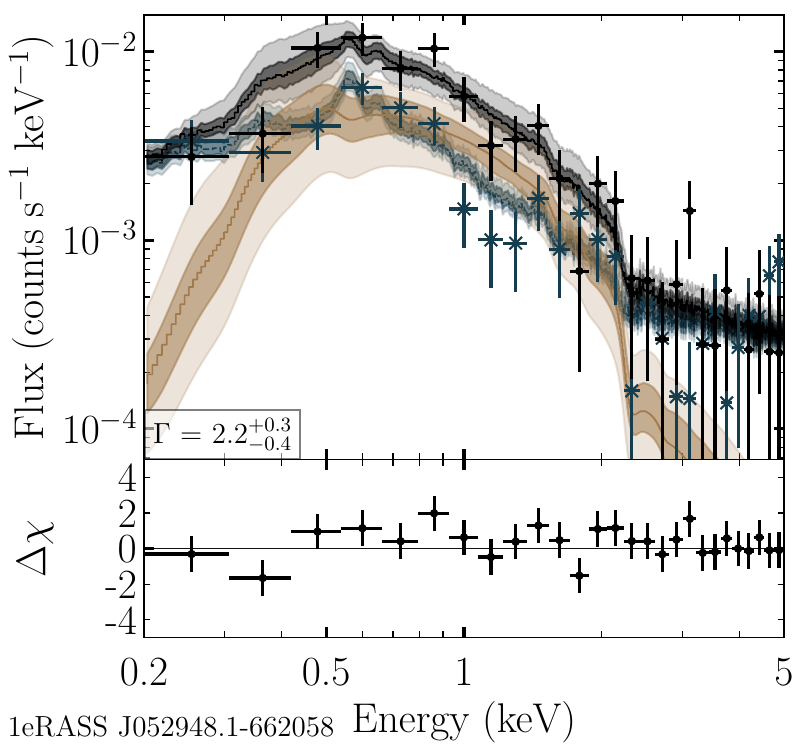}}
            \resizebox{0.33\hsize}{!}{\includegraphics{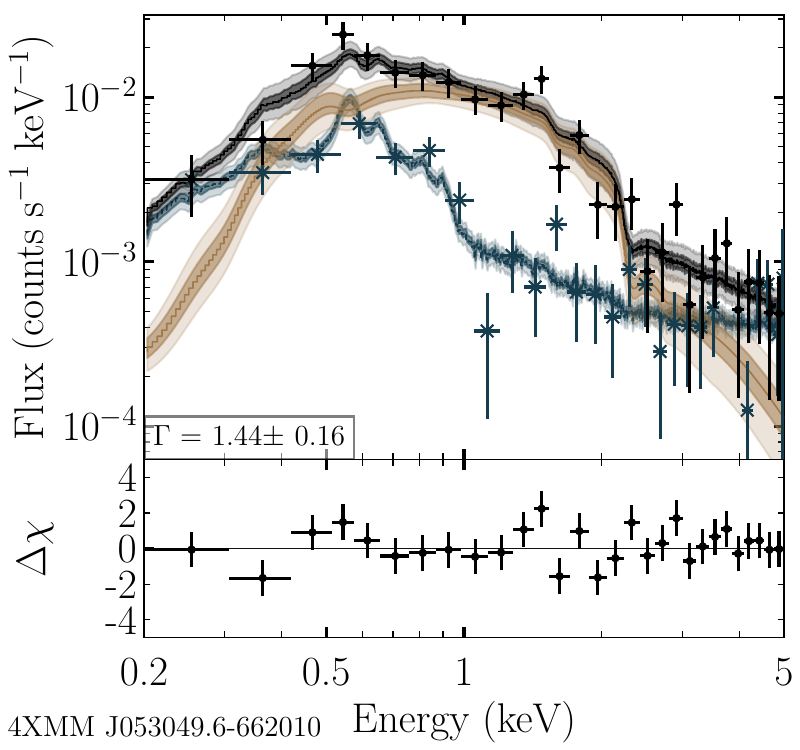}}
            \resizebox{0.33\hsize}{!}{\includegraphics{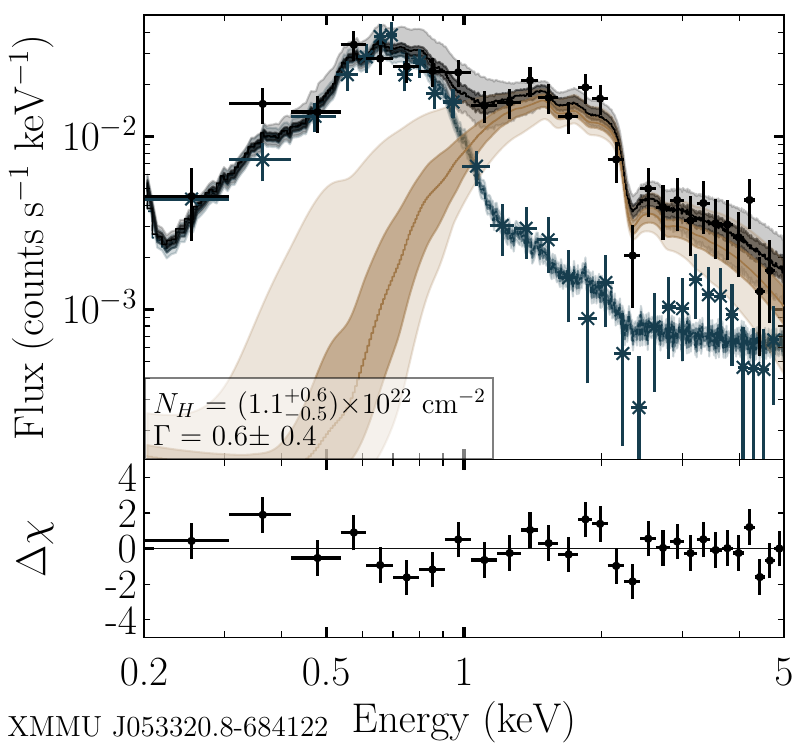}}
            \resizebox{0.33\hsize}{!}{\includegraphics{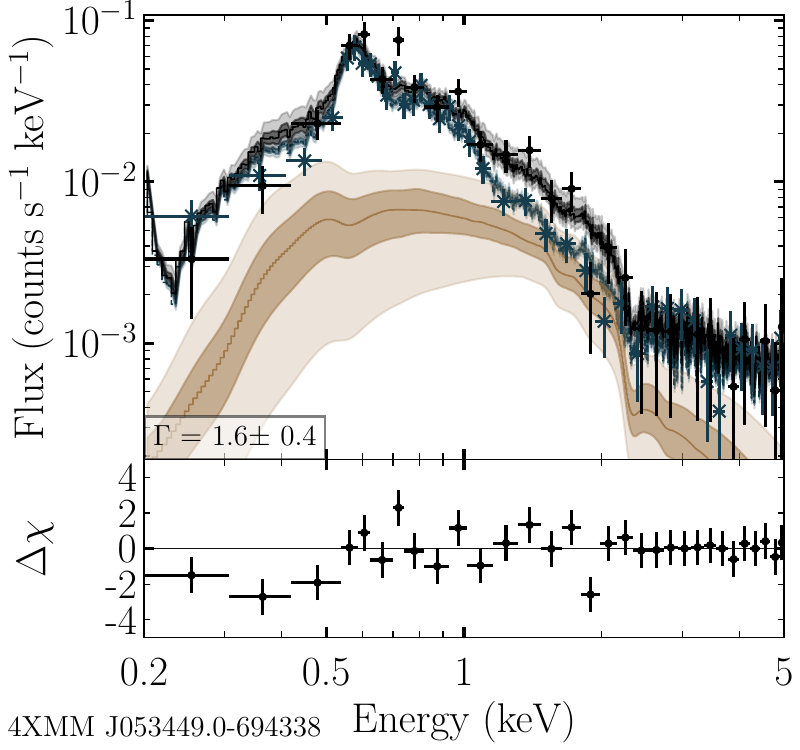}}
            \resizebox{0.33\hsize}{!}{\includegraphics{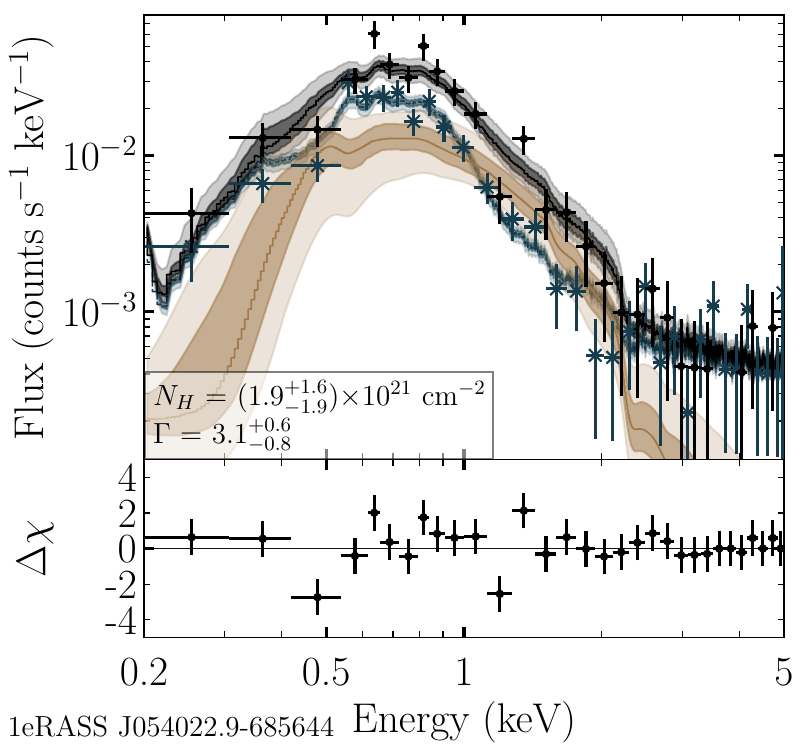}}
            \resizebox{0.33\hsize}{!}{\includegraphics{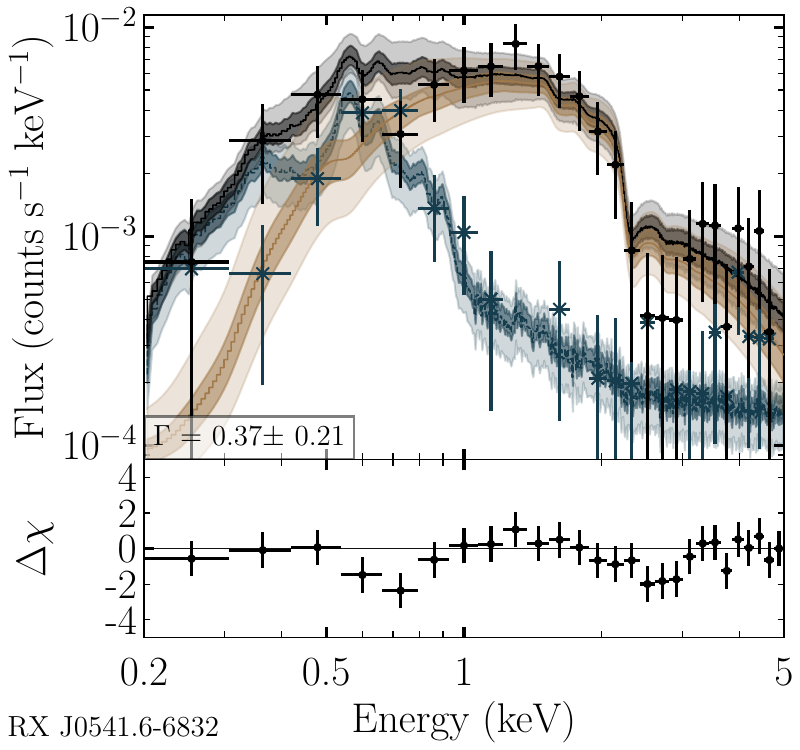}}
            \resizebox{0.33\hsize}{!}{\includegraphics{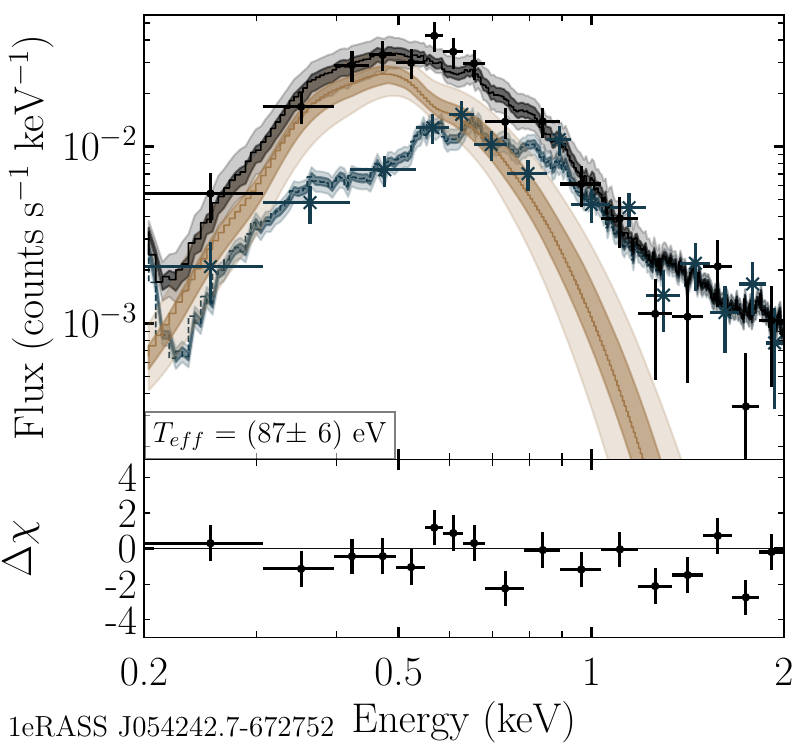}}
            \resizebox{0.33\hsize}{!}{\includegraphics{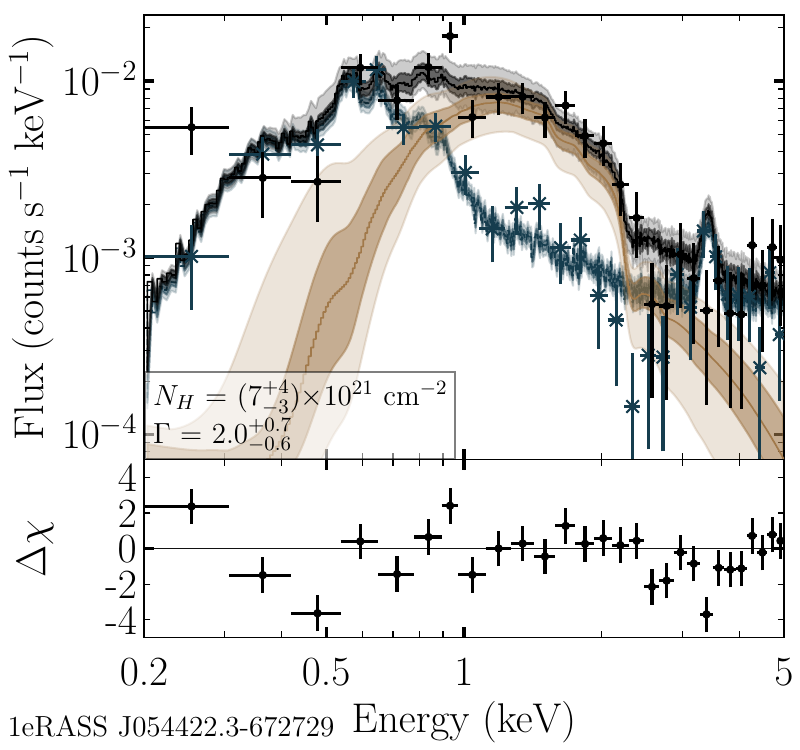}}
            \resizebox{0.33\hsize}{!}{\includegraphics{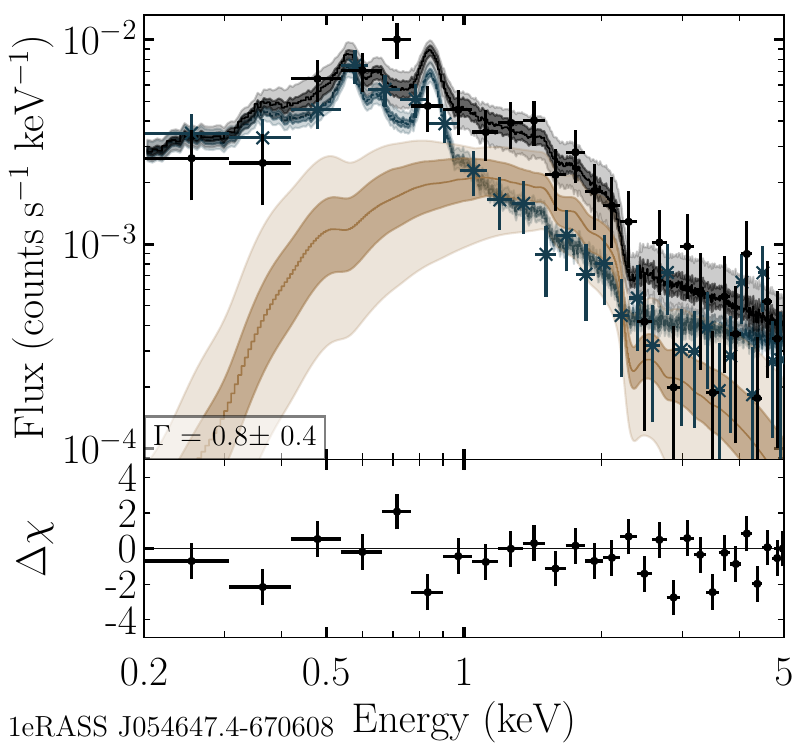}}
            \caption{continued for \#37 to \#48}
        \end{figure*}
        \addtocounter{figure}{-1}\begin{figure*}
            \centering
            \resizebox{0.33\hsize}{!}{\includegraphics{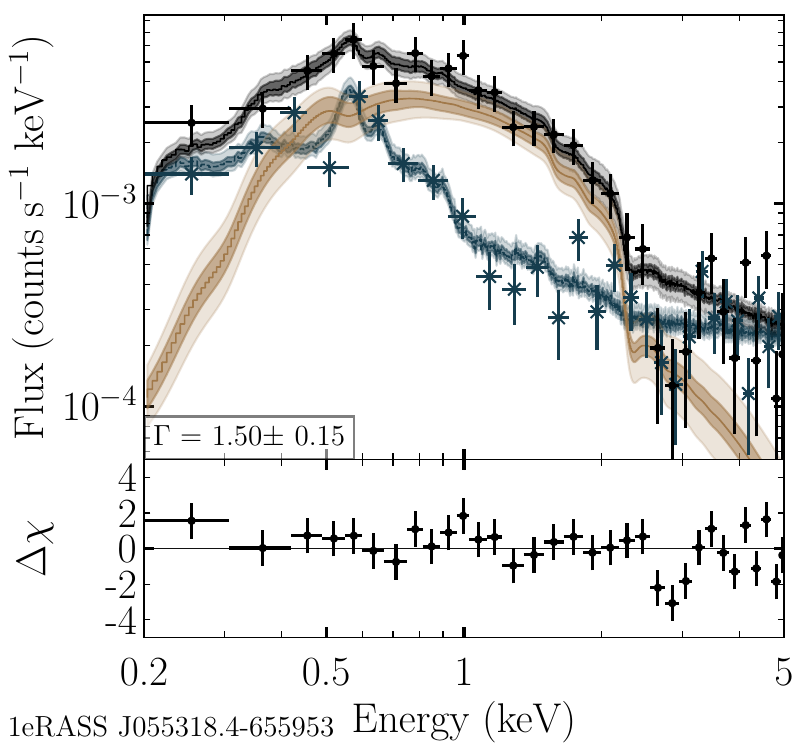}}
            \resizebox{0.33\hsize}{!}{\includegraphics{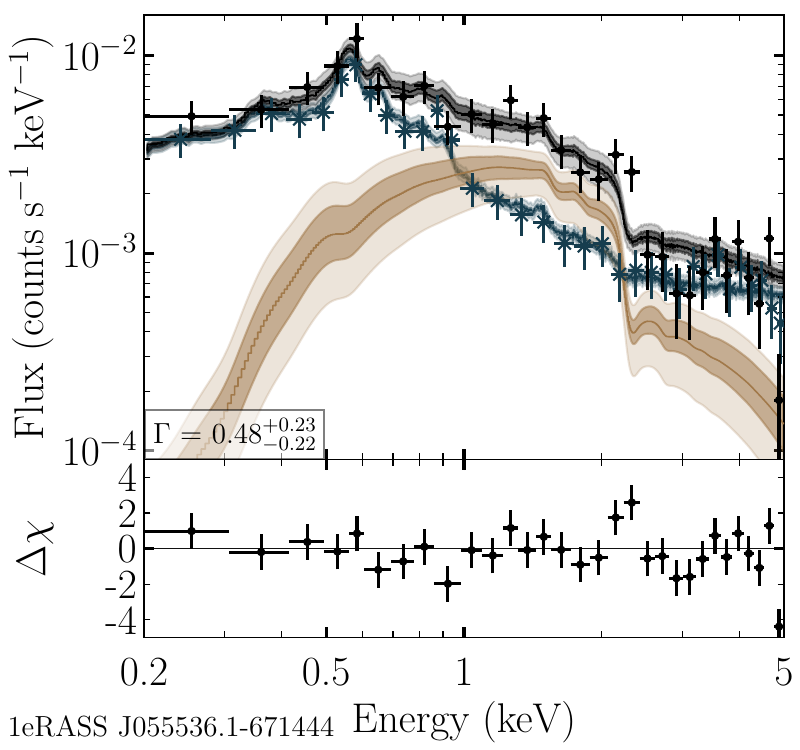}}
            \resizebox{0.33\hsize}{!}{\includegraphics{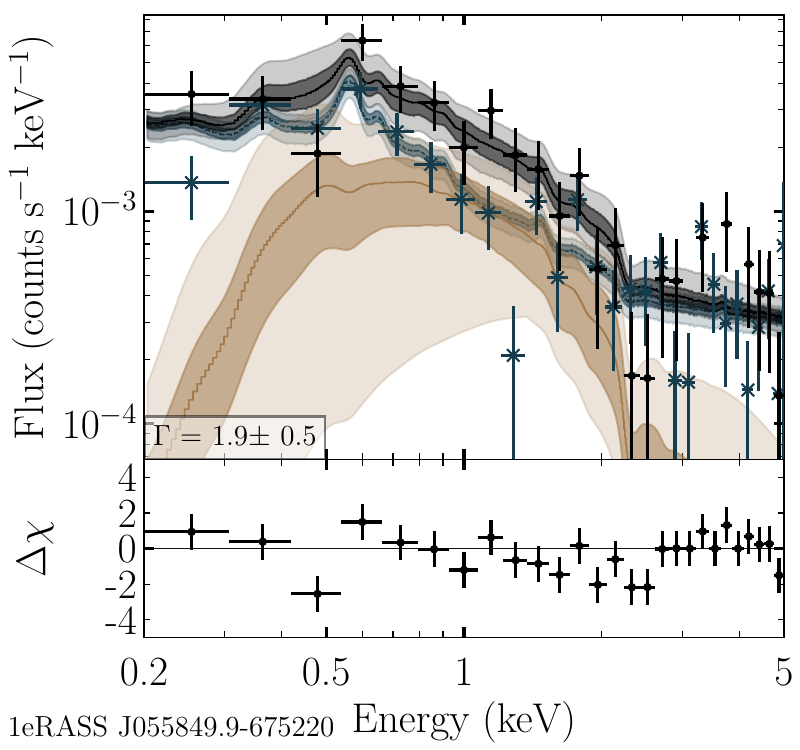}}
            \resizebox{0.33\hsize}{!}{\includegraphics{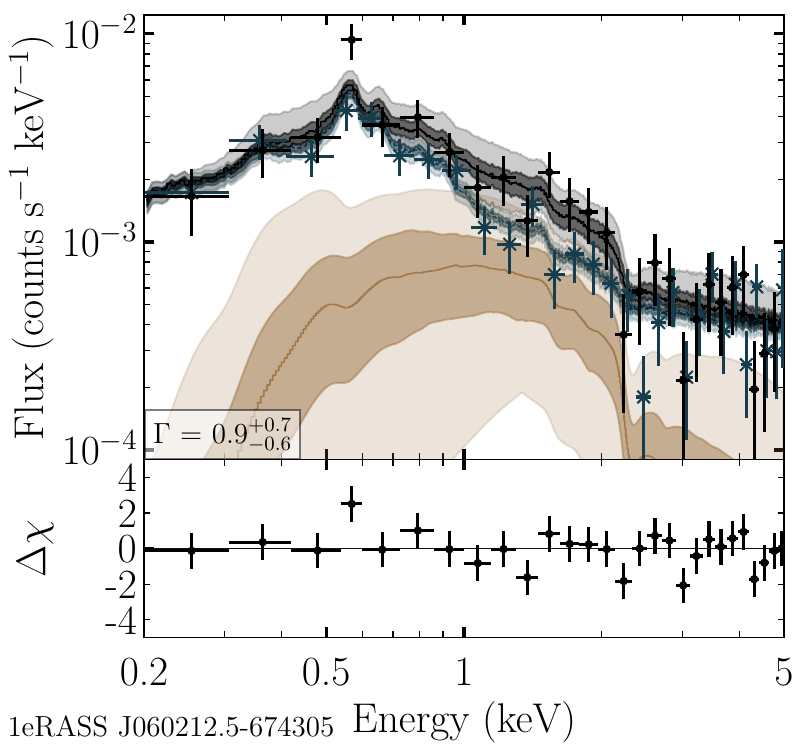}}
            \resizebox{0.33\hsize}{!}{\includegraphics{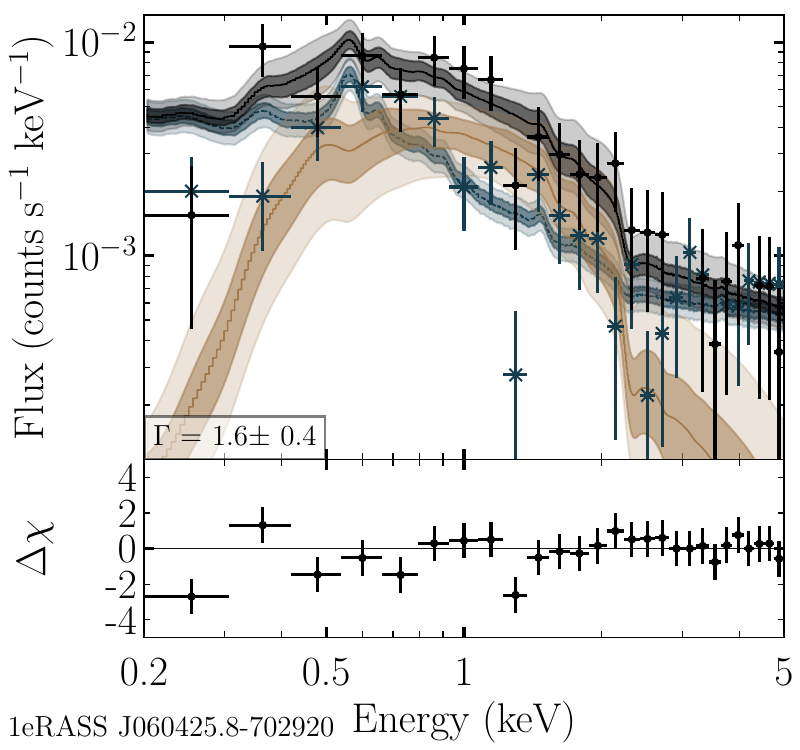}}
            \caption{continued for \#49 to \#53}
        \end{figure*}
        \clearpage
        
        \begin{figure*}
            \centering
            \resizebox{0.495\hsize}{!}{\includegraphics{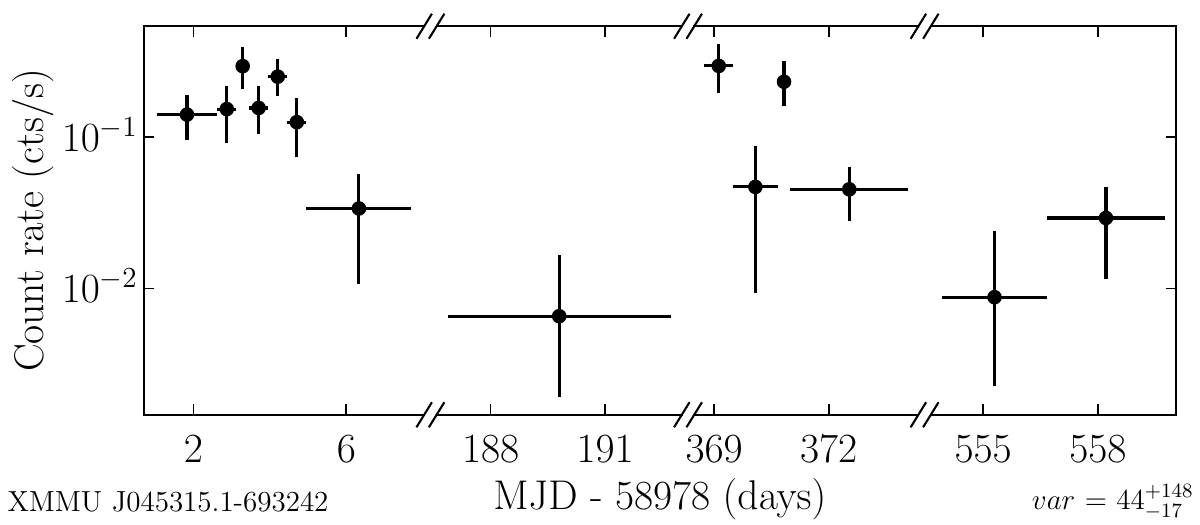}}
            \resizebox{0.495\hsize}{!}{\includegraphics{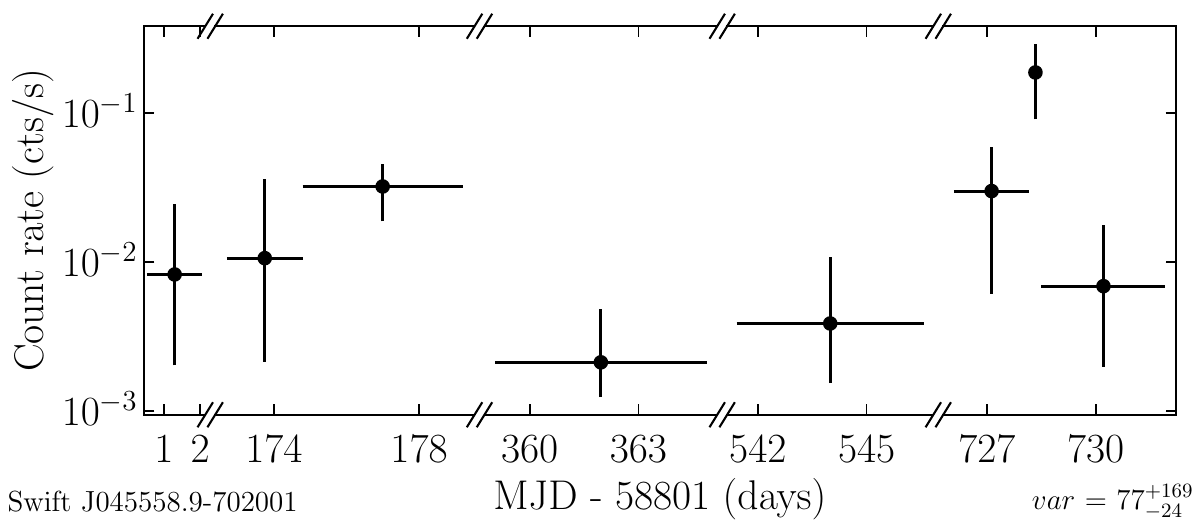}}
            \resizebox{0.495\hsize}{!}{\includegraphics{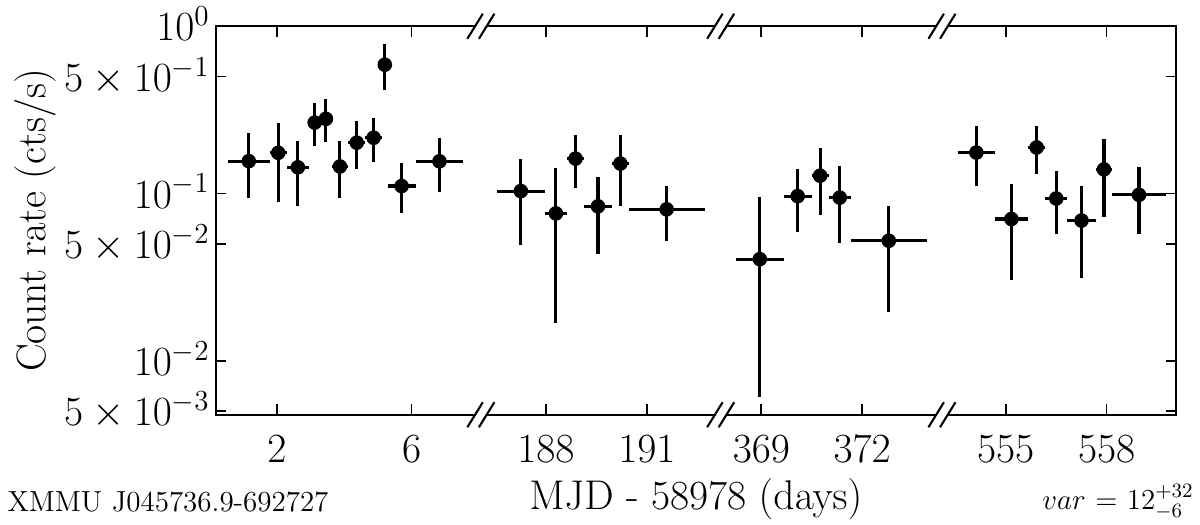}}
            \resizebox{0.495\hsize}{!}{\includegraphics{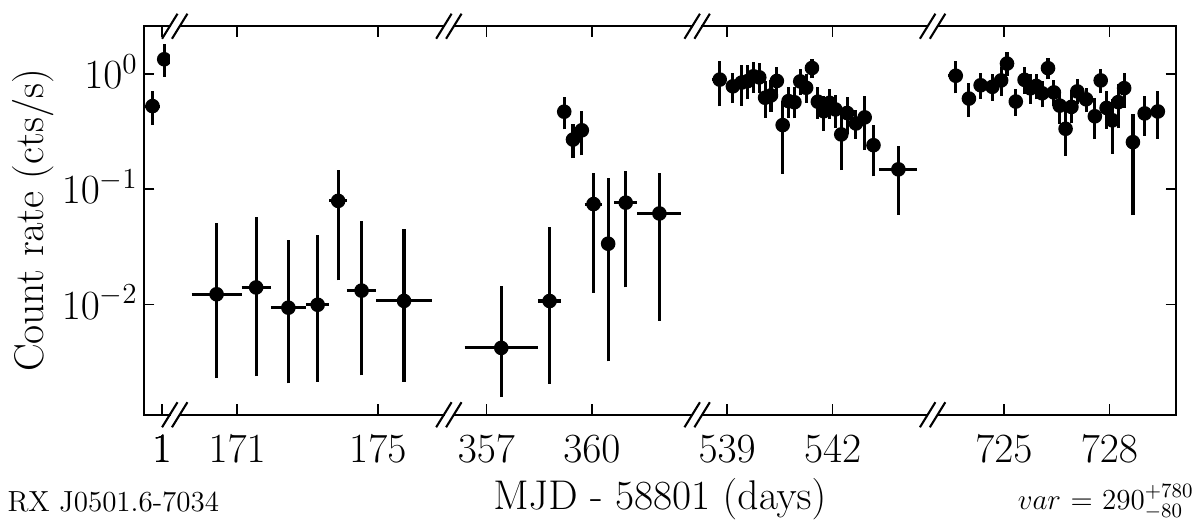}}
            \resizebox{0.495\hsize}{!}{\includegraphics{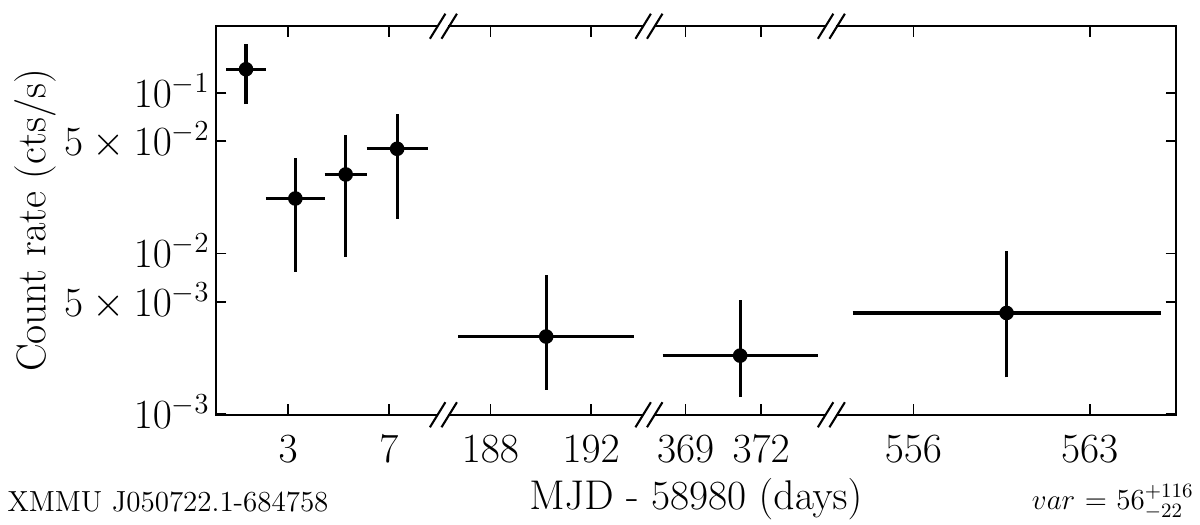}}
            \resizebox{0.495\hsize}{!}{\includegraphics{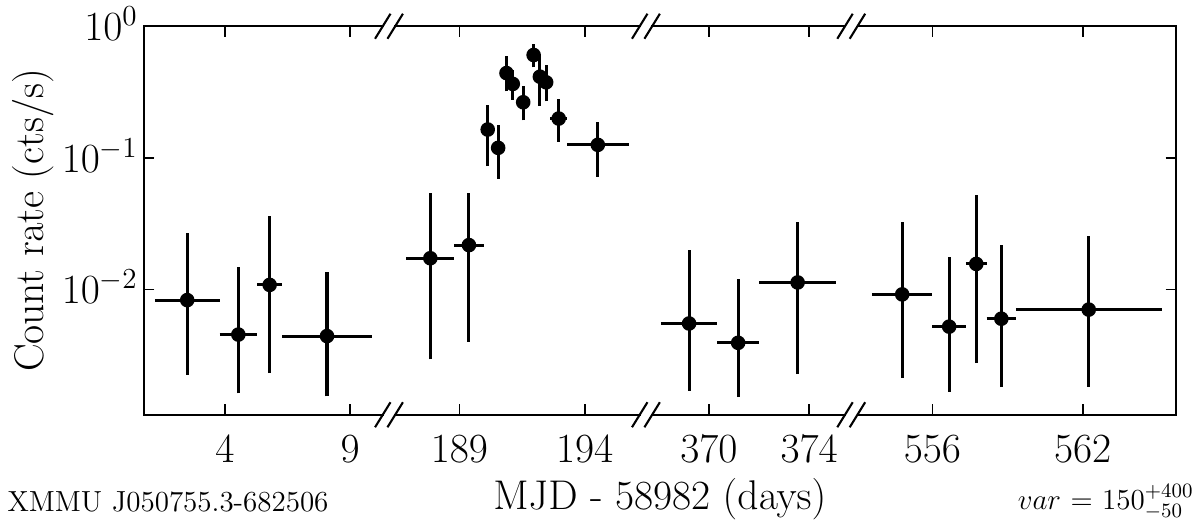}}
            \resizebox{0.495\hsize}{!}{\includegraphics{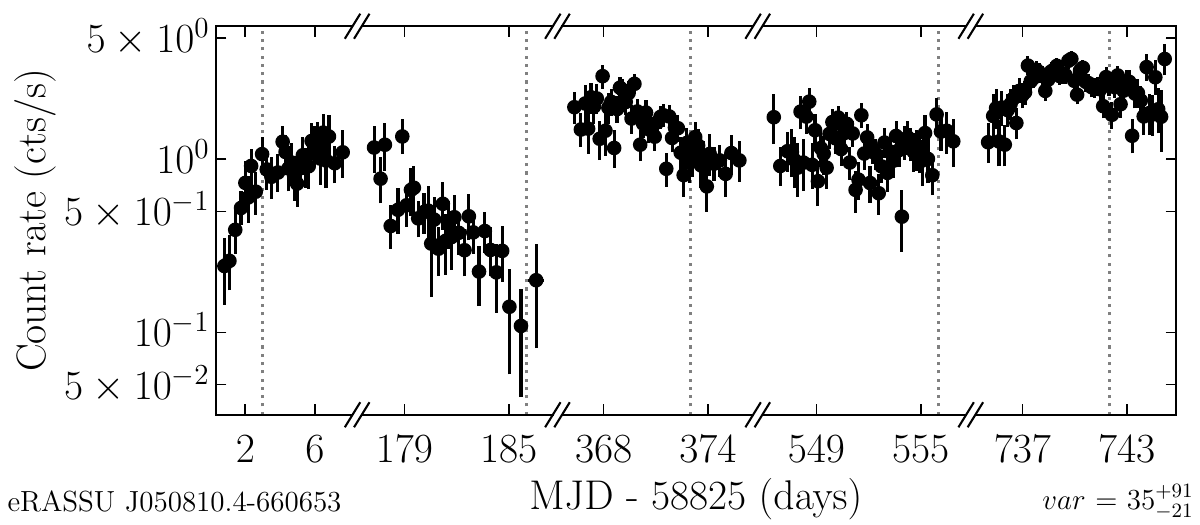}}
            \resizebox{0.495\hsize}{!}{\includegraphics{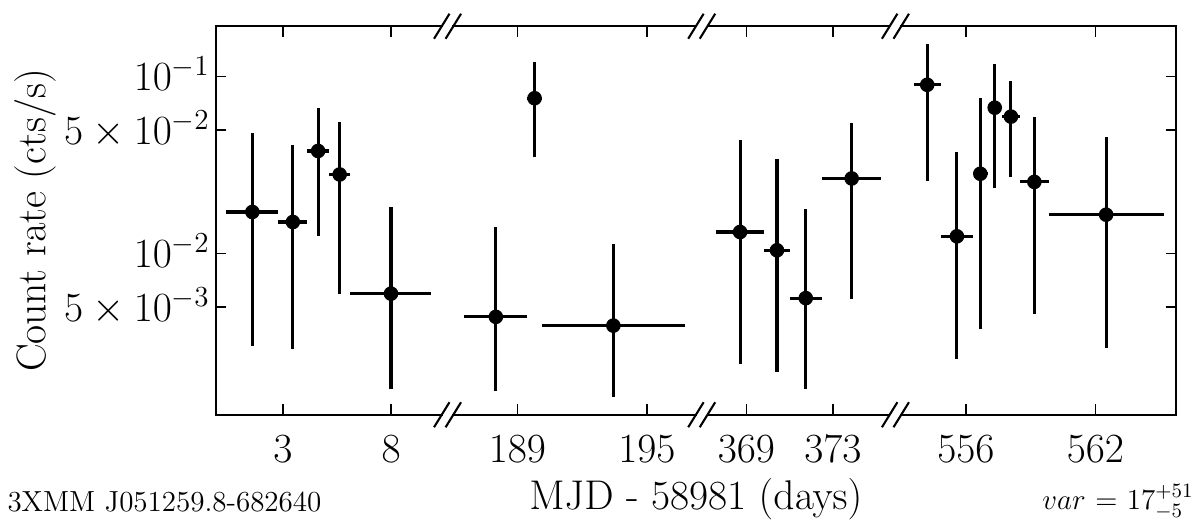}}
            \resizebox{0.495\hsize}{!}{\includegraphics{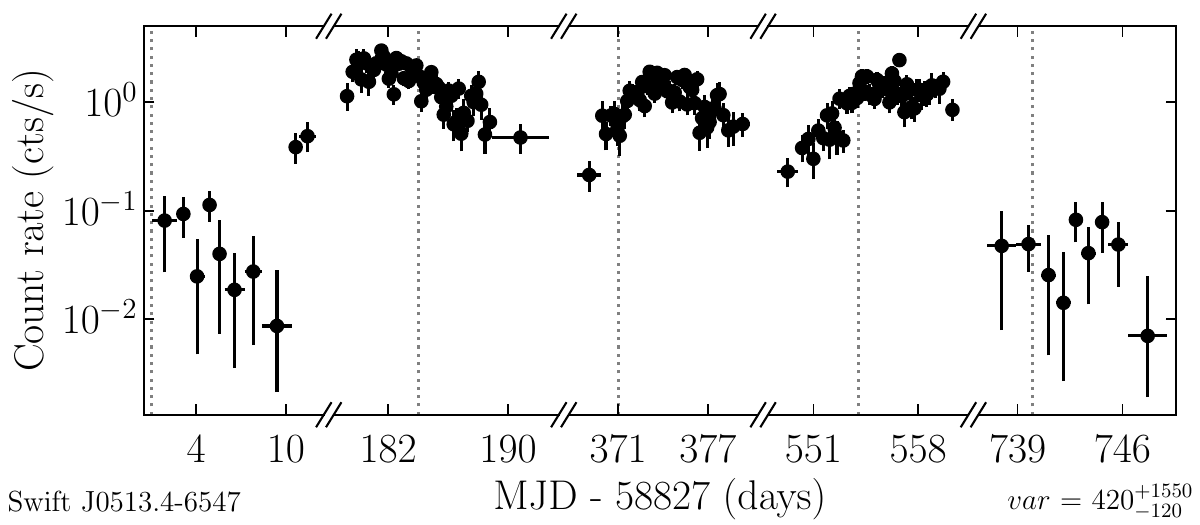}}
            \resizebox{0.495\hsize}{!}{\includegraphics{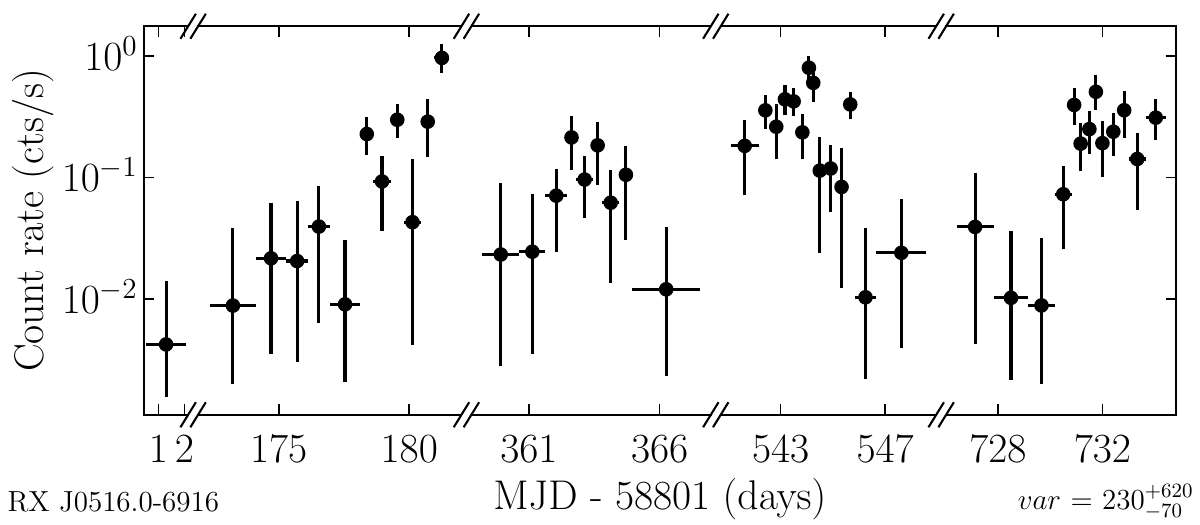}}
            \caption{\ero lightcurves for \#1 to \#10. For readability white spaces between the individual eRASSs are removed. The lightcurves are rebinned to have a minimum number of 10 net source counts per time bin as described in Sect. \ref{sec:LCs}. The variability $var$ (found in the bottom right of each figure) is calculated as described in Sect \ref{sec:MAV}.}
            \label{fig:eRO_LCs}
        \end{figure*}
        \addtocounter{figure}{-1}\begin{figure*}
            \centering
            \resizebox{0.495\hsize}{!}{\includegraphics{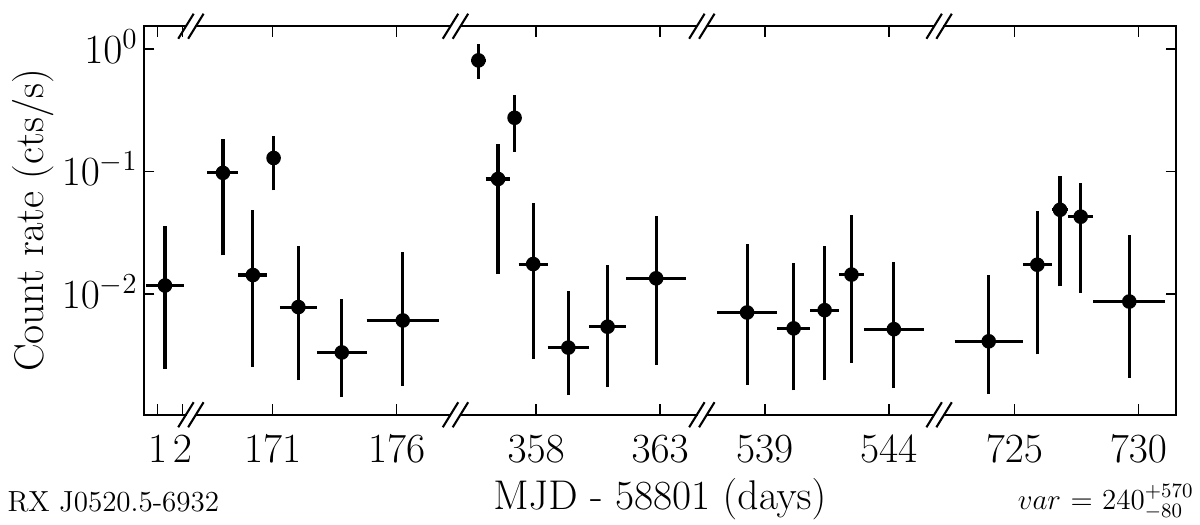}}
            \resizebox{0.495\hsize}{!}{\includegraphics{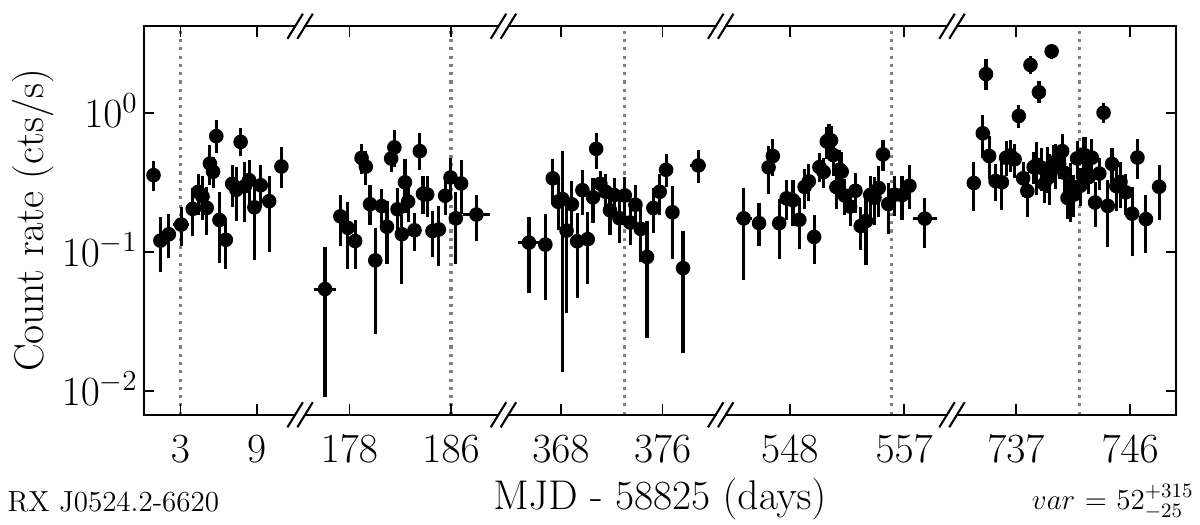}}
            \resizebox{0.495\hsize}{!}{\includegraphics{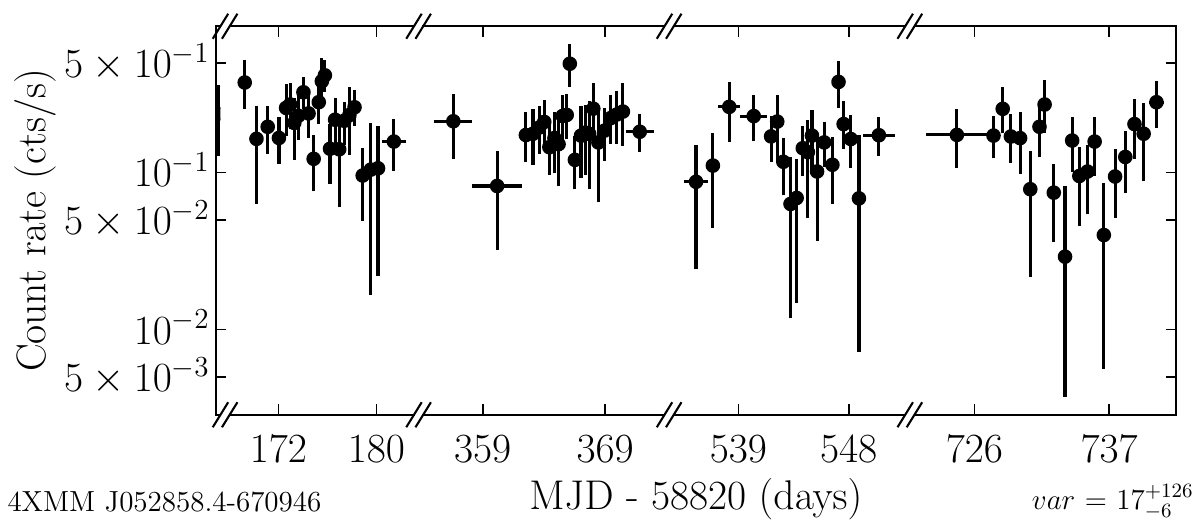}}
            \resizebox{0.495\hsize}{!}{\includegraphics{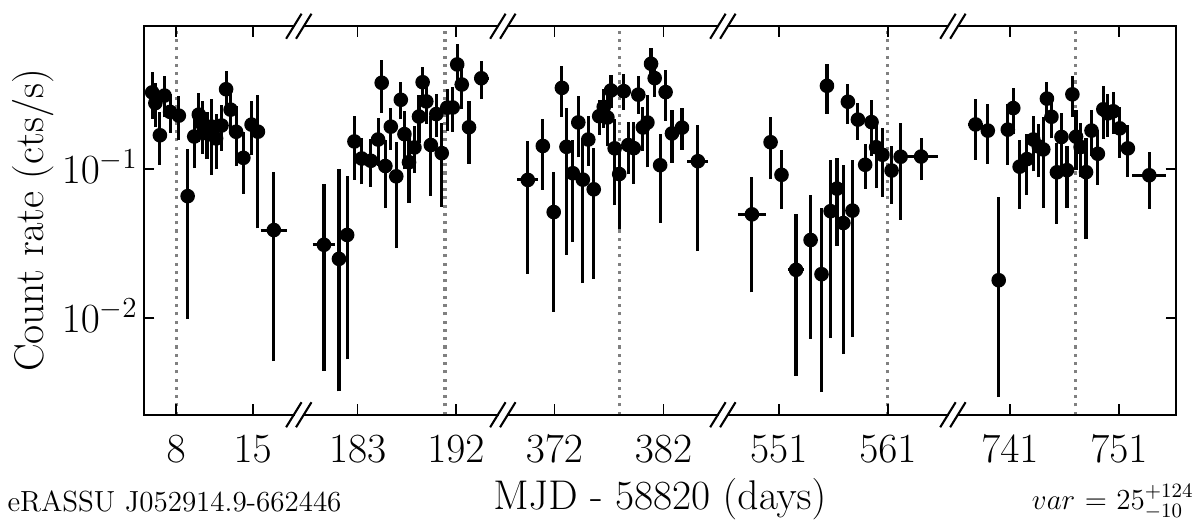}}
            \resizebox{0.495\hsize}{!}{\includegraphics{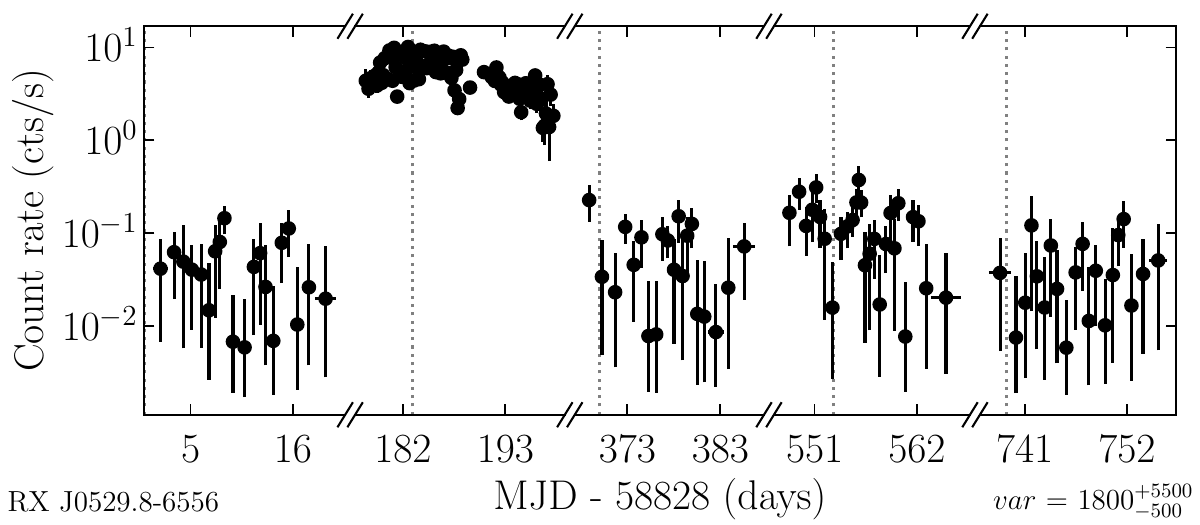}}
            \resizebox{0.495\hsize}{!}{\includegraphics{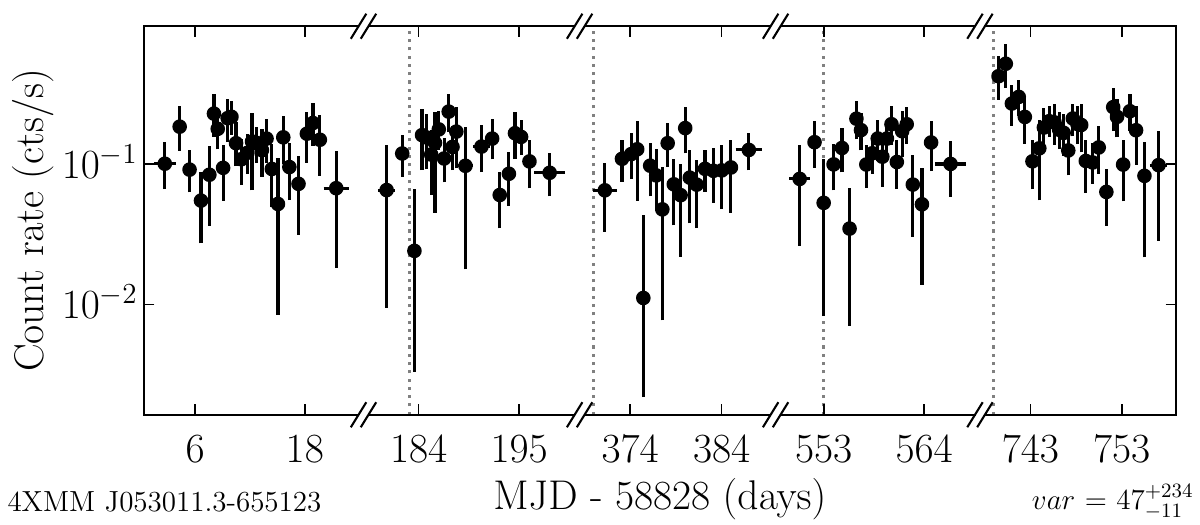}}
            \resizebox{0.495\hsize}{!}{\includegraphics{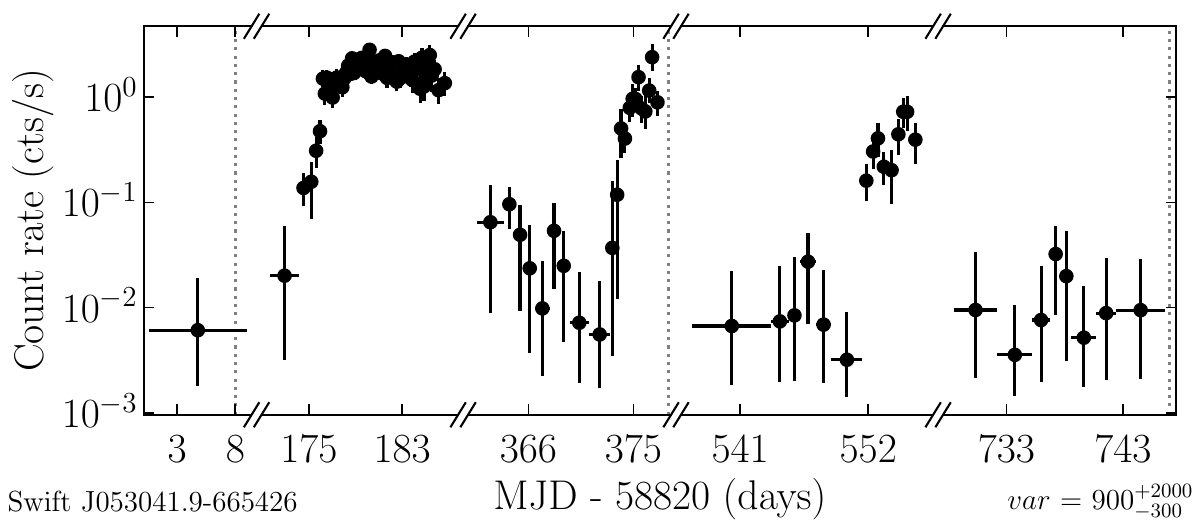}}
            \resizebox{0.495\hsize}{!}{\includegraphics{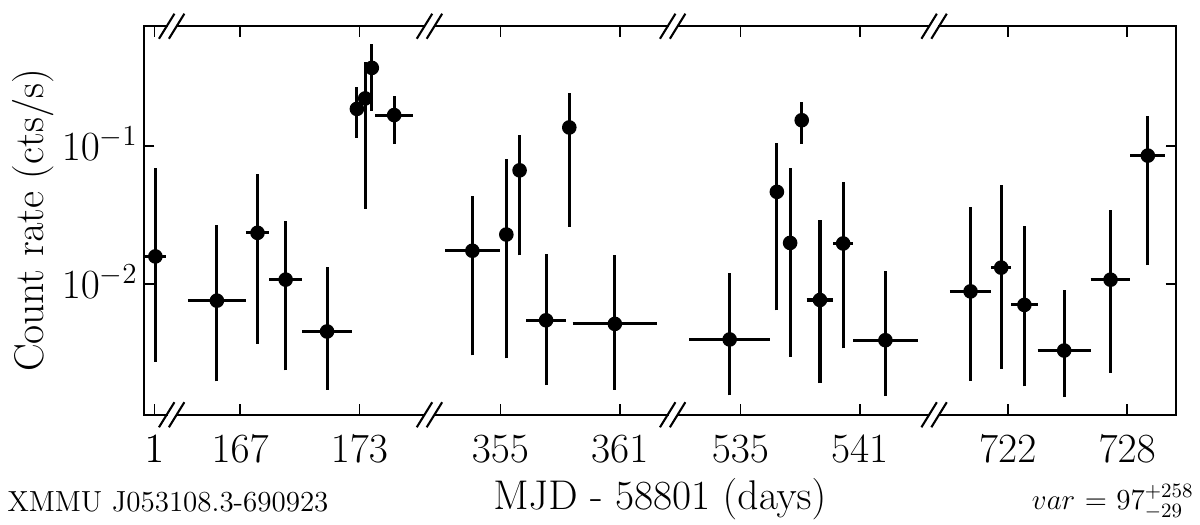}}
            \resizebox{0.495\hsize}{!}{\includegraphics{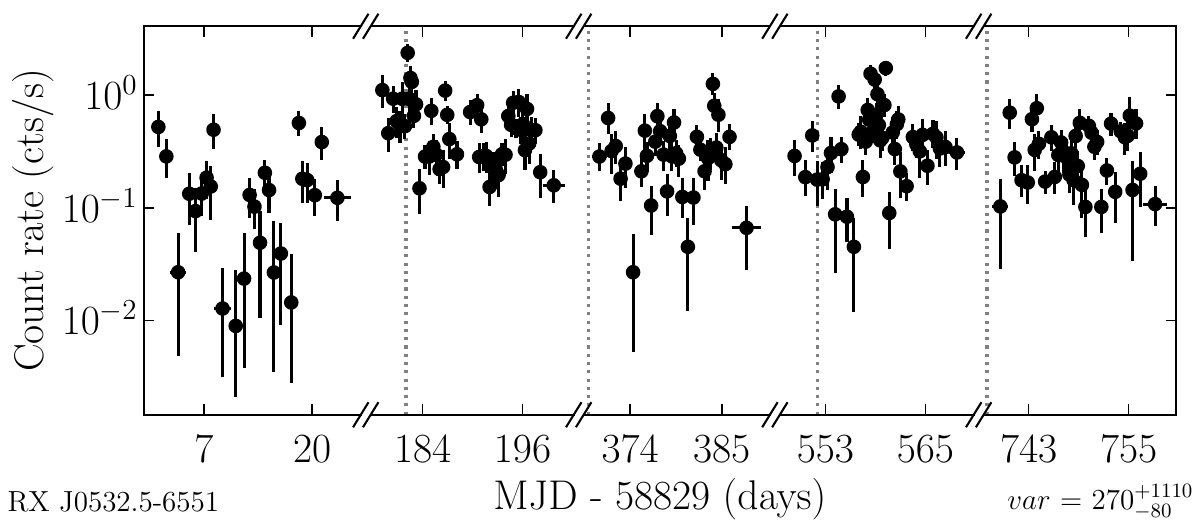}}
            \resizebox{0.495\hsize}{!}{\includegraphics{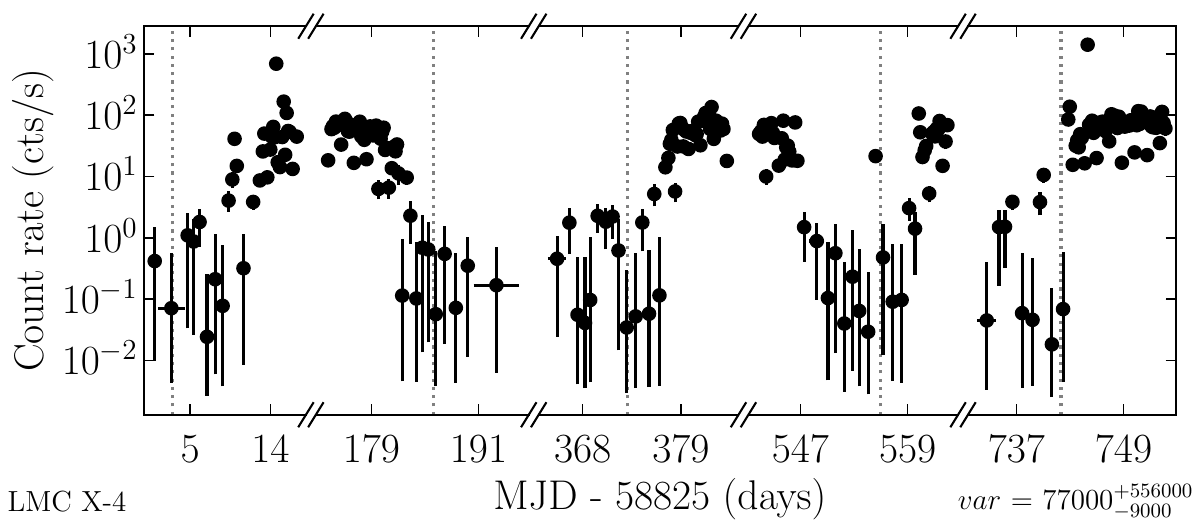}}
            \resizebox{0.495\hsize}{!}{\includegraphics{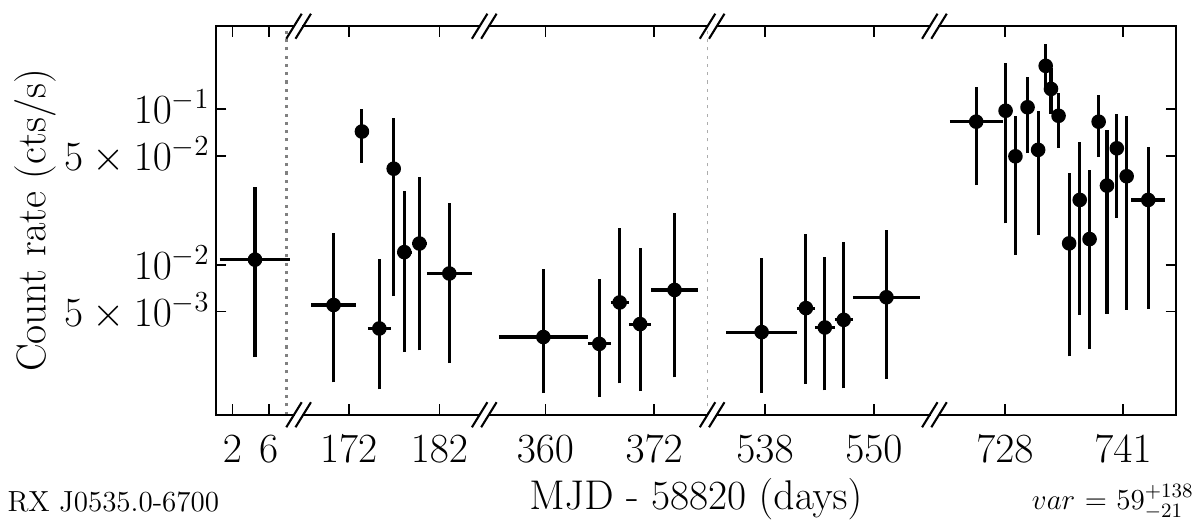}}
            \resizebox{0.495\hsize}{!}{\includegraphics{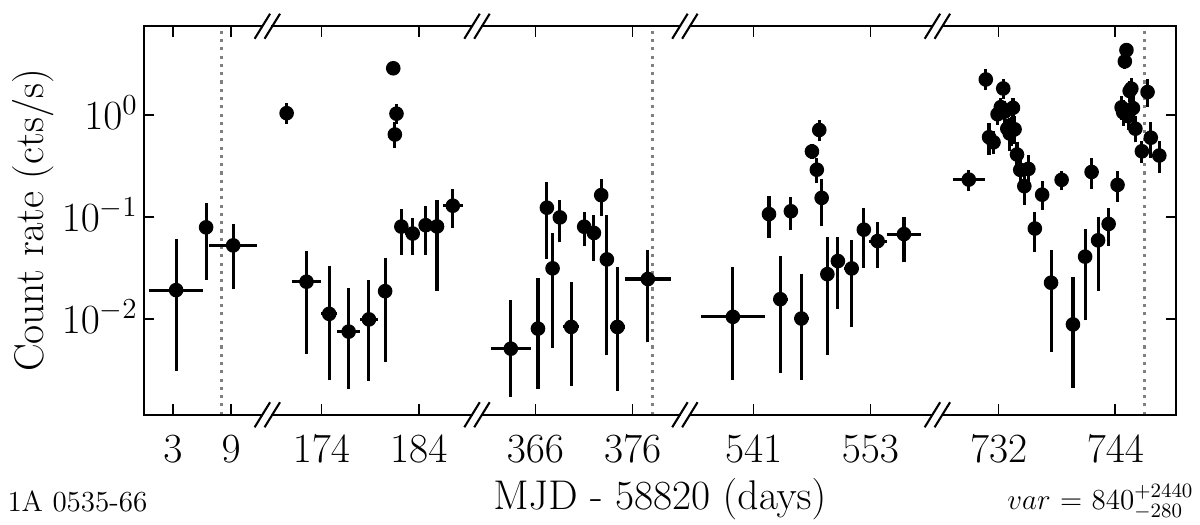}}
            \caption{continued for \#11 to \#22}
        \end{figure*}
        \addtocounter{figure}{-1}\begin{figure*}
            \centering
            \resizebox{0.495\hsize}{!}{\includegraphics{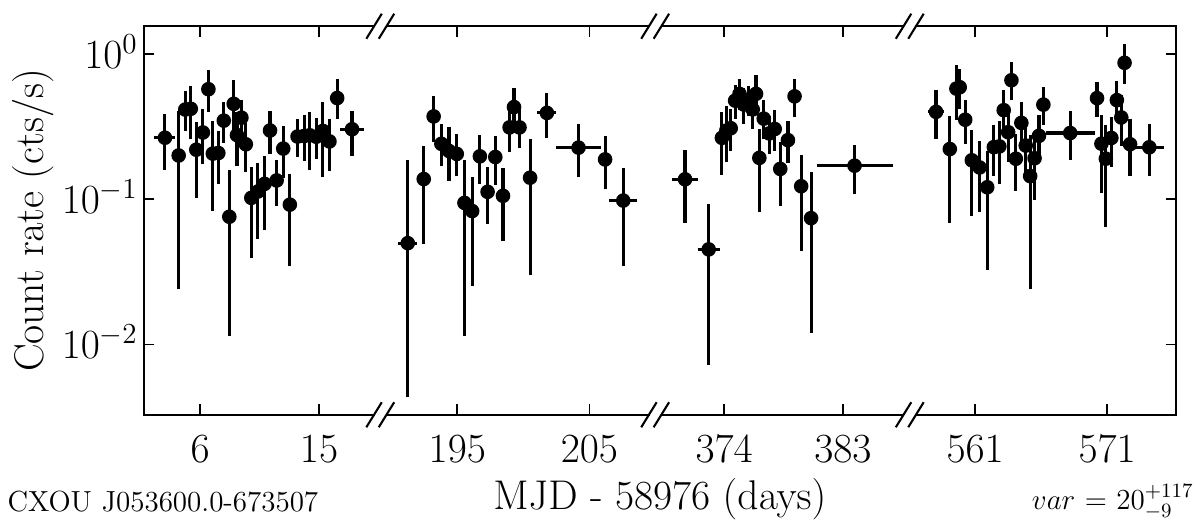}}
            \resizebox{0.495\hsize}{!}{\includegraphics{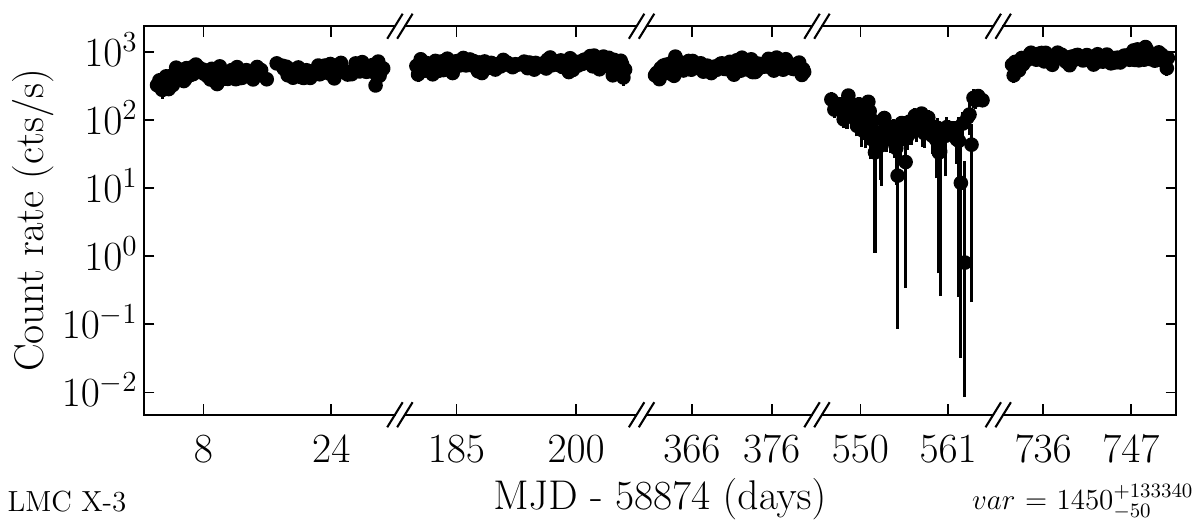}}
            \resizebox{0.495\hsize}{!}{\includegraphics{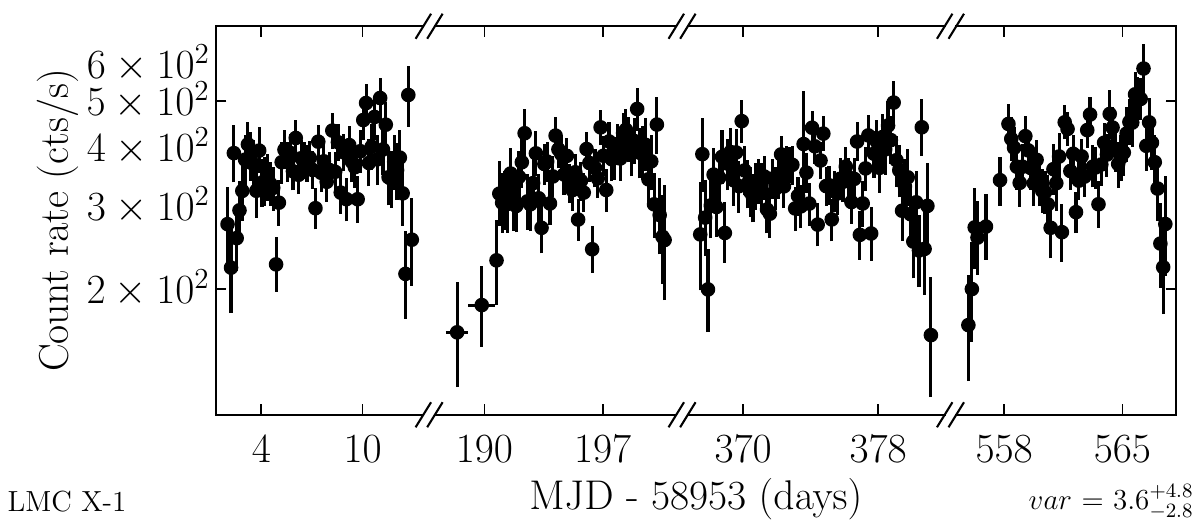}}
            \resizebox{0.495\hsize}{!}{\includegraphics{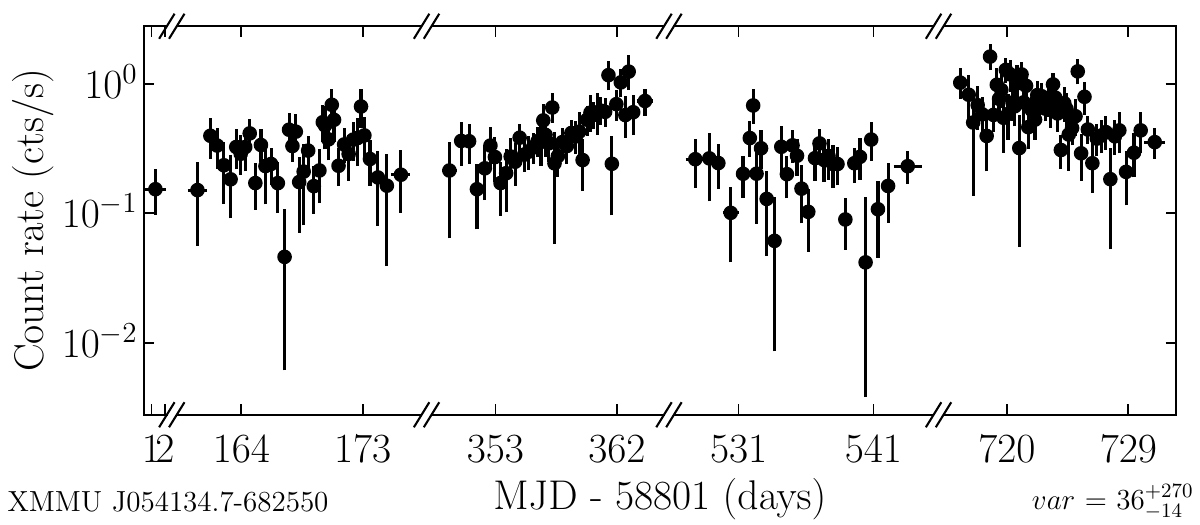}}
            \resizebox{0.495\hsize}{!}{\includegraphics{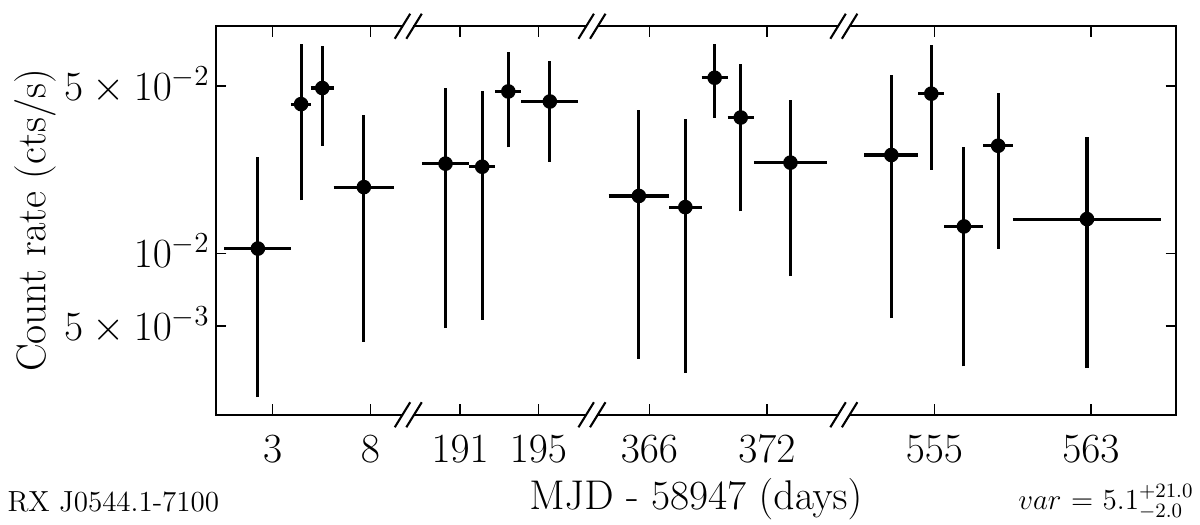}}
            \resizebox{0.495\hsize}{!}{\includegraphics{plots/eROSITA_LCs/seq28_lc_small.pdf}}
            \resizebox{0.495\hsize}{!}{\includegraphics{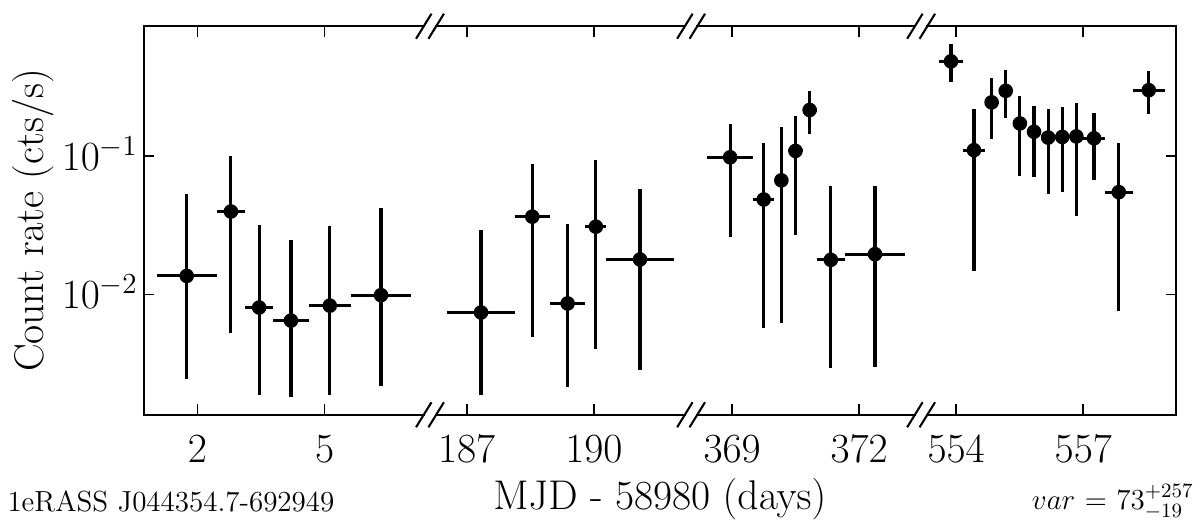}}
            \resizebox{0.495\hsize}{!}{\includegraphics{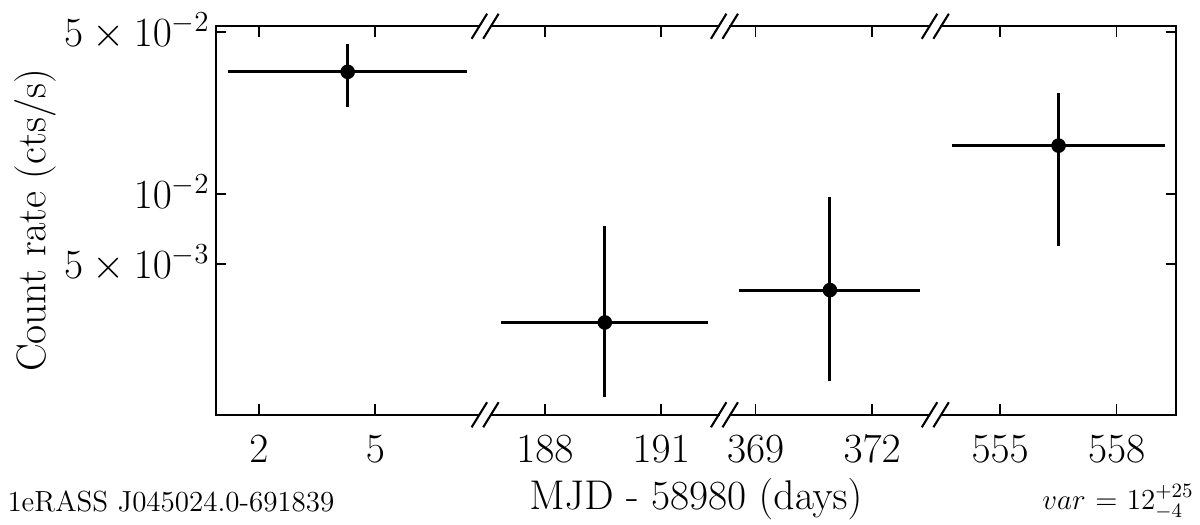}}
            \resizebox{0.495\hsize}{!}{\includegraphics{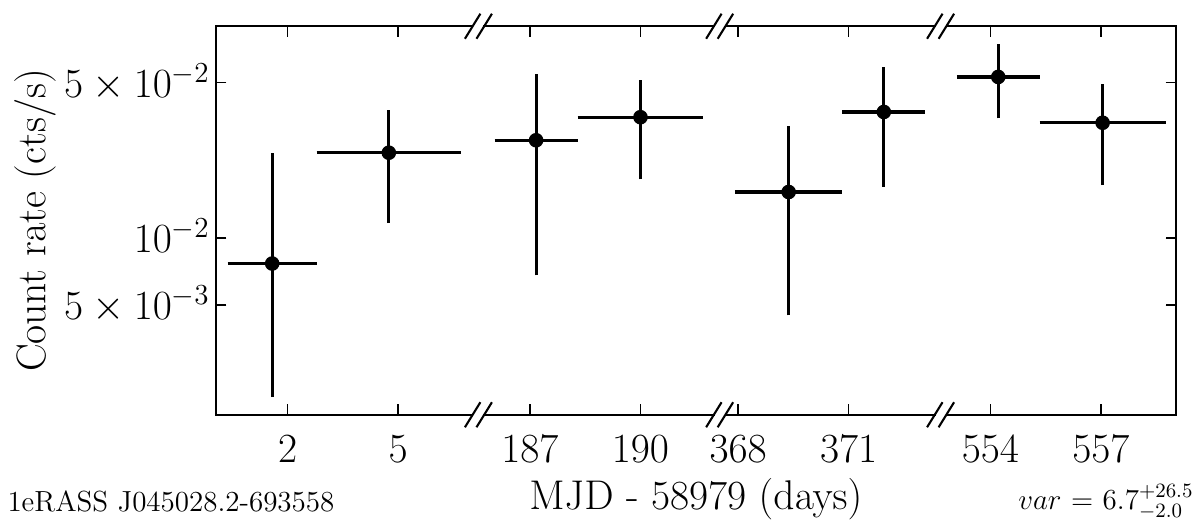}}
            \resizebox{0.495\hsize}{!}{\includegraphics{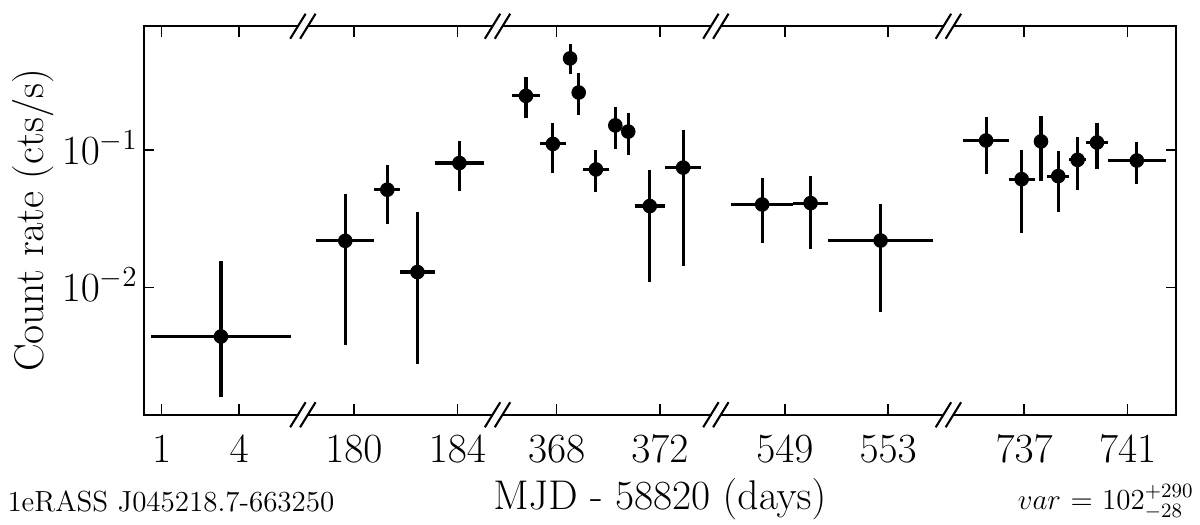}}
            \resizebox{0.495\hsize}{!}{\includegraphics{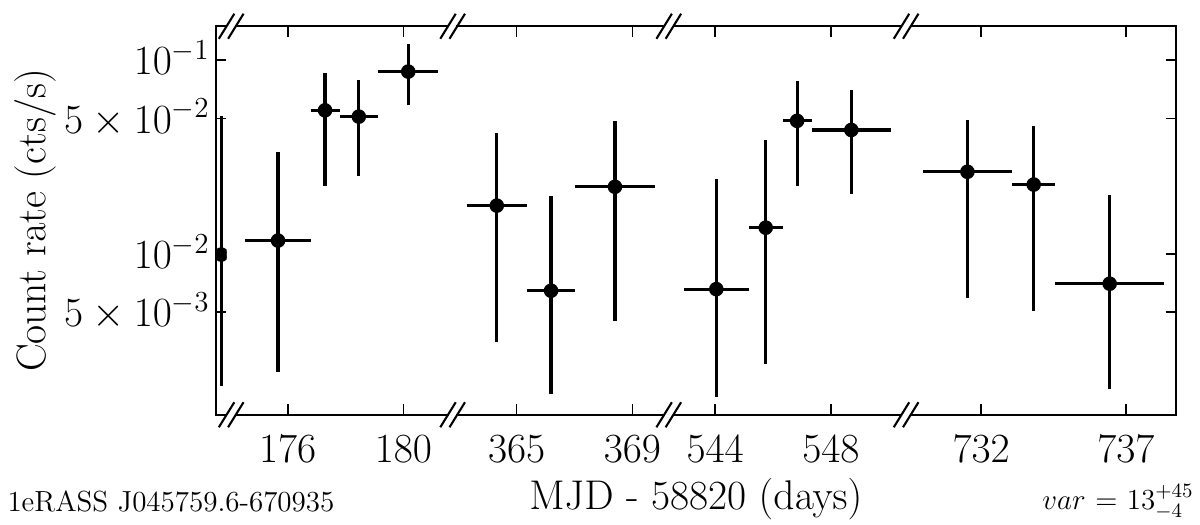}}
            \resizebox{0.495\hsize}{!}{\includegraphics{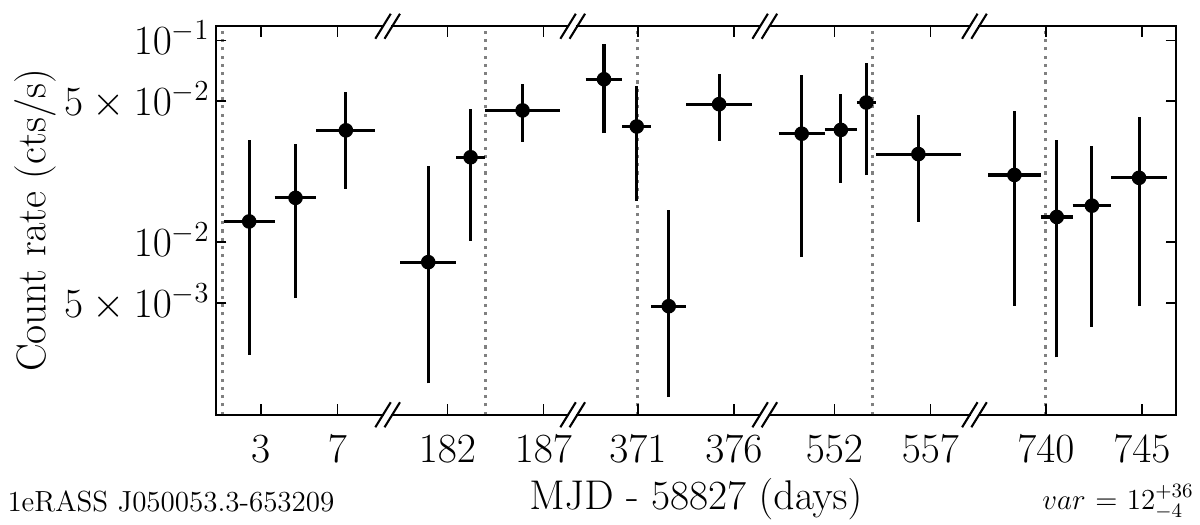}}
            \caption{continued for \#23 to \#34}
        \end{figure*}
        \addtocounter{figure}{-1}\begin{figure*}
            \centering
            \resizebox{0.495\hsize}{!}{\includegraphics{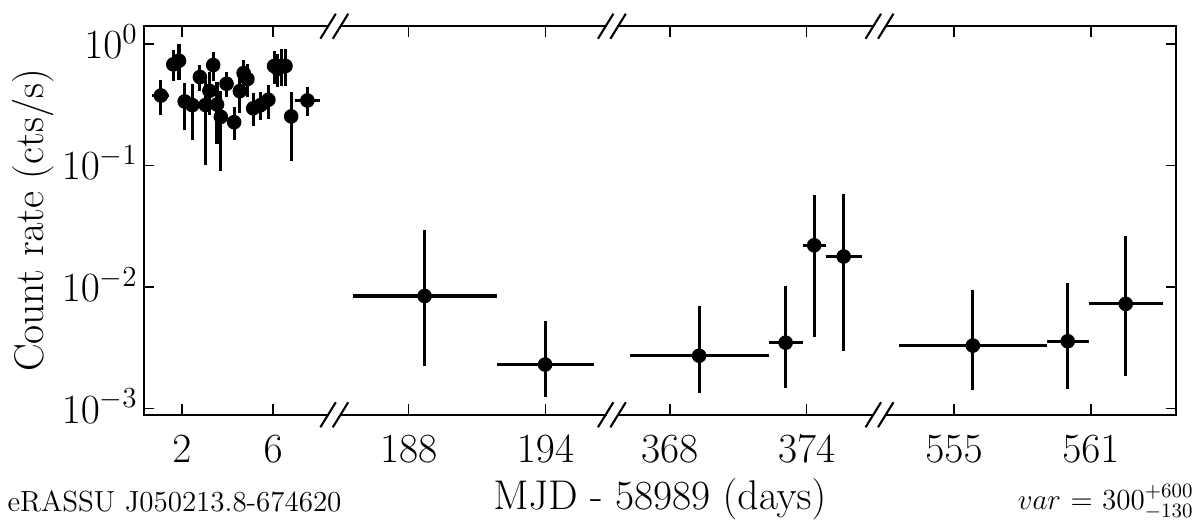}}
            \resizebox{0.495\hsize}{!}{\includegraphics{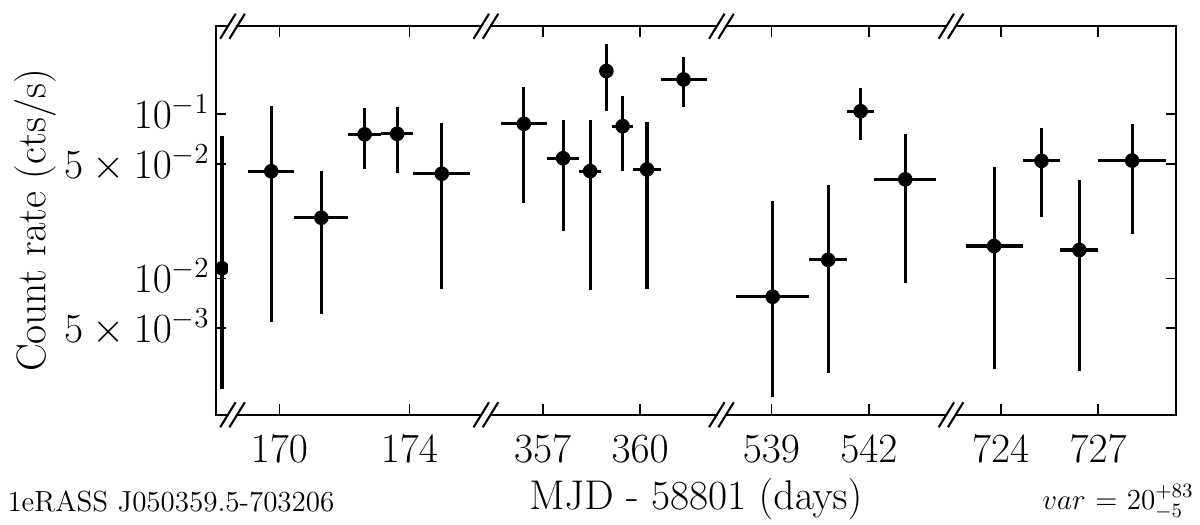}}
            \resizebox{0.495\hsize}{!}{\includegraphics{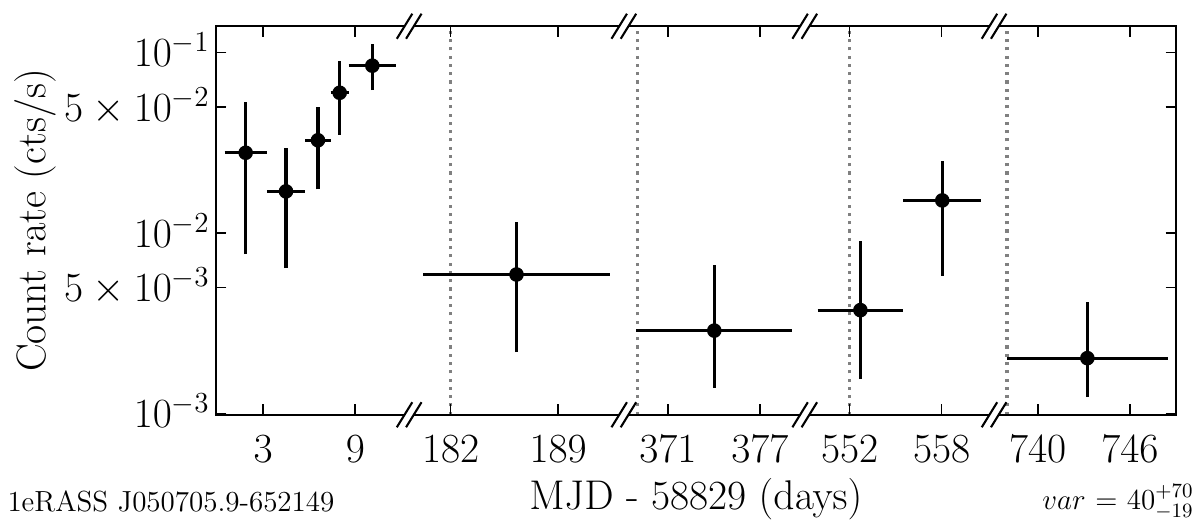}}
            \resizebox{0.495\hsize}{!}{\includegraphics{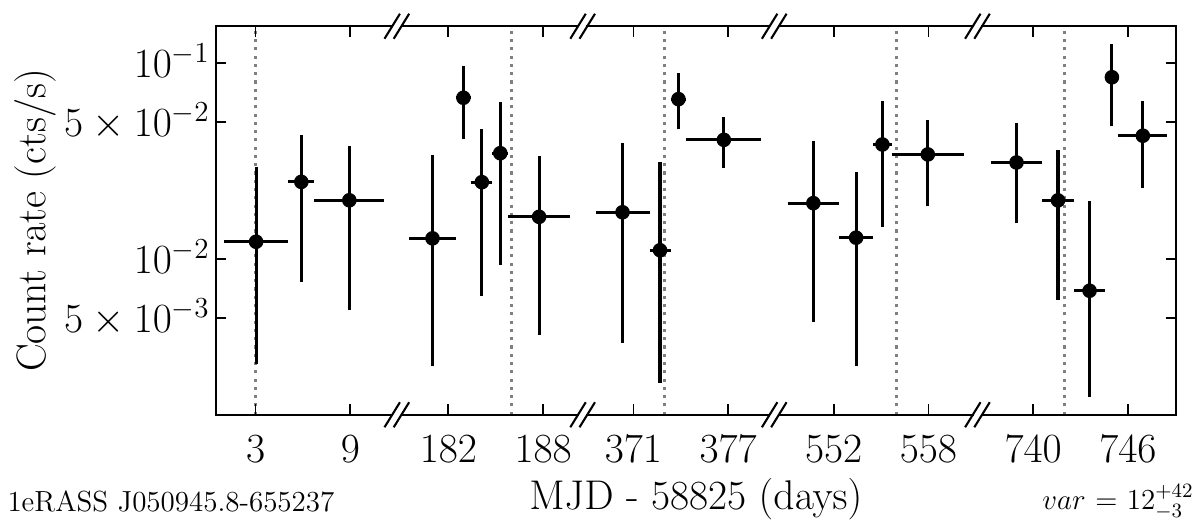}}
            \resizebox{0.495\hsize}{!}{\includegraphics{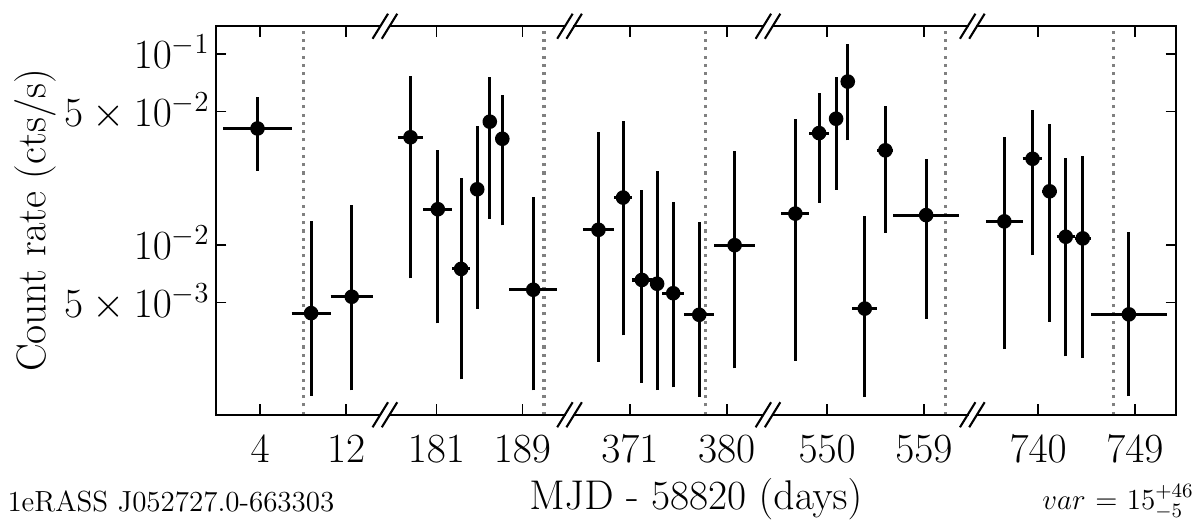}}
            \resizebox{0.495\hsize}{!}{\includegraphics{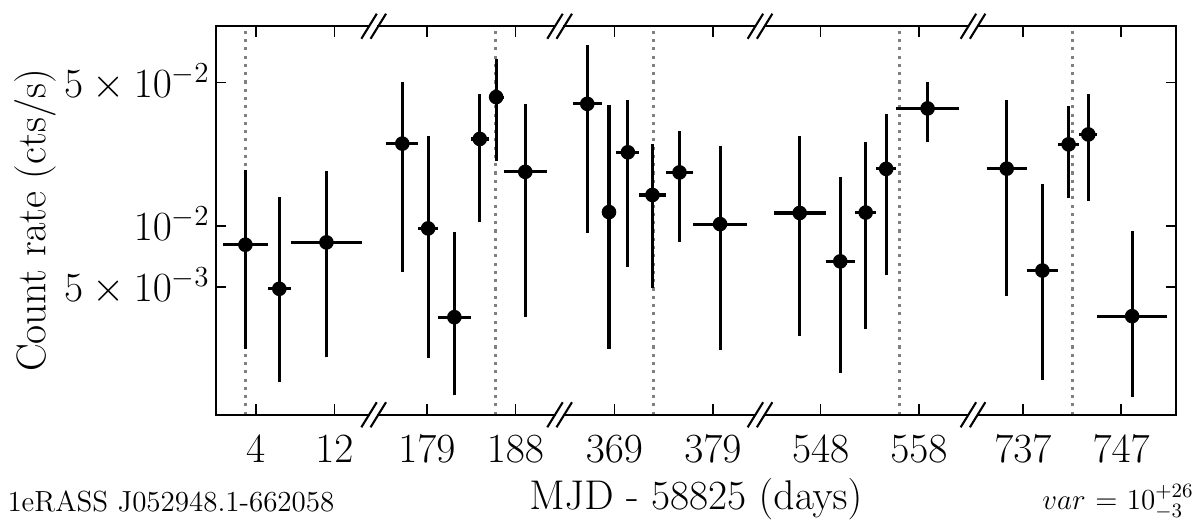}}
            \resizebox{0.495\hsize}{!}{\includegraphics{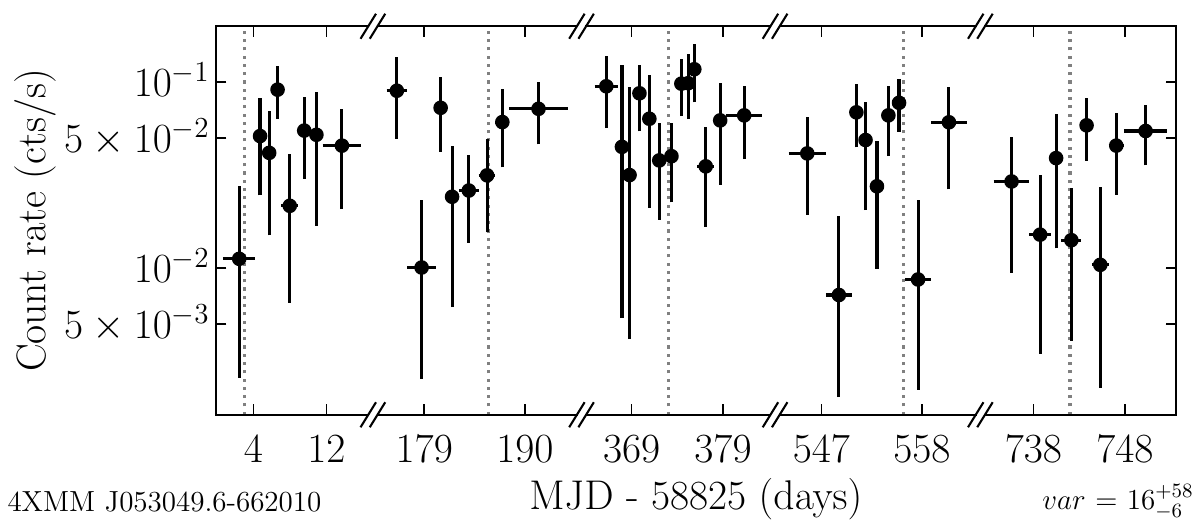}}
            \resizebox{0.495\hsize}{!}{\includegraphics{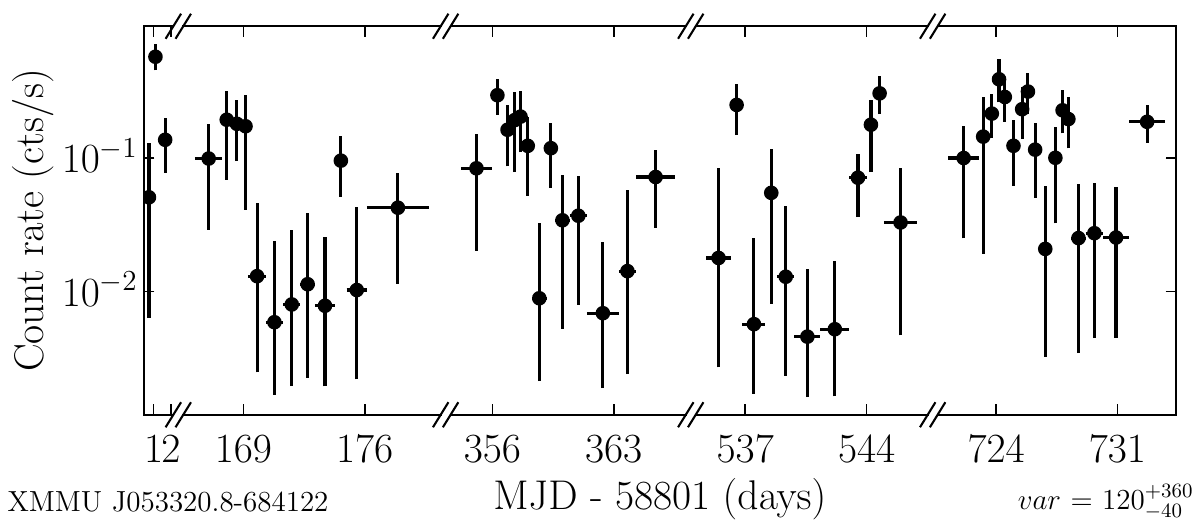}}
            \resizebox{0.495\hsize}{!}{\includegraphics{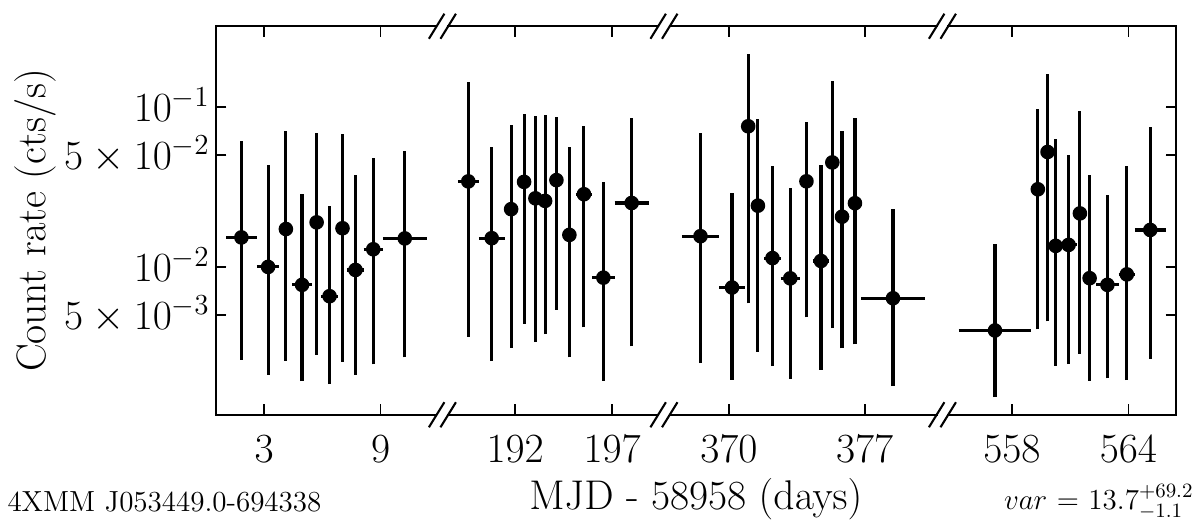}}
            \resizebox{0.495\hsize}{!}{\includegraphics{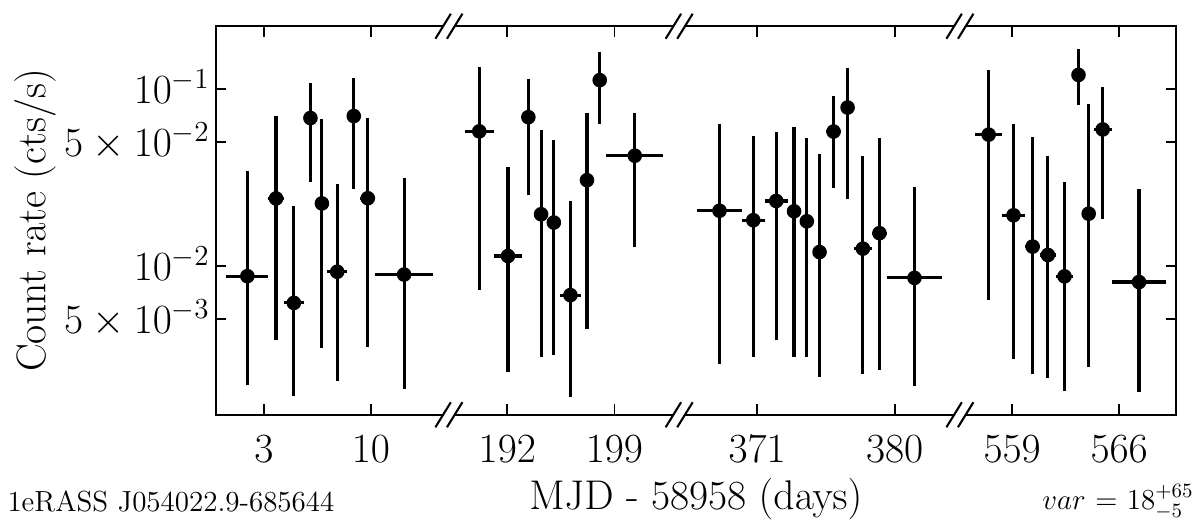}}
            \resizebox{0.495\hsize}{!}{\includegraphics{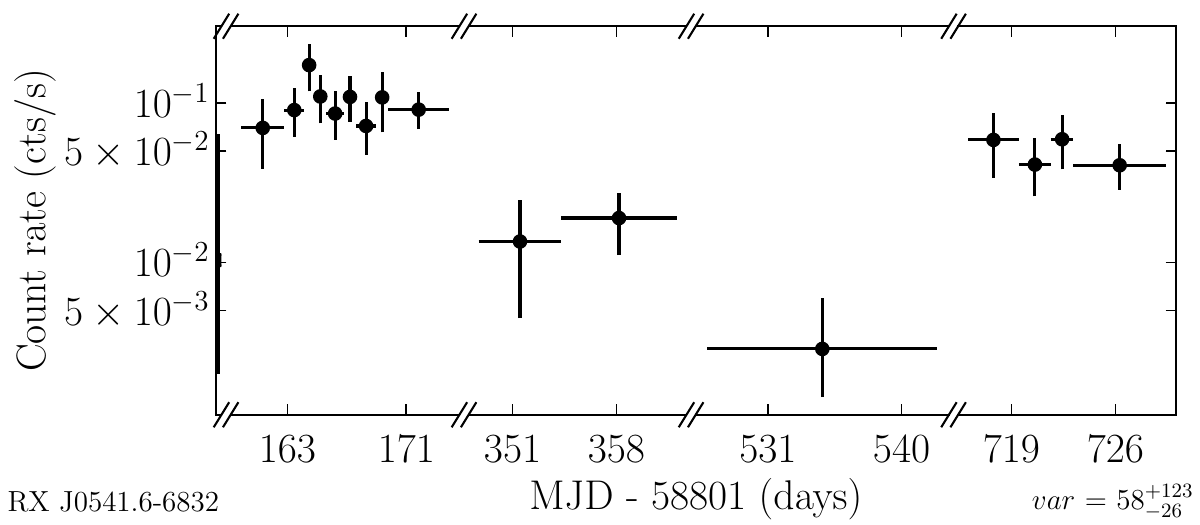}}
            \resizebox{0.495\hsize}{!}{\includegraphics{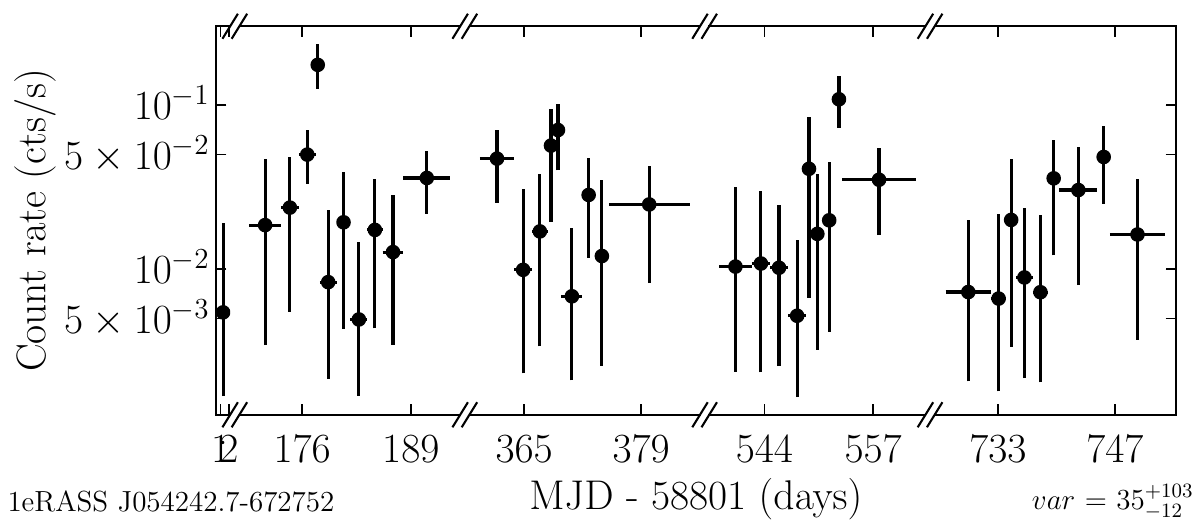}}
            \caption{continued for \#35 to \#46}
        \end{figure*}
        \addtocounter{figure}{-1}\begin{figure*}
            \centering
            \resizebox{0.495\hsize}{!}{\includegraphics{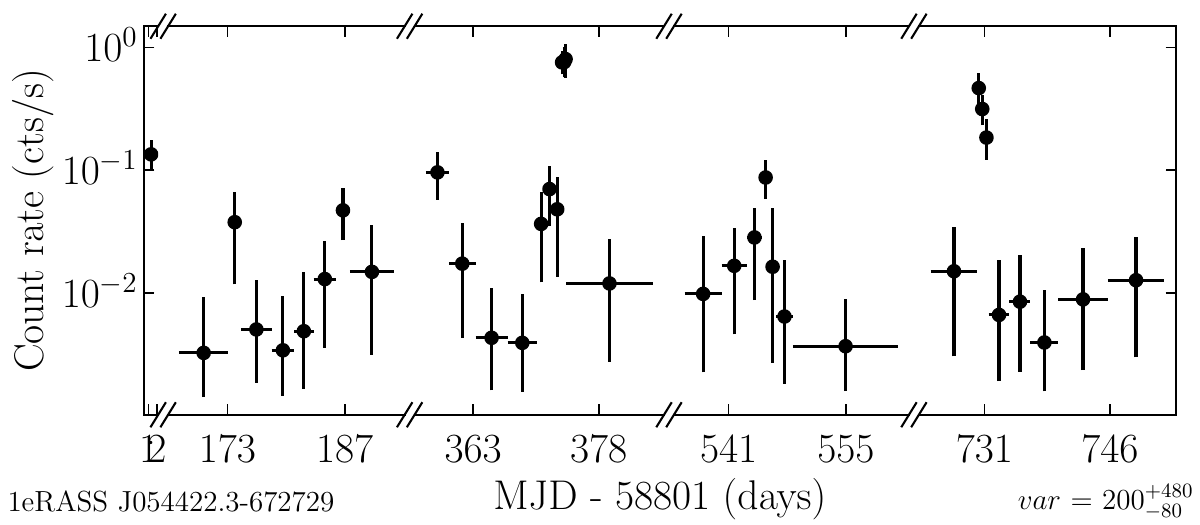}}
            \resizebox{0.495\hsize}{!}{\includegraphics{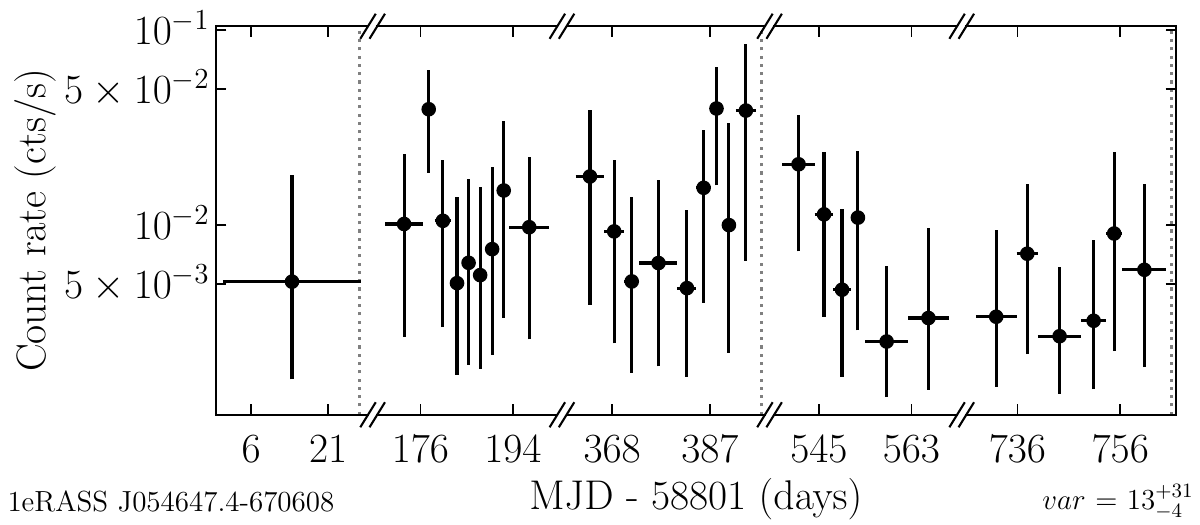}}
            \resizebox{0.495\hsize}{!}{\includegraphics{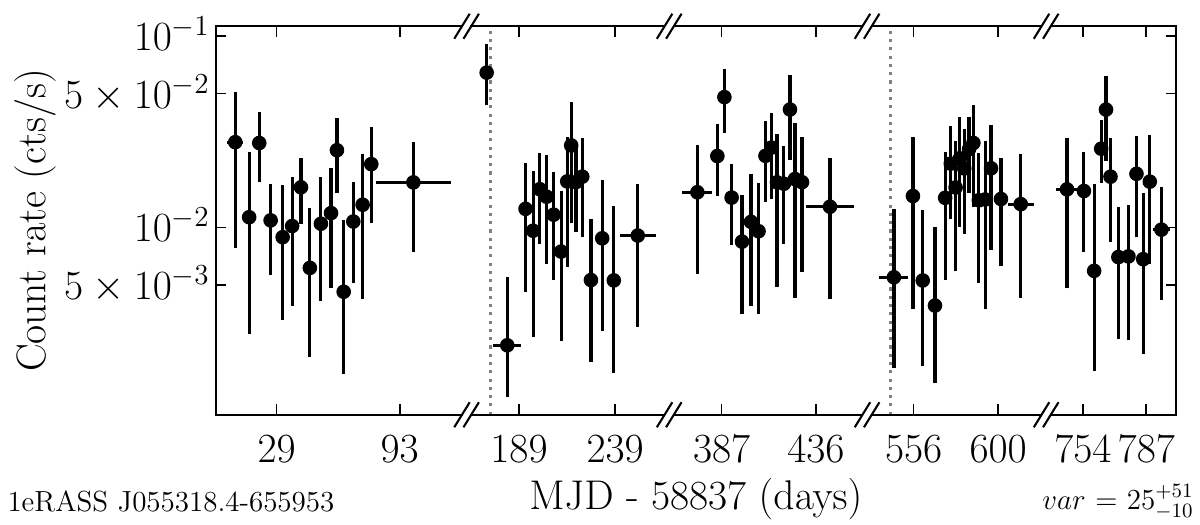}}
            \resizebox{0.495\hsize}{!}{\includegraphics{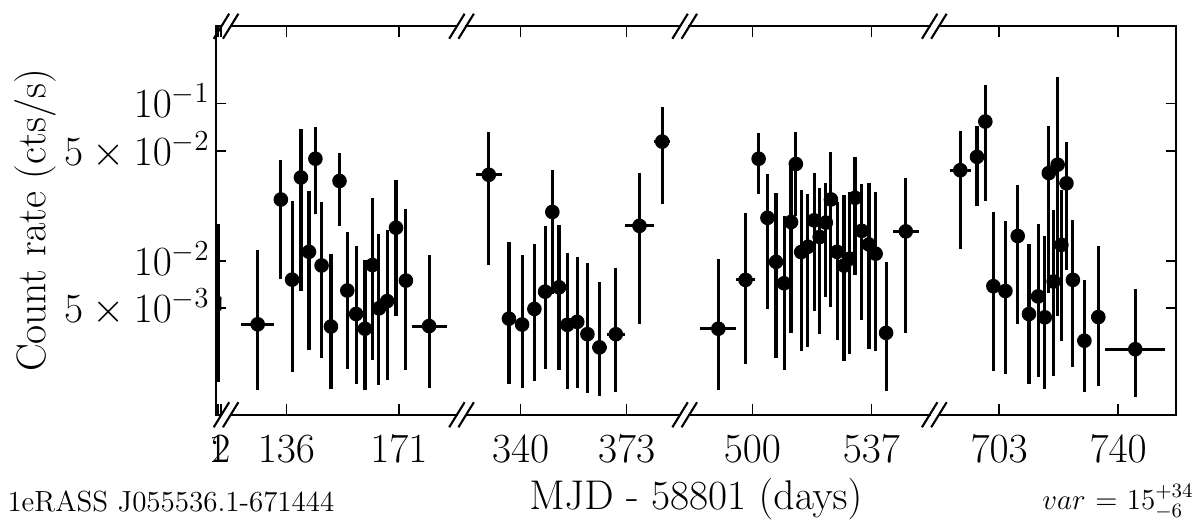}}
            \resizebox{0.495\hsize}{!}{\includegraphics{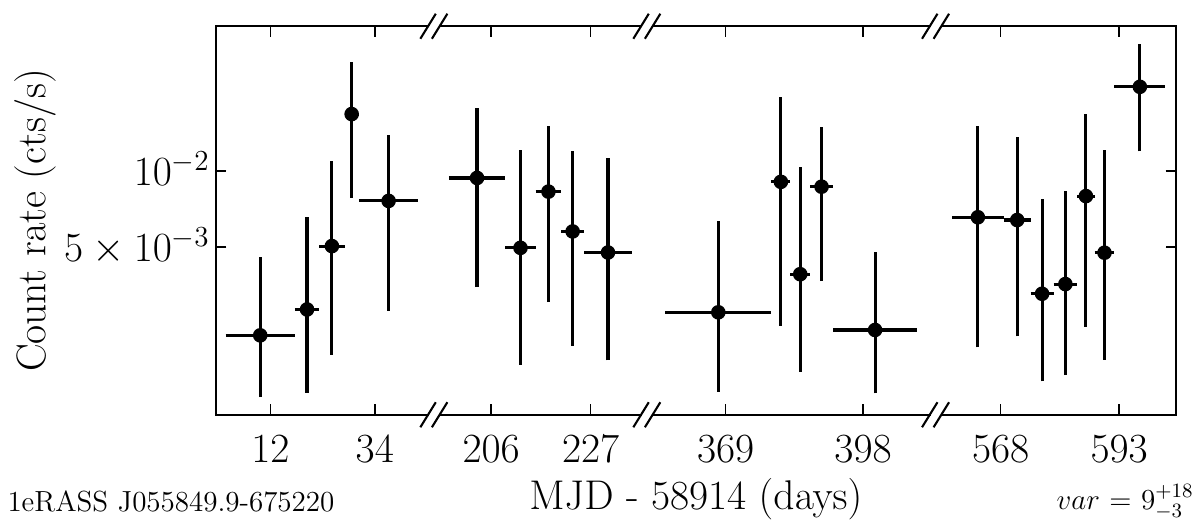}}
            \resizebox{0.495\hsize}{!}{\includegraphics{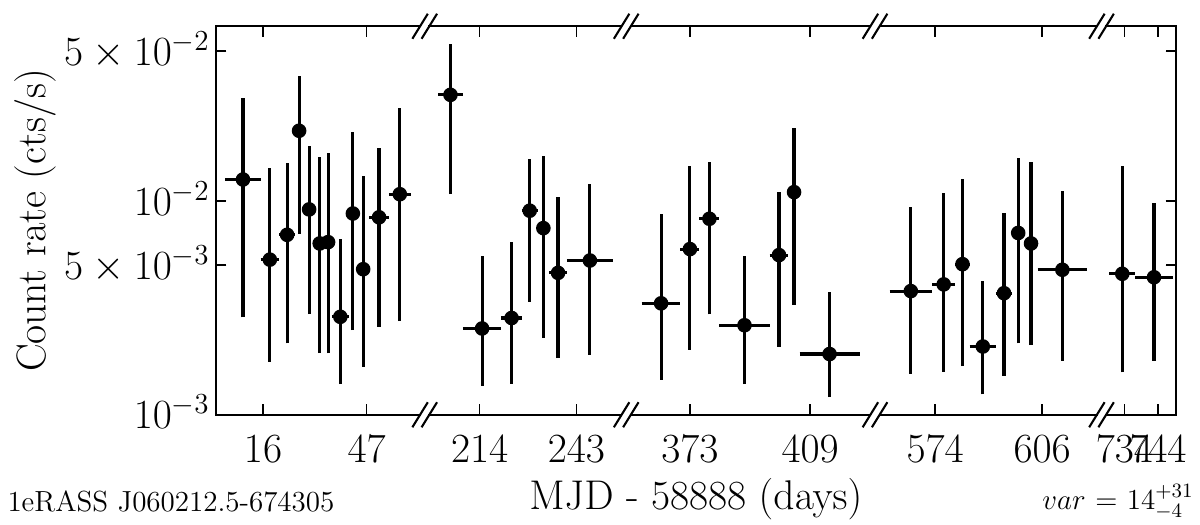}}
            \resizebox{0.495\hsize}{!}{\includegraphics{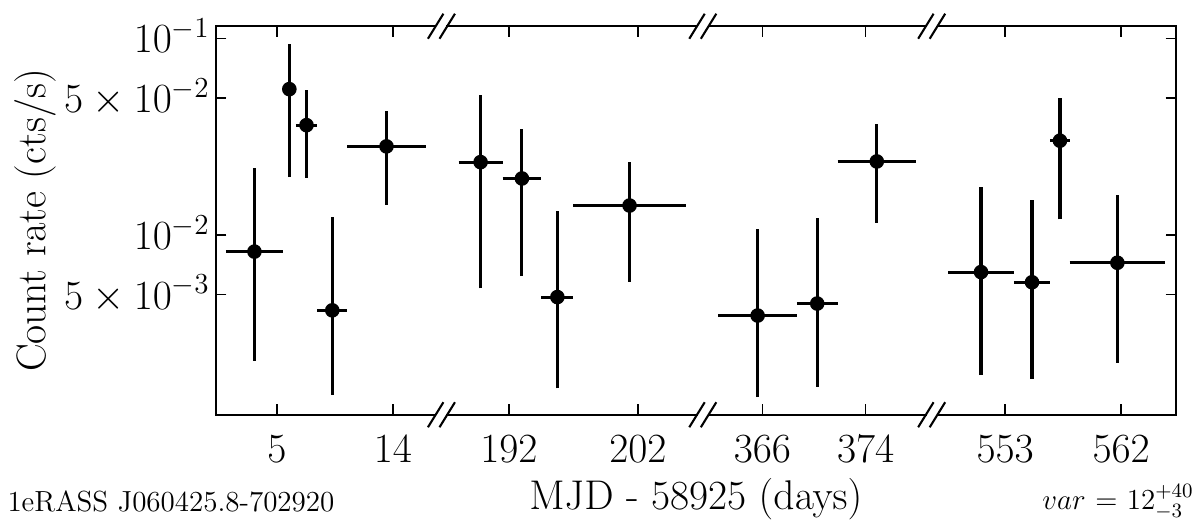}}
            \caption{continued for \#47 to \#53}
        \end{figure*}
        \clearpage

        \begin{figure*}
            \centering
            \resizebox{0.495\hsize}{!}{\includegraphics{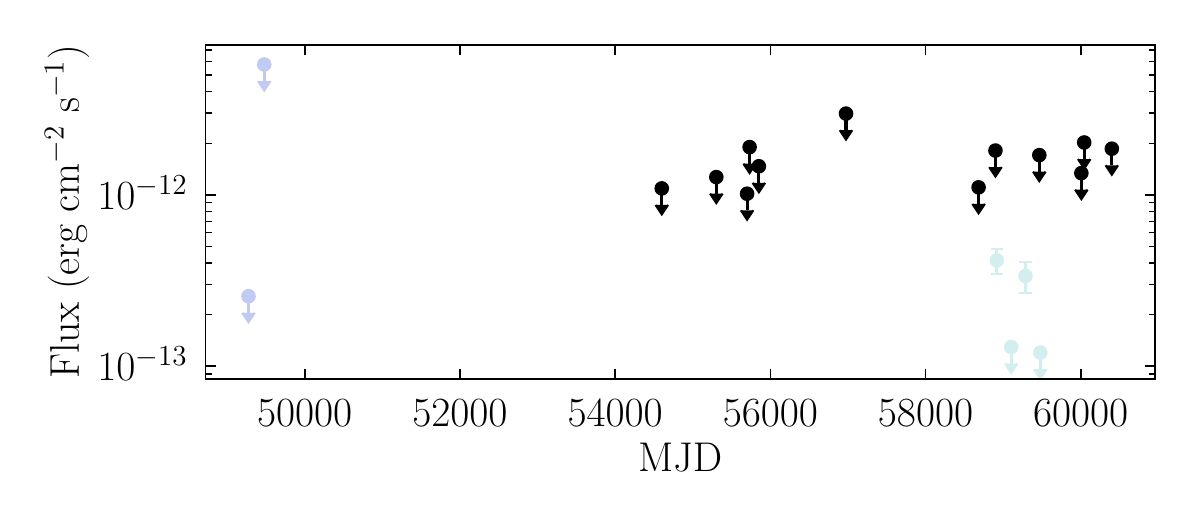}}
            \resizebox{0.495\hsize}{!}{\includegraphics{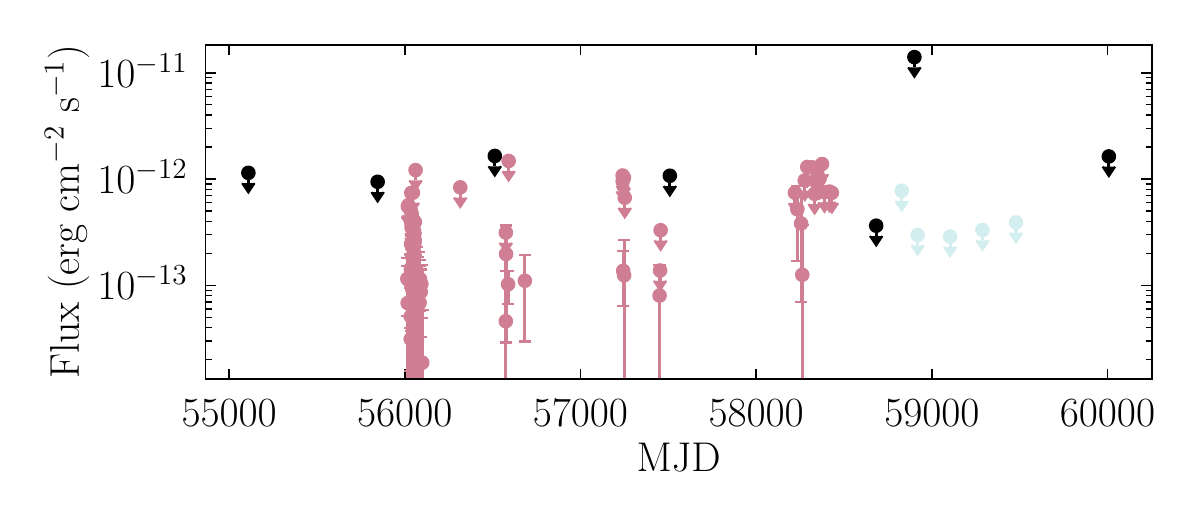}}
            \resizebox{0.495\hsize}{!}{\includegraphics{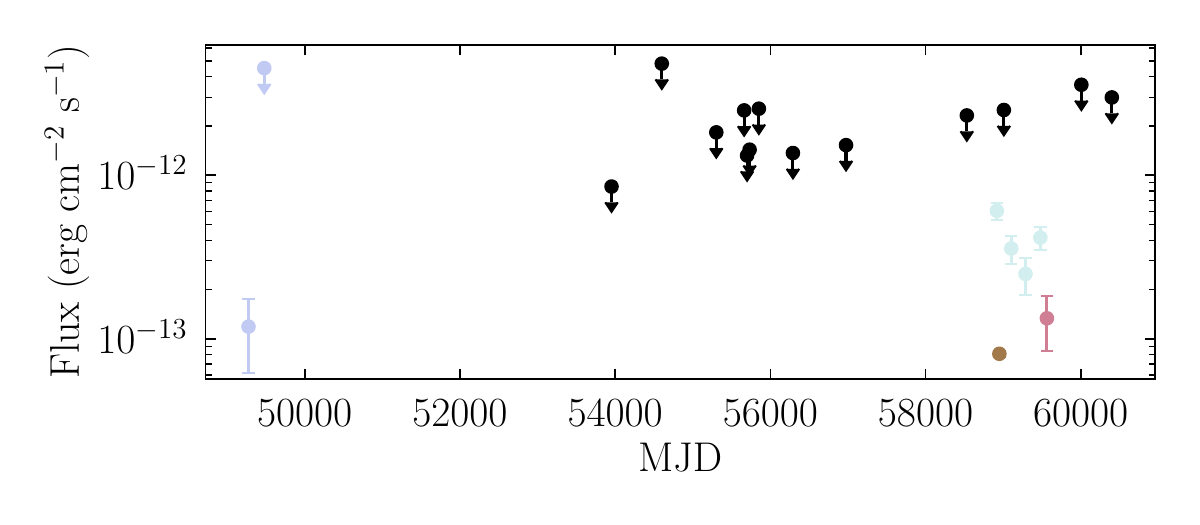}}
            \resizebox{0.495\hsize}{!}{\includegraphics{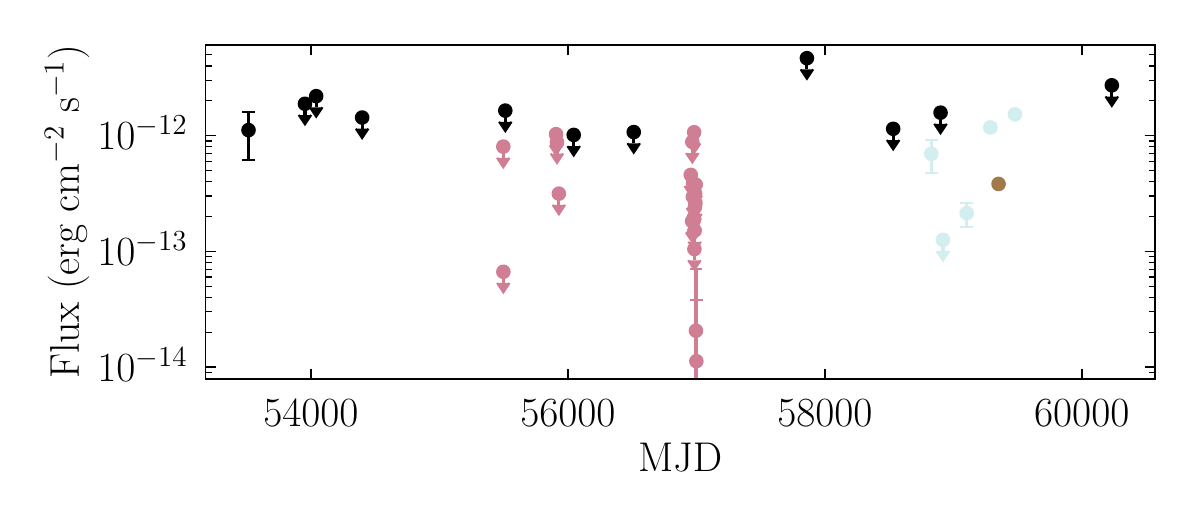}}
            \resizebox{0.495\hsize}{!}{\includegraphics{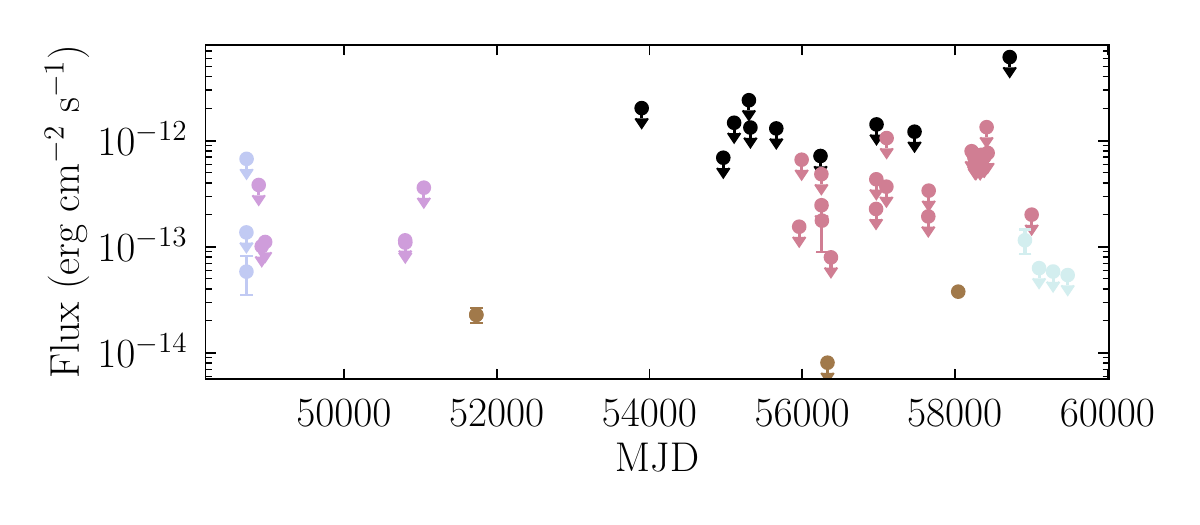}}
            \resizebox{0.495\hsize}{!}{\includegraphics{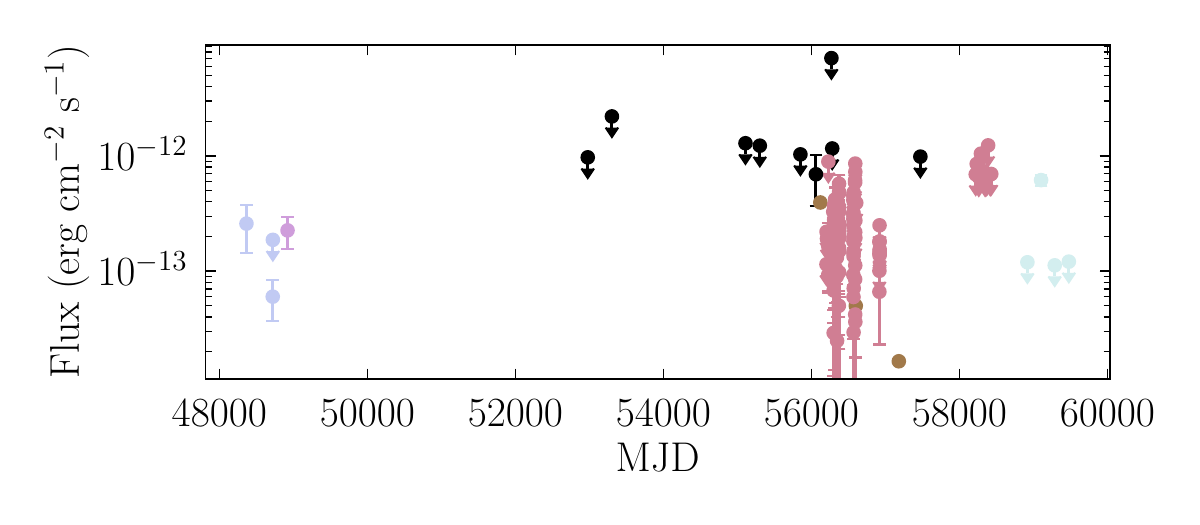}}
            \resizebox{0.495\hsize}{!}{\includegraphics{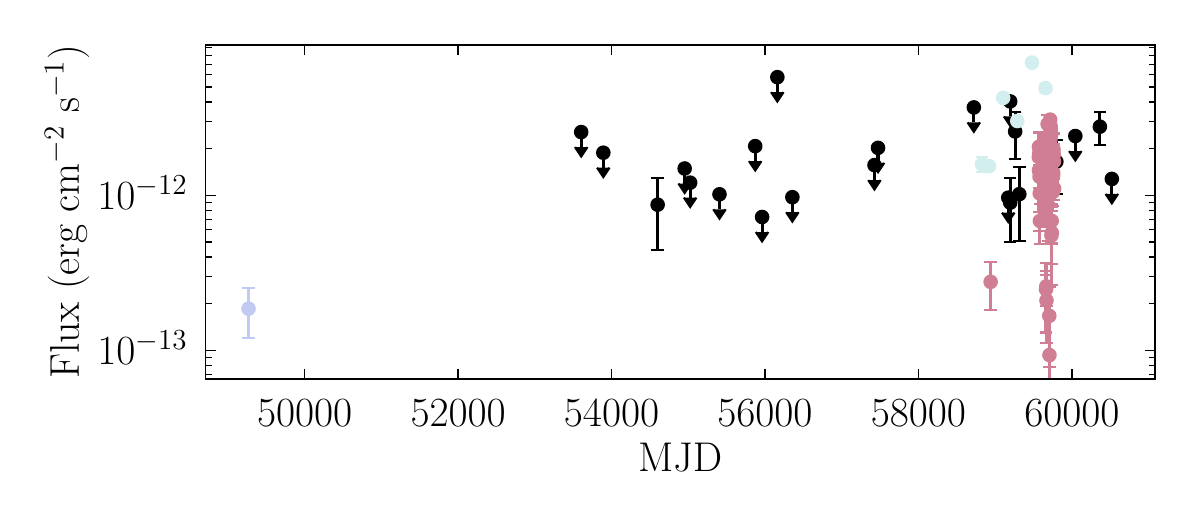}}
            \resizebox{0.495\hsize}{!}{\includegraphics{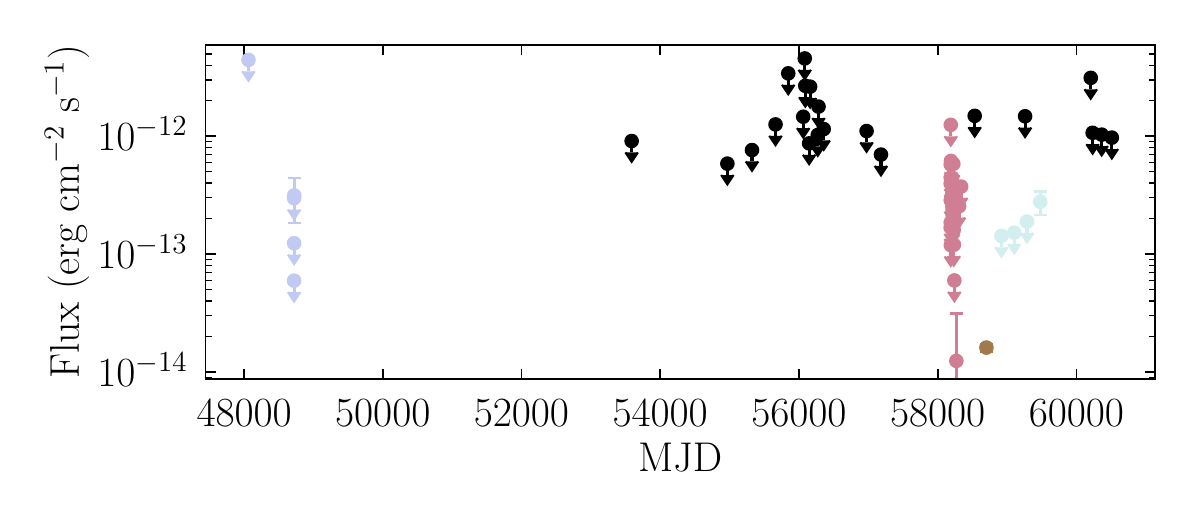}}
            \resizebox{0.495\hsize}{!}{\includegraphics{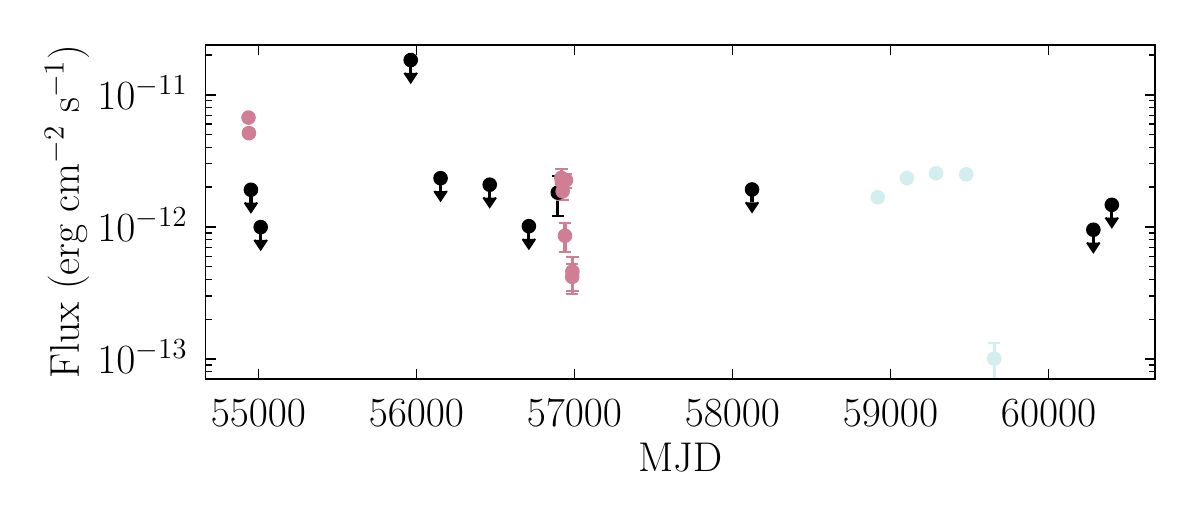}}
            \resizebox{0.495\hsize}{!}{\includegraphics{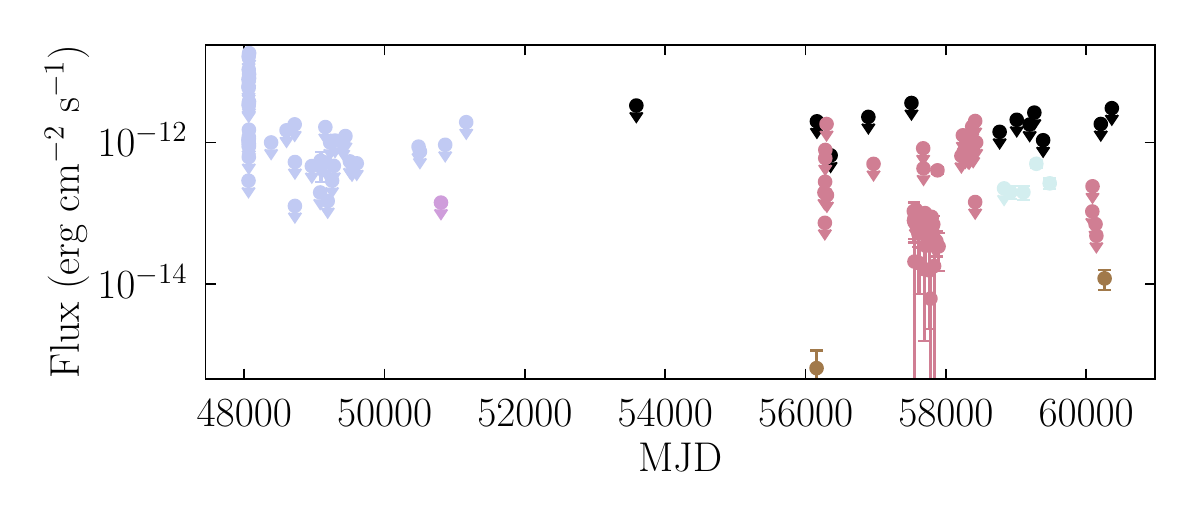}}
            \caption{Long-term light curves extracted as explained in Sect. \ref{sec:long_var} for \#1 to \#10. The flux values and upper limits are shown for \xmm Slew (black), \xmm Pointed (dark yellow), \swift-XRT (salmon), \ROSAT Pointed-HRI (plum), \ROSAT Pointed-PSPC (light steel blue) and \ero (azure).}
            \label{fig:UL_lightcurves}
        \end{figure*}
        \addtocounter{figure}{-1}
        \begin{figure*}
            \centering
            \resizebox{0.495\hsize}{!}{\includegraphics{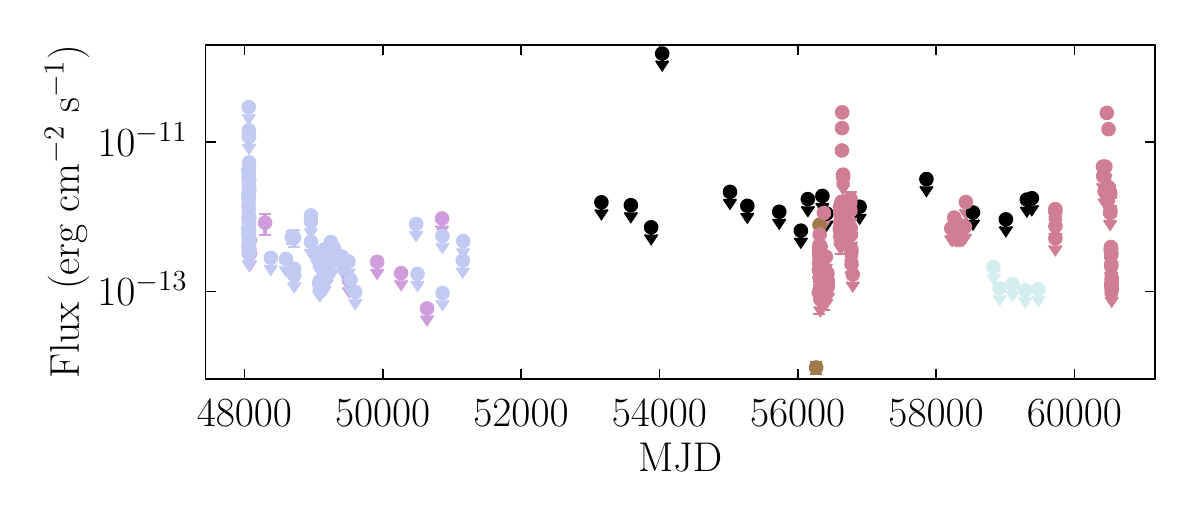}}
            \resizebox{0.495\hsize}{!}{\includegraphics{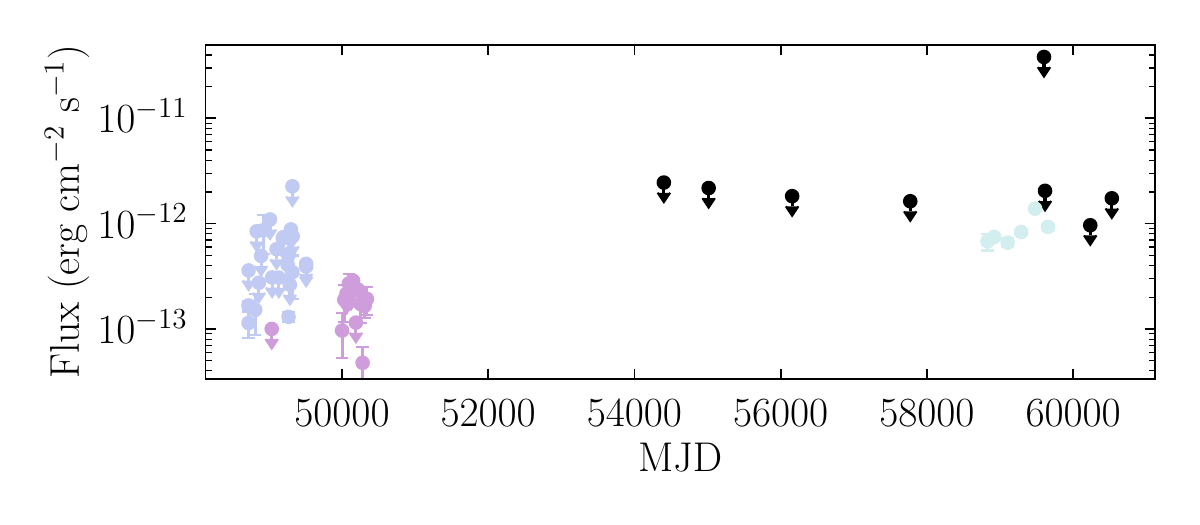}}
            \resizebox{0.495\hsize}{!}{\includegraphics{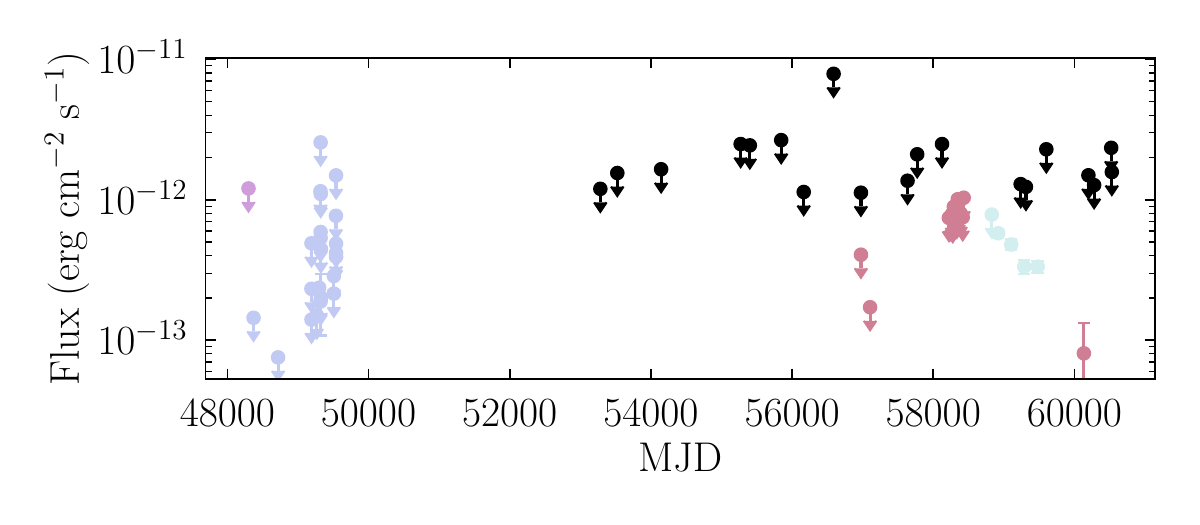}}
            \resizebox{0.495\hsize}{!}{\includegraphics{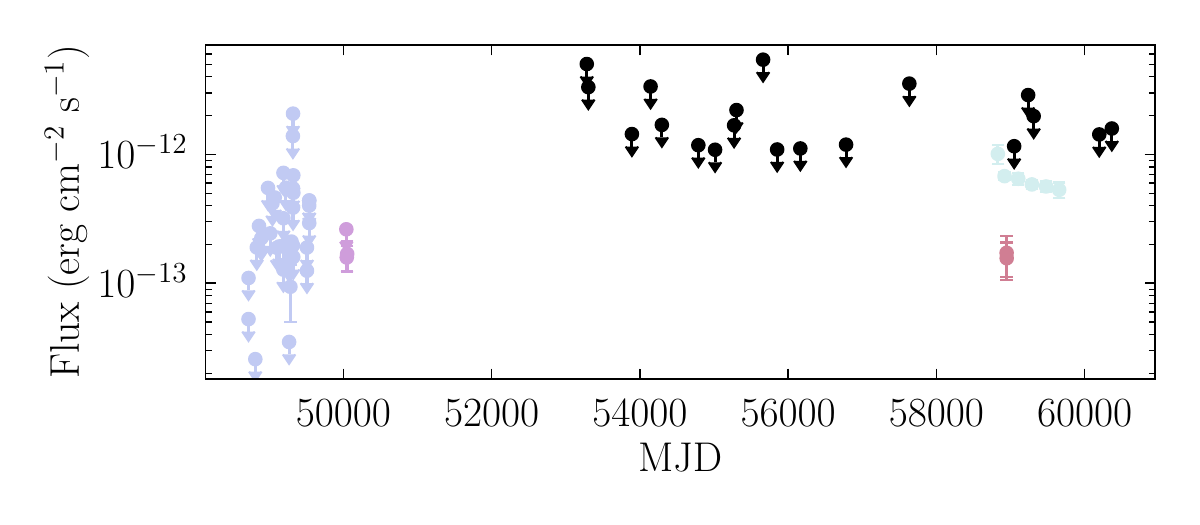}}
            \resizebox{0.495\hsize}{!}{\includegraphics{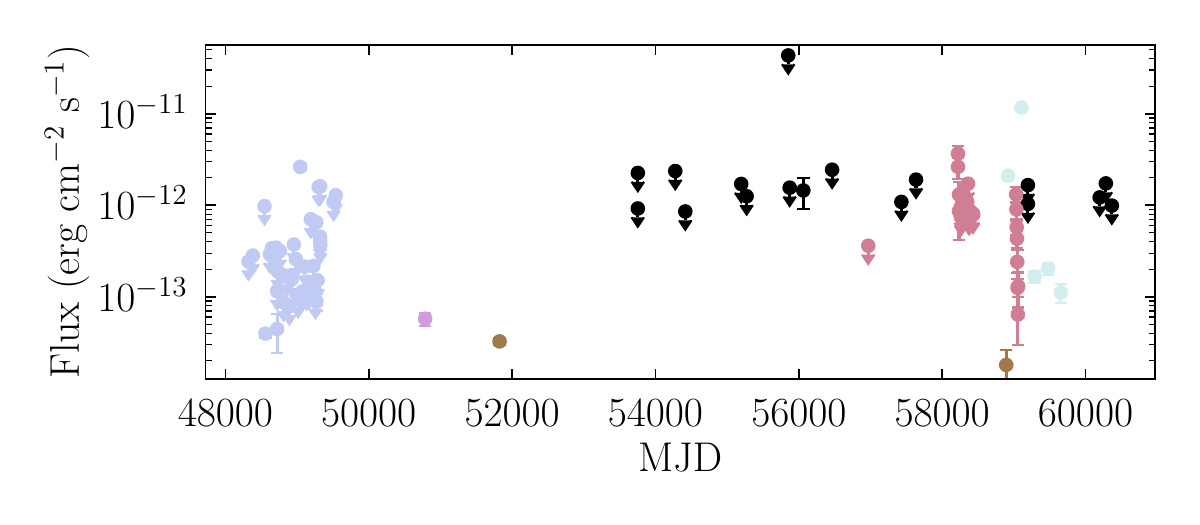}}
            \resizebox{0.495\hsize}{!}{\includegraphics{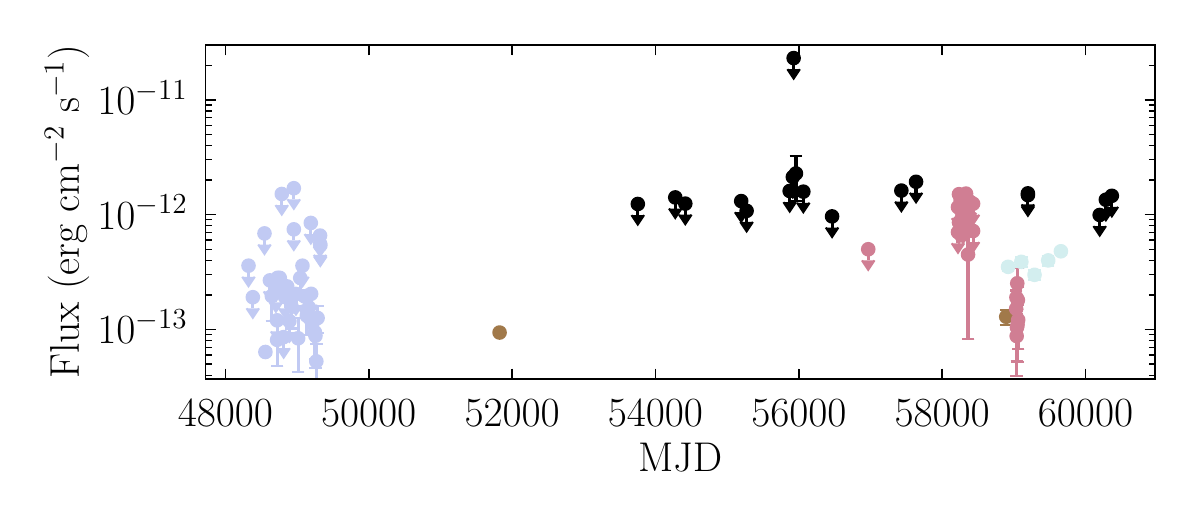}}
            \resizebox{0.495\hsize}{!}{\includegraphics{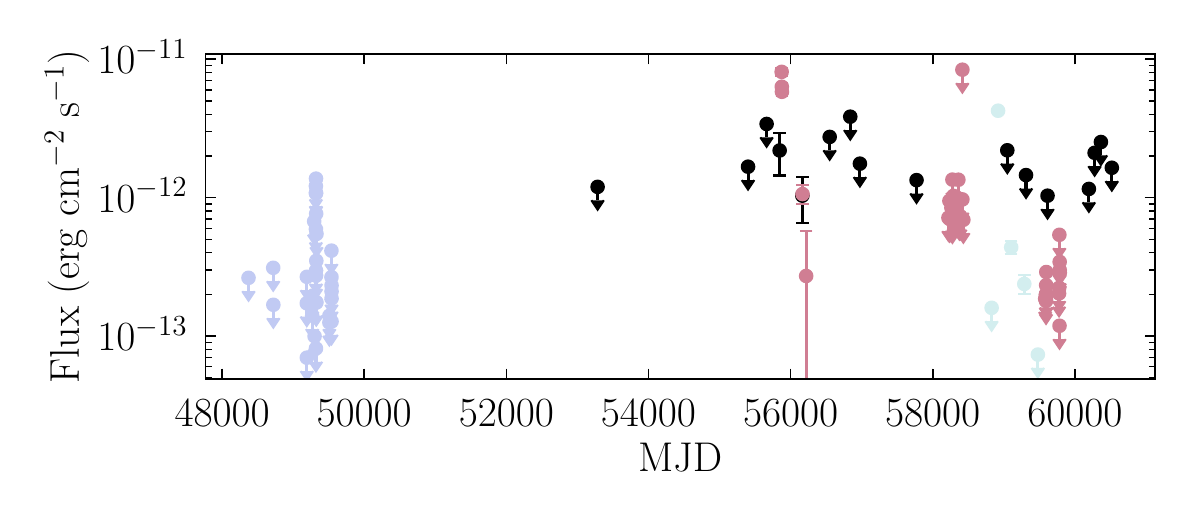}}
            \resizebox{0.495\hsize}{!}{\includegraphics{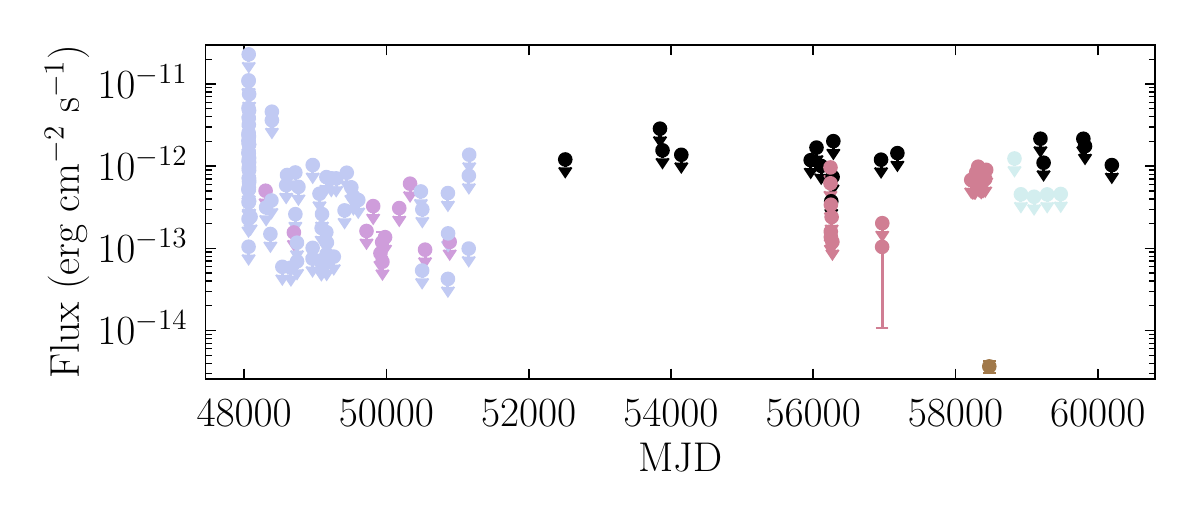}}
            \resizebox{0.495\hsize}{!}{\includegraphics{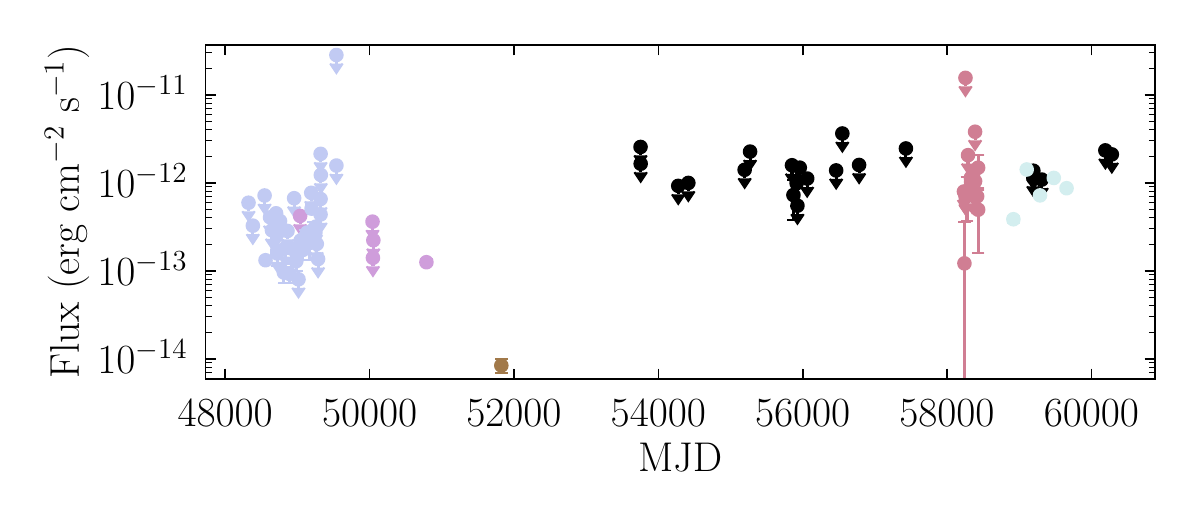}}
            \resizebox{0.495\hsize}{!}{\includegraphics{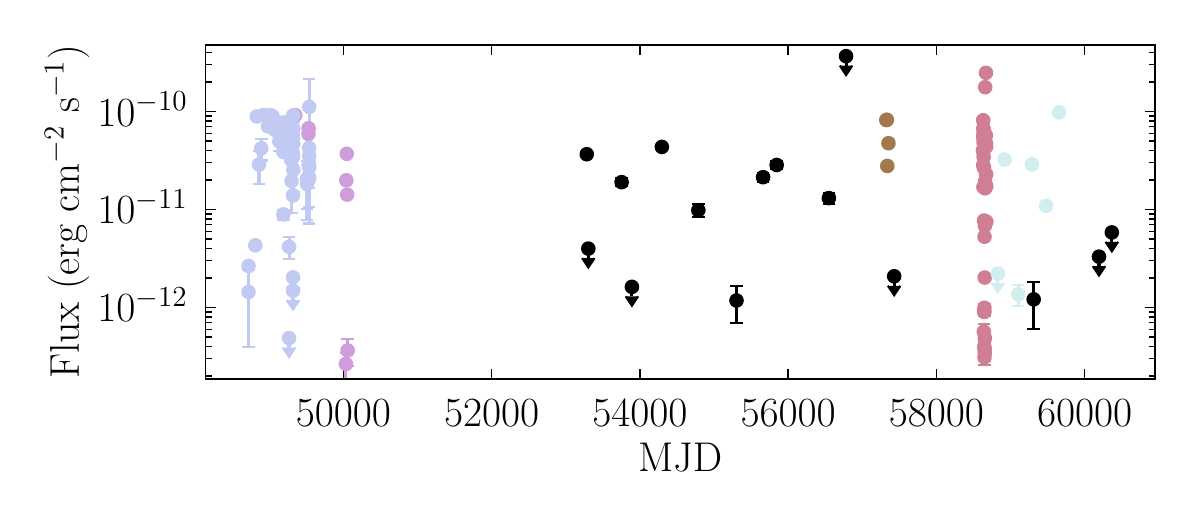}}
            \resizebox{0.495\hsize}{!}{\includegraphics{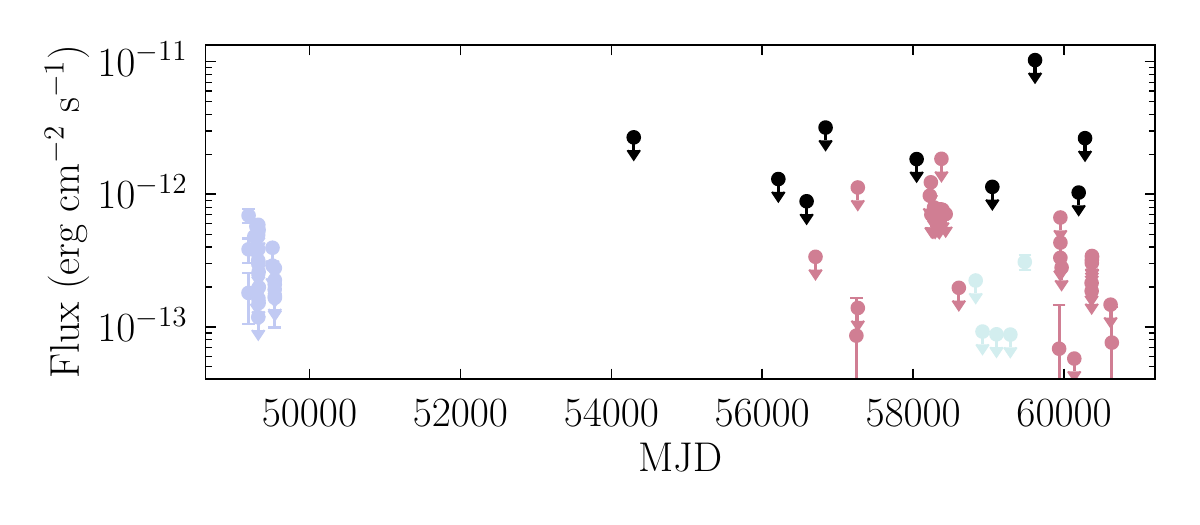}}
            \resizebox{0.495\hsize}{!}{\includegraphics{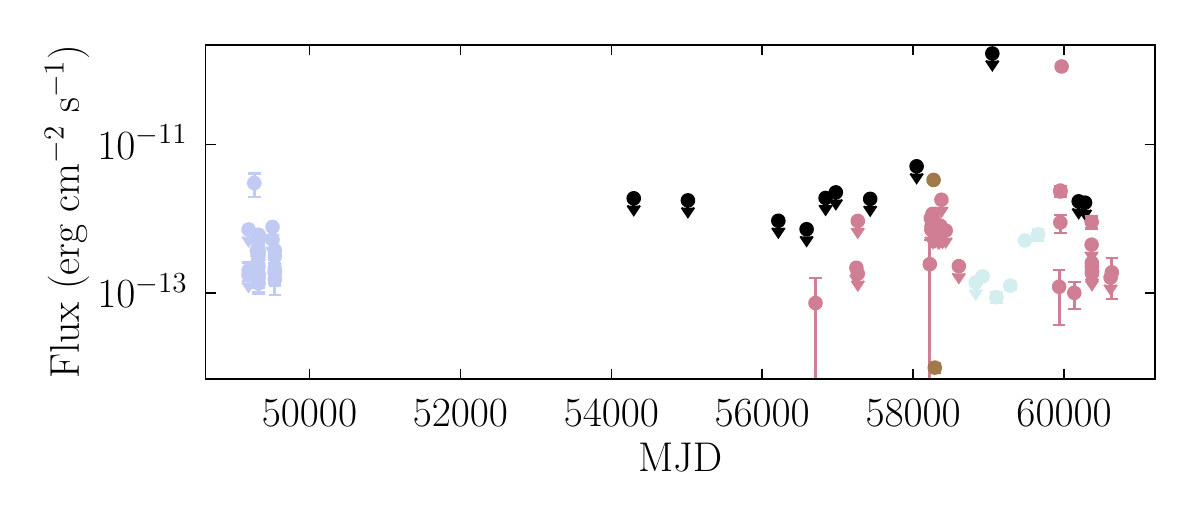}}
            \caption{continued for \#11 to \#23}
        \end{figure*}
        \addtocounter{figure}{-1}
        \begin{figure*}
            \centering
            \resizebox{0.495\hsize}{!}{\includegraphics{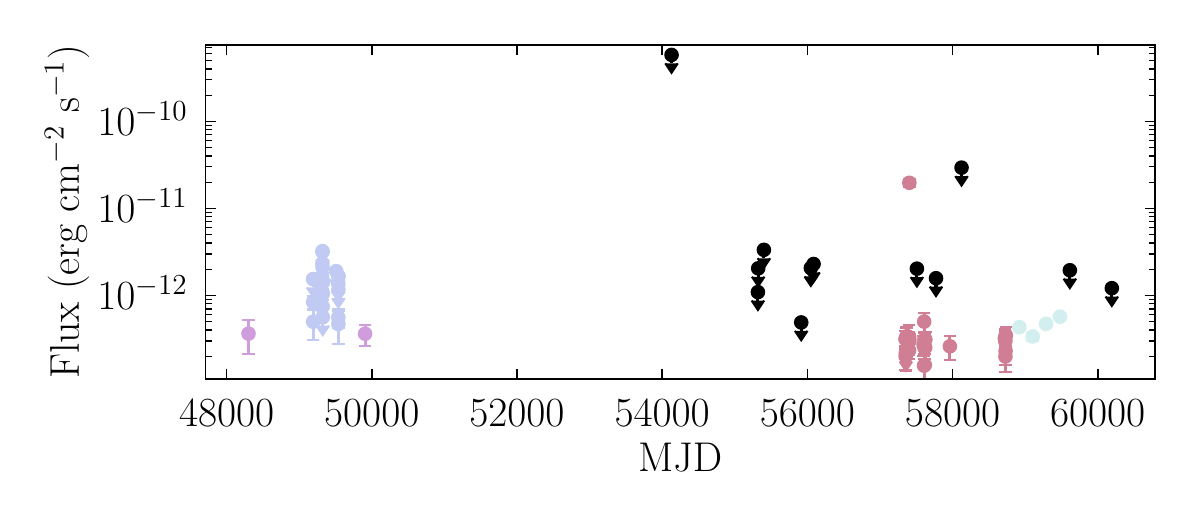}}
            \resizebox{0.495\hsize}{!}{\includegraphics{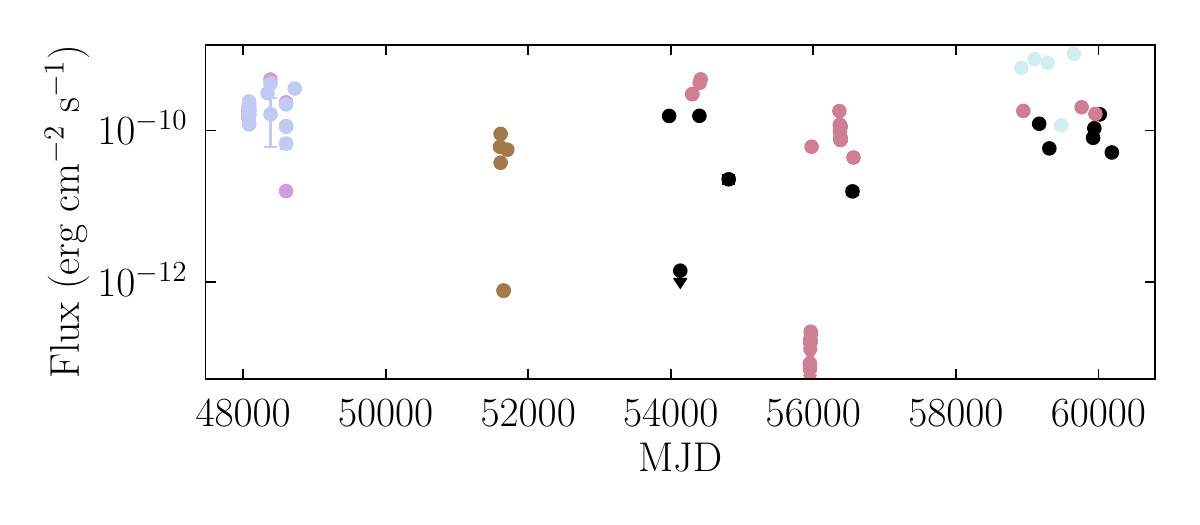}}
            \resizebox{0.495\hsize}{!}{\includegraphics{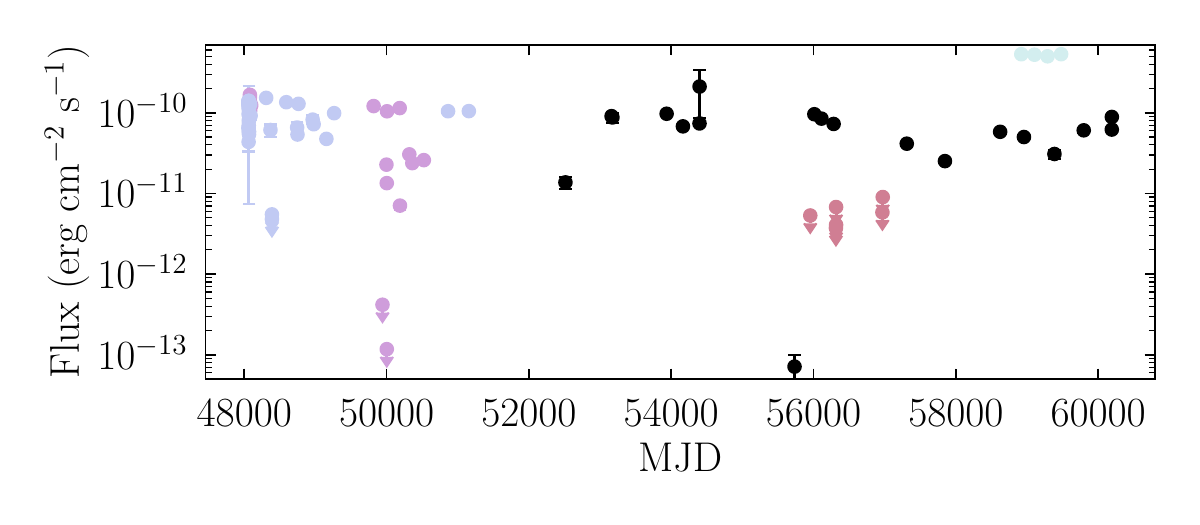}}
            \resizebox{0.495\hsize}{!}{\includegraphics{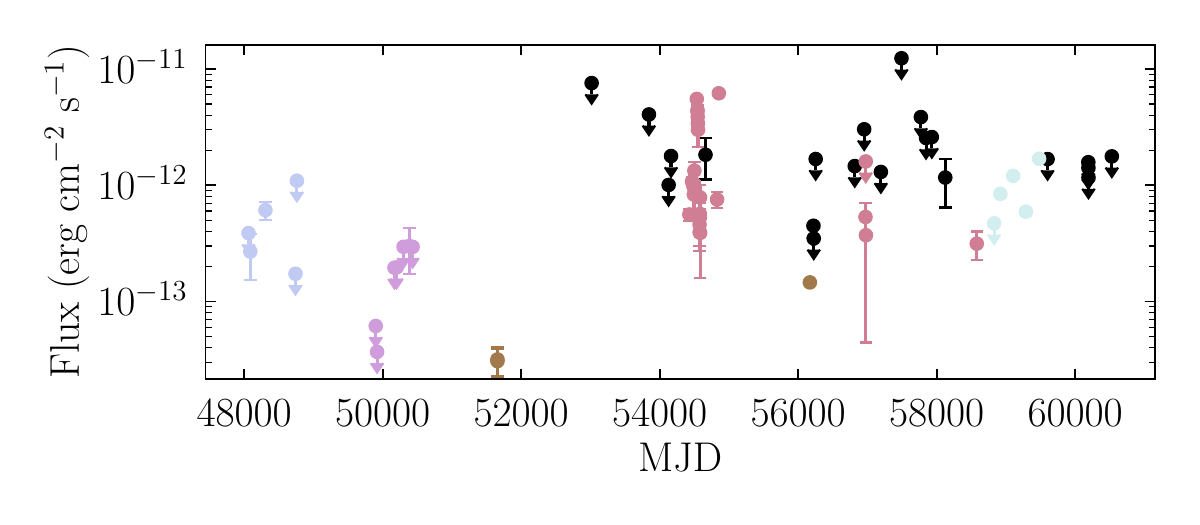}}
            \resizebox{0.495\hsize}{!}{\includegraphics{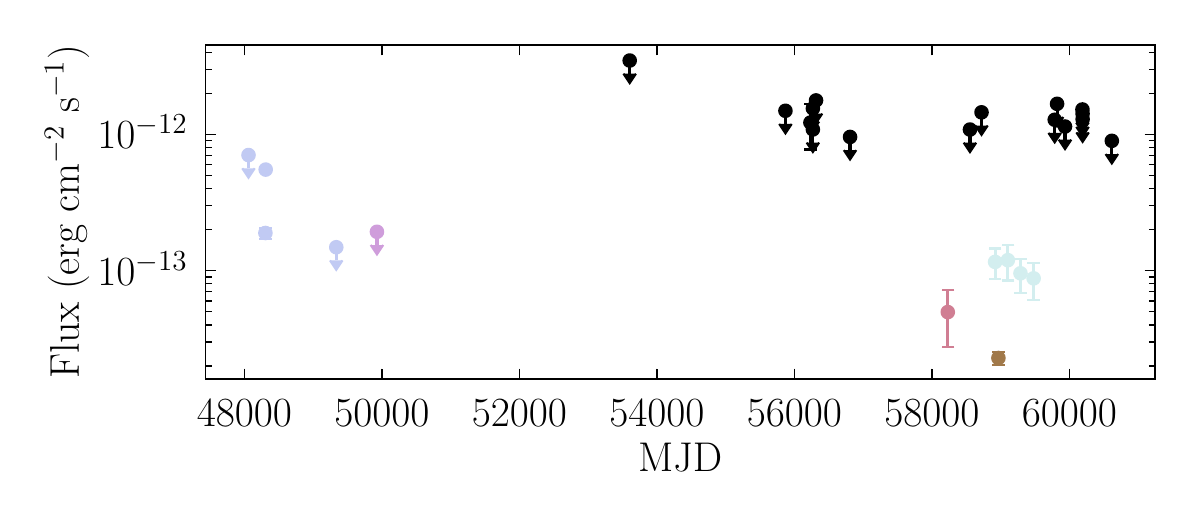}}
            \resizebox{0.495\hsize}{!}{\includegraphics{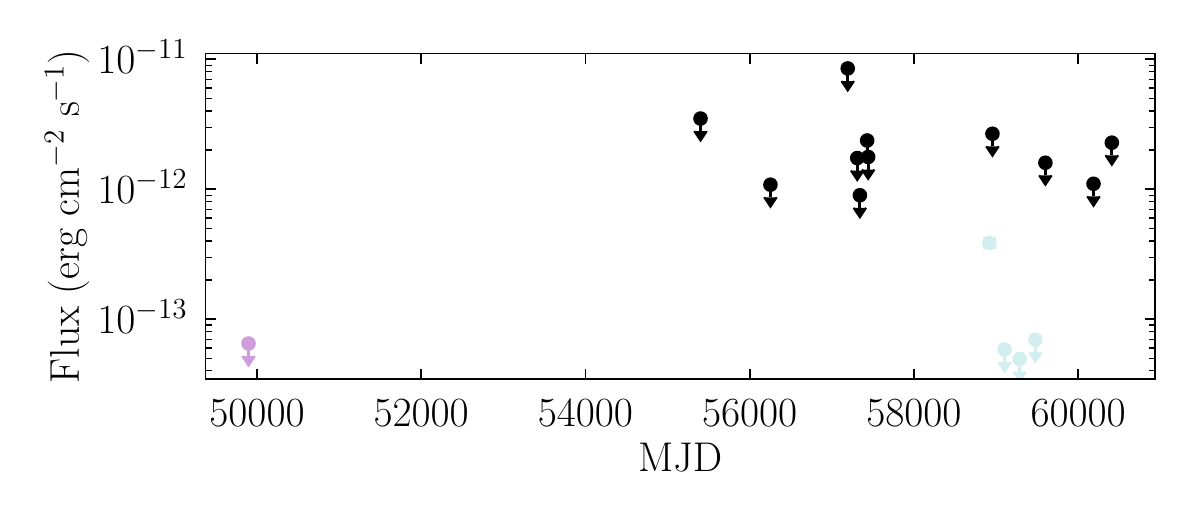}}
            \resizebox{0.495\hsize}{!}{\includegraphics{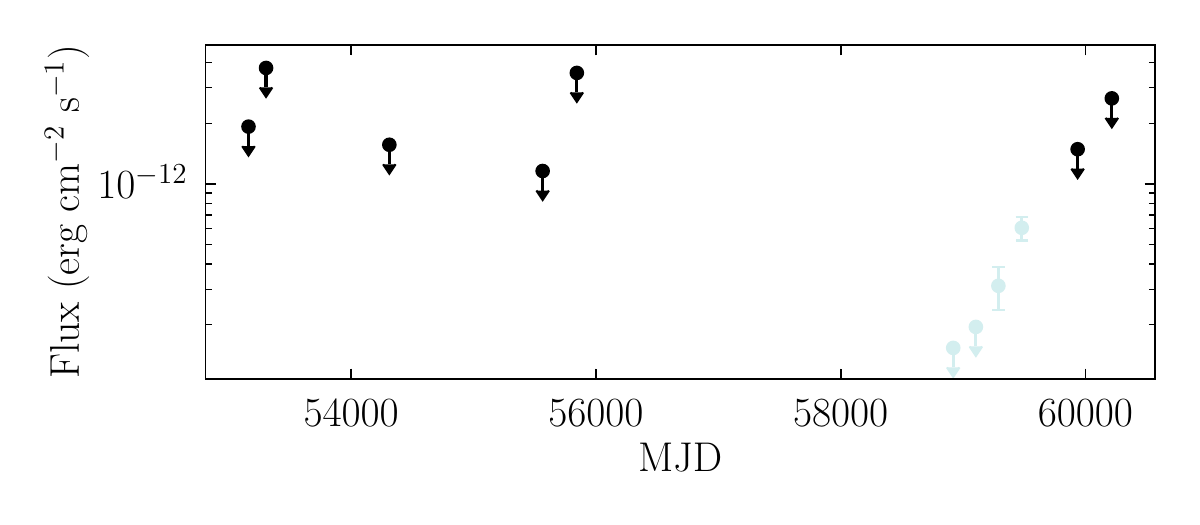}}
            \resizebox{0.495\hsize}{!}{\includegraphics{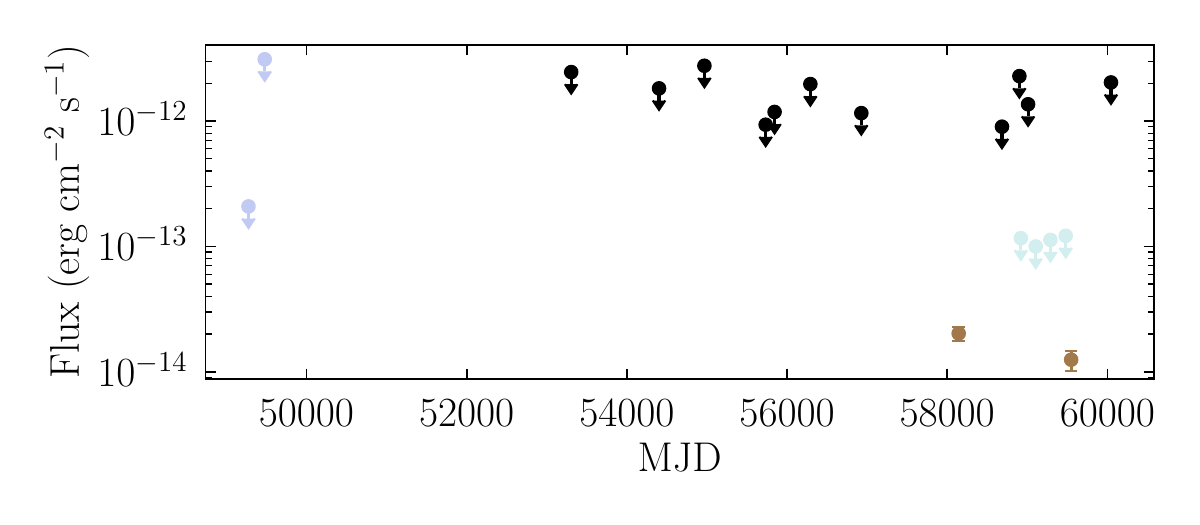}}
            \resizebox{0.495\hsize}{!}{\includegraphics{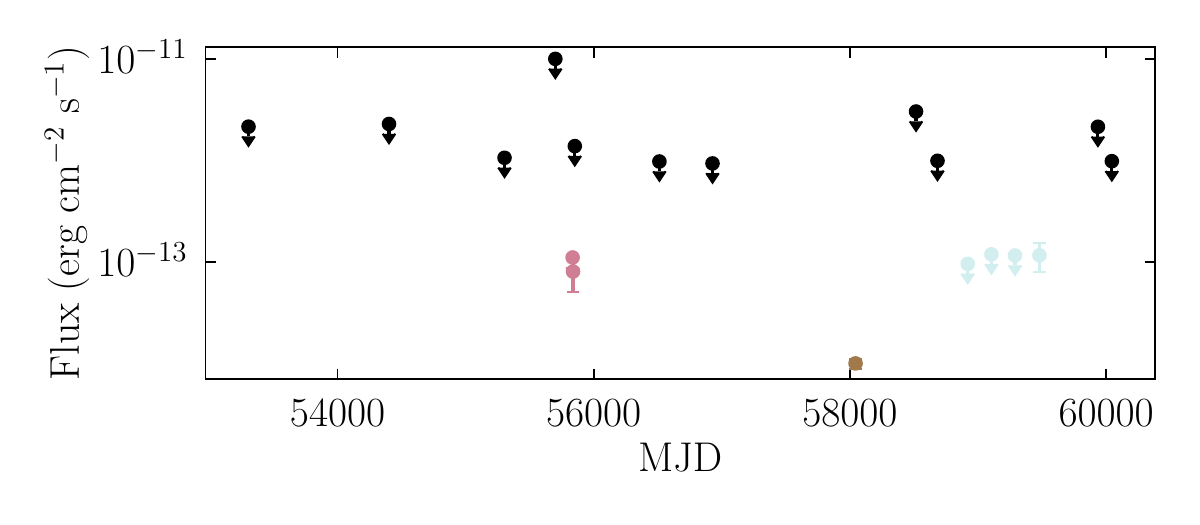}}
            \resizebox{0.495\hsize}{!}{\includegraphics{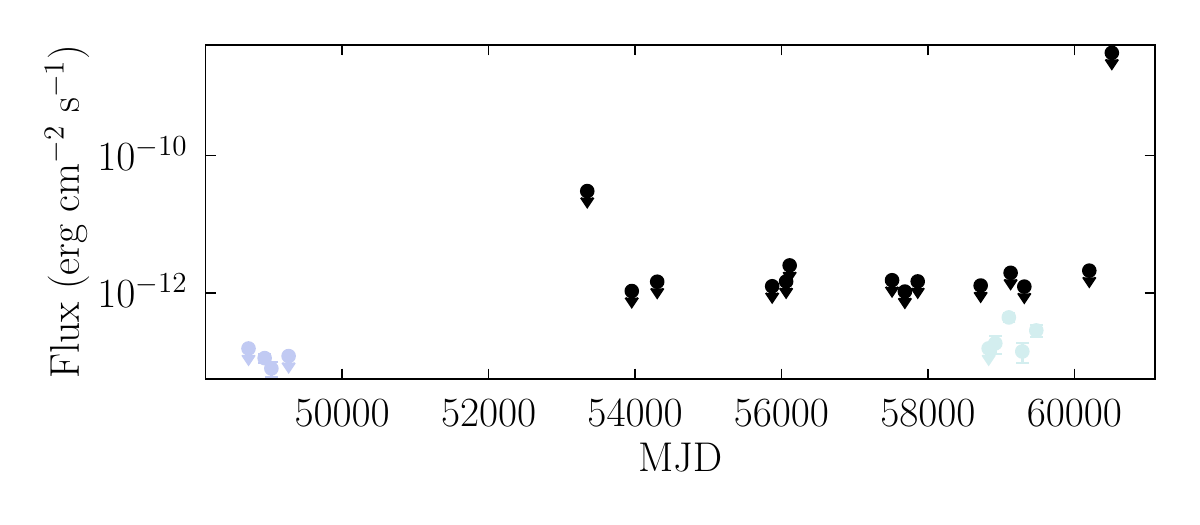}}
            \resizebox{0.495\hsize}{!}{\includegraphics{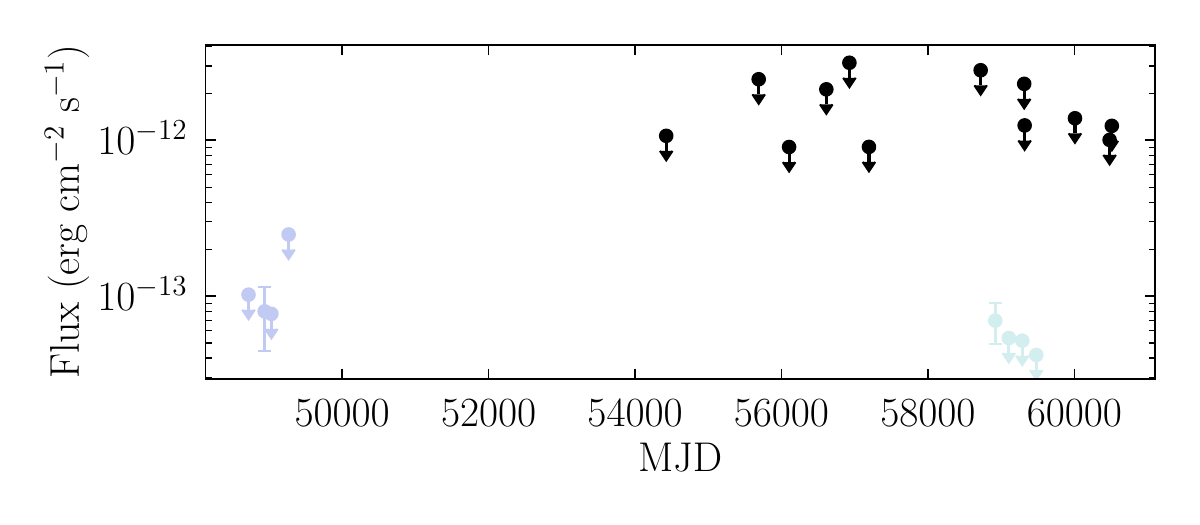}}
            \resizebox{0.495\hsize}{!}{\includegraphics{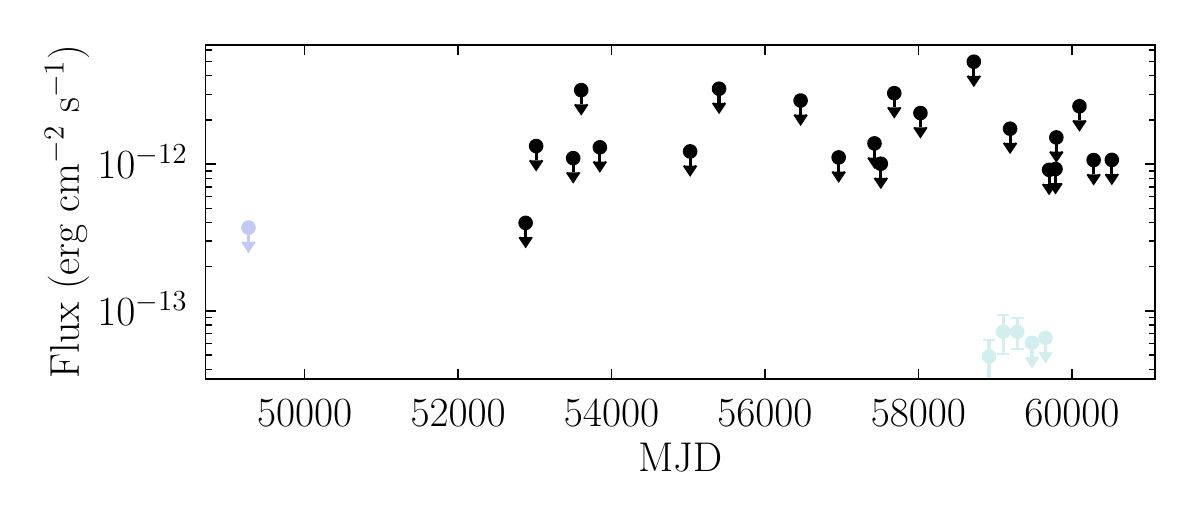}}
            \caption{continued for \#23 to \#34.}
        \end{figure*}
        \addtocounter{figure}{-1}
        \begin{figure*}
            \centering
            \resizebox{0.495\hsize}{!}{\includegraphics{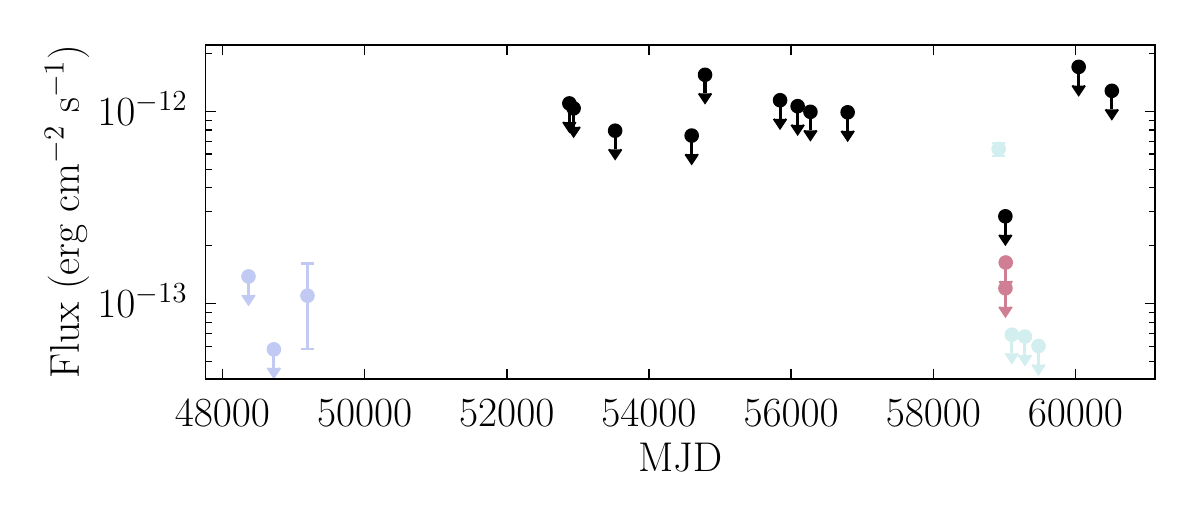}}
            \resizebox{0.495\hsize}{!}{\includegraphics{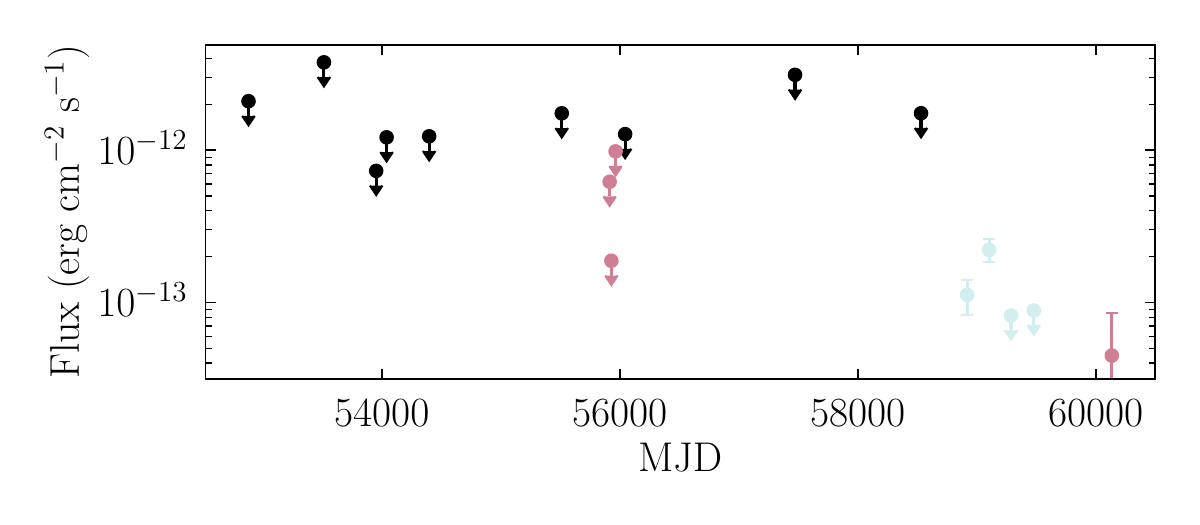}}
            \resizebox{0.495\hsize}{!}{\includegraphics{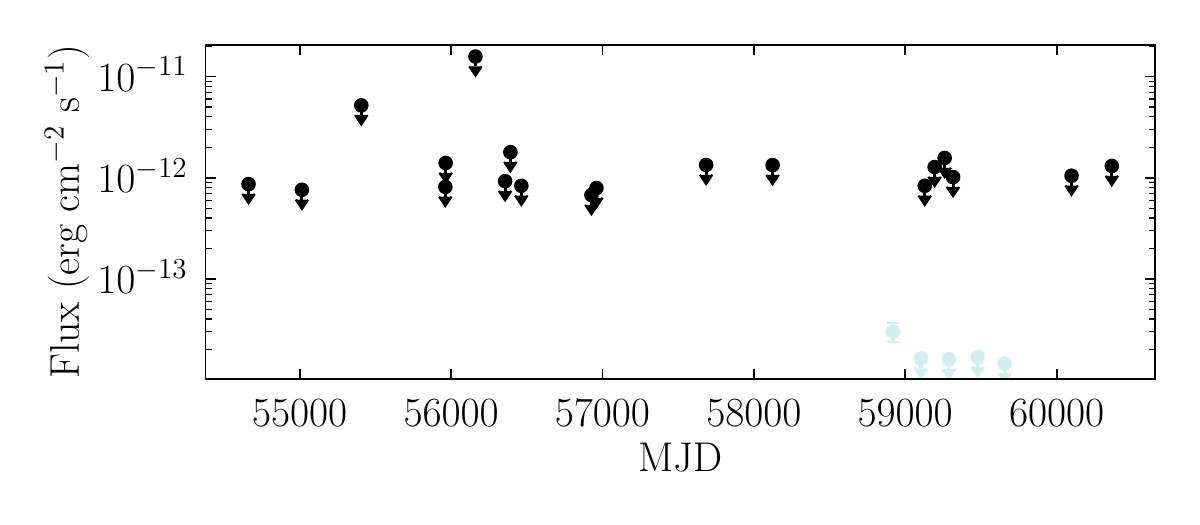}}
            \resizebox{0.495\hsize}{!}{\includegraphics{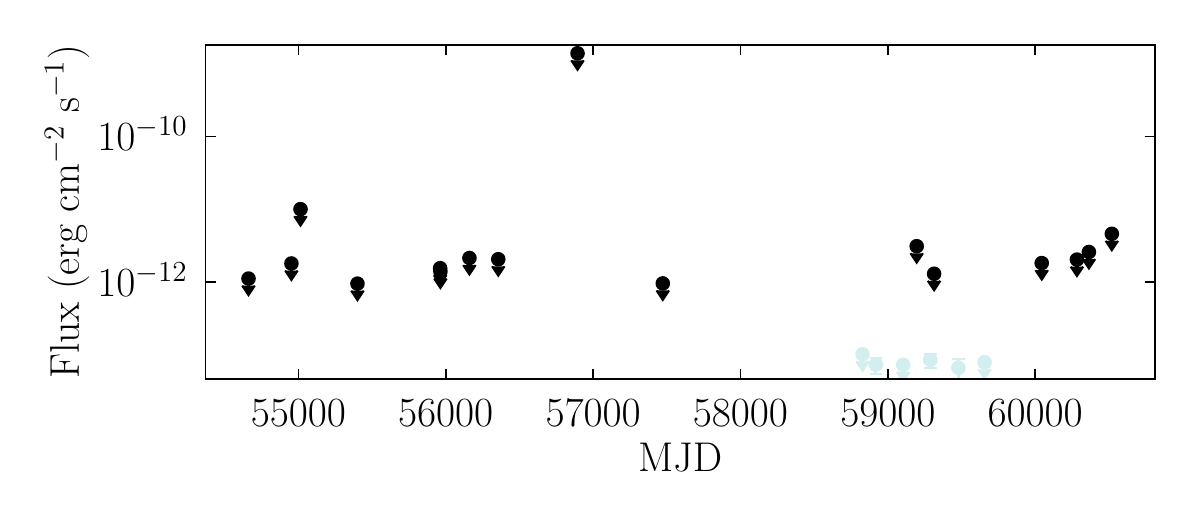}}
            \resizebox{0.495\hsize}{!}{\includegraphics{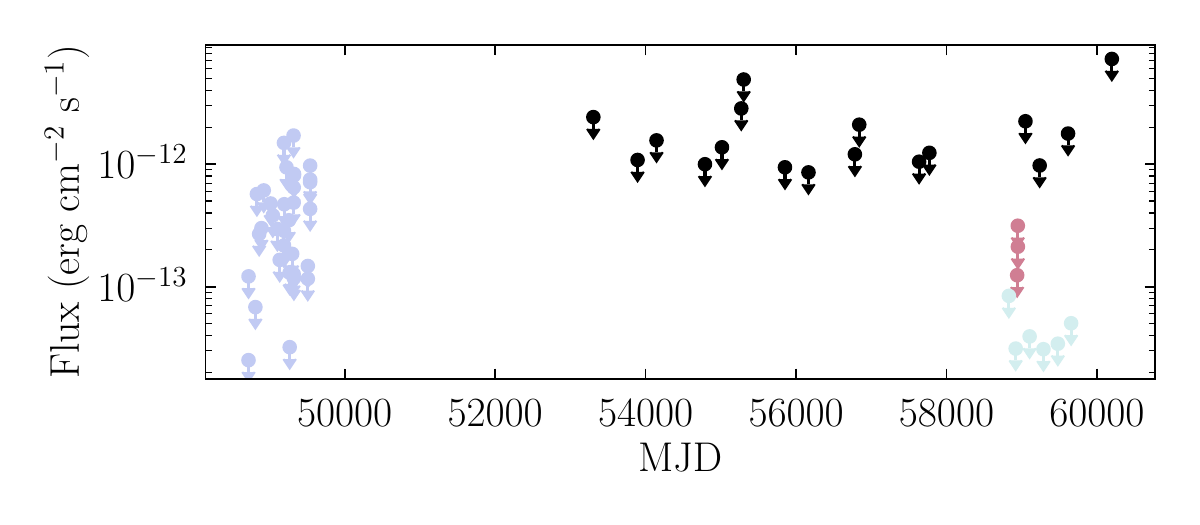}}
            \resizebox{0.495\hsize}{!}{\includegraphics{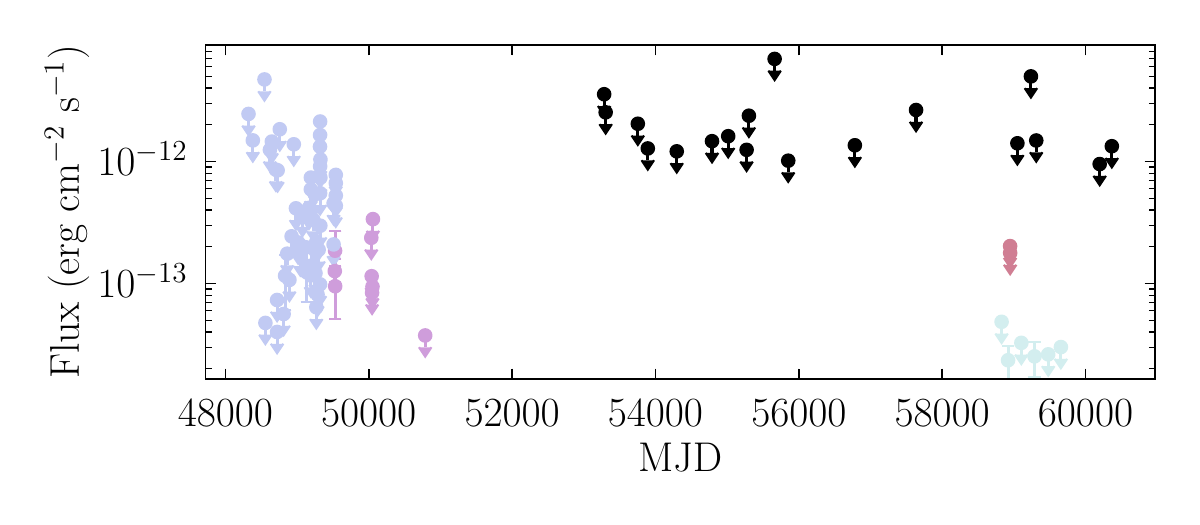}}
            \resizebox{0.495\hsize}{!}{\includegraphics{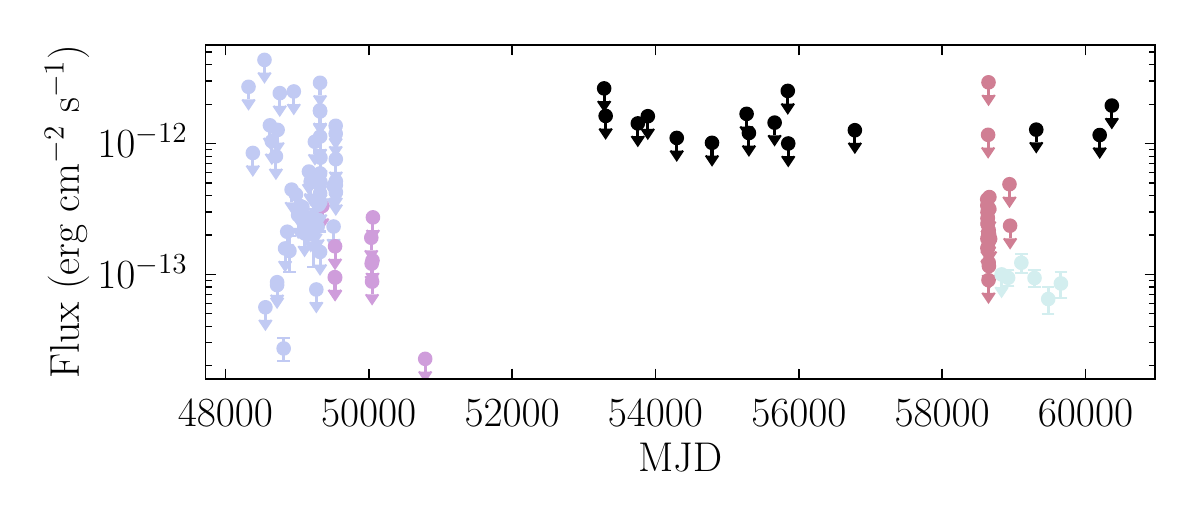}}
            \resizebox{0.495\hsize}{!}{\includegraphics{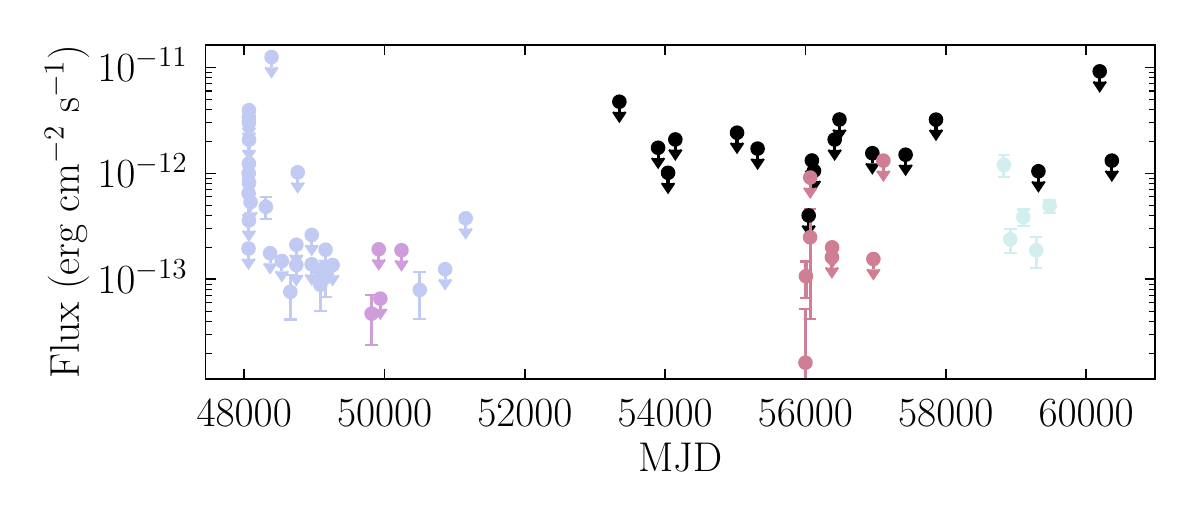}}
            \resizebox{0.495\hsize}{!}{\includegraphics{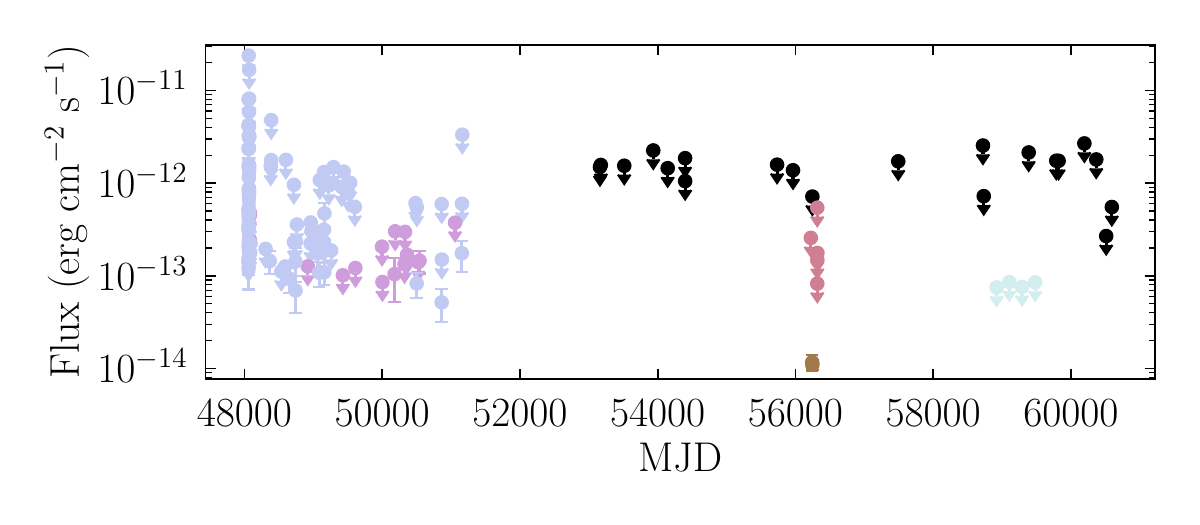}}
            \resizebox{0.495\hsize}{!}{\includegraphics{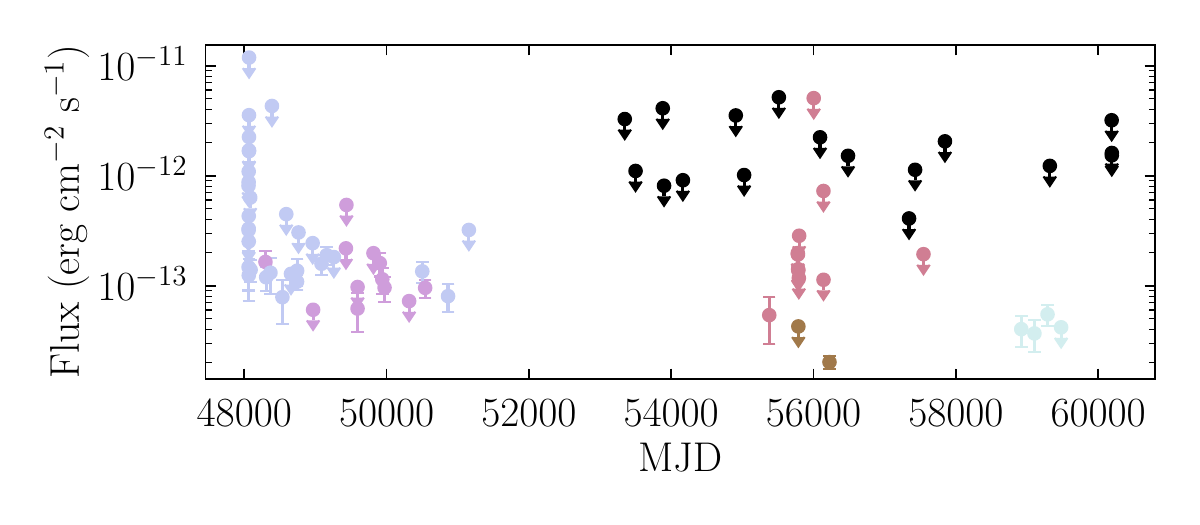}}
            \resizebox{0.495\hsize}{!}{\includegraphics{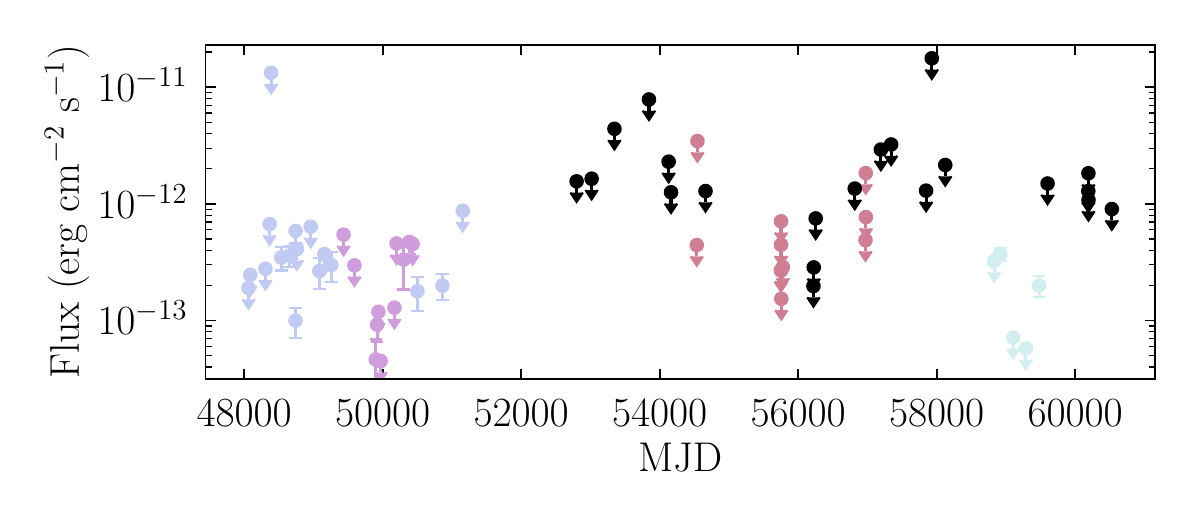}}
            \resizebox{0.495\hsize}{!}{\includegraphics{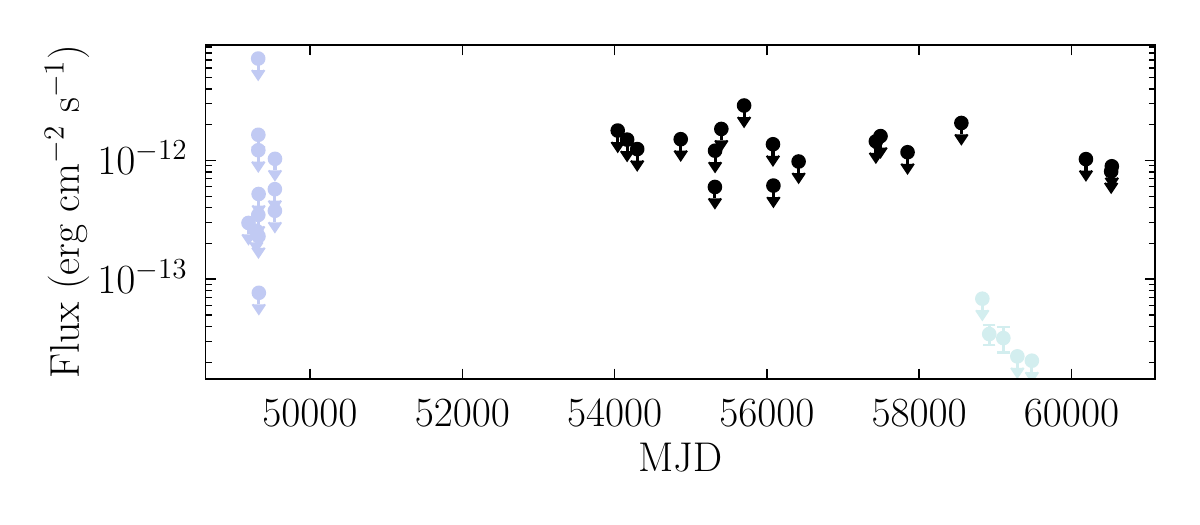}}
            \caption{continued for \#35 to \#46.}
        \end{figure*}
        \addtocounter{figure}{-1}
        \begin{figure*}
            \centering
            \resizebox{0.495\hsize}{!}{\includegraphics{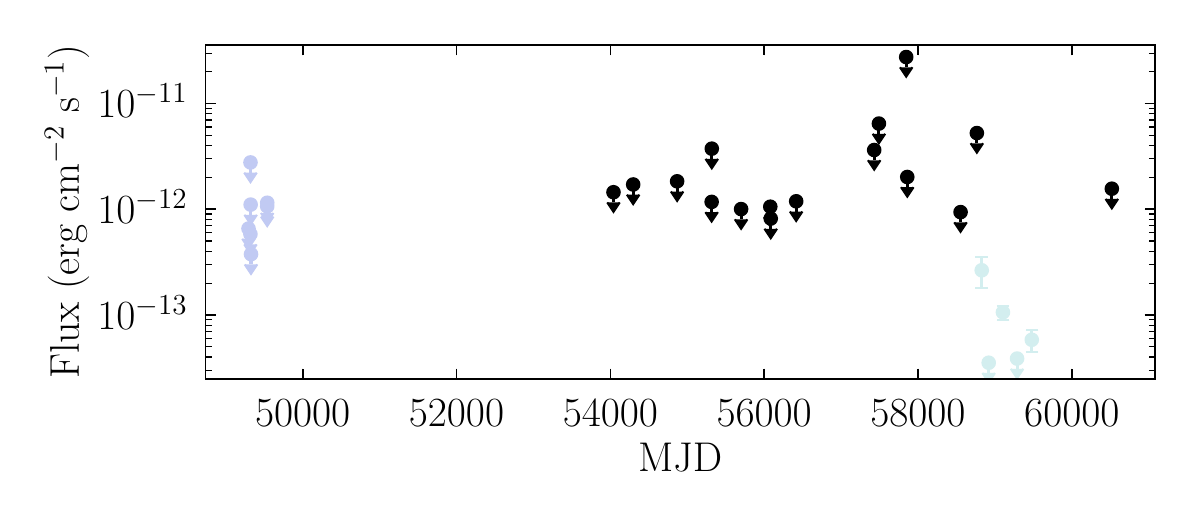}}
            \resizebox{0.495\hsize}{!}{\includegraphics{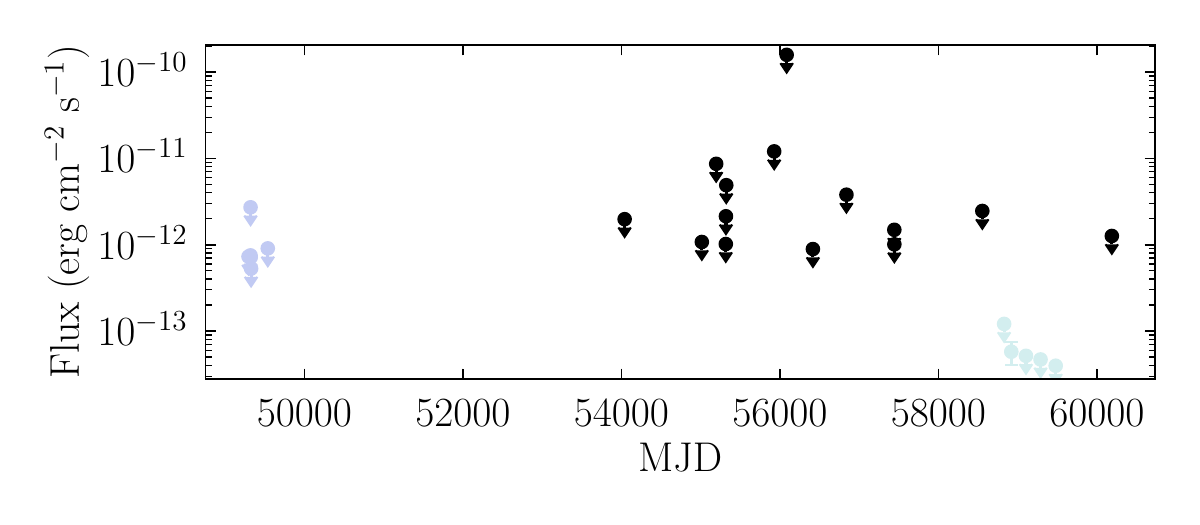}}
            \resizebox{0.495\hsize}{!}{\includegraphics{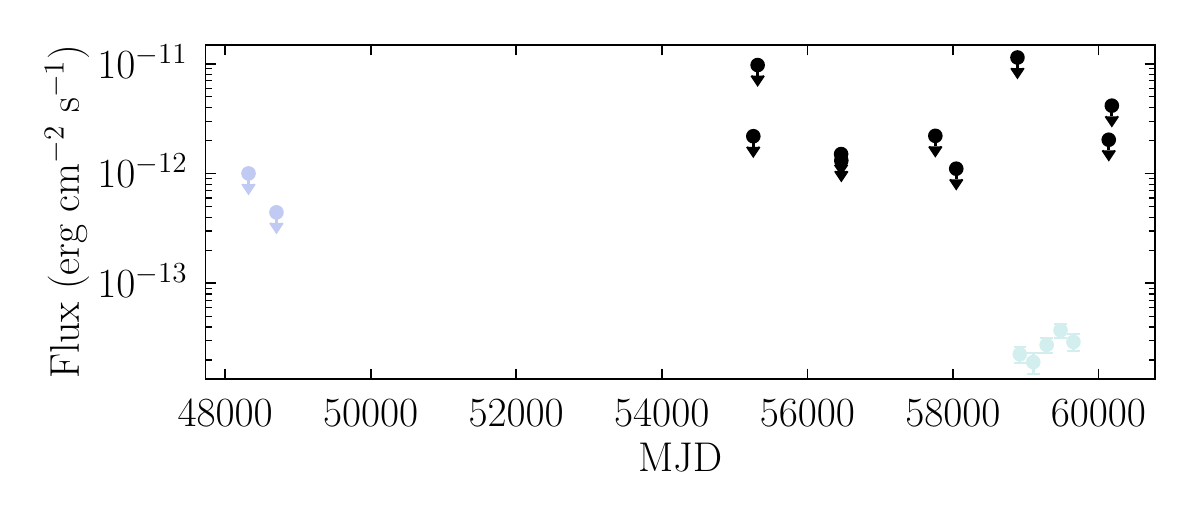}}
            \resizebox{0.495\hsize}{!}{\includegraphics{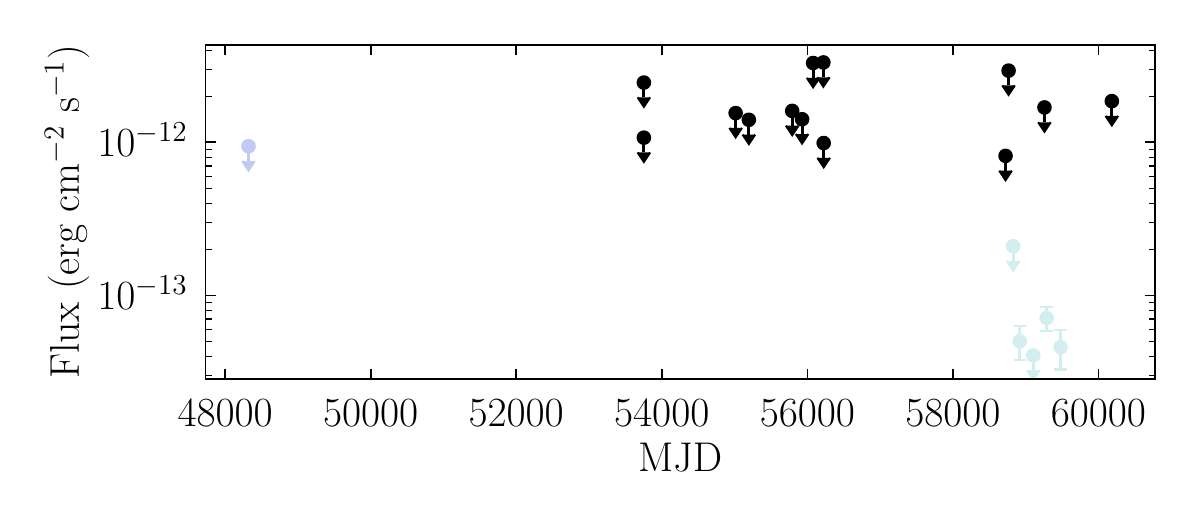}}
            \resizebox{0.495\hsize}{!}{\includegraphics{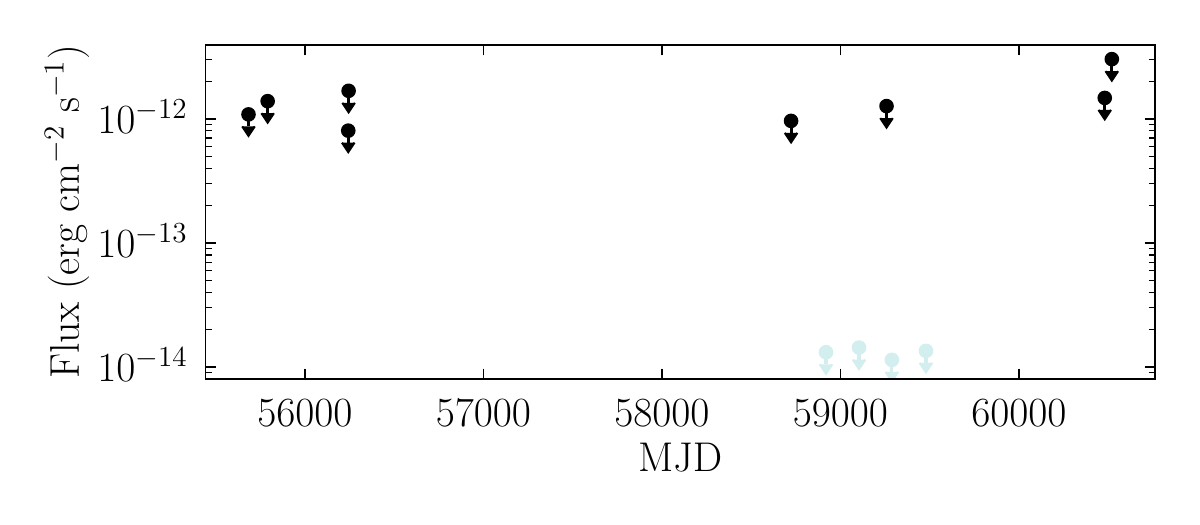}}
            \resizebox{0.495\hsize}{!}{\includegraphics{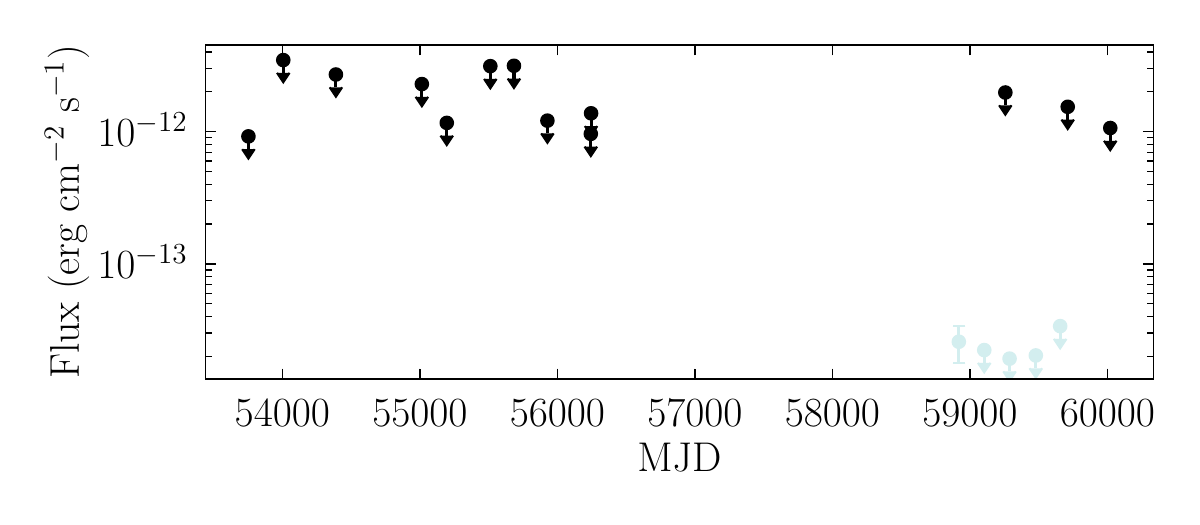}}
            \resizebox{0.495\hsize}{!}{\includegraphics{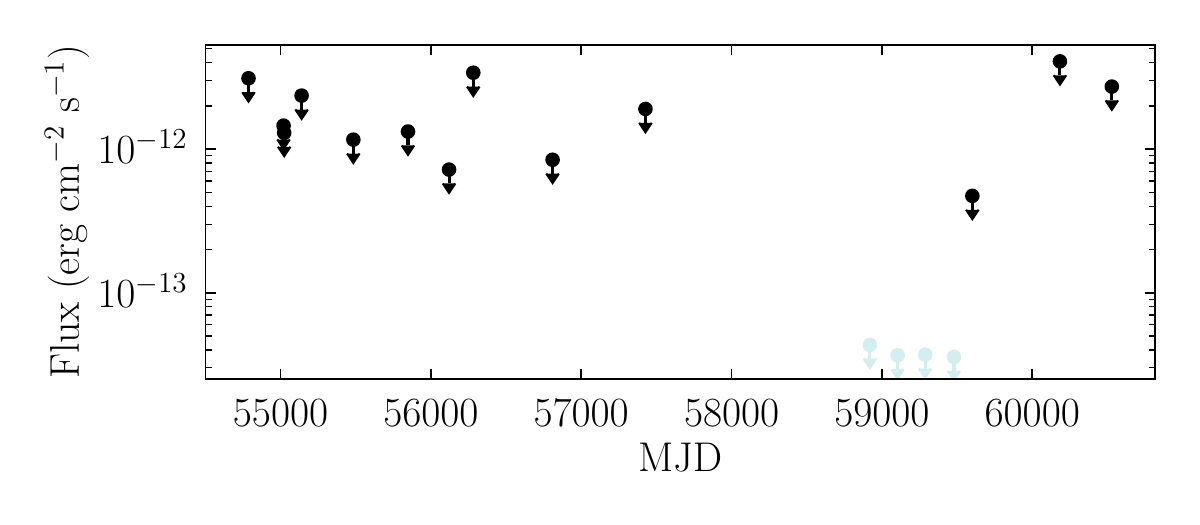}}
            \caption{continued for \#49 to \#53.}
        \end{figure*}
        \clearpage
        
        \section{SED fitting}
        
        Figure\,\ref{fig:SED-fitting} shows the results from SED fitting as explained in Sect.\,\ref{sec:SED-fitting} for all sources listed as candidates in our sample. Sources we flag to have an IR excess fit well for wavelengths below 10 microns and show overluminosity compared to the model prediction above. Table\,\ref{tab:SED_results} summarises the best-fit results we obtained. For 1eRASS\,J050945.8$-$655237 (\#38), 1eRASS\,J055318.4$-$655953 (\#49), 1eRASS\,J055536.1$-$671444 (\#50), and 1eRASS\,J060212.5$-$674305 (\#52), spectral fitting was unsuccessful, with inferred temperatures lower than expected for Be stars. This makes their classification as HMXBs uncertain. In the cases of 1eRASS\,J045028.2$-$693558 (\#31) and 1eRASS\,J050053.3$-$653209 (\#34), two distinct sources fall within the VizieR selection region and cannot be distinguished based on position alone. The spectral parameters presented in Table\,\ref{tab:SED_results} correspond to the object whose SED is consistent with that of a Be star, though the results -- particularly the parameter uncertainties -- should be taken with a grain of salt.
        
        For sources for which optical spectra with \salt or LCO/FLOYDS were taken, Fig.\,\ref{fig:SED_LCO} shows a comparison of the best-fit result from SED fitting and the observed spectrum. This nicely shows the enhanced \Halpha flux.
        
        \renewcommand{\arraystretch}{1.5} 
        \begin{table}[!htbp]
                \centering
                \caption{SED fitting results.} 
                \label{tab:SED_results} 
                \begin{tabular}{llll} 
                        \hline\hline\noalign{\smallskip}
                        \# & $T_{eff}$ & E(B$-$V) & $R$ \\
                        & K & mag & R$_{\odot}$ \\
                        \noalign{\smallskip}\hline\noalign{\smallskip}
                        16 &  <17100.0 & >0.0 & 15.0$^{+1.4}_{-2.8}$ \\ 
                        18 &  <16200.0 & >0.0 & 40.0$^{+1.6}_{-9.2}$ \\ 
                        28 &  18000$^{+6900}_{-2600}$ & >0.0 & 15.0$^{+5.3}_{-4.0}$ \\ 
                        29 &  >15300.0 & $-$ & 5.0$^{+8.7}_{-0.9}$ \\ 
                        30 &  15000$^{+14000}_{-9700}$ & $-$ & 12.0$^{+3.2}_{-5.8}$ \\ 
                        31 &  20000$^{+24500}_{-17800}$ & $-$ & <12.2 \\ 
                        32 &  <17300.0 & >0.0 & 15.0$^{+1.7}_{-3.1}$ \\ 
                        33 &  <23400.0 & $-$ & 5.0$^{+2.7}_{-0.9}$ \\ 
                        34 &  15000$^{+14600}_{-1900}$ & $-$ & 8.0$^{+0.3}_{-3.6}$ \\ 
                        35 &  20000$^{+9400}_{-4100}$ & $-$ & 20.0$^{+8.1}_{-6.0}$ \\ 
                        36 &  <17400.0 & >0.0 & 15.0$^{+2.6}_{-2.8}$ \\ 
                        37 &  <17400.0 & $-$ & 15.0$^{+2.6}_{-2.5}$ \\ 
                        38 &  <18900.0 & >0.0 & 5.0$^{+1.2}_{-0.6}$ \\ 
                        39 &  19000$^{+21100}_{-4600}$ & $-$ & 8.0$^{+3.0}_{-3.6}$ \\ 
                        40 &  21000$^{+2200}_{-5900}$ & $-$ & 5.0$^{+2.4}_{-0.6}$ \\ 
                        41 &  21000$^{+2800}_{-4700}$ & $-$ & 5.0$^{+2.7}_{-0.6}$ \\ 
                        42 &  <17800.0 & >0.0 & 40.0$^{+3.4}_{-7.1}$ \\ 
                        43 &  <16200.0 & >0.0 & 15.0$^{+1.1}_{-2.8}$ \\ 
                        44 &  15000$^{+22100}_{-3500}$ & 0.127$^{+0.228}_{-0.136}$ & 8.0$^{+0.6}_{-3.3}$ \\ 
                        45 &  20000$^{+7600}_{-3900}$ & >0.0 & 15.0$^{+4.7}_{-3.7}$ \\ 
                        46 &  <15900.0 & >0.0 & 20.0$^{+0.9}_{-3.9}$ \\ 
                        47 &  25000$^{+1600}_{-1300}$ & >0.0 & 20.0$^{+9.6}_{-7.2}$ \\ 
                        48 &  16000$^{+17400}_{-2700}$ & >0.0 & 5.0$^{+0.9}_{-3.0}$ \\ 
                        49 &  <21600.0 & $-$ & 5.0$^{+2.1}_{-0.6}$ \\ 
                        50 &  <19100.0 & $-$ & 10.0$^{+3.4}_{-3.2}$ \\ 
                        51 &  <15500.0 & >0.0 & 8.0$^{+0.3}_{-1.5}$ \\ 
                        52 &  <24400.0 & $-$ & 15.0$^{+5.9}_{-4.9}$ \\ 
                        53 &  <16800.0 & >0.0 & 8.0$^{+0.6}_{-2.7}$ \\ 
                        
                        \noalign{\smallskip}\hline
                \end{tabular}
        \tablefoot{SED fitting results for the effective temperature of the optical counterpart, the absorption and the stellar radius. Upper limits denote parameters for which no lower uncertainty could be determined, and lower limits denote those with no upper uncertainty. Parameters for which neither uncertainty could be determined are indicated with ‘$-$’.}
        \end{table}
        \renewcommand{\arraystretch}{1.0}

        \begin{figure*}
            \centering
            \resizebox{0.497\hsize}{!}{\includegraphics{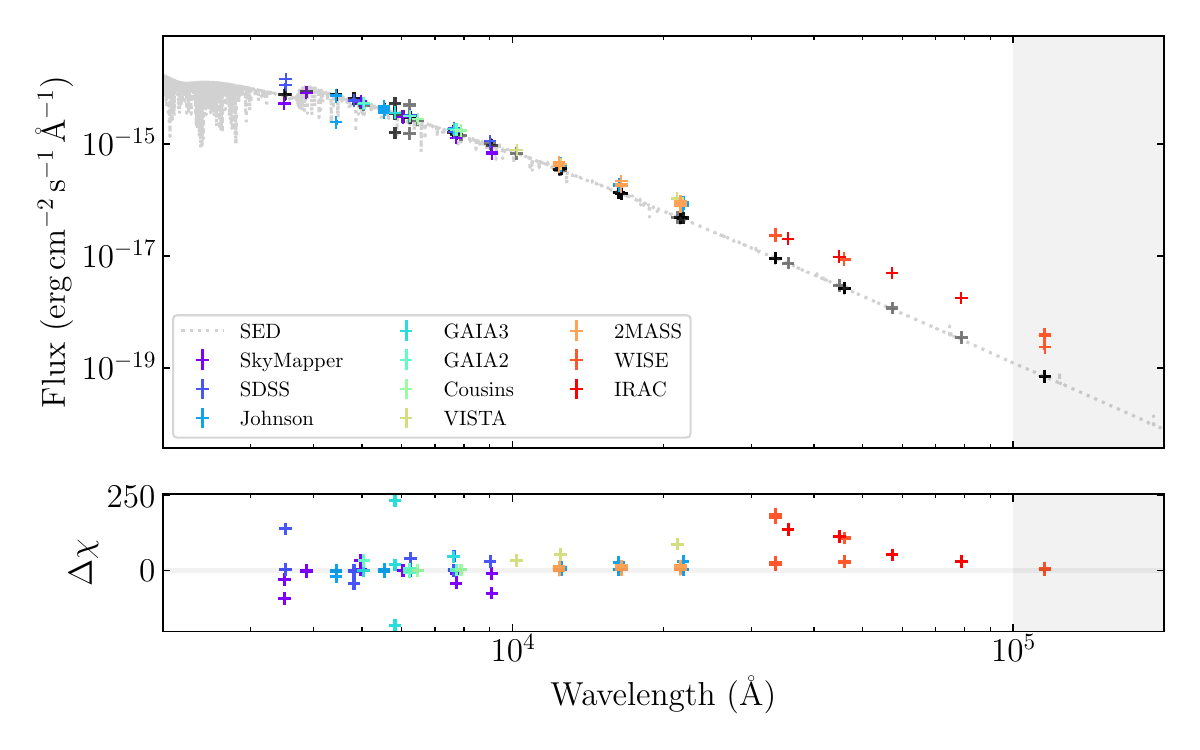}}
            \resizebox{0.497\hsize}{!}{\includegraphics{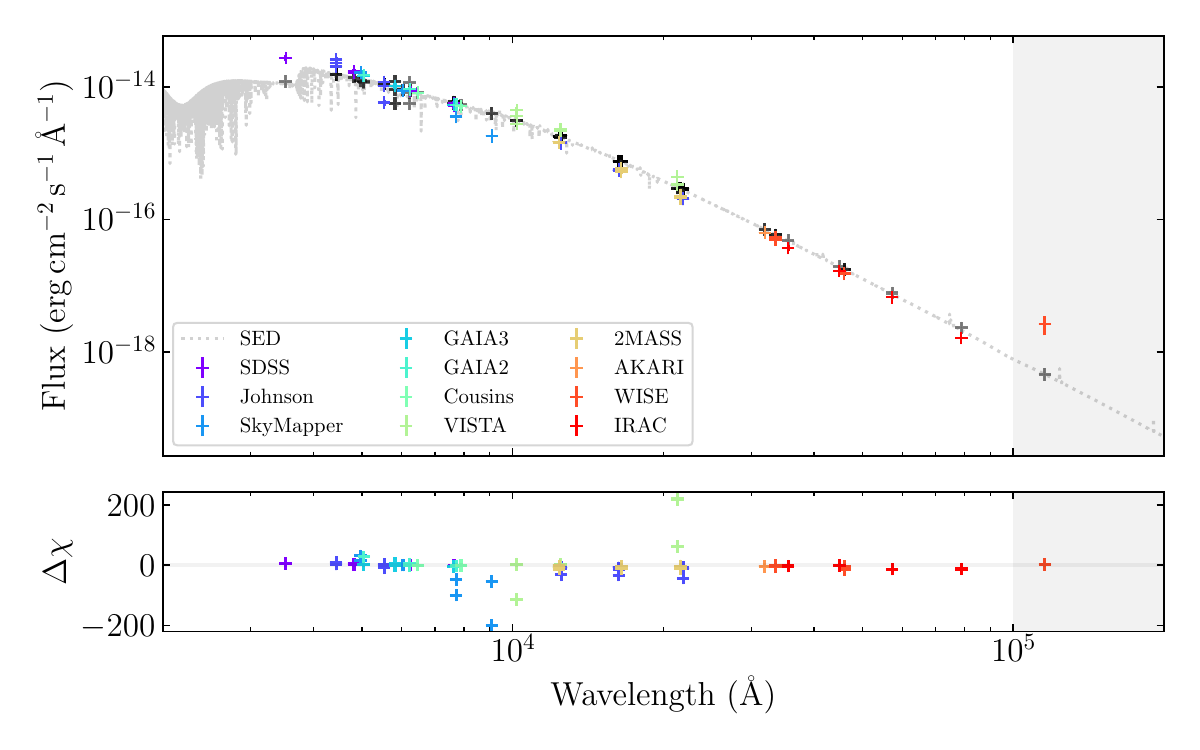}}
            \resizebox{0.497\hsize}{!}{\includegraphics{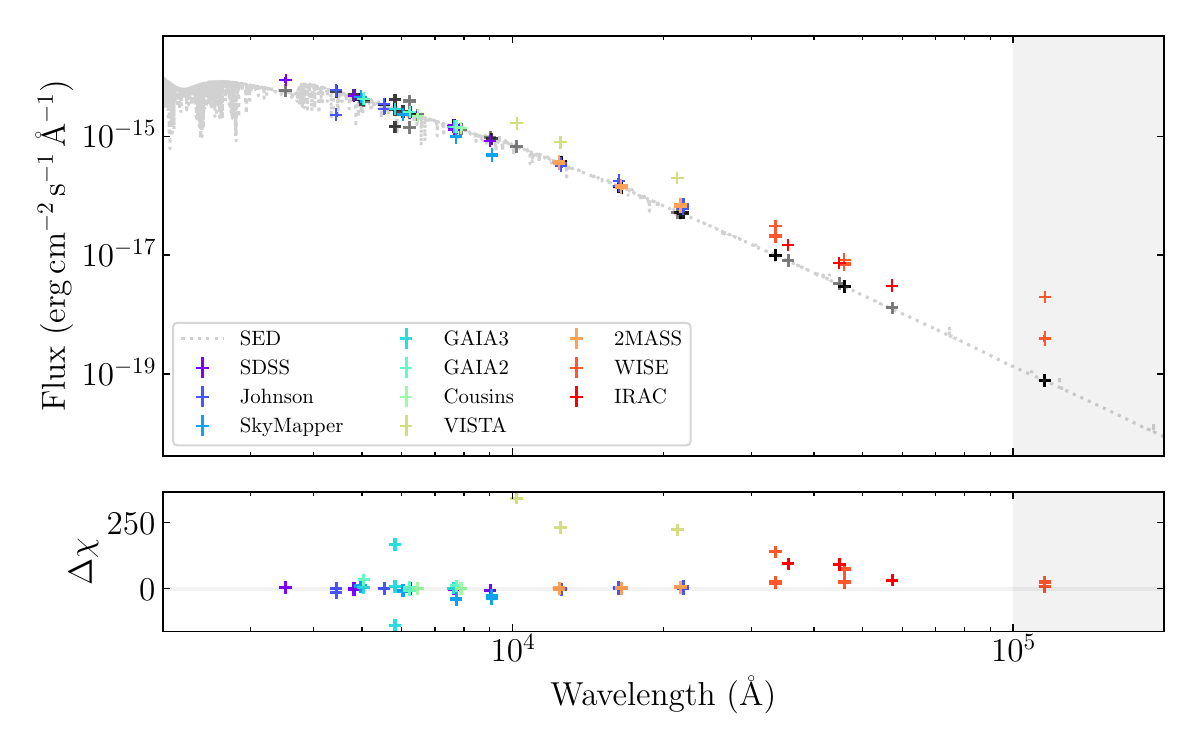}}
            \resizebox{0.497\hsize}{!}{\includegraphics{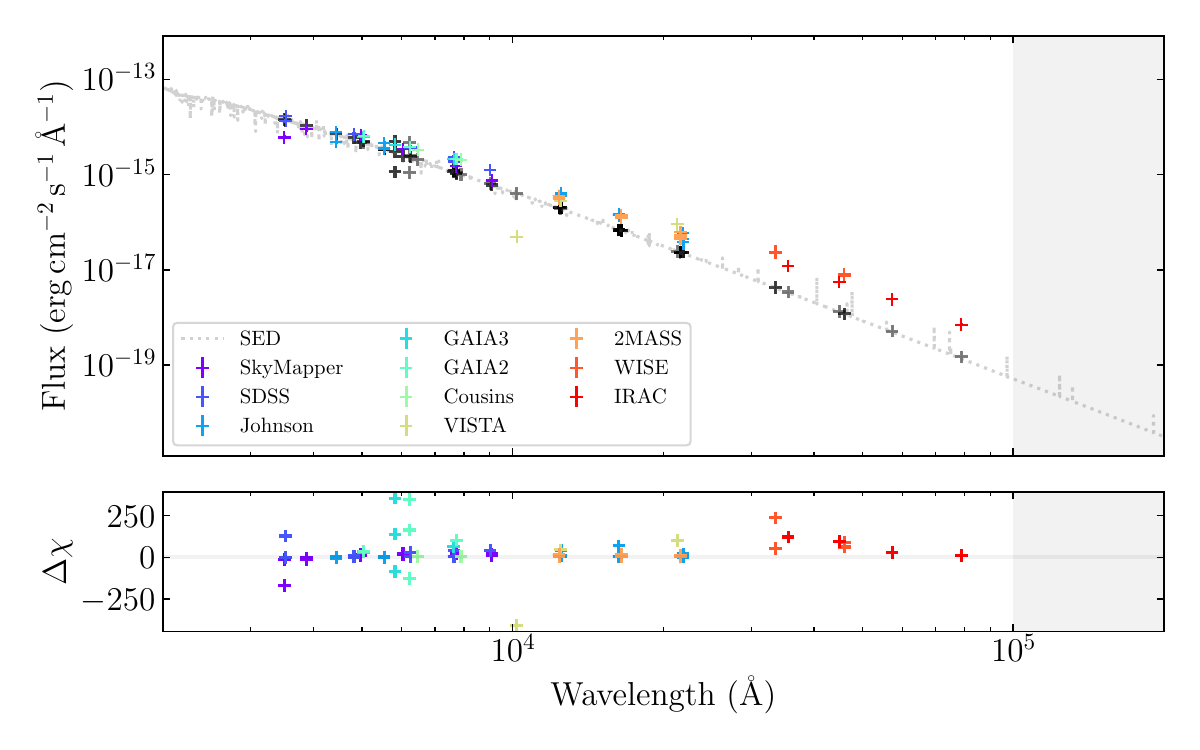}}
            \resizebox{0.497\hsize}{!}{\includegraphics{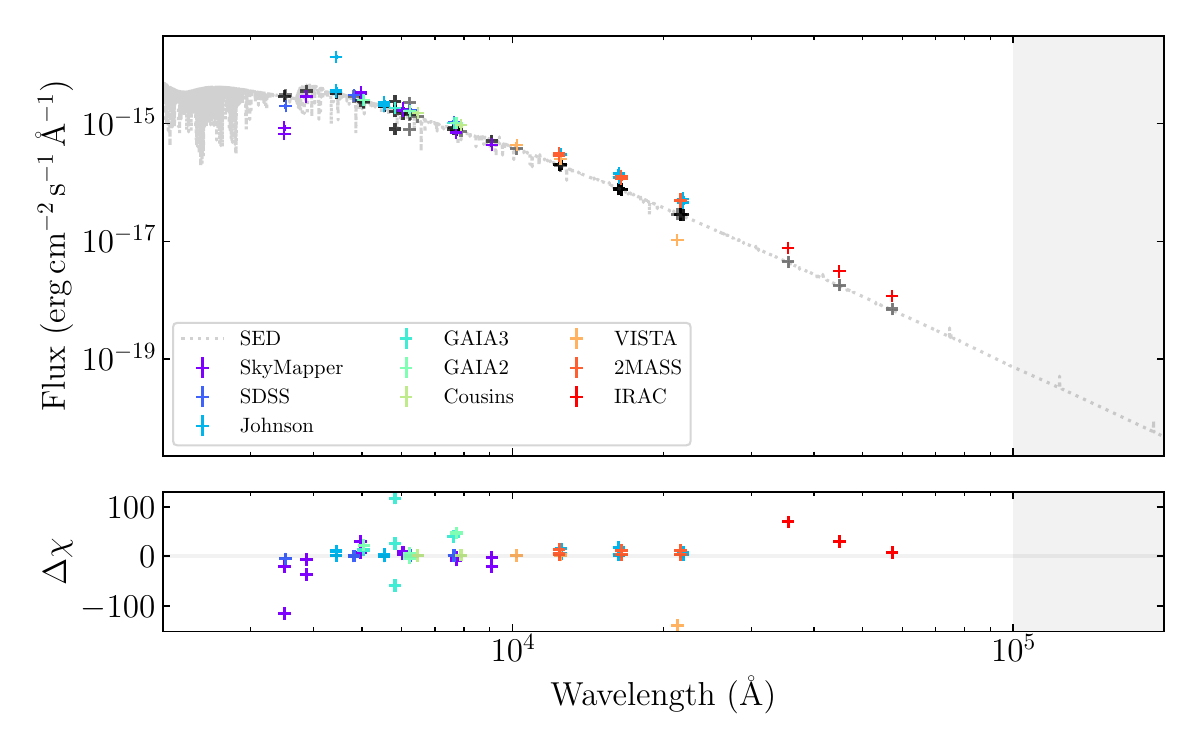}}
            \resizebox{0.497\hsize}{!}{\includegraphics{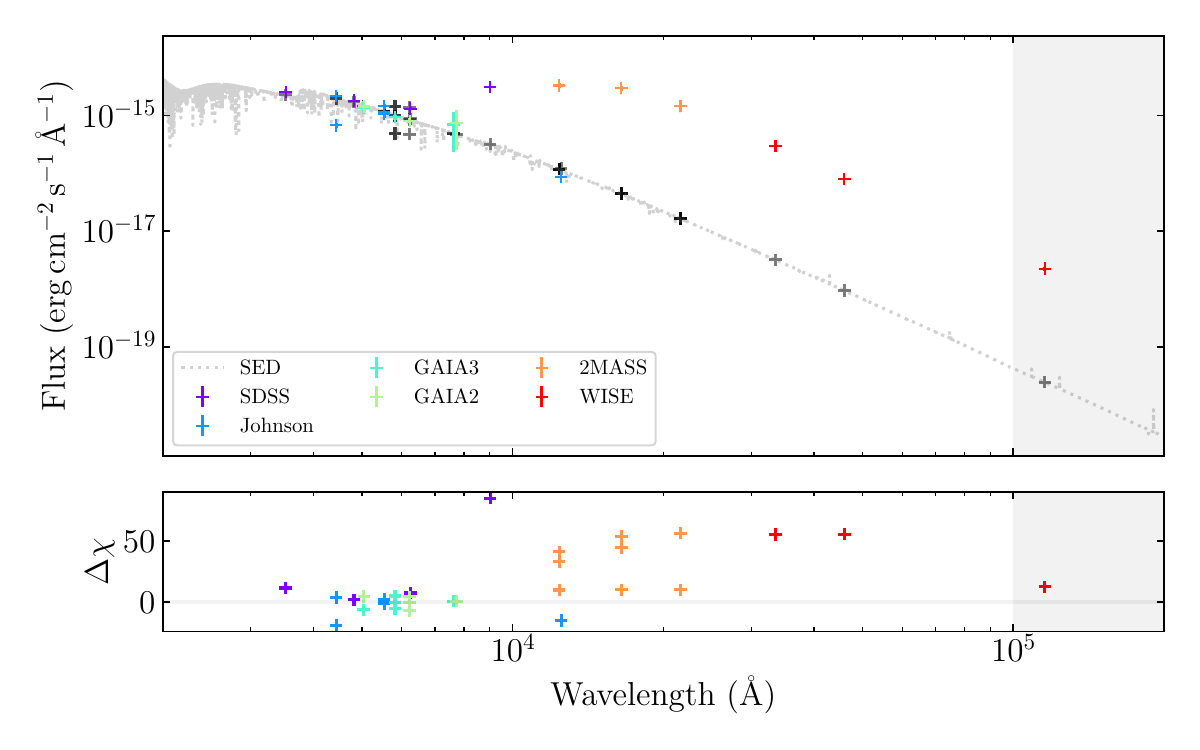}}
            \resizebox{0.497\hsize}{!}{\includegraphics{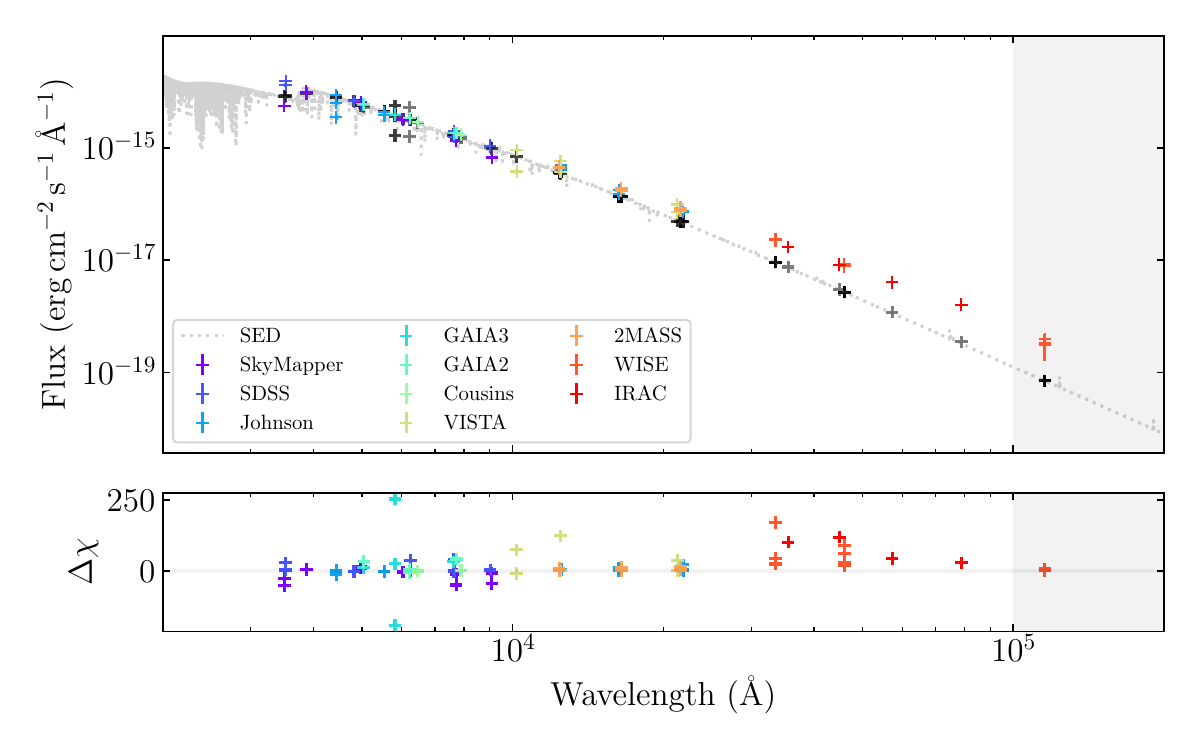}}
            \resizebox{0.497\hsize}{!}{\includegraphics{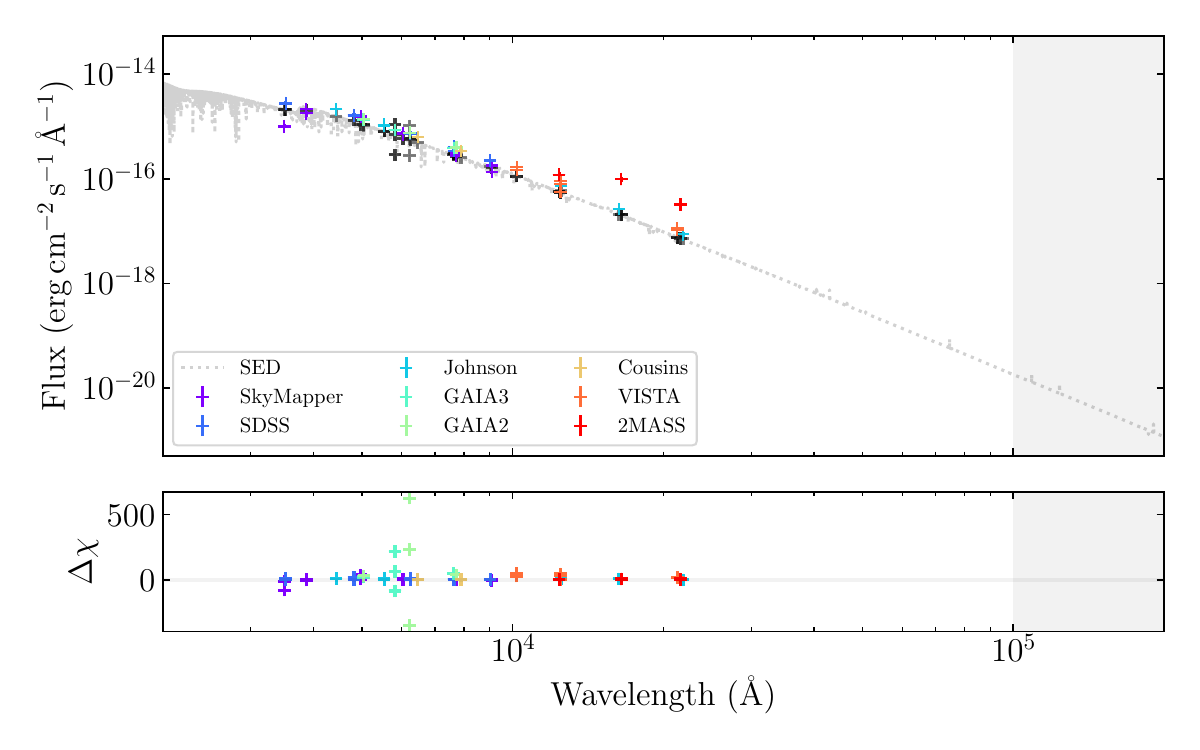}}
            \caption{SED fits as described in Sect.\,\ref{sec:SED-fitting} for \#16, \#18, and \#28 to \#33. Coloured points show data from the different instruments, black points are model predictions convolved with the corresponding filters. Only data below 10 micron were used for the fit.}
            \label{fig:SED-fitting} 
        \end{figure*}
        \addtocounter{figure}{-1}
        \begin{figure*}
            \centering
            \resizebox{0.497\hsize}{!}{\includegraphics{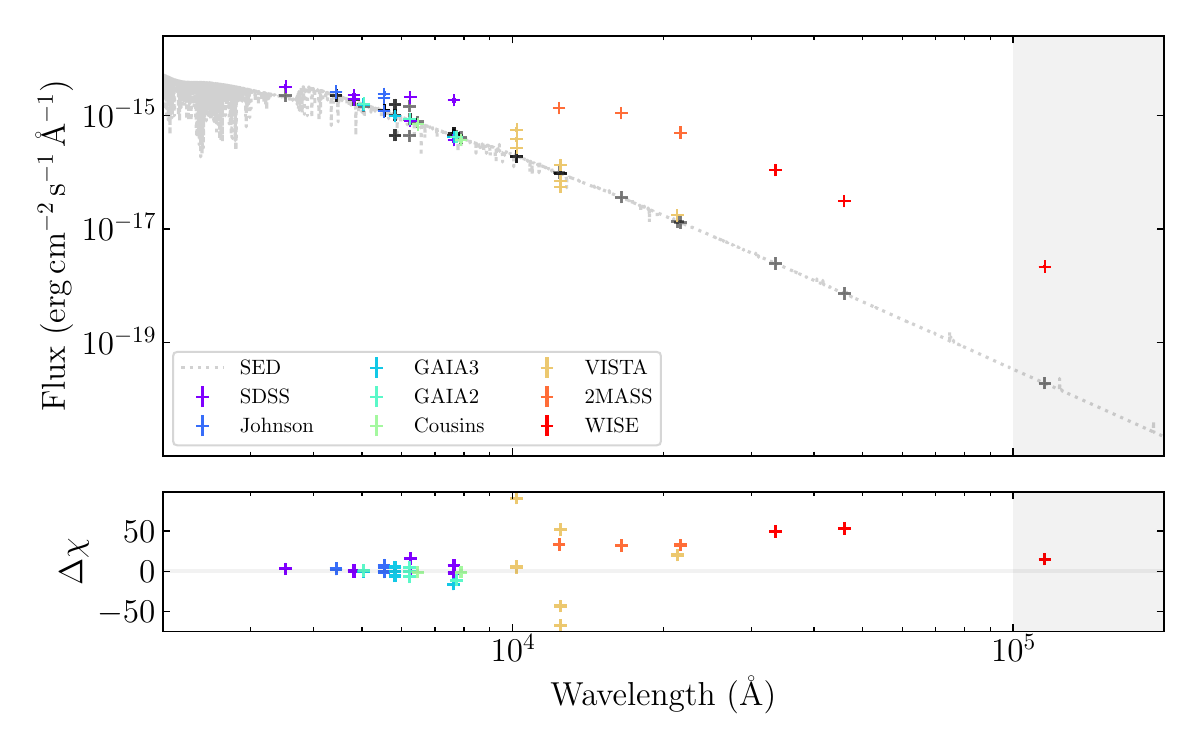}}
            \resizebox{0.497\hsize}{!}{\includegraphics{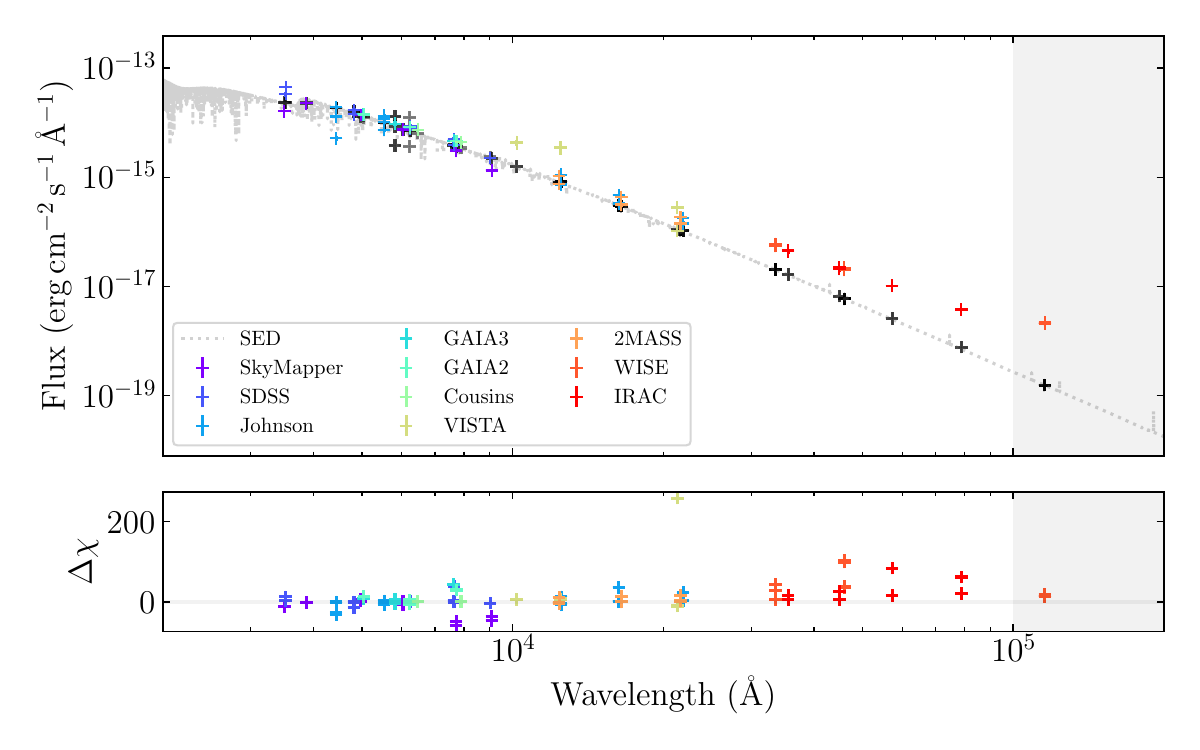}}
            \resizebox{0.497\hsize}{!}{\includegraphics{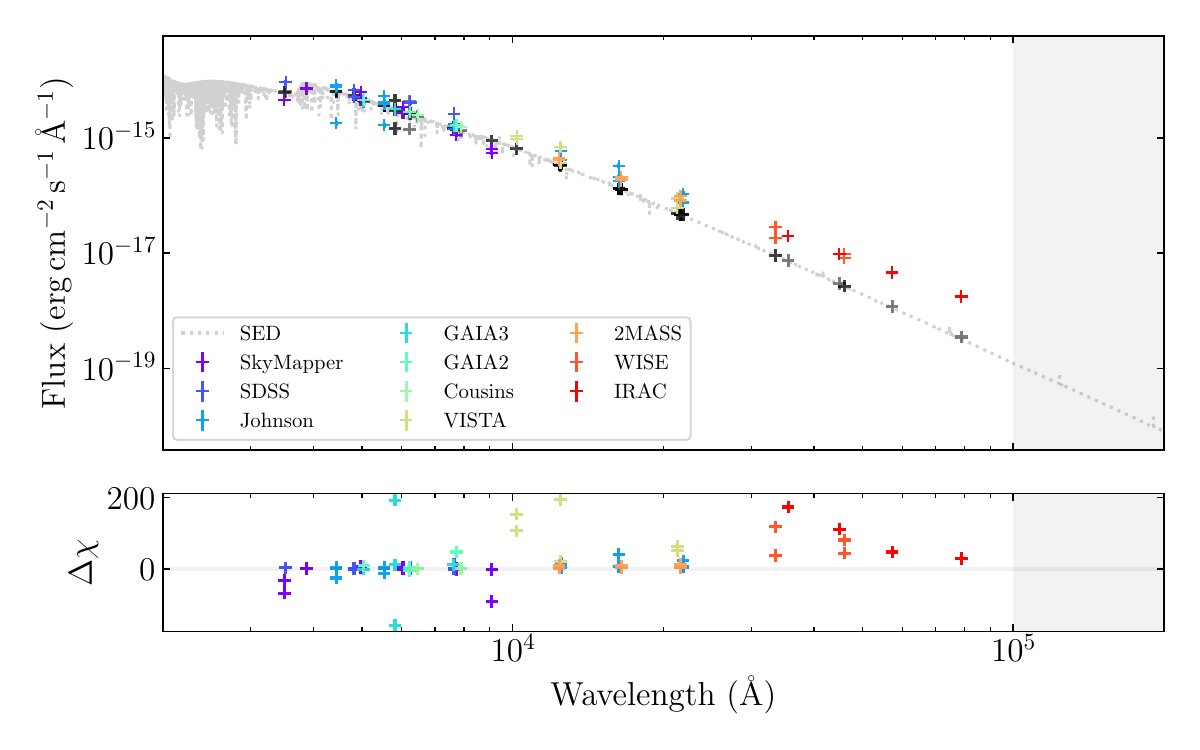}}
            \resizebox{0.497\hsize}{!}{\includegraphics{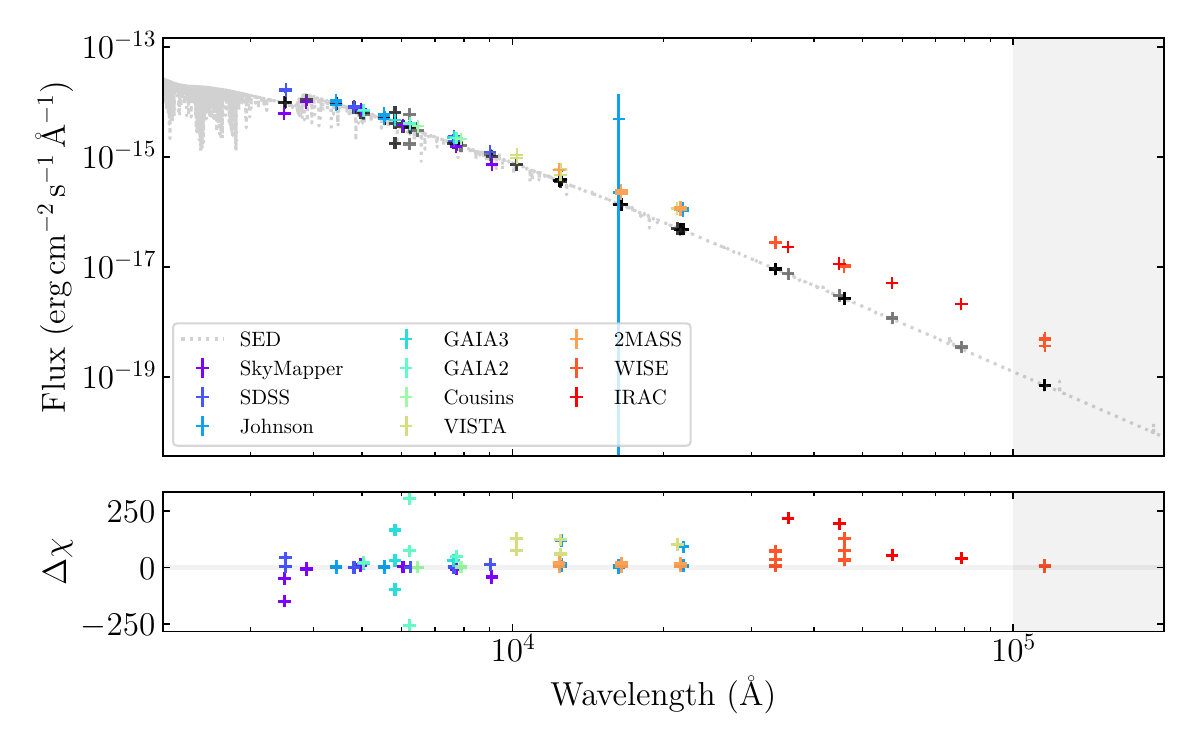}}
            \resizebox{0.497\hsize}{!}{\includegraphics{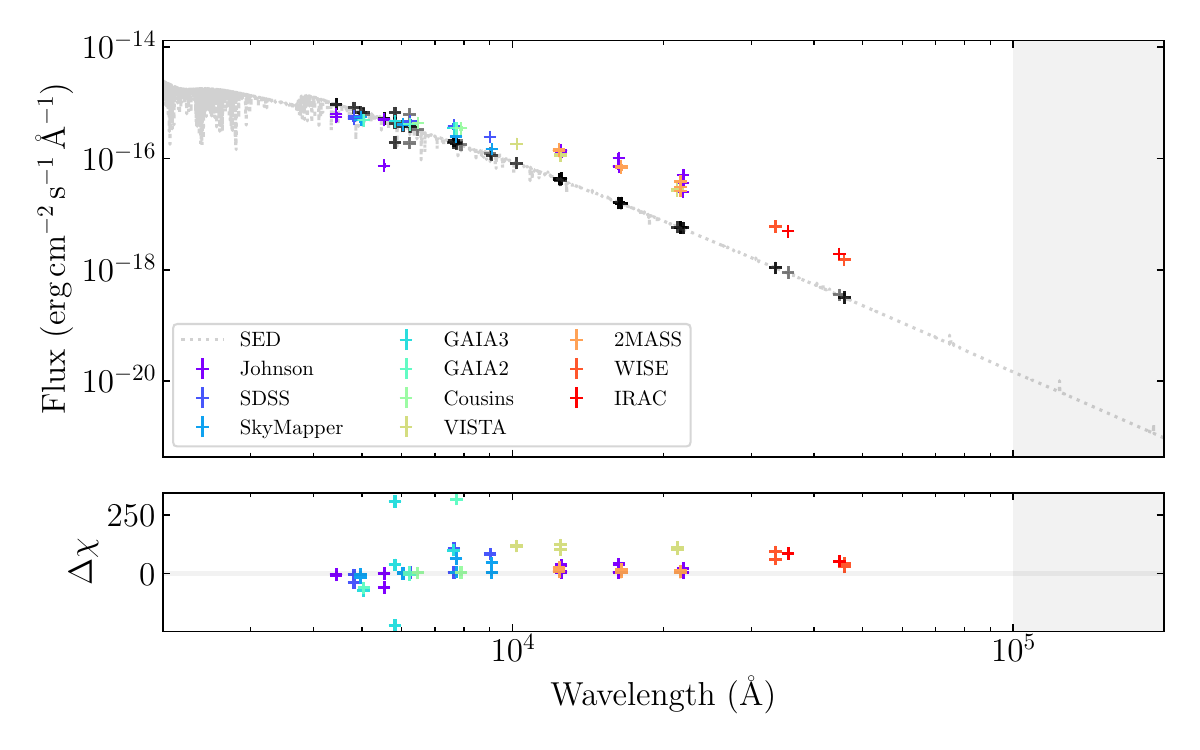}}
            \resizebox{0.497\hsize}{!}{\includegraphics{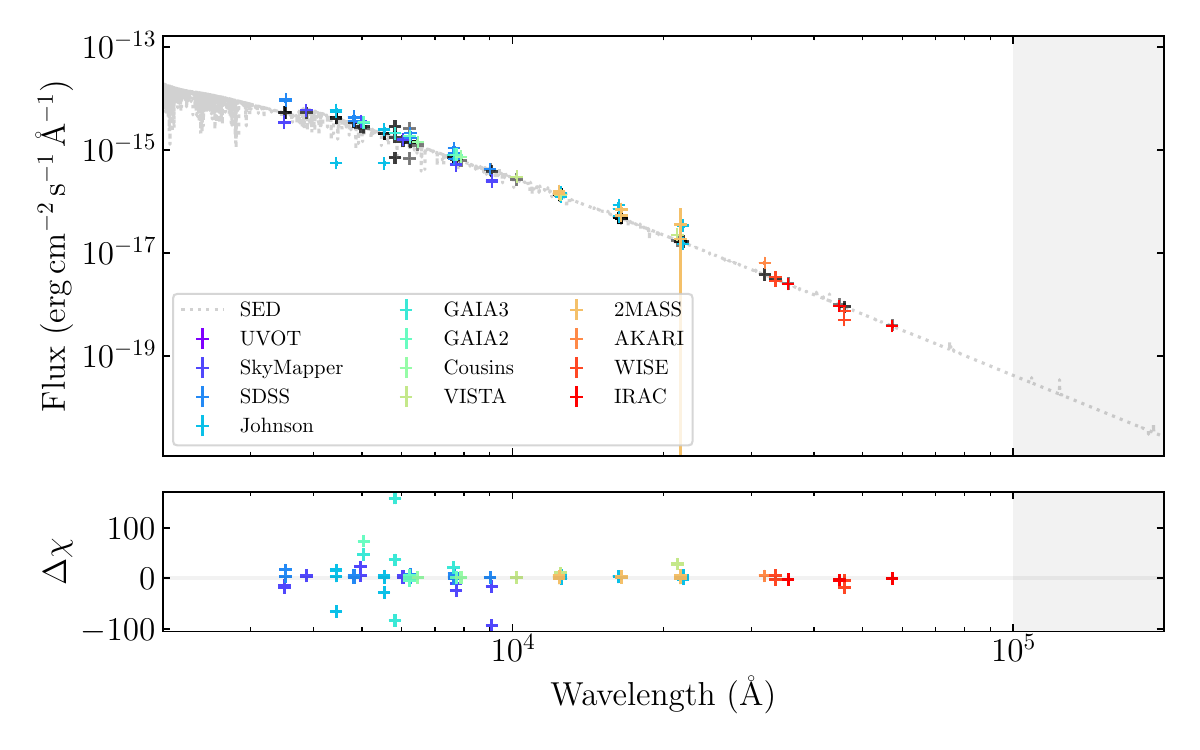}}
            \resizebox{0.497\hsize}{!}{\includegraphics{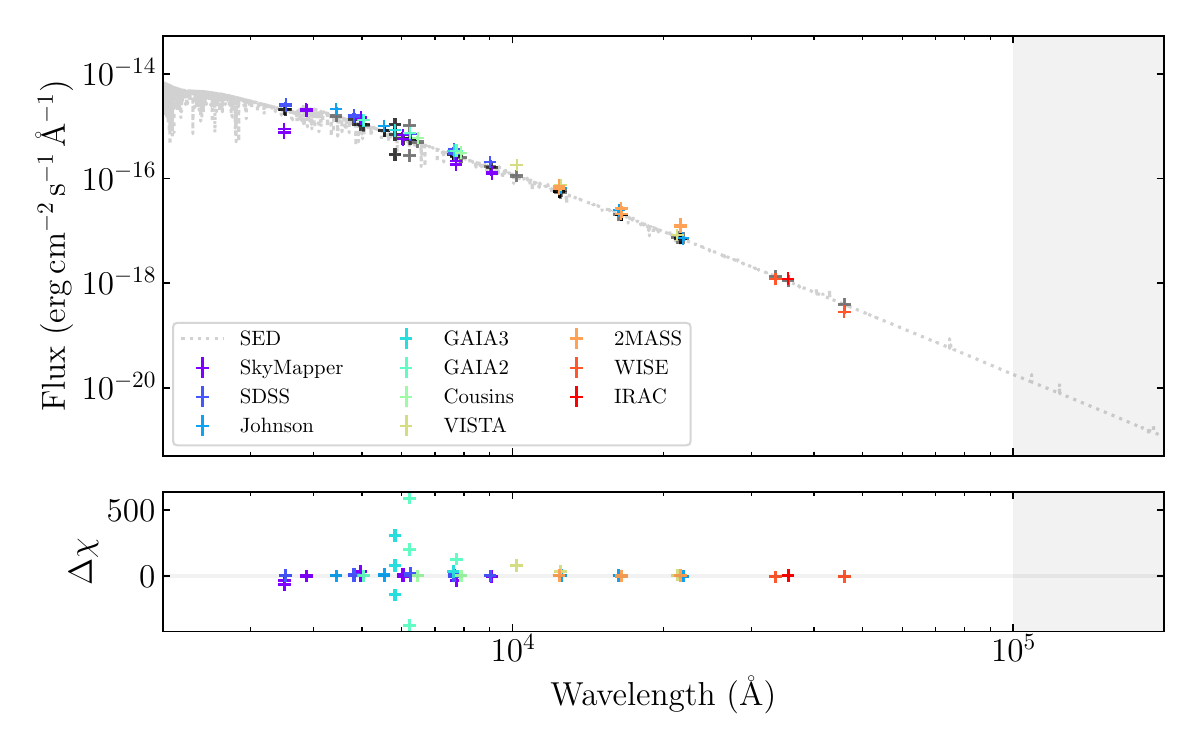}}
            \resizebox{0.497\hsize}{!}{\includegraphics{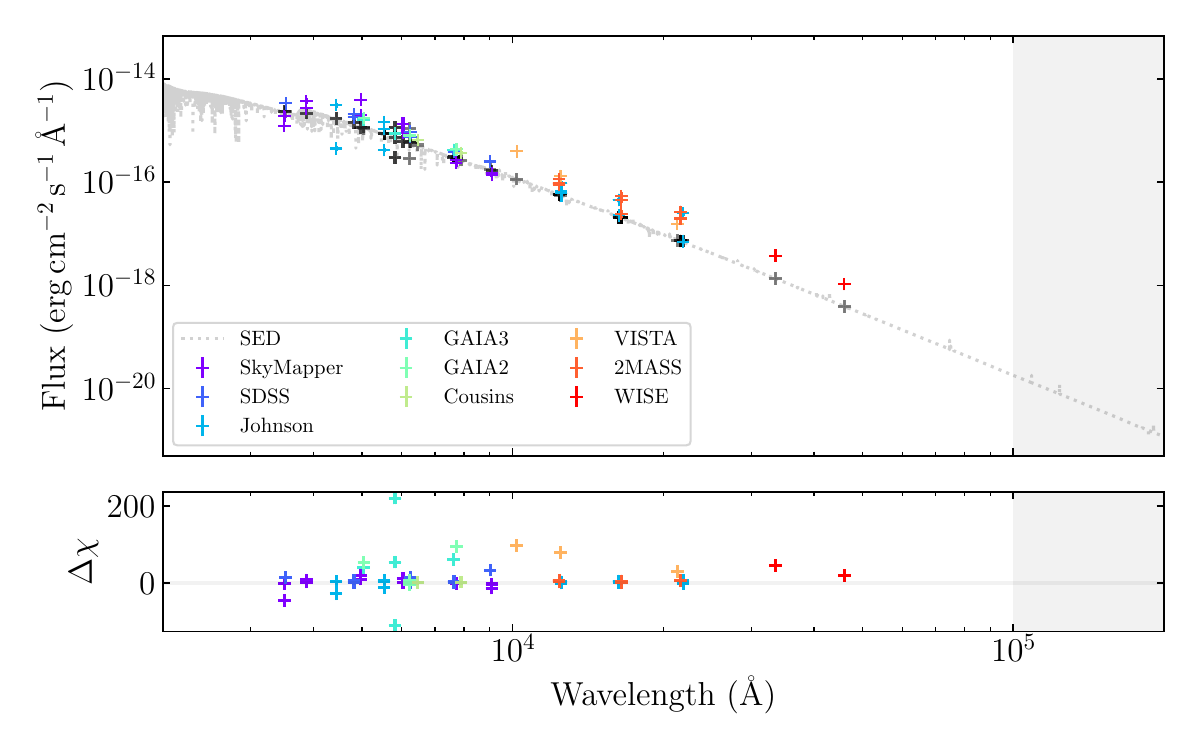}}
            \caption{continued for \#34 to \#41.}
        \end{figure*}
        \addtocounter{figure}{-1}
        \begin{figure*}
            \centering
            \resizebox{0.497\hsize}{!}{\includegraphics{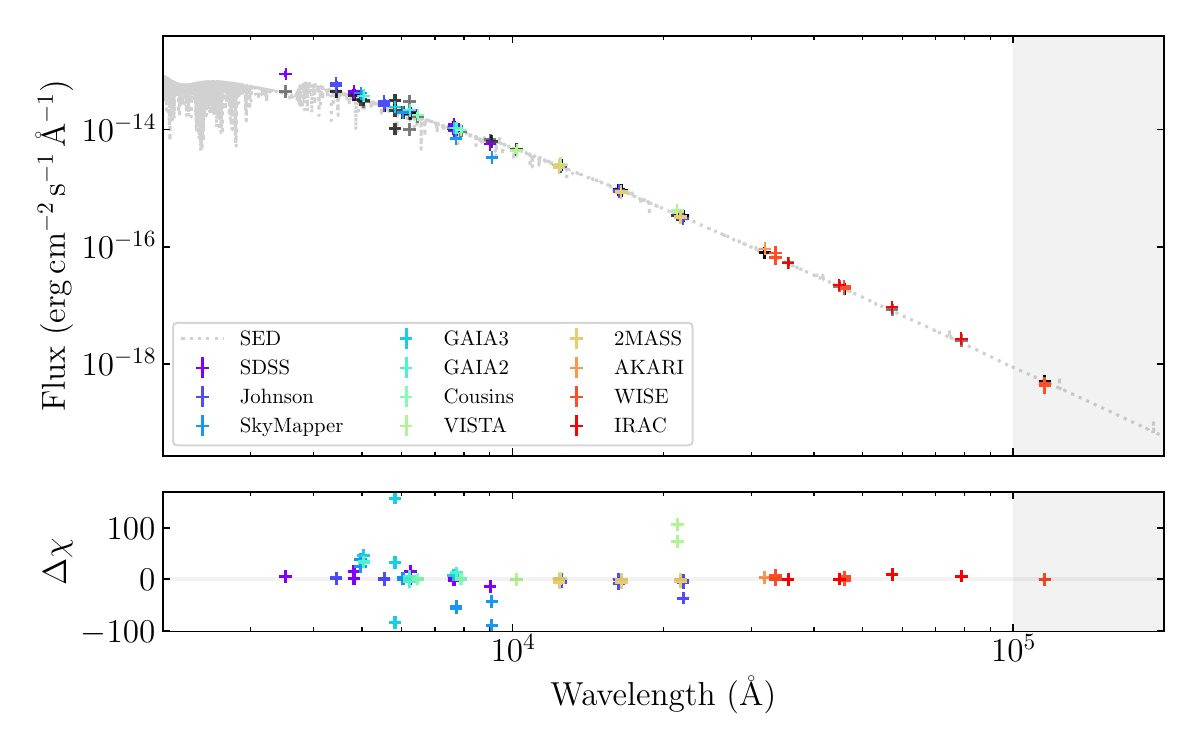}}
            \resizebox{0.497\hsize}{!}{\includegraphics{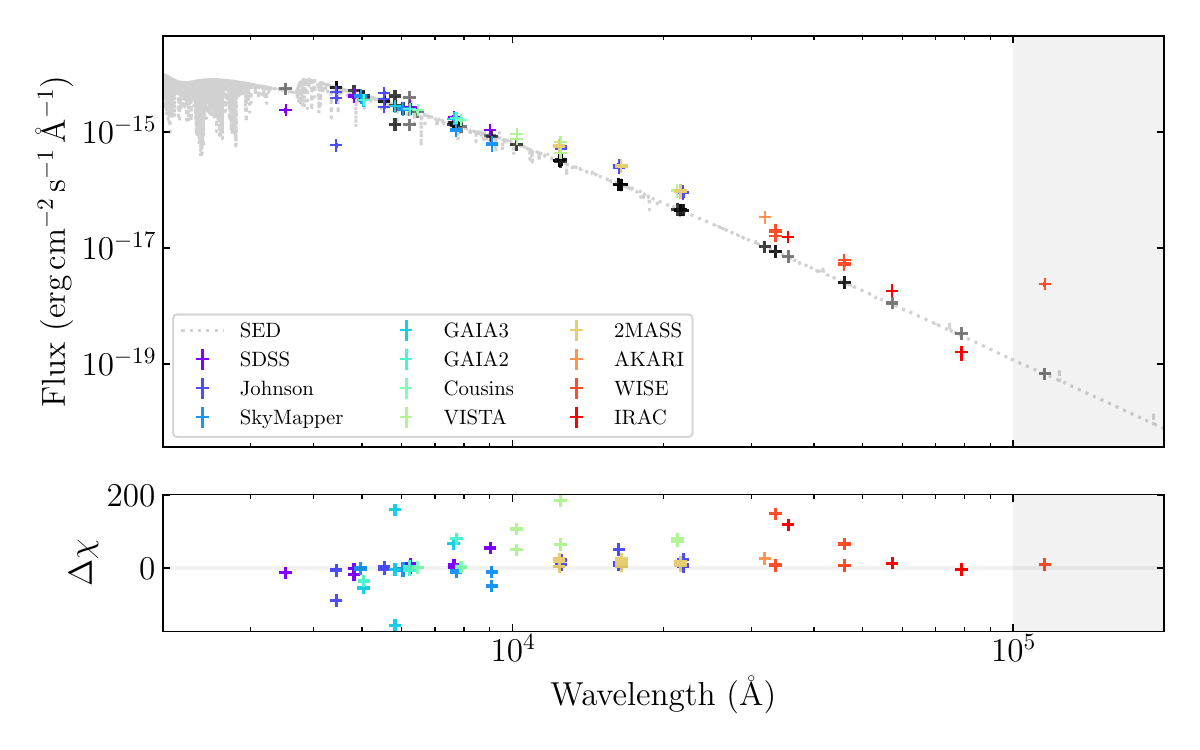}}
            \resizebox{0.497\hsize}{!}{\includegraphics{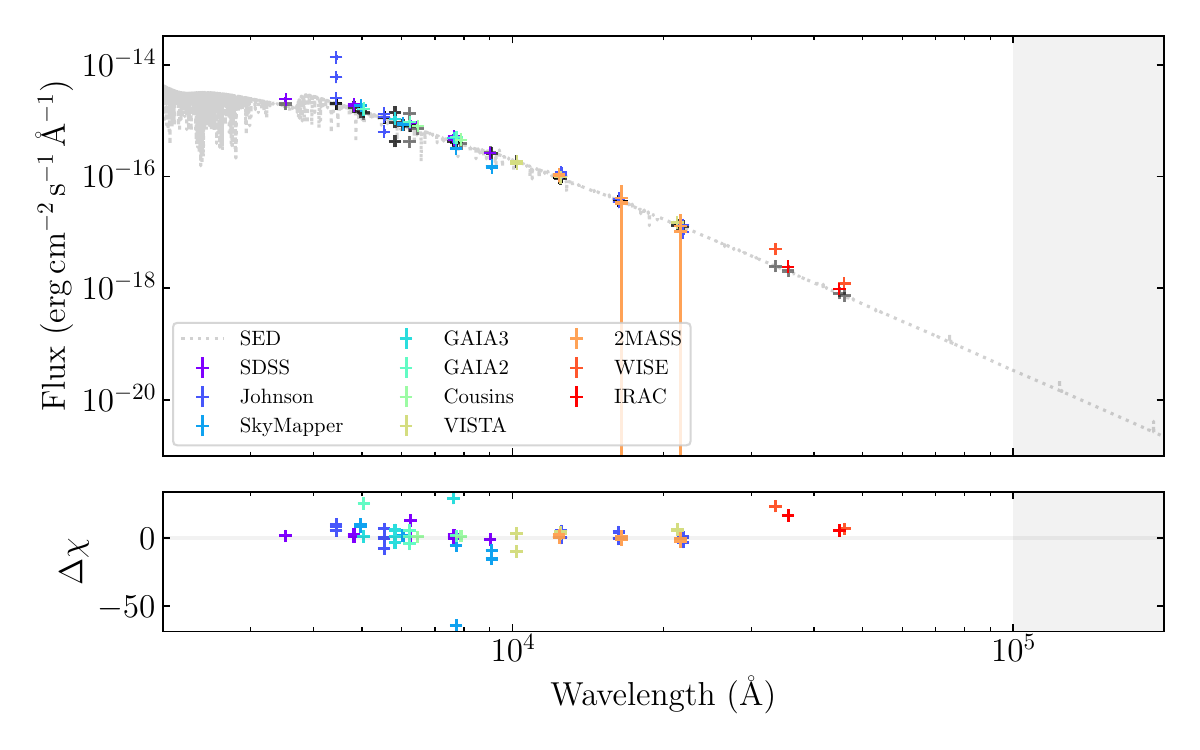}}
            \resizebox{0.497\hsize}{!}{\includegraphics{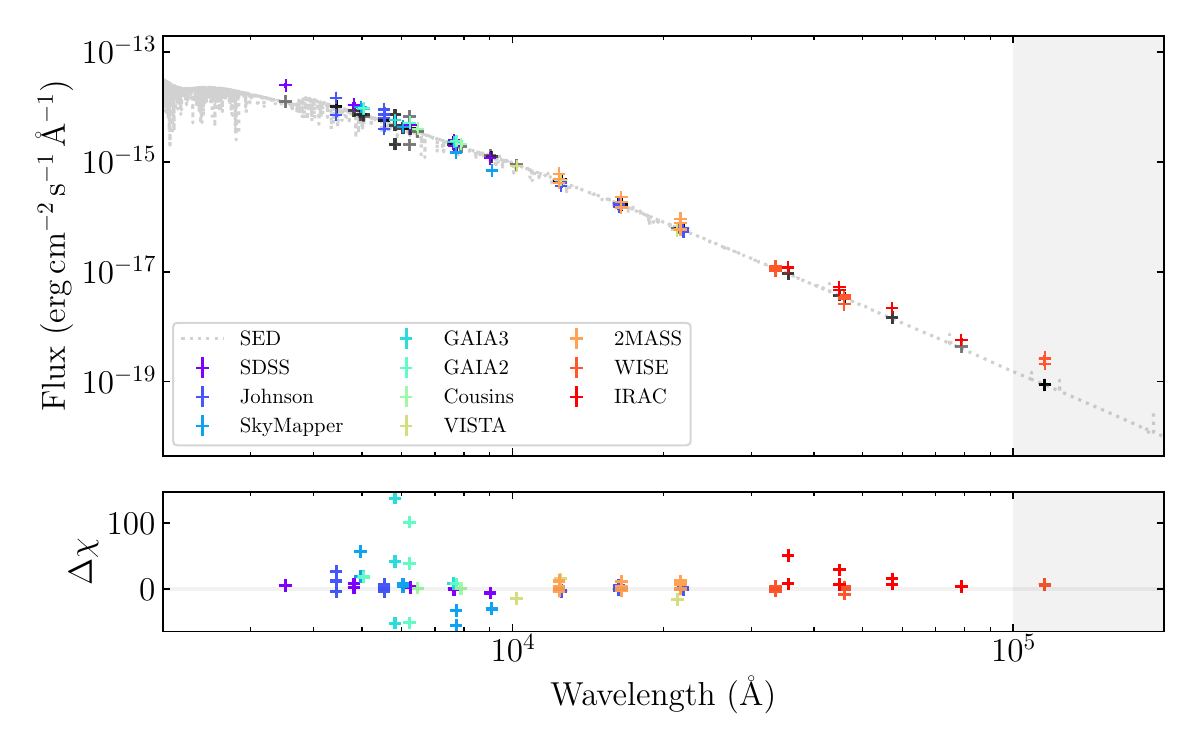}}
            \resizebox{0.497\hsize}{!}{\includegraphics{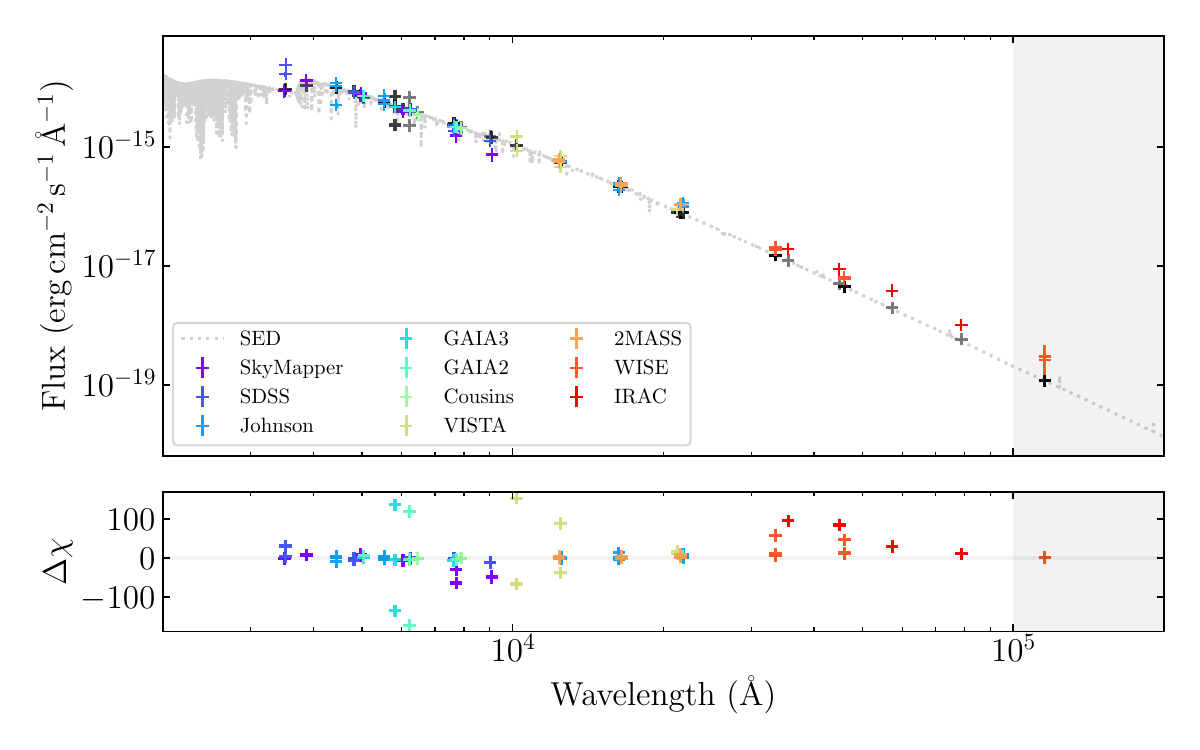}}
            \resizebox{0.497\hsize}{!}{\includegraphics{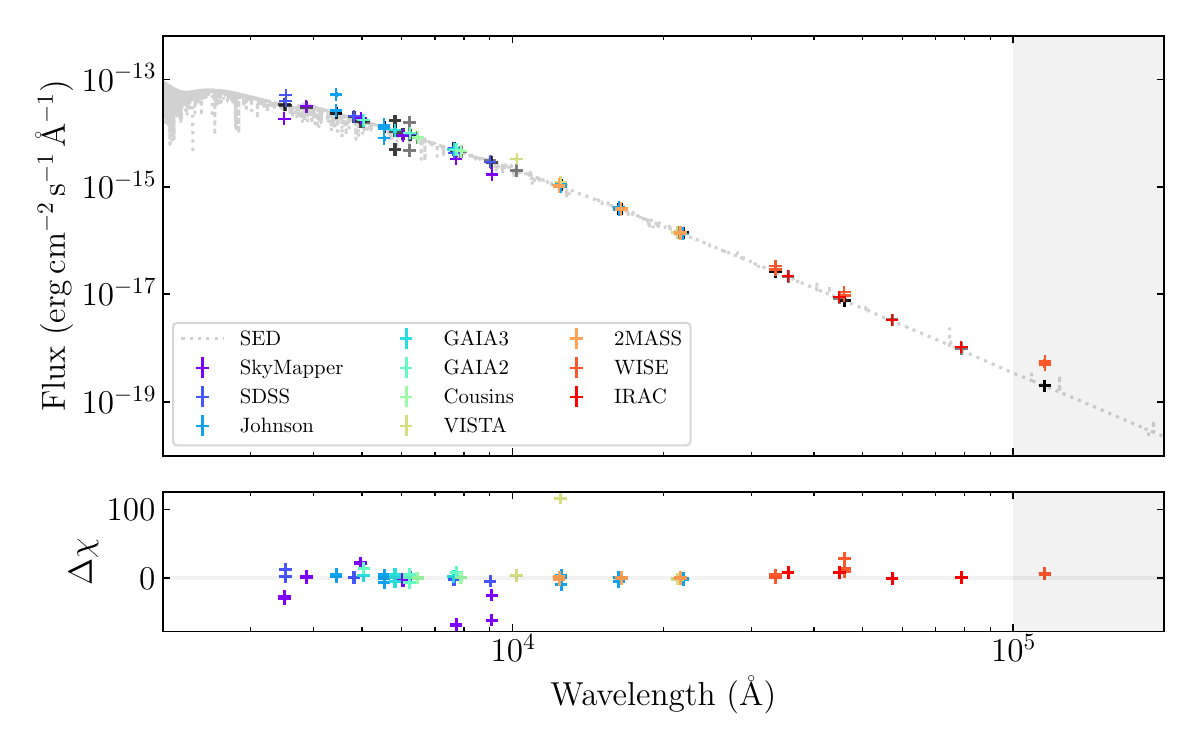}}
            \resizebox{0.497\hsize}{!}{\includegraphics{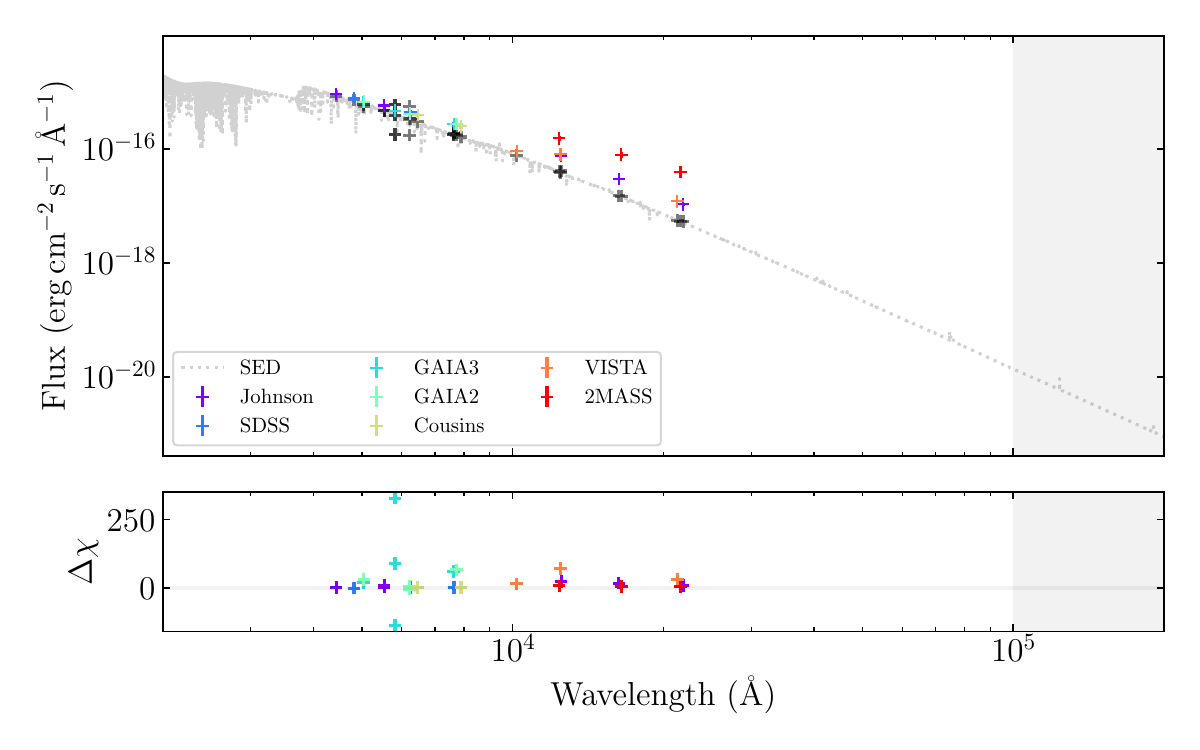}}
            \resizebox{0.497\hsize}{!}{\includegraphics{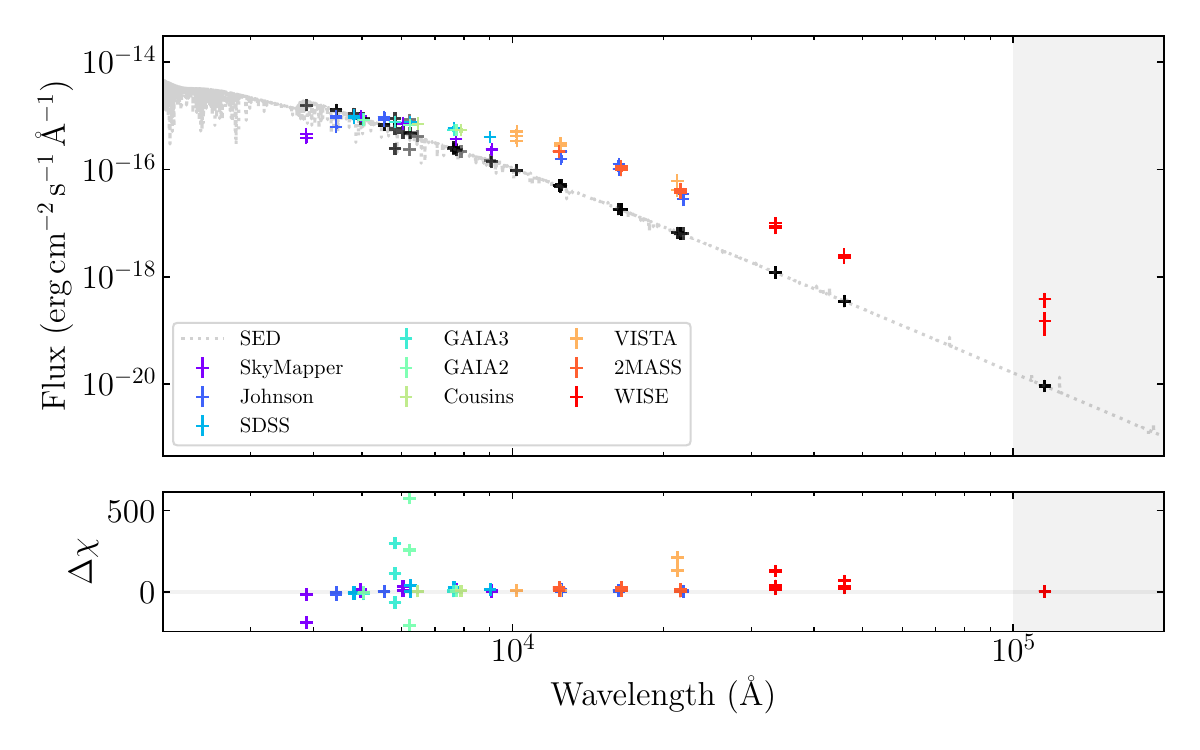}}
            \caption{continued for \#42 to \#49.}
        \end{figure*}
        \addtocounter{figure}{-1}
        \begin{figure*}
            \centering
            \resizebox{0.497\hsize}{!}{\includegraphics{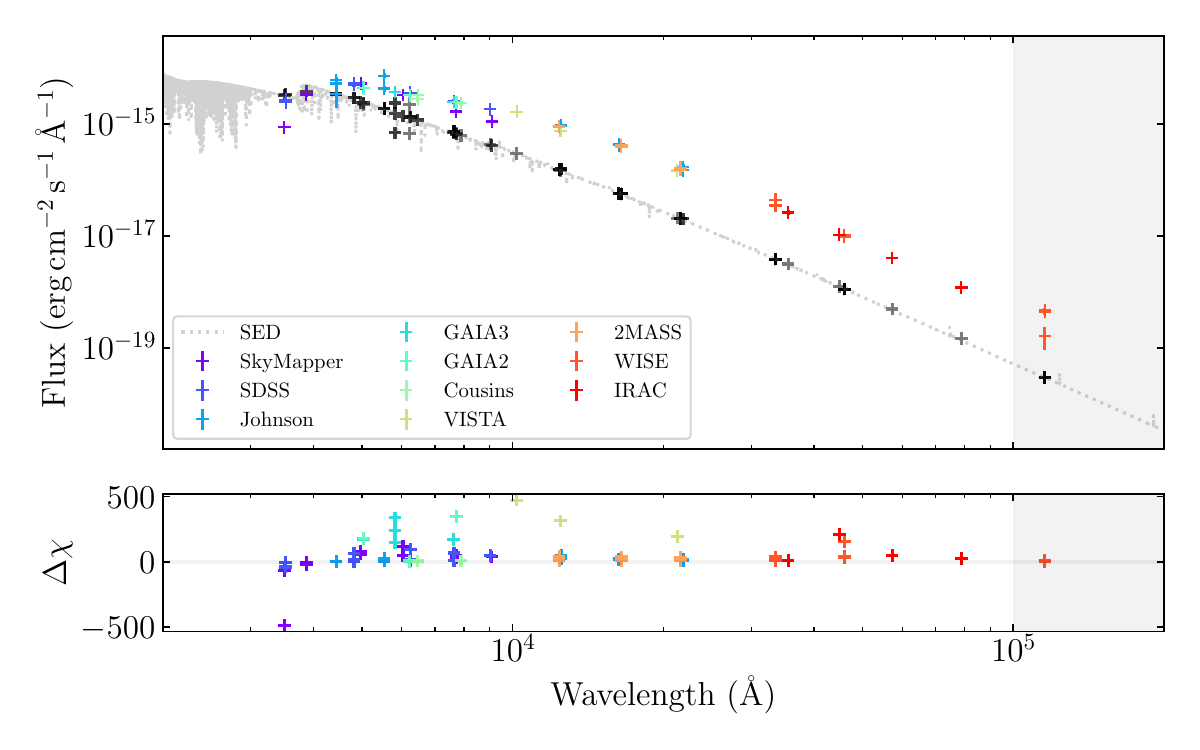}}
            \resizebox{0.497\hsize}{!}{\includegraphics{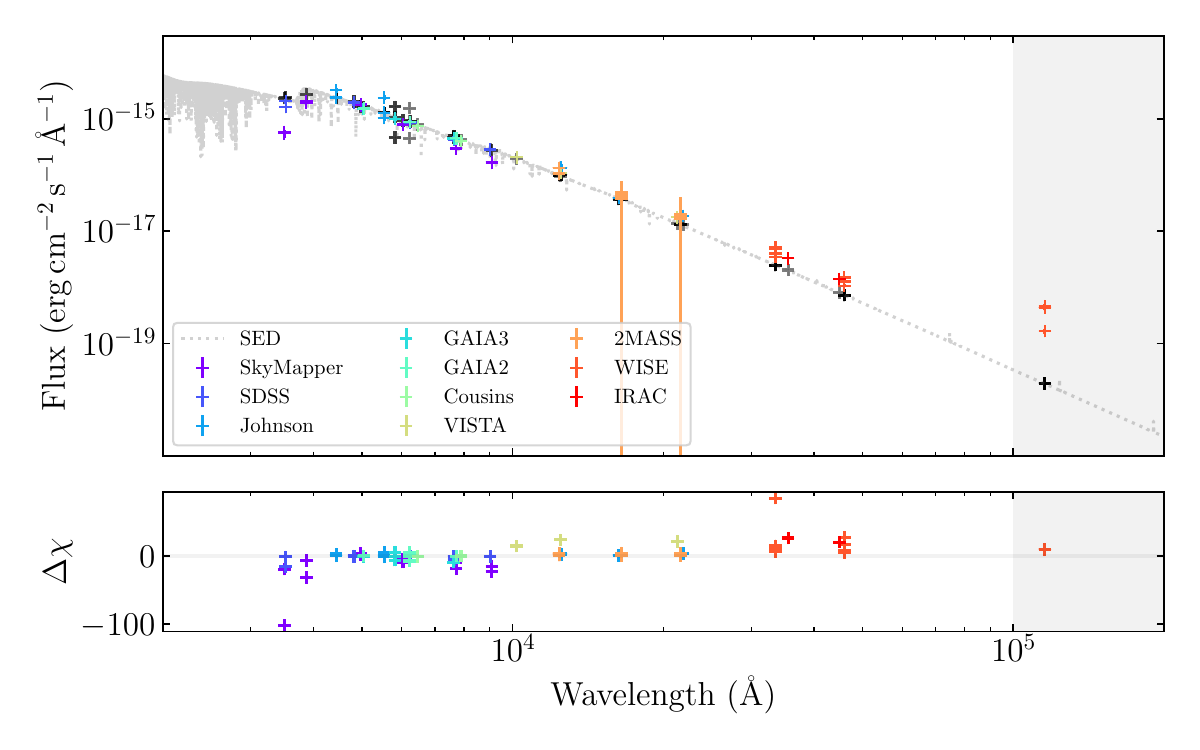}}
            \resizebox{0.497\hsize}{!}{\includegraphics{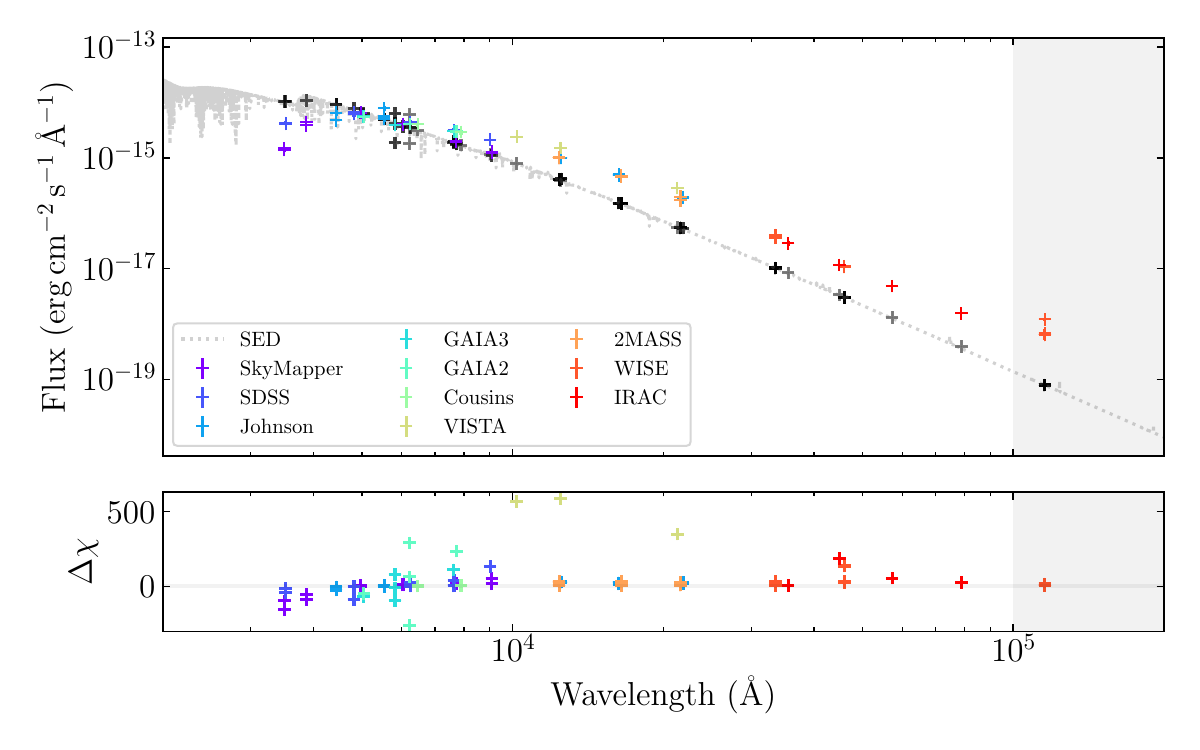}}
            \resizebox{0.497\hsize}{!}{\includegraphics{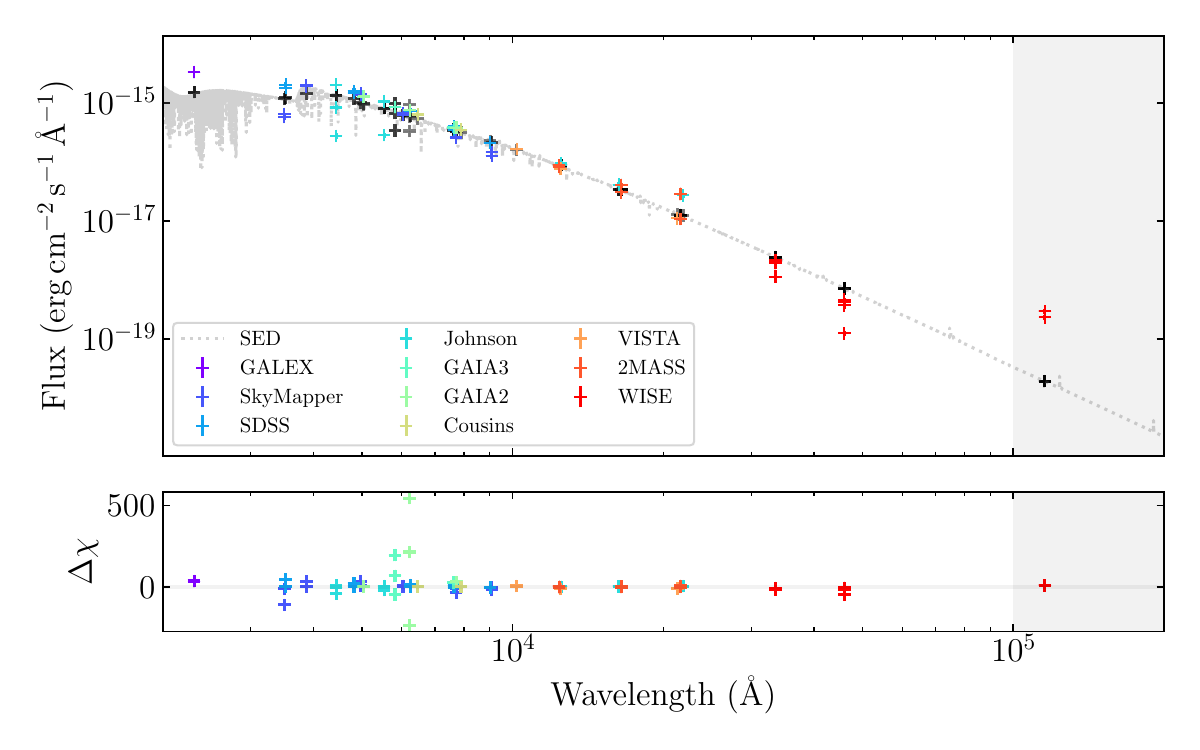}}
            \caption{continued for \#50 to \#53.}
        \end{figure*}
        
        \begin{figure*}
                \centering
                \resizebox{0.497\hsize}{!}{\includegraphics{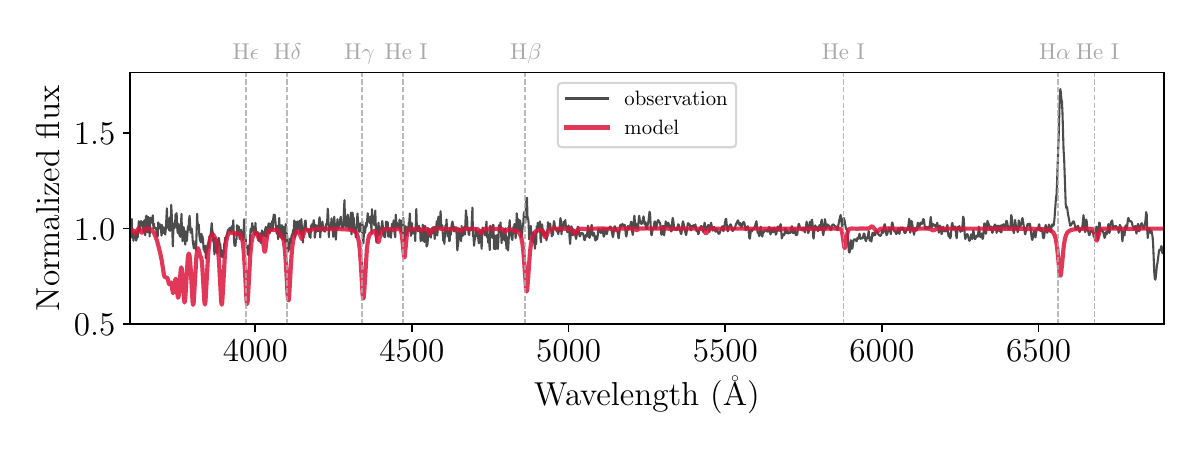}}
                \resizebox{0.497\hsize}{!}{\includegraphics{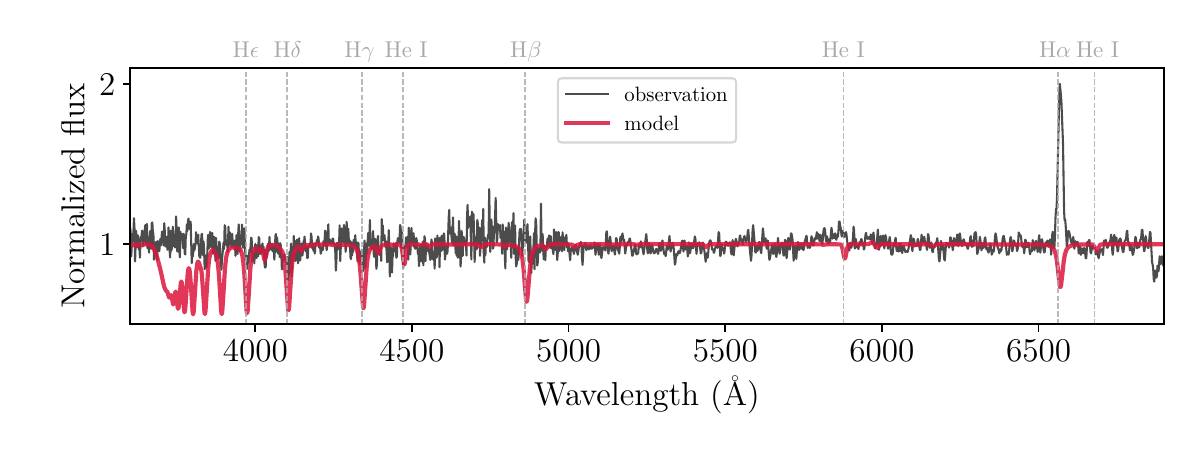}}
                \resizebox{0.497\hsize}{!}{\includegraphics{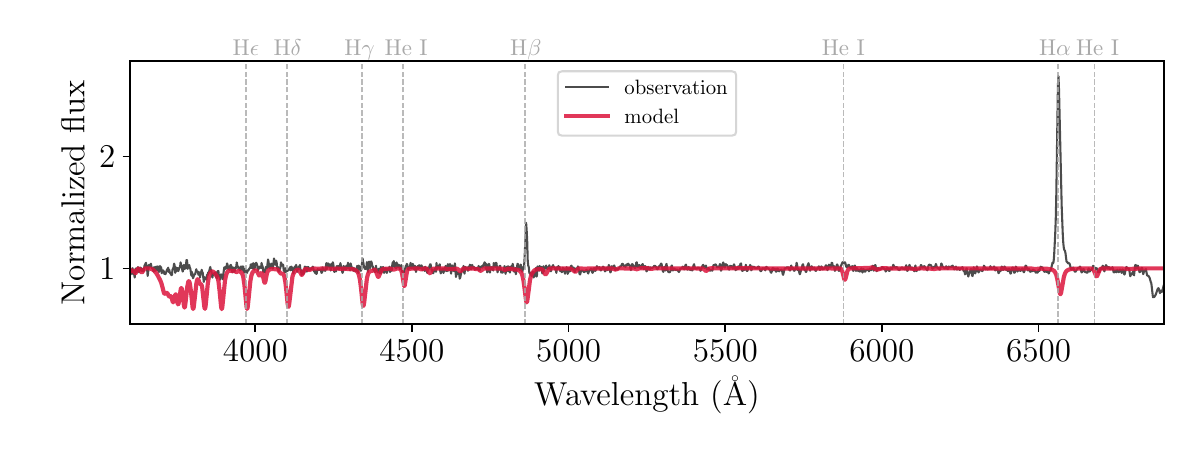}}
                \resizebox{0.497\hsize}{!}{\includegraphics{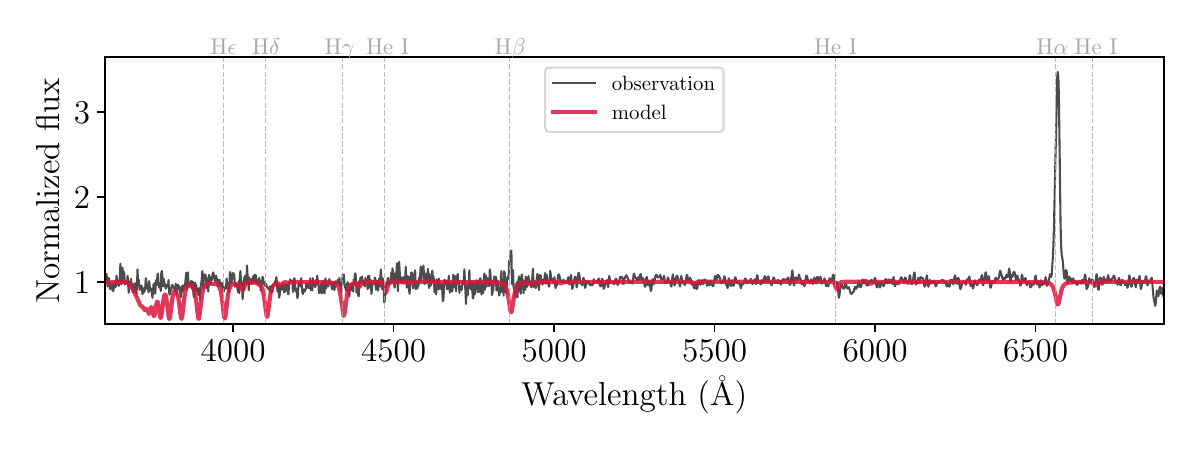}}
                \resizebox{0.497\hsize}{!}{\includegraphics{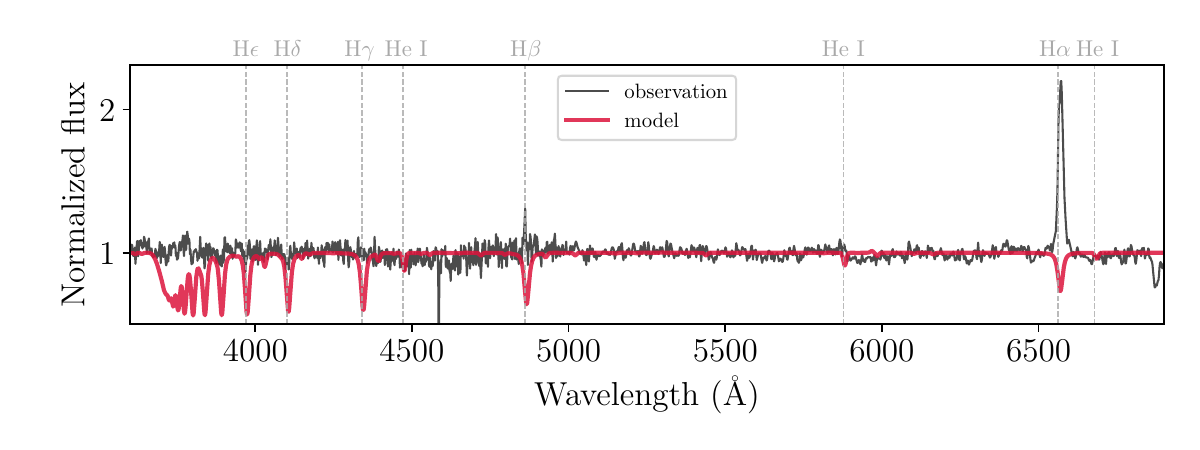}}
                \resizebox{0.497\hsize}{!}{\includegraphics{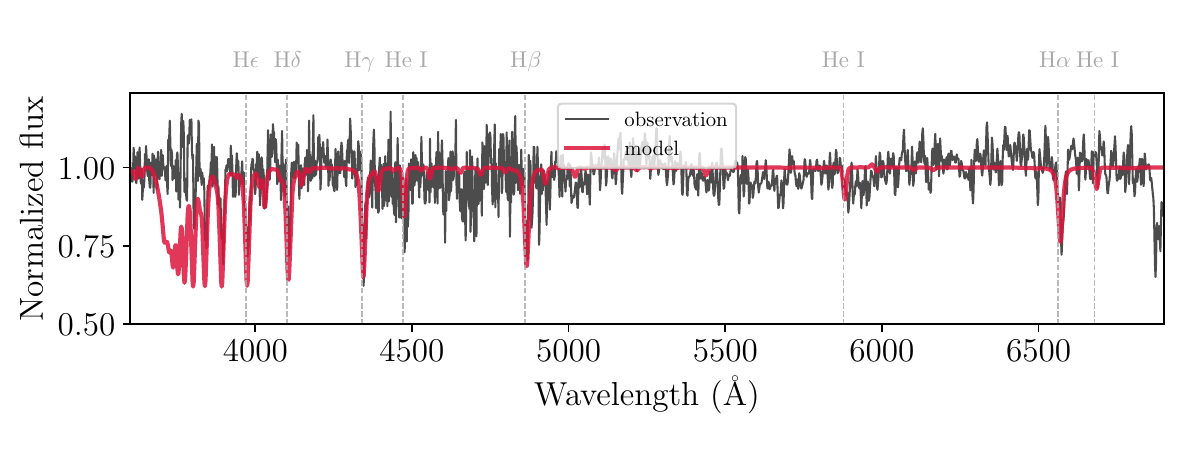}}
                \resizebox{0.497\hsize}{!}{\includegraphics{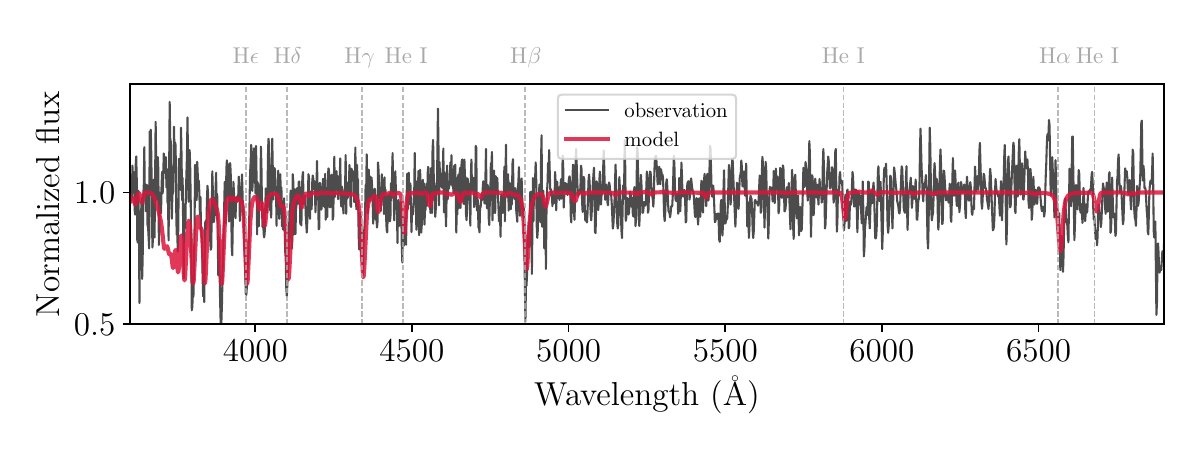}}
                \resizebox{0.497\hsize}{!}{\includegraphics{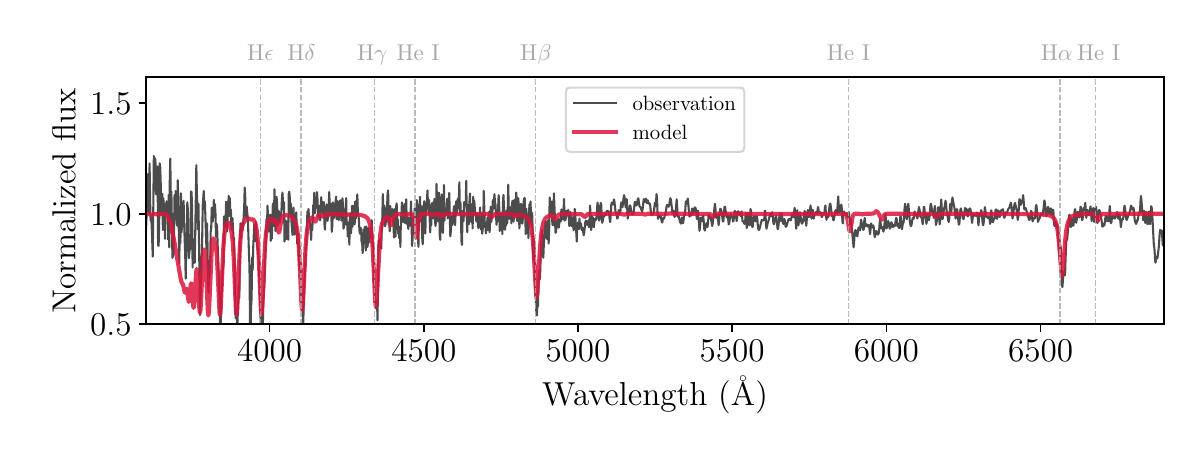}}
                \resizebox{0.497\hsize}{!}{\includegraphics{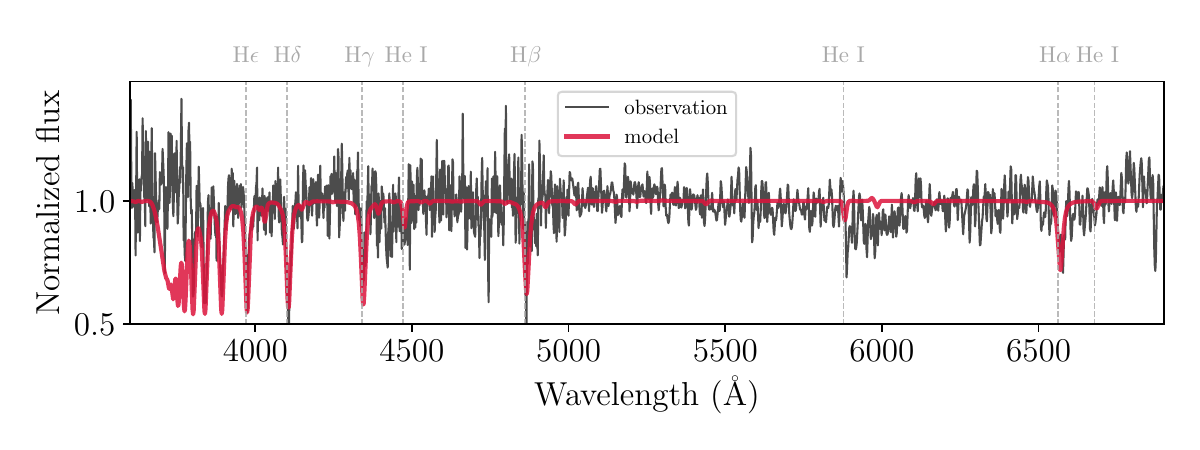}}
                \resizebox{0.497\hsize}{!}{\includegraphics{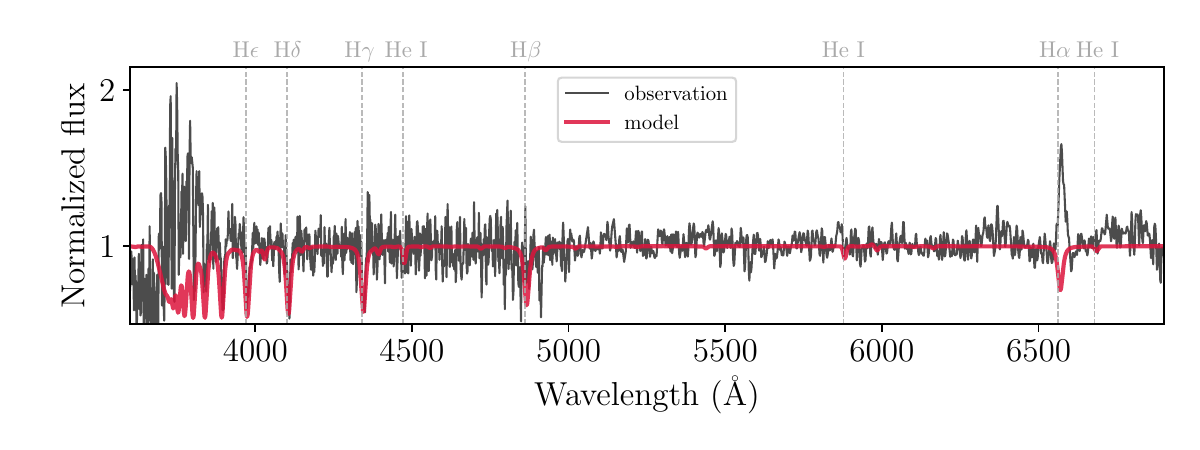}}
                \resizebox{0.497\hsize}{!}{\includegraphics{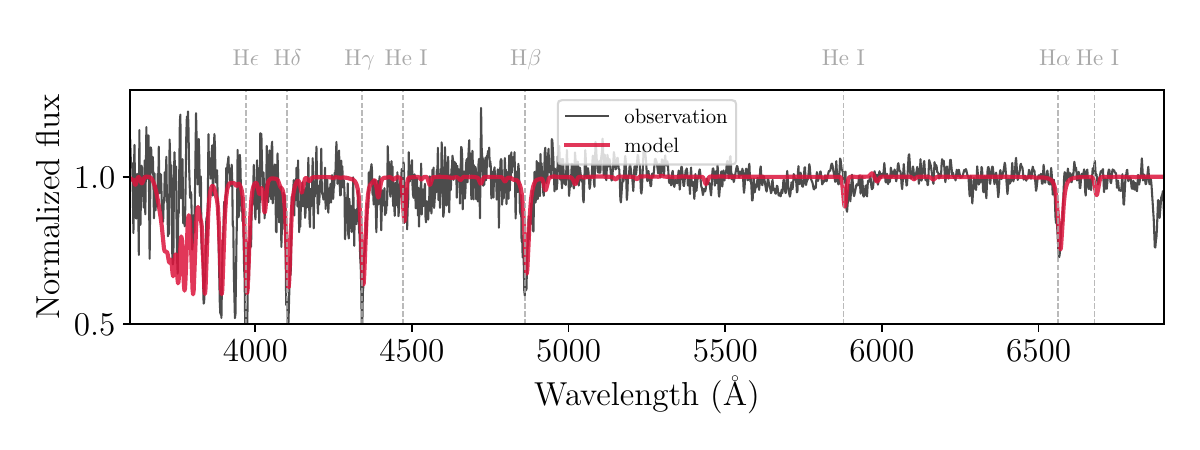}}
                \caption{Best-fit spectrum from SED fitting as described in Sect.\,\ref{sec:SED-fitting} compared to LCO/FLOYDS spectrum for \#28, \#32, \#35, \#36, \#37, \#39, \#41, \#43, \#44, \#51 and \#52.}
                \label{fig:SED_LCO}
        \end{figure*}
        \clearpage
        
        \section{\gaia counterparts}
        
        \begin{table*}[b]
                \centering
                \caption{\gaia DR3 counterparts of objects in our catalogue.} 
                \label{tab:Gaia_cp} 
                \begin{tabular}{lllllllllll} 
                        \hline\hline\noalign{\smallskip}
                        \#      & \gaia DR3& RA            & Dec & Dist.     & G  & e\_G & BP & e\_BP & RP & e\_RP \\
                        & name&  \multicolumn{2}{c}{J2016}    & arcsec & mag & mag & mag & mag & mag & mag  \\
                        \noalign{\smallskip}\hline\noalign{\smallskip}
                        1 & 4655356872255326464 & 04:53:15.1 & -69:32:42 & 2.0 & 15.164 & 0.004 & 15.120 & 0.012 & 15.217 & 0.011 \\ 
                        2 & 4655121130108671744 & 04:55:58.9 & -70:20:00 & 2.3 & 14.0511 & 0.0029 & 14.080 & 0.004 & 13.940 & 0.005 \\ 
                        3 & 4655263924844830848 & 04:57:37.0 & -69:27:28 & 4.2 & 13.682 & 0.003 & 13.691 & 0.004 & 13.627 & 0.006 \\ 
                        4 & 4655053441413348096 & 05:01:23.8 & -70:33:34 & 3.5 & 14.463 & 0.005 & 14.412 & 0.014 & 14.360 & 0.017 \\ 
                        5 & 4661267090898349824 & 05:07:22.2 & -68:48:00 & 0.9 & 15.849 & 0.003 & 15.819 & 0.006 & 15.752 & 0.008 \\ 
                        6 & 4661372884514746240 & 05:07:55.5 & -68:25:05 & 5.5 & 14.962 & 0.003 & 14.962 & 0.004 & 14.902 & 0.007 \\ 
                        7 & 4662048358263809408 & 05:08:09.9 & -66:06:52 & 2.1 & 14.0608 & 0.0028 & 14.046 & 0.003 & 14.028 & 0.005 \\ 
                        8 & 4658370908558254848 & 05:13:00.0 & -68:26:38 & 2.8 & 14.855 & 0.004 & 14.884 & 0.007 & 14.641 & 0.009 \\ 
                        9 & 4663520643026398848 & 05:13:28.3 & -65:47:19 & 2.7 & 14.892 & 0.003 & 14.866 & 0.005 & 14.855 & 0.005 \\ 
                        10 & 4658196459798969856 & 05:16:00.0 & -69:16:08 & 2.9 & 15.156 & 0.003 & 15.101 & 0.007 & 15.206 & 0.006 \\ 
                        11 & 4658173851121043200 & 05:20:29.9 & -69:31:56 & 3.8 & 14.3861 & 0.0029 & 14.367 & 0.004 & 14.342 & 0.005 \\ 
                        12 & 4660433042580352000 & 05:24:11.8 & -66:20:51 & 0.7 & 14.6477 & 0.0029 & 14.637 & 0.004 & 14.621 & 0.005 \\ 
                        13 & 4660184346803166848 & 05:28:58.5 & -67:09:46 & 1.3 & 14.989 & 0.003 & 14.914 & 0.004 & 15.092 & 0.007 \\ 
                        14 & 4660325634049868800 & 05:29:14.3 & -66:24:44 & 2.0 & 14.492 & 0.003 & 14.437 & 0.006 & 14.556 & 0.005 \\ 
                        15 & 4660363017447834112 & 05:29:47.9 & -65:56:44 & 3.1 & 14.5651 & 0.0028 & 14.499 & 0.003 & 14.661 & 0.004 \\ 
                        16 & 4660739153472610048 & 05:30:11.4 & -65:51:24 & 1.9 & 14.7305 & 0.0028 & 14.749 & 0.003 & 14.642 & 0.004 \\ 
                        17 & 4660214961332778880 & 05:30:42.1 & -66:54:30 & 0.8 & 14.979 & 0.003 & 14.961 & 0.004 & 14.974 & 0.005 \\ 
                        18 & 4658429113912320000 & 05:31:08.4 & -69:09:23 & 2.1 & 13.6018 & 0.0029 & 13.645 & 0.004 & 13.481 & 0.006 \\ 
                        19 & 4660694932493354752 & 05:32:32.6 & -65:51:41 & 1.6 & 13.0418 & 0.0028 & 12.945 & 0.003 & 13.205 & 0.004 \\ 
                        20 & 4660300345280168192 & 05:32:49.6 & -66:22:13 & 0.9 & 14.071 & 0.003 & 13.942 & 0.005 & 14.287 & 0.005 \\ 
                        21 & 4660245713290392448 & 05:35:06.0 & -67:00:16 & 7.5 & 14.7826 & 0.0030 & 14.740 & 0.004 & 14.811 & 0.006 \\ 
                        22 & 4660250867252215424 & 05:35:41.0 & -66:51:54 & 3.4 & 15.026 & 0.004 & 14.84 & 0.01 & 15.103 & 0.007 \\ 
                        23 & 4660111538511490944 & 05:36:00.0 & -67:35:07 & 0.7 & 13.4073 & 0.0028 & 13.339 & 0.003 & 13.511 & 0.004 \\ 
                        24 & 4757068874690668160 & 05:38:56.6 & -64:05:03 & 6.6 & 16.996 & 0.011 & 16.98 & 0.03 & 17.02 & 0.04 \\ 
                        25 & 4657637156283982336 & 05:39:38.8 & -69:44:36 & 6.0 & 14.5561 & 0.0030 & 14.678 & 0.004 & 14.164 & 0.005 \\ 
                        26 & 4657822114771262848 & 05:41:34.3 & -68:25:48 & 1.3 & 13.9931 & 0.0029 & 13.996 & 0.004 & 13.947 & 0.006 \\ 
                        27 & 4657045928301192320 & 05:44:05.2 & -71:00:51 & 2.2 & 15.308 & 0.008 & 15.332 & 0.017 & 15.24 & 0.03 \\ 
                        28 & 4659106859733759104 & 05:50:06.5 & -68:14:56 & 1.6 & 14.9763 & 0.0029 & 14.992 & 0.005 & 14.890 & 0.006 \\ 
                        29 & 4655407071838175744 & 04:43:54.7 & -69:29:46 & 2.8 & 14.561 & 0.003 & 14.579 & 0.005 & 14.473 & 0.007 \\ 
                        30 & 4655375701393865600 & 04:50:24.5 & -69:18:42 & 3.5 & 15.4728 & 0.0028 & 15.556 & 0.004 & 15.283 & 0.004 \\ 
                        31 & 4655344704637383296 & 04:50:28.1 & -69:35:50 & 8.4 & 16.207 & 0.003 & 16.218 & 0.011 & 15.7 & 0.7 \\ 
                        32 & 4661786335230690176 & 04:52:18.4 & -66:32:49 & 2.8 & 14.6628 & 0.0028 & 14.637 & 0.003 & 14.651 & 0.004 \\ 
                        33 & 4661704730845343872 & 04:58:00.5 & -67:09:26 & 11.3 & 16.2952 & 0.0029 & 16.220 & 0.005 & 16.332 & 0.006 \\ 
                        34 & 4662277920008206720 & 05:00:51.9 & -65:32:06 & 9.2 & 16.1617 & 0.0028 & 16.080 & 0.003 & 16.248 & 0.005 \\ 
                        35 & 4661465209126624512 & 05:02:14.1 & -67:46:18 & 3.0 & 13.678 & 0.004 & 13.670 & 0.010 & 13.619 & 0.005 \\ 
                        36 & 4655045603122201216 & 05:03:59.7 & -70:32:10 & 3.5 & 14.885 & 0.003 & 14.913 & 0.005 & 14.761 & 0.009 \\ 
                        37 & 4663605271066598784 & 05:07:06.2 & -65:21:47 & 2.7 & 14.451 & 0.003 & 14.441 & 0.006 & 14.405 & 0.007 \\ 
                        38 & 4663551635512438784 & 05:09:46.4 & -65:52:33 & 5.7 & 16.9651 & 0.0028 & 17.326 & 0.004 & 16.441 & 0.004 \\ 
                        39 & 4660238944432296192 & 05:27:26.1 & -66:33:08 & 6.8 & 15.3273 & 0.0028 & 15.222 & 0.003 & 15.517 & 0.004 \\ 
                        40 & 4660327489475880576 & 05:29:47.3 & -66:20:59 & 4.9 & 16.3413 & 0.0028 & 16.286 & 0.003 & 16.470 & 0.004 \\ 
                        41 & 4660347521204963712 & 05:30:49.6 & -66:20:11 & 1.8 & 16.275 & 0.003 & 16.016 & 0.004 & 16.231 & 0.005 \\ 
                        42 & 4658459900242918912 & 05:33:20.7 & -68:41:23 & 3.8 & 12.679 & 0.003 & 12.632 & 0.004 & 12.731 & 0.005 \\ 
                        43 & 4657274450578946048 & 05:34:48.9 & -69:43:39 & 4.8 & 15.0373 & 0.0028 & 15.198 & 0.003 & 14.729 & 0.004 \\ 
                        44 & 4657778477899251968 & 05:40:21.9 & -68:56:46 & 5.5 & 16.0581 & 0.0028 & 16.056 & 0.004 & 16.079 & 0.005 \\ 
                        45 & 4657768547932332544 & 05:41:37.5 & -68:32:33 & 1.9 & 14.239 & 0.005 & 14.149 & 0.012 & 14.393 & 0.012 \\ 
                        46 & 4659421732399012736 & 05:42:41.6 & -67:27:55 & 6.7 & 14.442 & 0.004 & 14.413 & 0.008 & 14.434 & 0.013 \\ 
                        47 & 4659423759624086784 & 05:44:22.1 & -67:27:33 & 3.7 & 13.5011 & 0.0030 & 13.457 & 0.004 & 13.562 & 0.005 \\ 
                        48 & 4659447364765216512 & 05:46:47.6 & -67:06:05 & 3.2 & 16.9646 & 0.0029 & 17.014 & 0.007 & 16.718 & 0.006 \\ 
                        49 & 4659796150440969984 & 05:53:19.1 & -65:59:56 & 5.1 & 16.4192 & 0.0028 & 16.736 & 0.004 & 15.956 & 0.005 \\ 
                        50 & 4659594080826426752 & 05:55:36.8 & -67:14:51 & 7.8 & 14.7255 & 0.0028 & 14.943 & 0.003 & 14.330 & 0.005 \\ 
                        51 & 4659162831750602112 & 05:58:50.6 & -67:52:27 & 7.0 & 16.1306 & 0.0028 & 16.106 & 0.004 & 16.173 & 0.005 \\ 
                        52 & 5283852012332458624 & 06:02:13.1 & -67:43:06 & 3.5 & 14.652 & 0.008 & 14.6858 & 0.0030 & 14.100 & 0.004 \\ 
                        53 & 5279359888851182464 & 06:04:27.4 & -70:29:16 & 9.1 & 16.2995 & 0.0029 & 16.268 & 0.004 & 16.330 & 0.005 \\ 
                        
                        \noalign{\smallskip}\hline
                \end{tabular}
        \end{table*}
        \clearpage
        
        \section{\ogle photometry}
        
        Our Fig.\,\ref{fig:ogle_IVlc} and Fig\,\ref{fig:ogle_Ilc} present \ogle light curves of the systems investigated in this work.

        \begin{figure*}[b]
        \centering
        \resizebox{0.495\hsize}{!}{\includegraphics{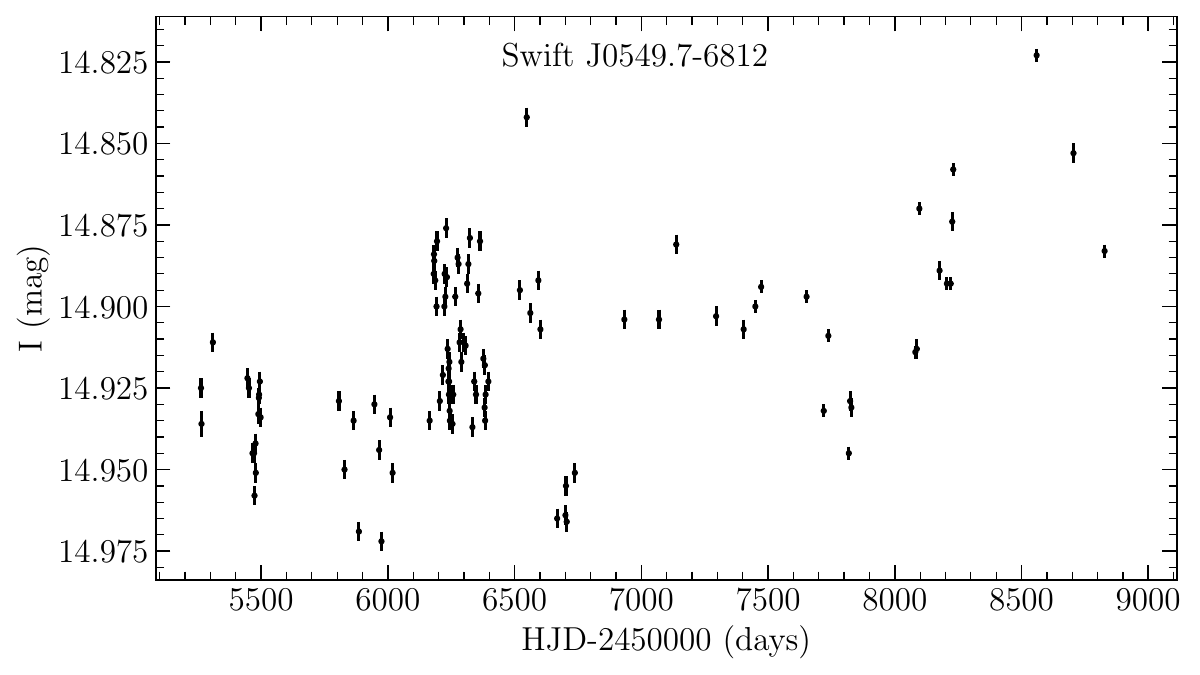}}
        \resizebox{0.495\hsize}{!}{\includegraphics{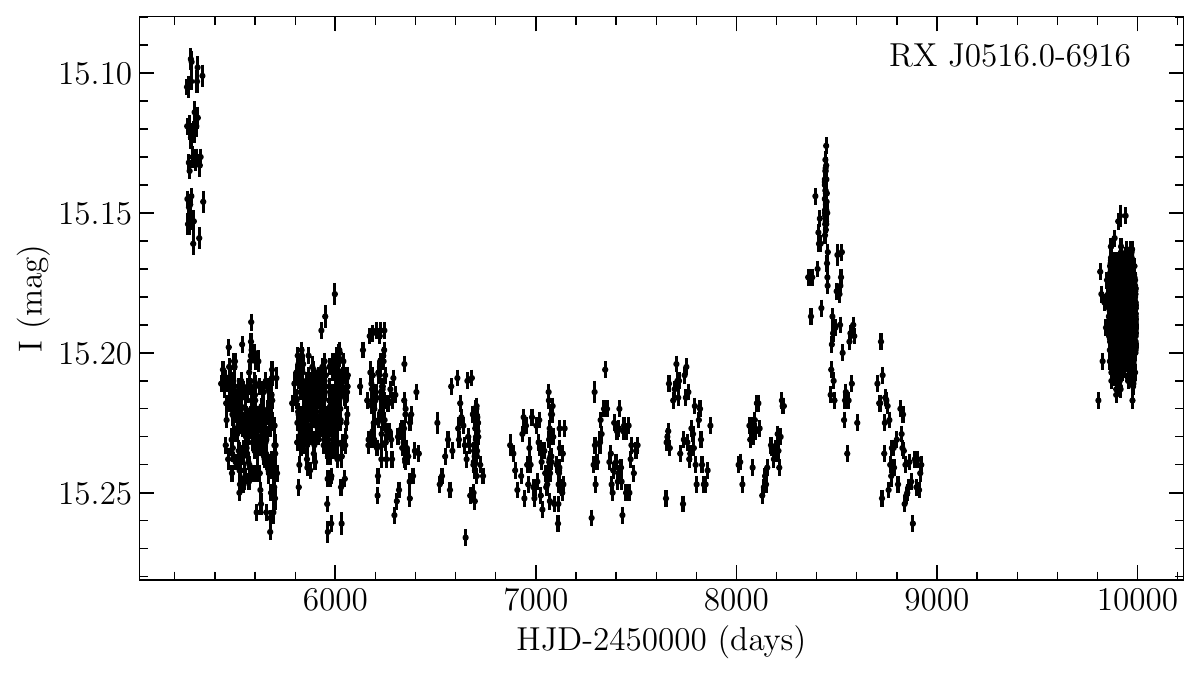}}
        \resizebox{0.495\hsize}{!}{\includegraphics{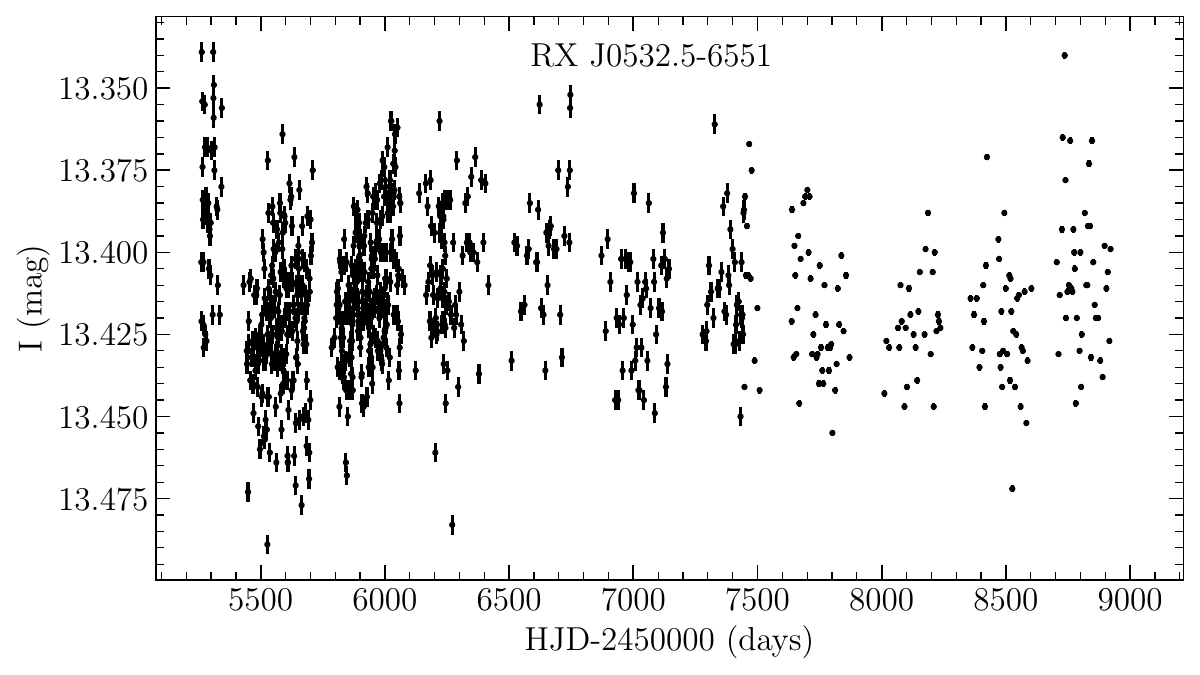}}
        \resizebox{0.495\hsize}{!}{\includegraphics{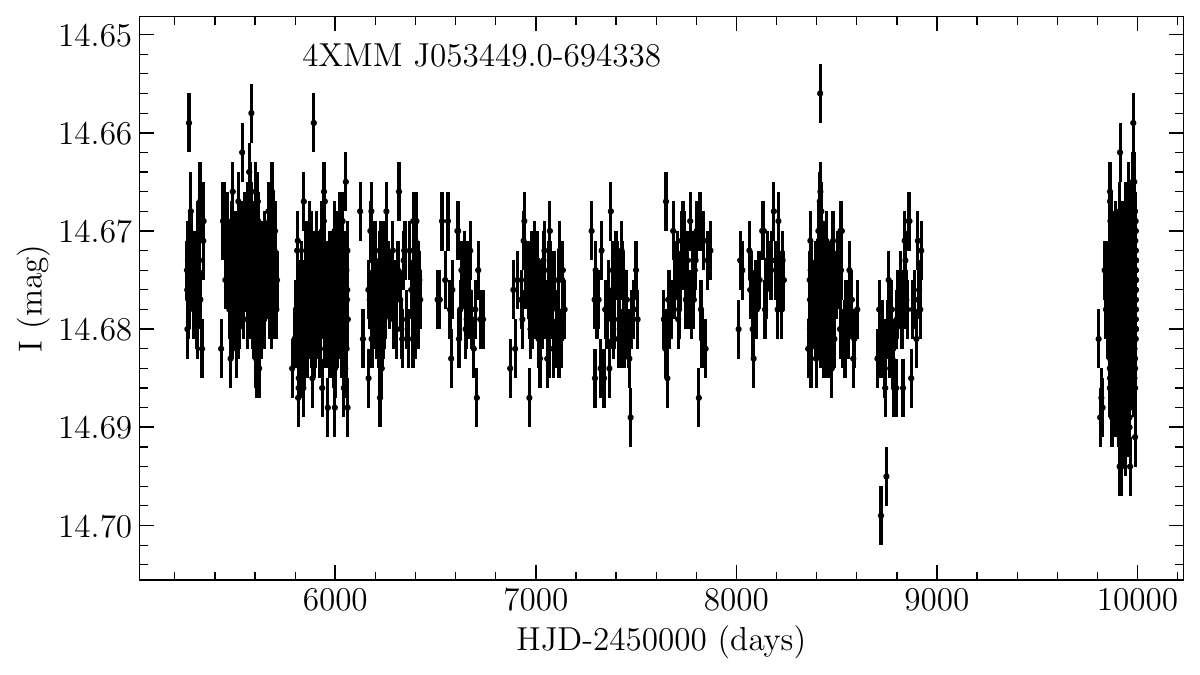}}
        \resizebox{0.495\hsize}{!}{\includegraphics{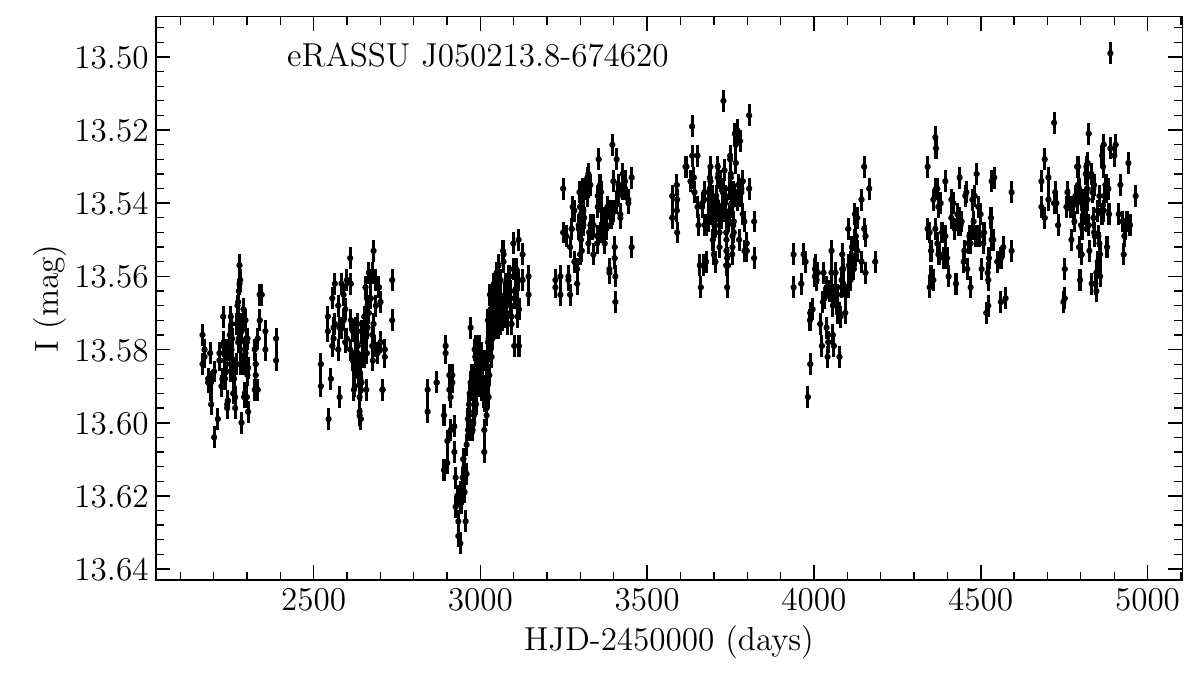}}
        \resizebox{0.495\hsize}{!}{\includegraphics{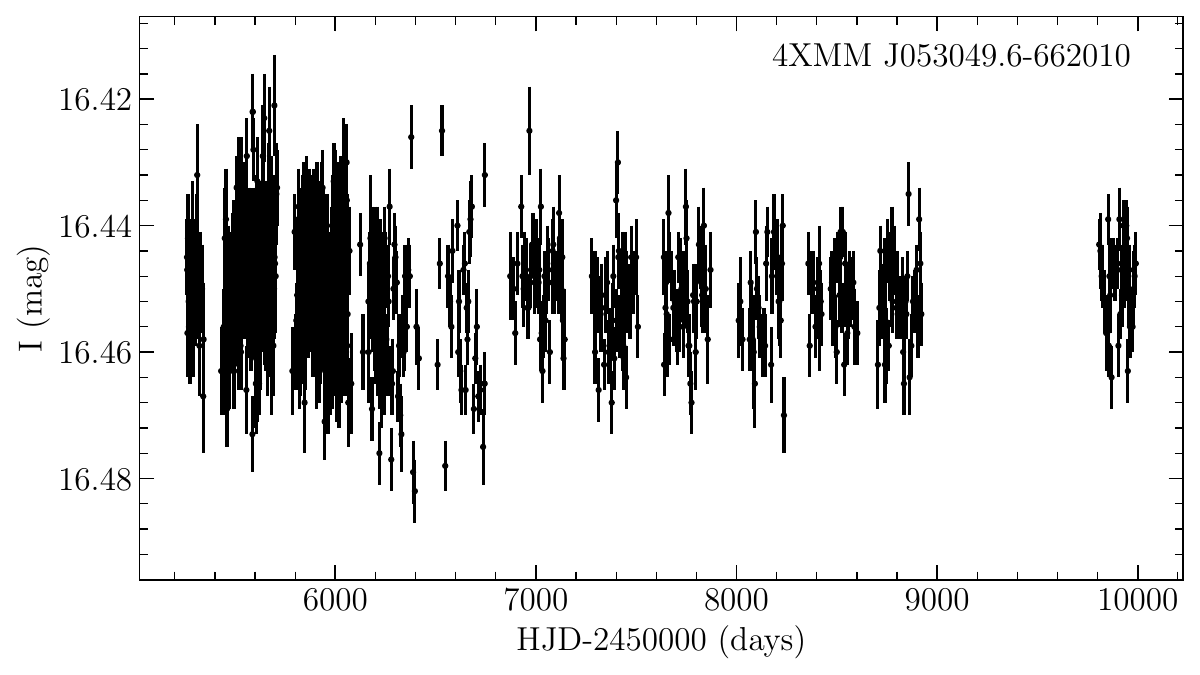}}
        \resizebox{0.495\hsize}{!}{\includegraphics{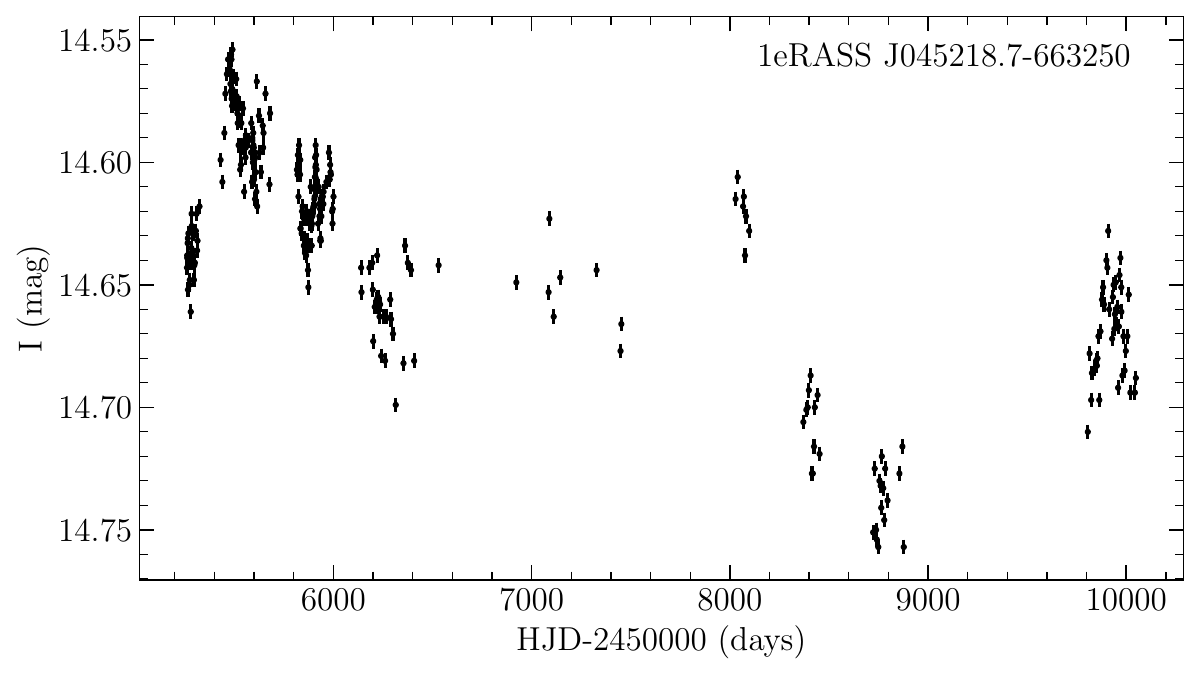}}
        \resizebox{0.495\hsize}{!}{\includegraphics{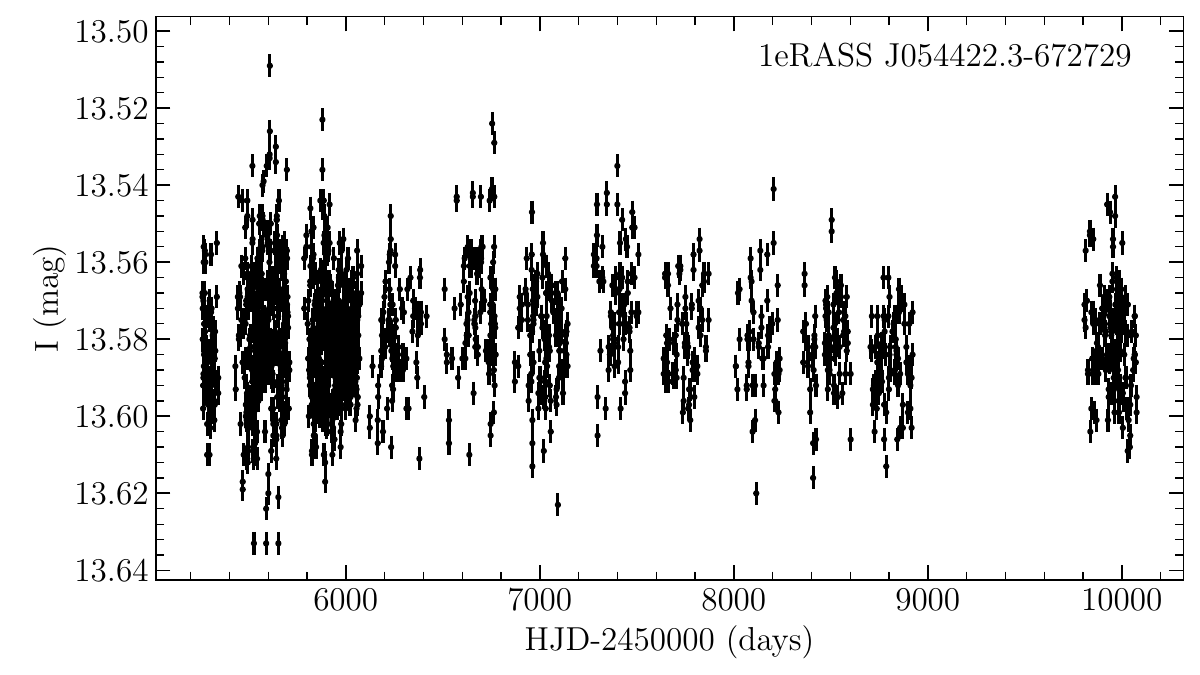}}
        \caption{\ogle I-band light curves of stars with low-amplitude ($<$0.3 mag) variability.}
        \end{figure*}
        \addtocounter{figure}{-1}
        \begin{figure*}
        \centering
        \resizebox{0.495\hsize}{!}{\includegraphics{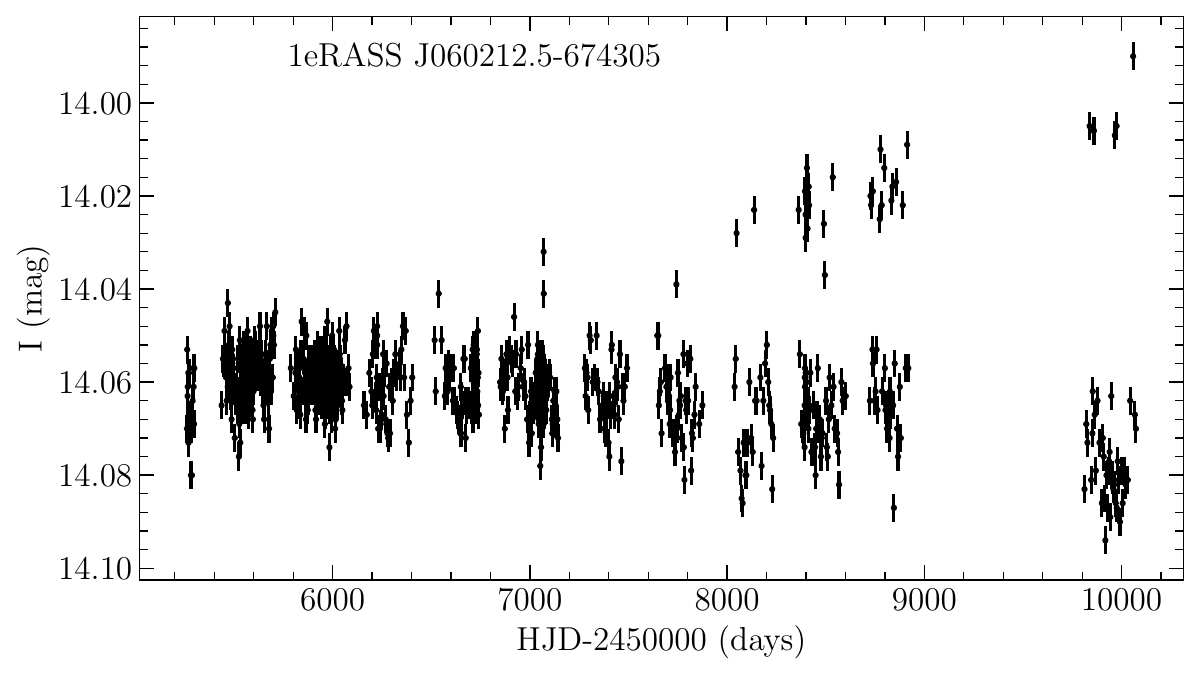}}
        \resizebox{0.495\hsize}{!}{\includegraphics{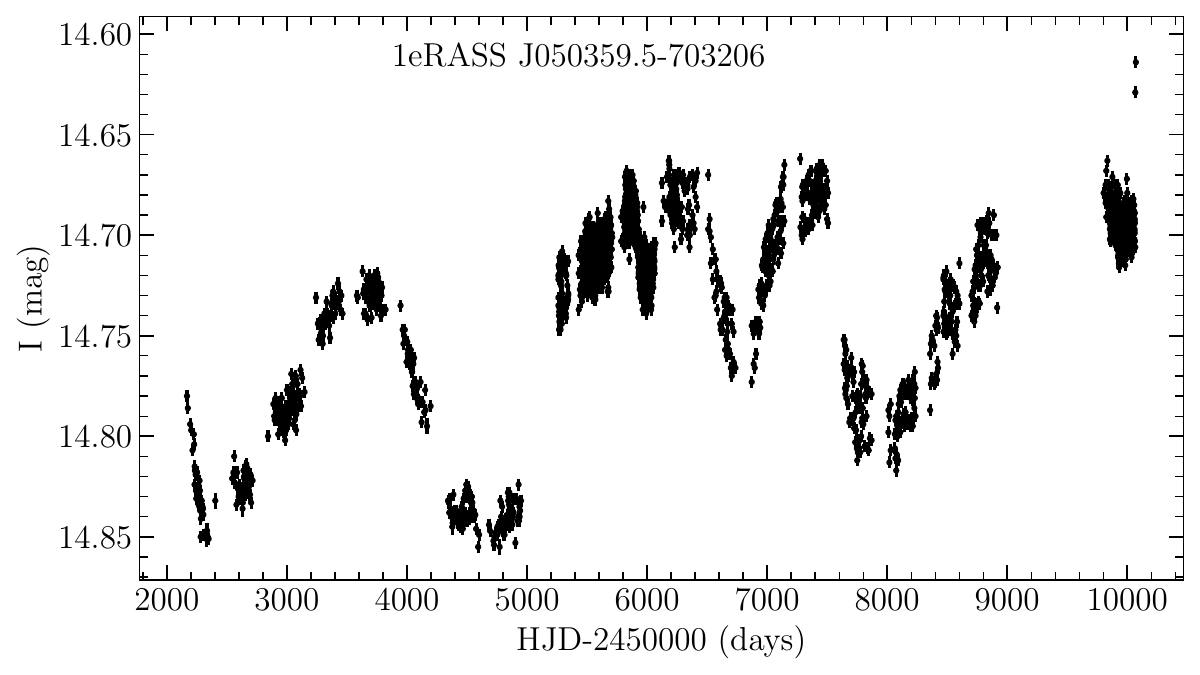}}
        \resizebox{0.495\hsize}{!}{\includegraphics{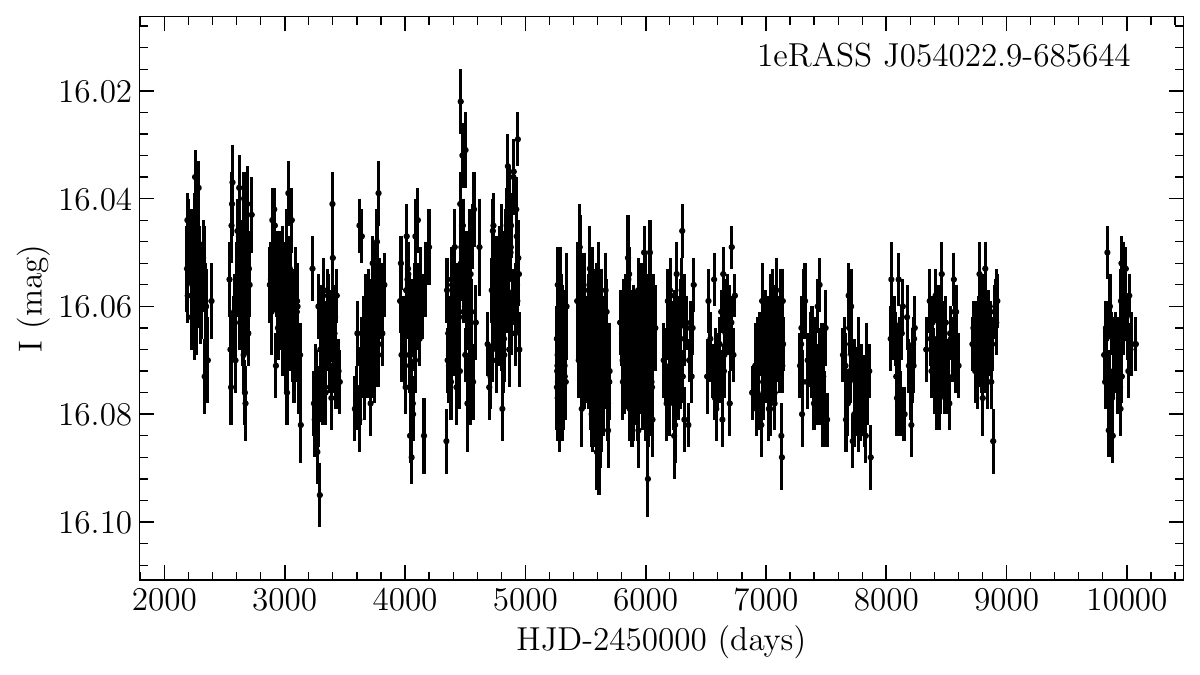}}
        \resizebox{0.495\hsize}{!}{\includegraphics{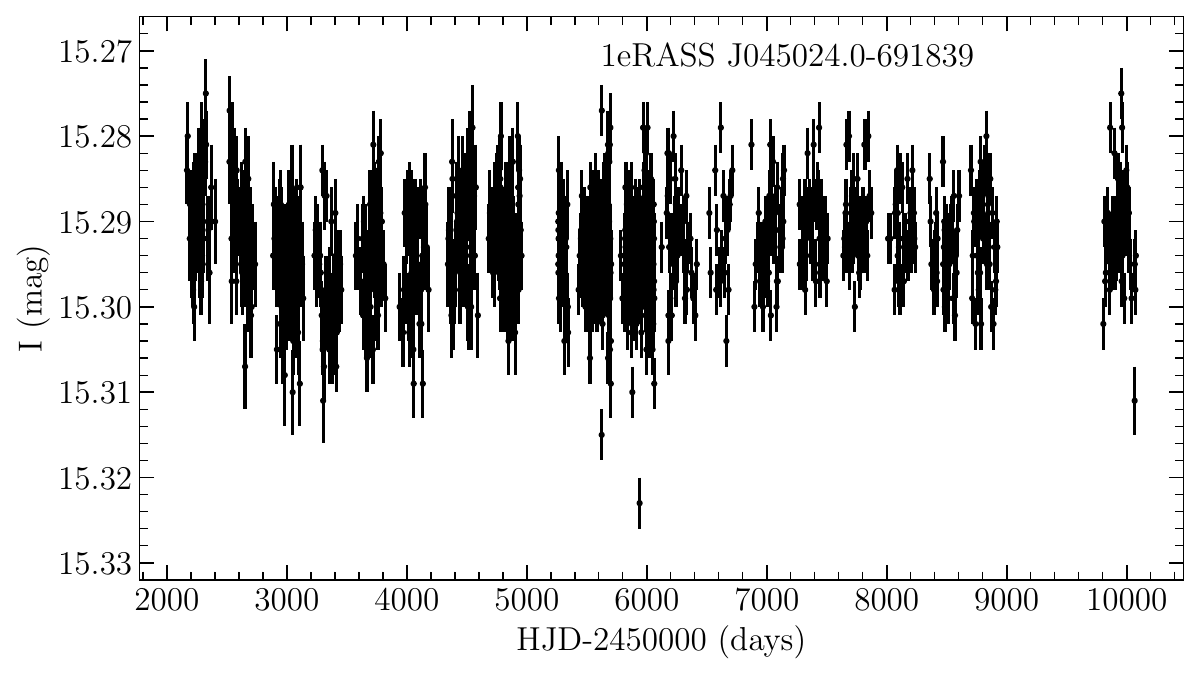}}
        \resizebox{0.495\hsize}{!}{\includegraphics{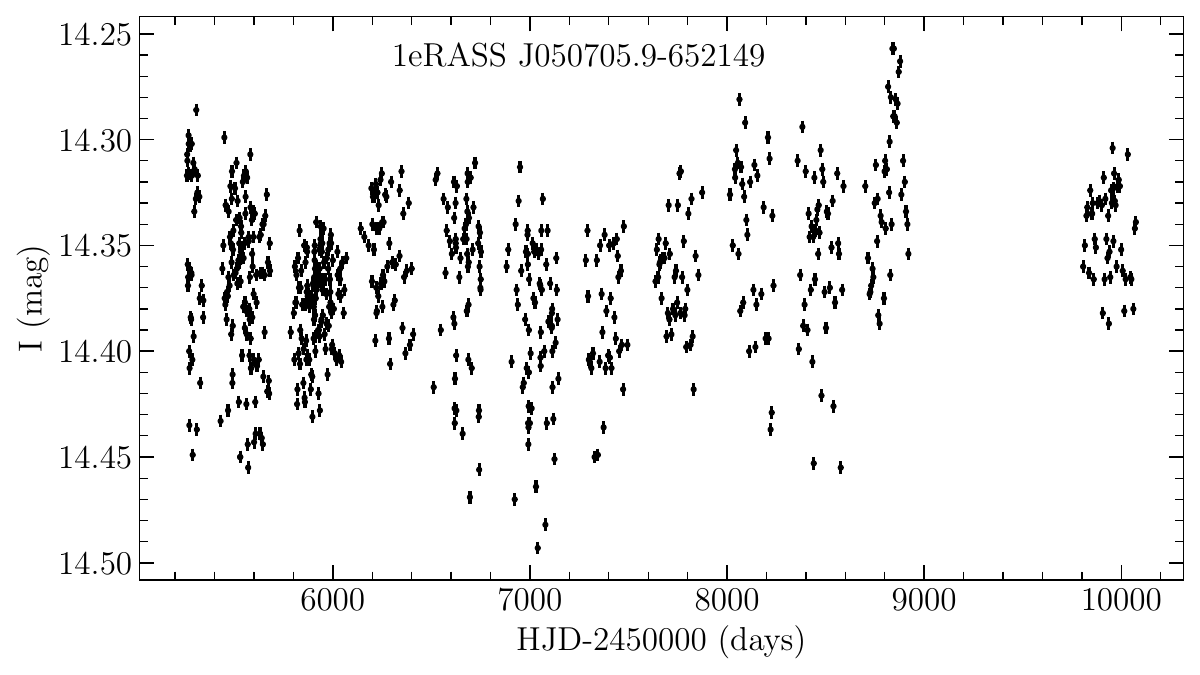}}
        \resizebox{0.495\hsize}{!}{\includegraphics{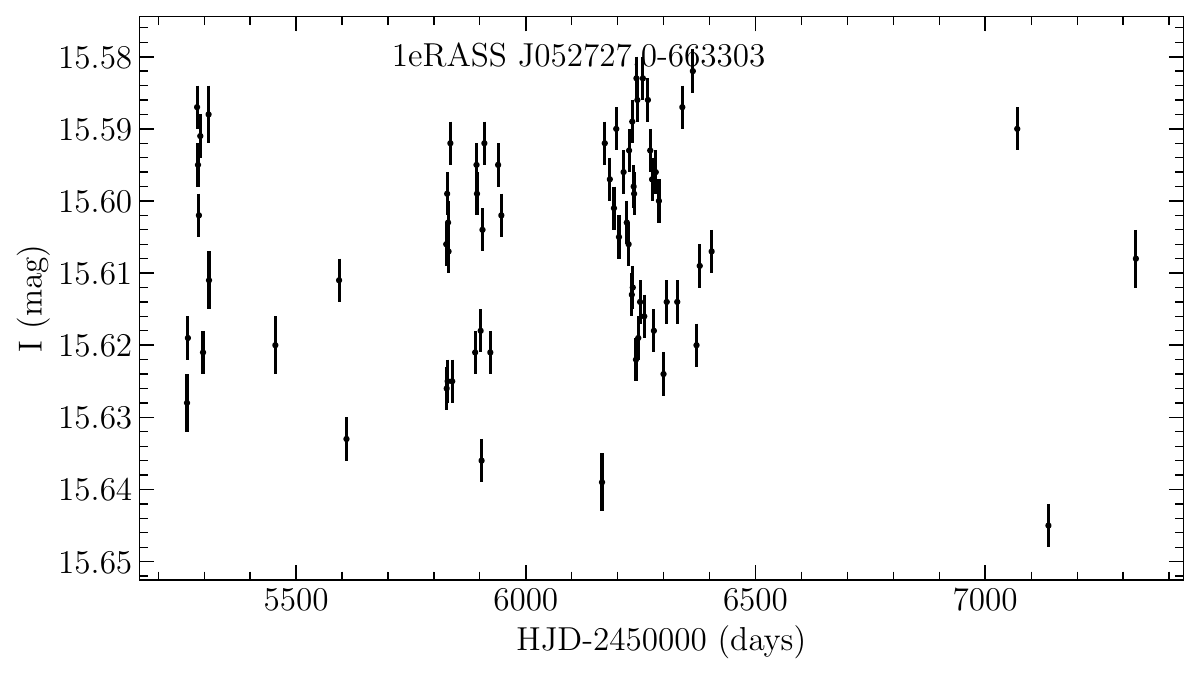}}
        \resizebox{0.495\hsize}{!}{\includegraphics{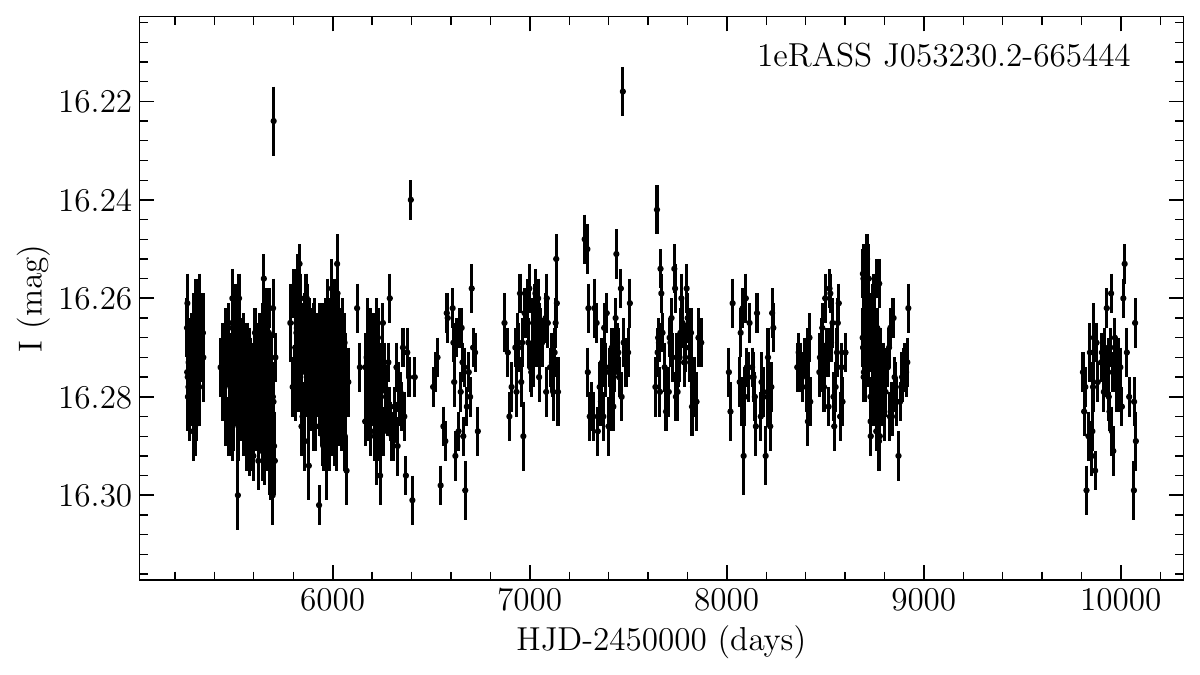}}
        \resizebox{0.495\hsize}{!}{\includegraphics{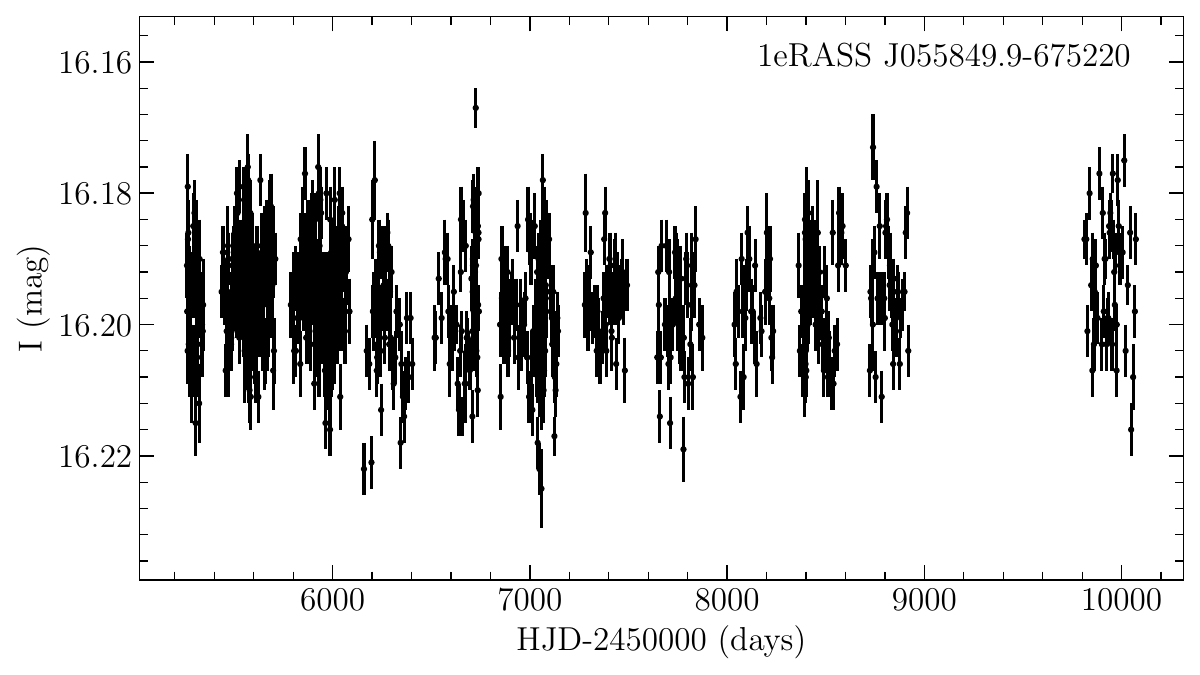}}
        \caption{Continued.}
        \end{figure*}
        \addtocounter{figure}{-1}
        \begin{figure*}
        \centering
        \resizebox{0.495\hsize}{!}{\includegraphics{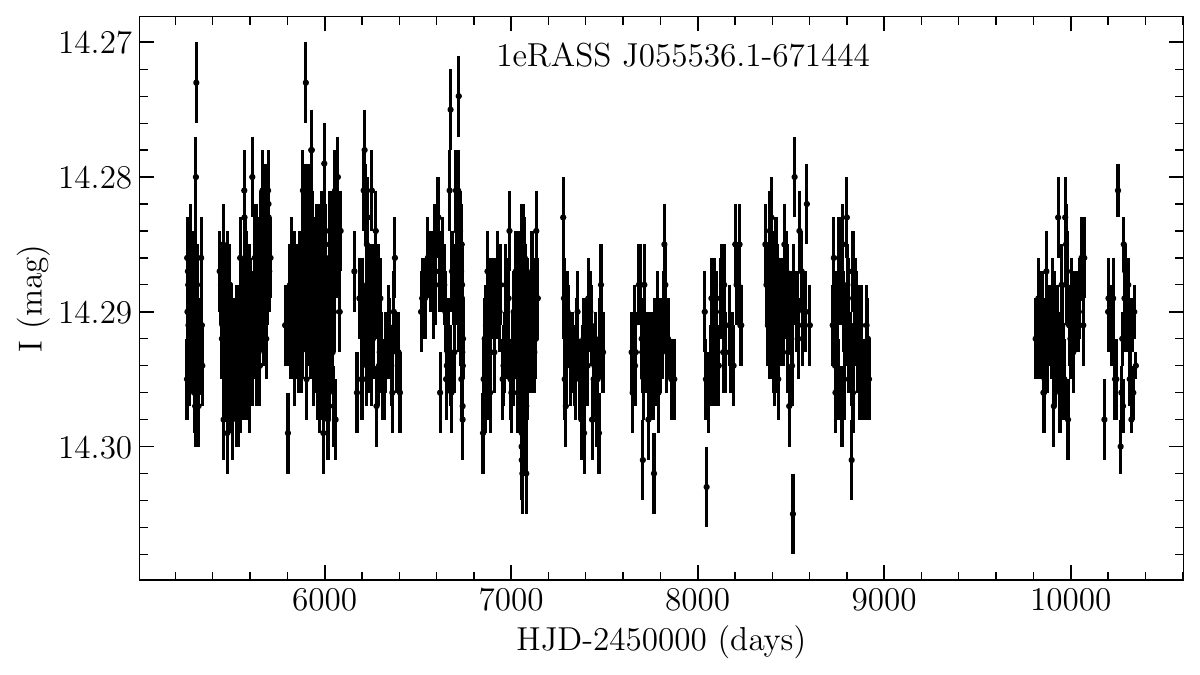}}
        \resizebox{0.495\hsize}{!}{\includegraphics{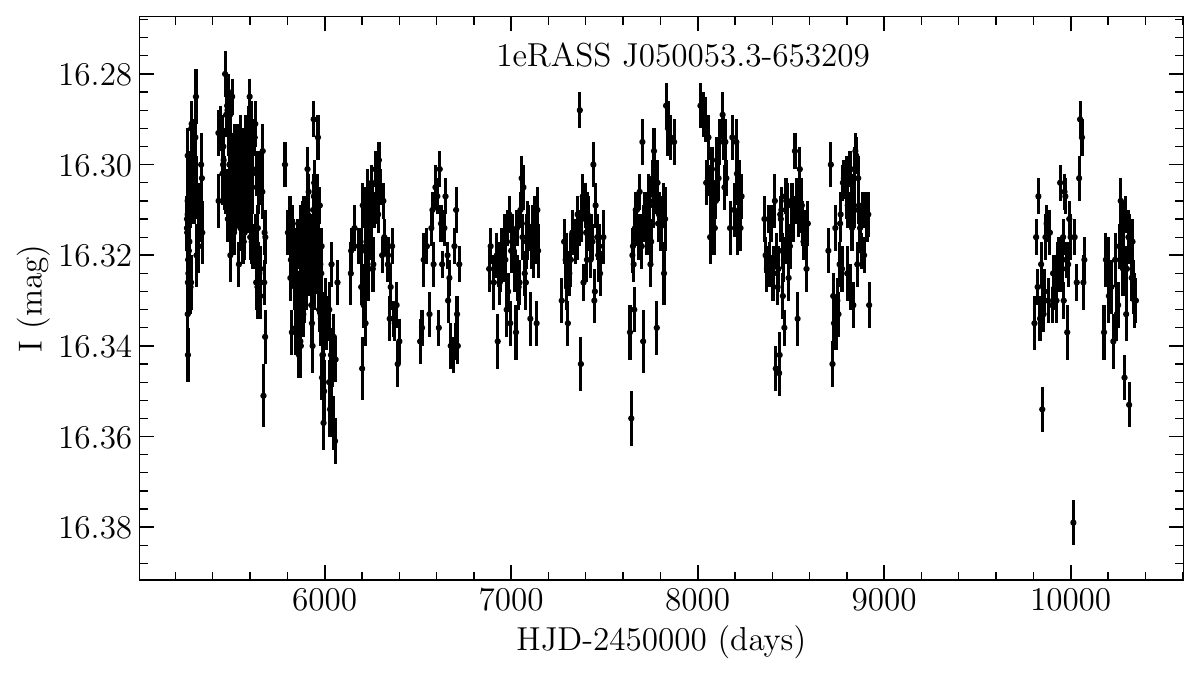}}
        \resizebox{0.495\hsize}{!}{\includegraphics{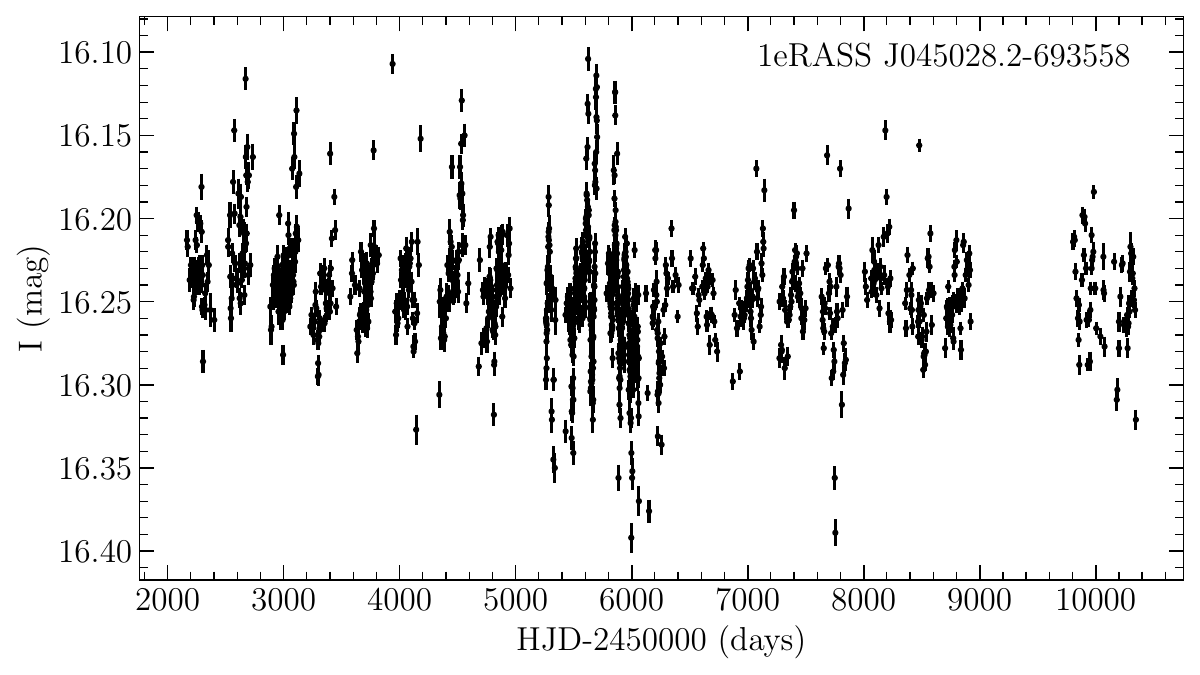}}
        \resizebox{0.495\hsize}{!}{\includegraphics{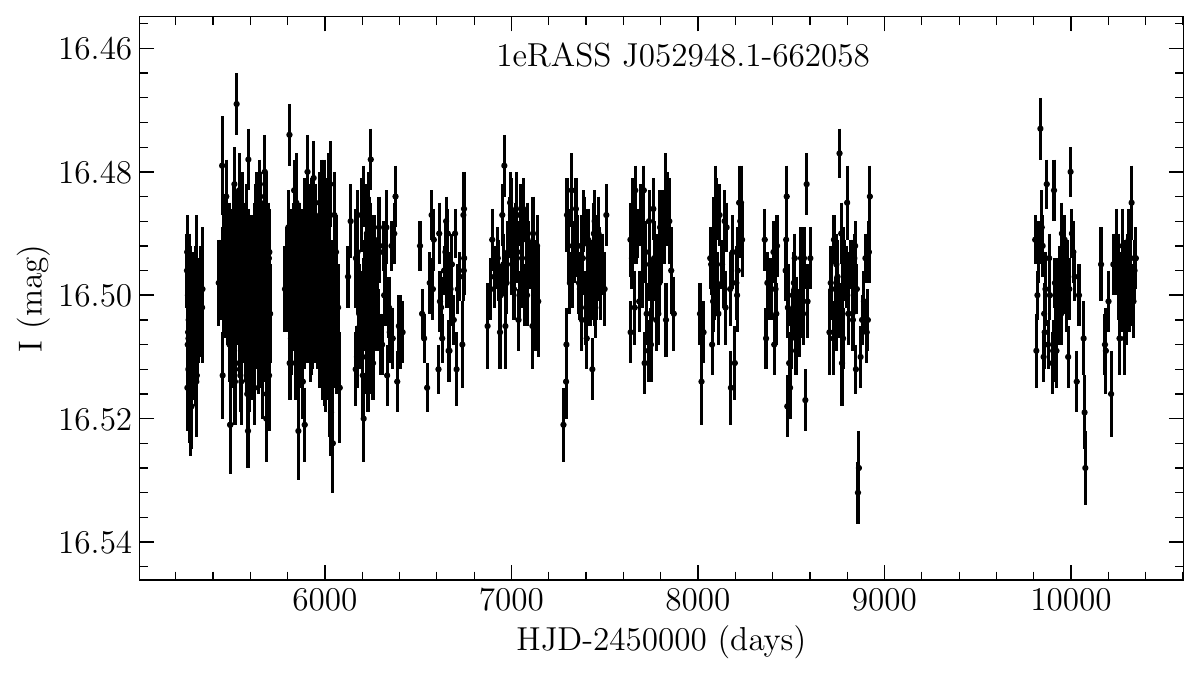}}
        \resizebox{0.495\hsize}{!}{\includegraphics{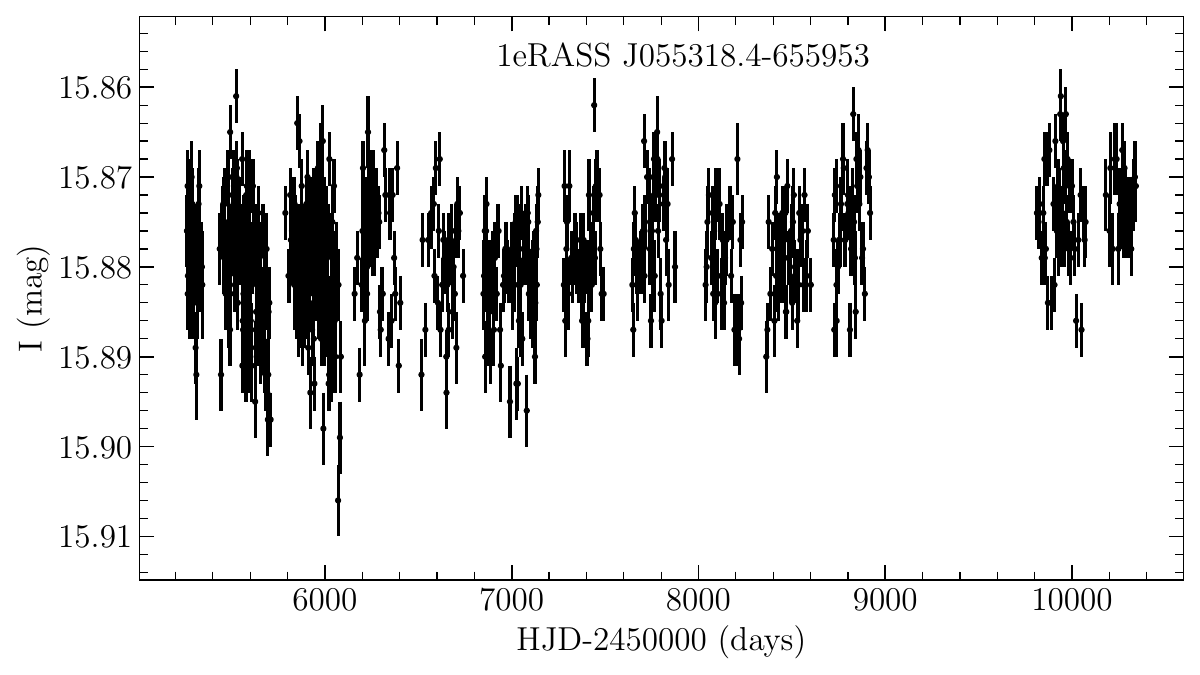}}
        \resizebox{0.495\hsize}{!}{\includegraphics{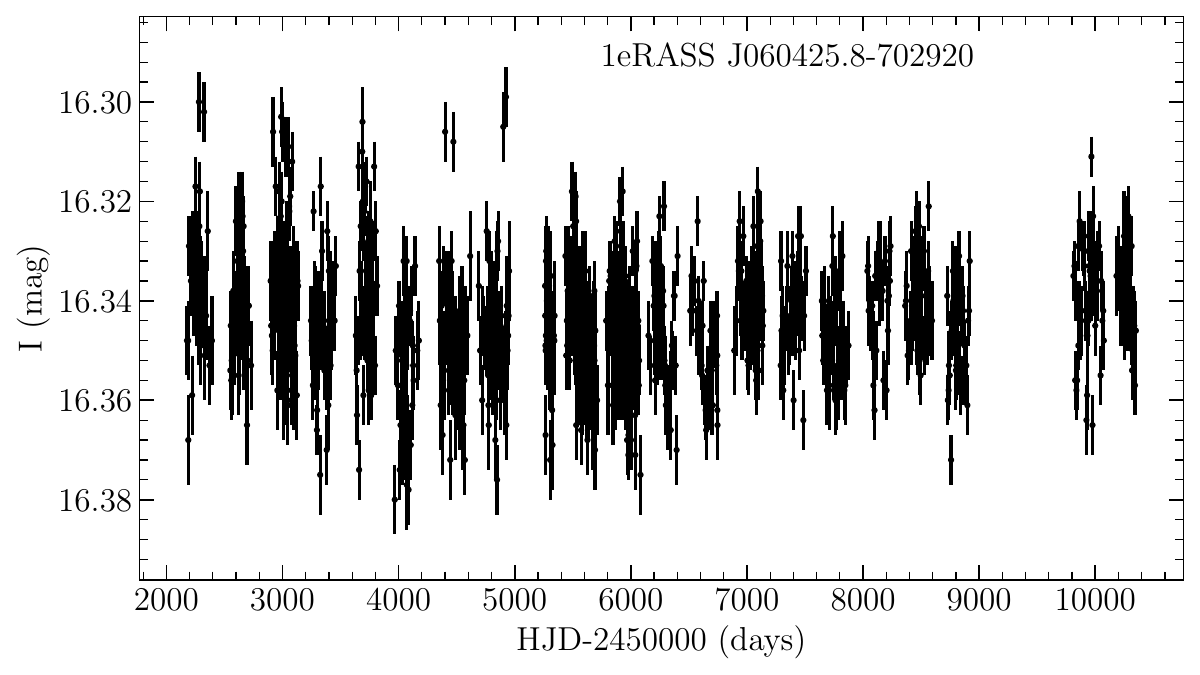}}
        \resizebox{0.495\hsize}{!}{\includegraphics{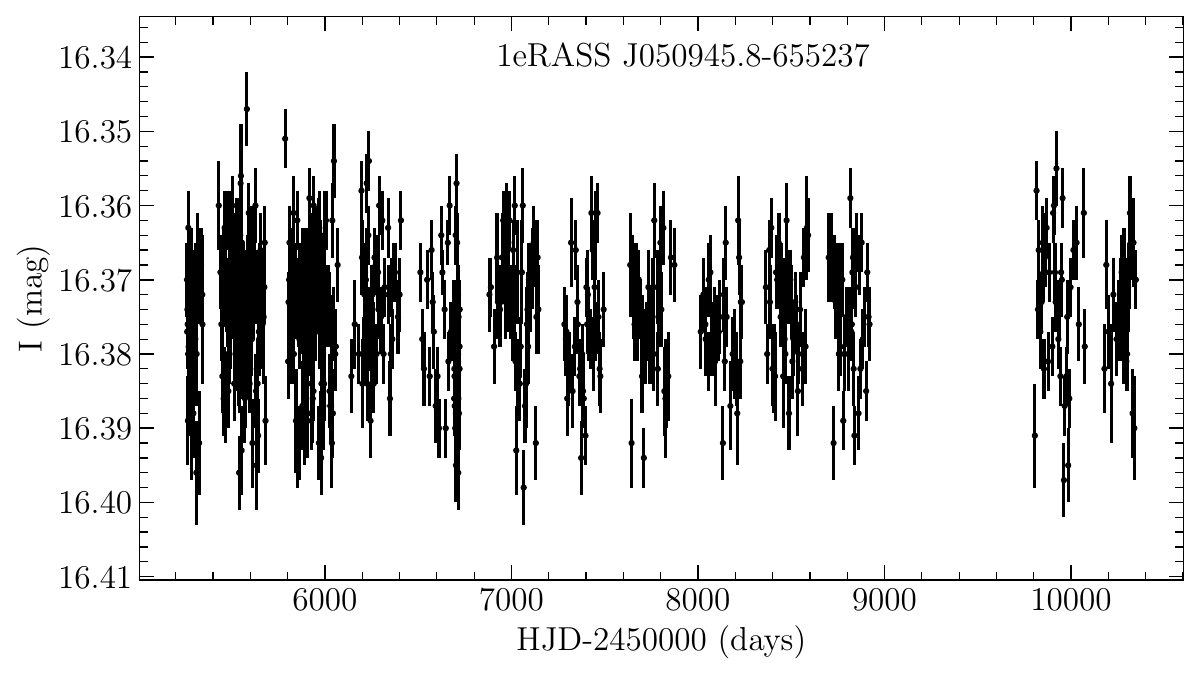}}
        \resizebox{0.495\hsize}{!}{\includegraphics{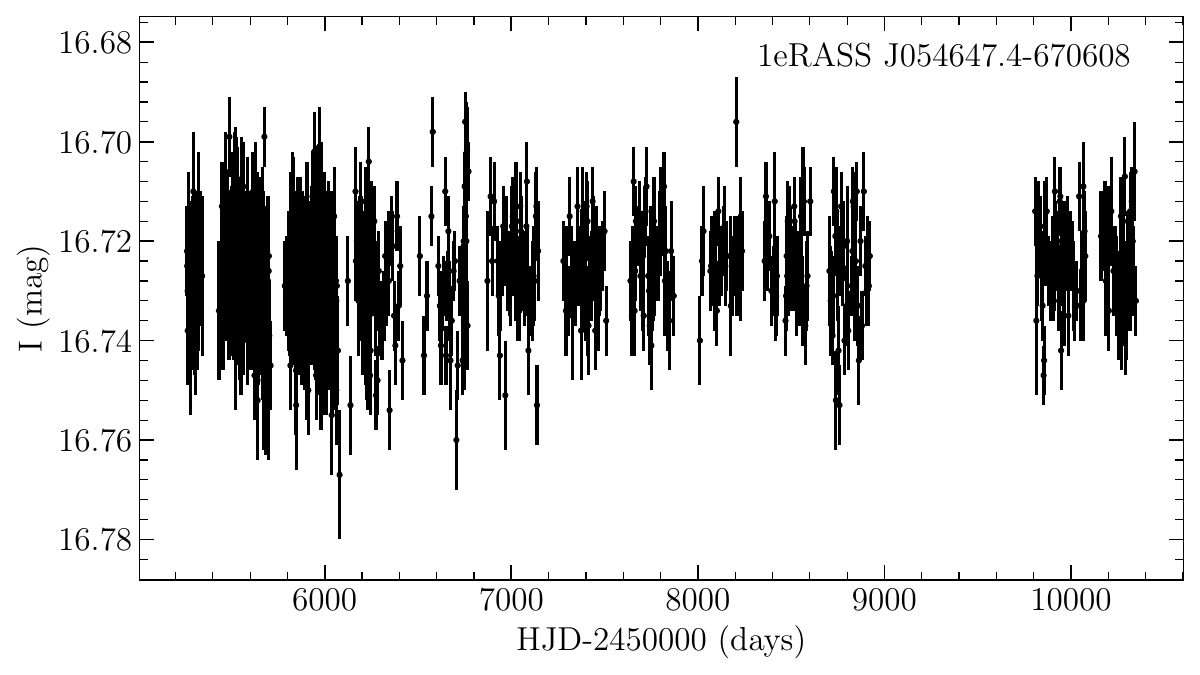}}
        \caption{Continued.}
        \label{fig:ogle_Ilc}
        \end{figure*}
        
        \begin{table*}[th]
                \centering
                \caption[]{Periodicities in \ogle data.}
                \label{tab:ogle_data}
                \begin{tabular}{rcll}
                        \hline\hline\noalign{\smallskip}
                        \multicolumn{1}{c}{\#} &
                        \multicolumn{1}{c}{Period (days)} &
                        \multicolumn{1}{l}{References} &
                        \multicolumn{1}{l}{OGLE-IDs (LMC...)} \\
                        \noalign{\smallskip}\hline\noalign{\smallskip}
                        1  &  49.6          & I, V (HMV22)             & \\  
                        2  &                & I (VSH13); XROM          & \\  
                        3  &                & I, V (HMV22)             & \\  
                        4  &  451           & I, V (HMK23)             & \\
                        5  &  40.16         & I (vJBM18); I, V (MHM21) & \\
                        6  &  262           & I (vJBM18); XROM         & \\
                        7  &                & I, V (HMK23)             & \\
                        8  &  1350          & I, V (HIR17); XROM       & \\
                        9  &  27.4          & I (vJBM18); XROM         & \\
                        10  &                & I (TW)                   & 503.23.76   \\
                        11  &  24.5          & I, V (VHS14); XROM       & \\
                        12  &                & I, V (HMV22)             & \\
                        13  &  193           & I (vJBM18)               & \\
                        14  &  151           & I, V (MKH23)             & \\
                        15  &                & I, V (TVB21); XROM       & \\
                        16  &  74            & I (vJBM18); XROM         & \\
                        17  &                & V (VMH13)                & \\
                        18  &                & I (vJBM18)               & \\
                        19  & 3.37/7.19 (TW) & I (TW)                   & 519.28.286D\\
                        20  &  1.40841       & Roche-lobe overflow      & \\
                        21  &  161 (TW)      & I, V (TW; see also RGC88) & 518.27.2345 518.27.v.99\\
                        22  &  16.651501     & I (DMS22); XROM          & \\
                        23  &  10.3          & I (CCC16)                & \\
                        24  &  1.7048        & BH               & \\
                        25  &  3.909         & BH               & \\
                        26  &  31.5          & I (vJBM18); XROM         & \\
                        27  &                & I (CNB01); XROM          & \\
                        28  &  45.7 (TW)     & I (TW)                   & 554.08.234D 554.09.806D\\
                        29  &                & I, V (TW)                & 538.26.19 142.8.11 538.26.v.7\\
                        30  &                & I (TW)                   & 531.13.36290 135.7.17289\\
                        31  &  83.87 (TW)    & I (TW)                   & 531.13.22231\\
                        32  &                & I (TW)                   & 533.23.7302\\
                        33  &                & no data                  & \\
                        34  &  160 (TW)      & I (TW)                   & 534.18.80\\
                        35  &                & I (TW) & 124.1.4920 125.4.33202 124.1.v.6401 125.4.v.25878\\
                        36  &                & I (TW)                   & 508.29.20314 509.05.2 121.7.11219\\
                        37  & 7.61/7.77 (TW) & I (TW)                   & 513.04.11386\\
                        38  &                & I (TW)                   & 512.27.106\\
                        39  &                & I (TW)                   & 519.06.4885D\\
                        40  &                & I (TW)                   & 519.13.8356\\
                        41  &                & I (TW)                   & 519.12.24706\\
                        42  &                & I (vJBM18)               & \\
                        43  &  2.125 (TW)    & I (TW)                   & 516.20.21135\\
                        44  &                & I (TW)                   & 553.32.17530 174.1.18879\\
                        45  &                & I, V (TW)                & 554.06.18667 554.06.v.30622\\
                        46  & 6.49/6.61 (TW) & I, V (TW)                & 554.30.12948 555.04.14513 554.30.v.23929 555.04.v.2\\
                        47  &  5.24 (TW)     & I (TW)                   & 555.03.13254 554.29.14293\\
                        48  &                & I (TW)                   & 555.10.68\\
                        49  &                & I (TW)                   & 563.21.7043\\
                        50  &                & I (TW)                   & 562.21.9149\\
                        51  &  170.3 (TW)    & I (TW)                   & 562.02.7990\\
                        52  &  122 (TW)      & I (TW)                   & 562.08.5 \\
                        53  &                & I (TW)                   & 560.01.54\\
                        \noalign{\smallskip}\hline
                \end{tabular}
                \tablefoot{
                        Column 1 refers to source numbers from Table\,\ref{tab:MasterTable_known}.
                        Periods are reported from literature or proposed in this work (TW).
                        References to previously published \ogle I- and V-band light curves:
                        CCC16 \citep{2016ApJ...829..105C}; 
                        CNB01 \citep{2001MNRAS.324..623C};
                        DMS22 \citep{2022A&A...661A..22D};
                        HIR17 \citep{2017A&A...598A..69H};
                        HMK23 \citep{2023A&A...671A..90H}; 
                        HMV22 \citep{2022A&A...662A..22H};
                        MHM21 \citep{2021MNRAS.504..326M};
                        MKH23 \citep{2023A&A...669A..30M};
                        RGC88 \citep[241 days was found by][from a UK Schmidt telescope photographic survey]{1988MNRAS.232...53R};
                        TVB21 a\citep{2021MNRAS.503.6187T};
                        vJBM18 \citep{2018MNRAS.475.3253V};
                        VSH13 \citep{2013ATel.5540....1V};
                        VHS14 \citep{2014A&A...567A.129V};
                        VMH13 \citep[MACHO data,][]{2013A&A...558A..74V};
                        TW (this work);
                        XROM\footnote{X-Ray variables \ogle Monitoring with regularly updated \ogle-IV I-band light curves: https://ogle.astrouw.edu.pl/ogle4/xrom/xrom.html} \citep[][]{2008AcA....58..187U}.
                }
        \end{table*}
        
        \begin{figure*}
                \centering
                \resizebox{0.495\hsize}{!}{\includegraphics{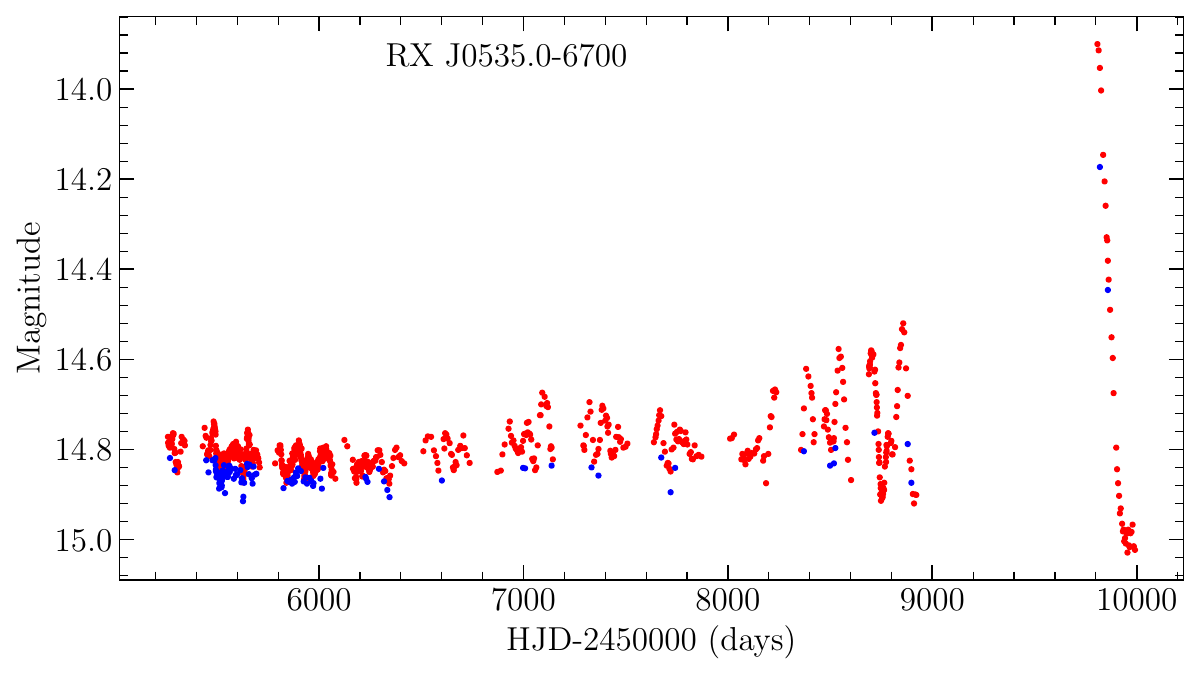}}
                \resizebox{0.495\hsize}{!}{\includegraphics{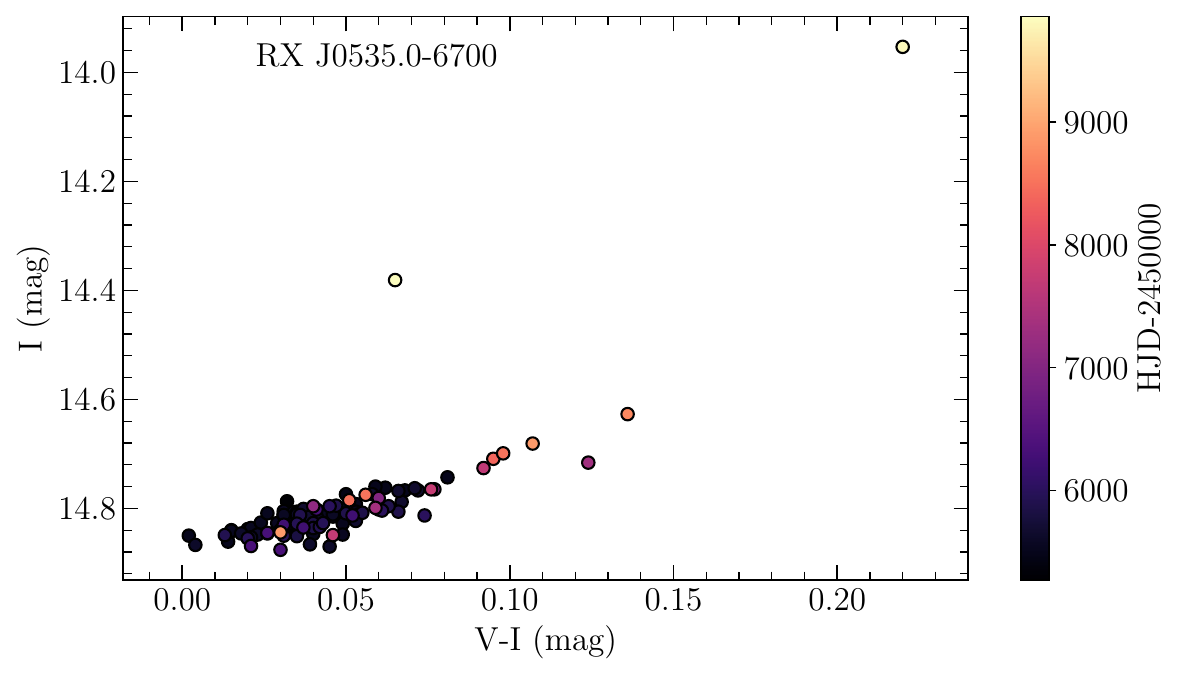}}
                \resizebox{0.495\hsize}{!}{\includegraphics{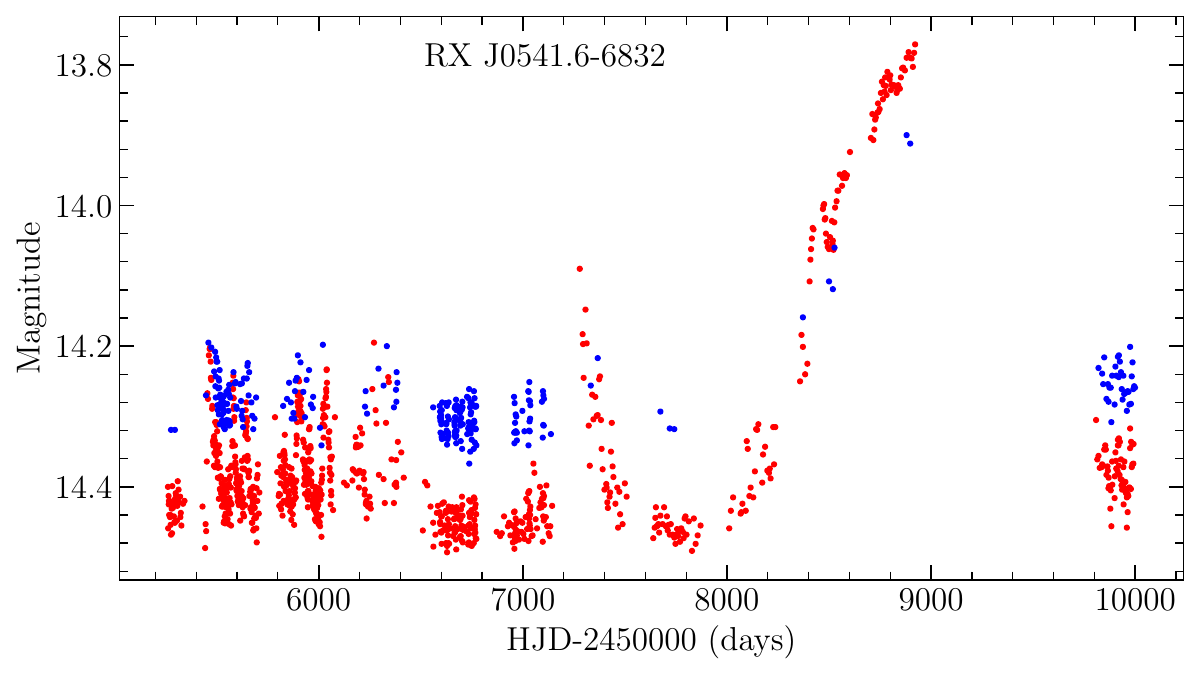}}
                \resizebox{0.495\hsize}{!}{\includegraphics{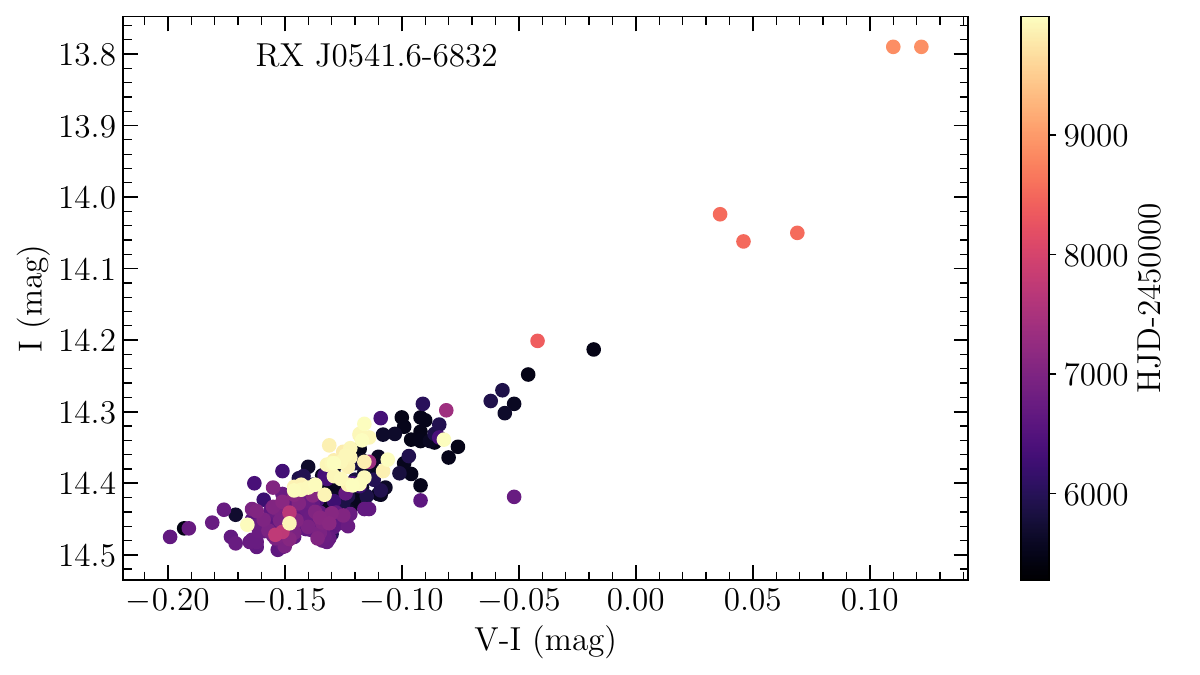}}
                \resizebox{0.495\hsize}{!}{\includegraphics{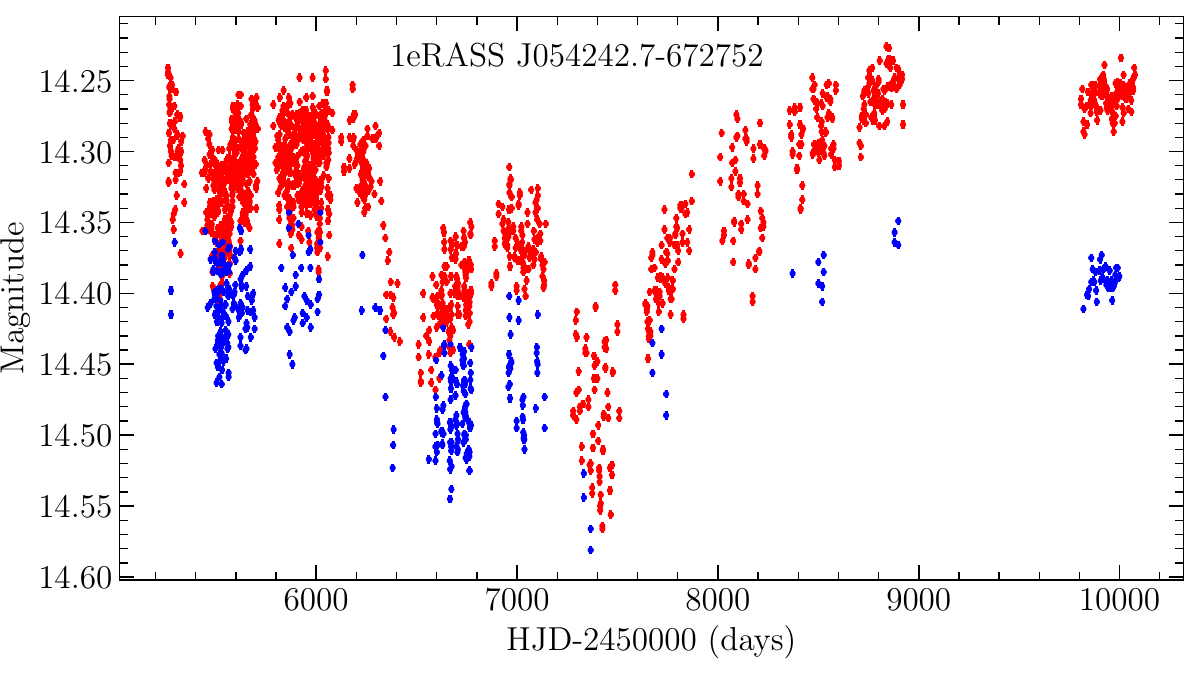}}
                \resizebox{0.495\hsize}{!}{\includegraphics{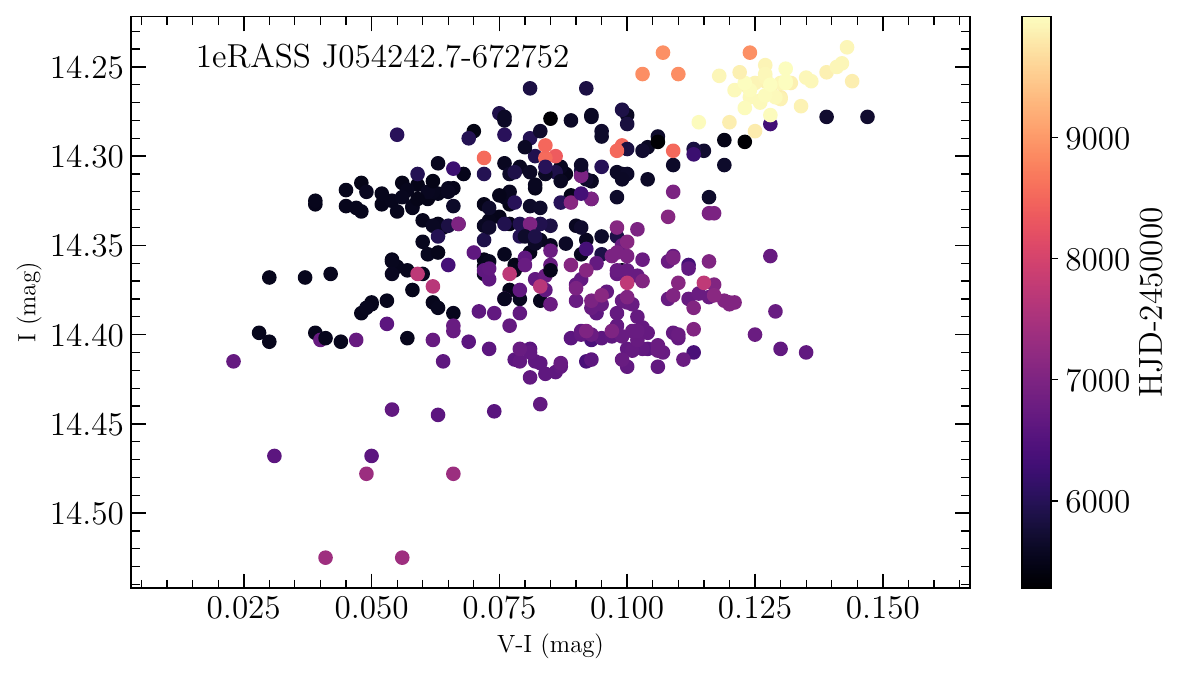}}
                \resizebox{0.495\hsize}{!}{\includegraphics{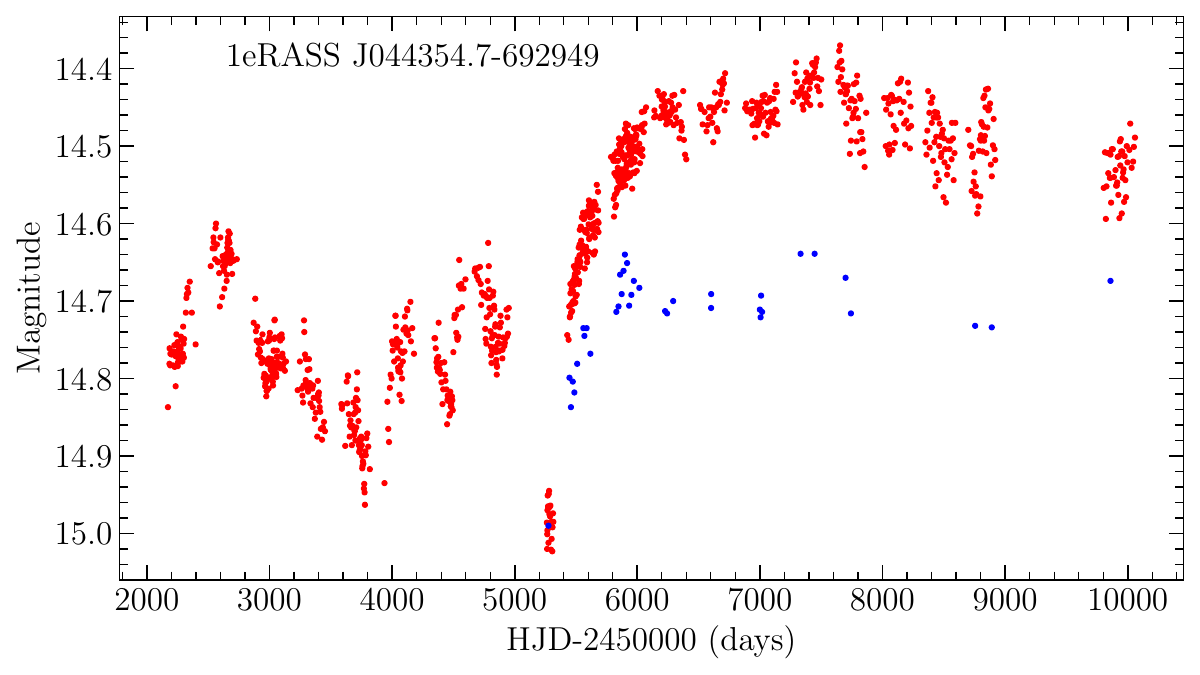}}
                \resizebox{0.495\hsize}{!}{\includegraphics{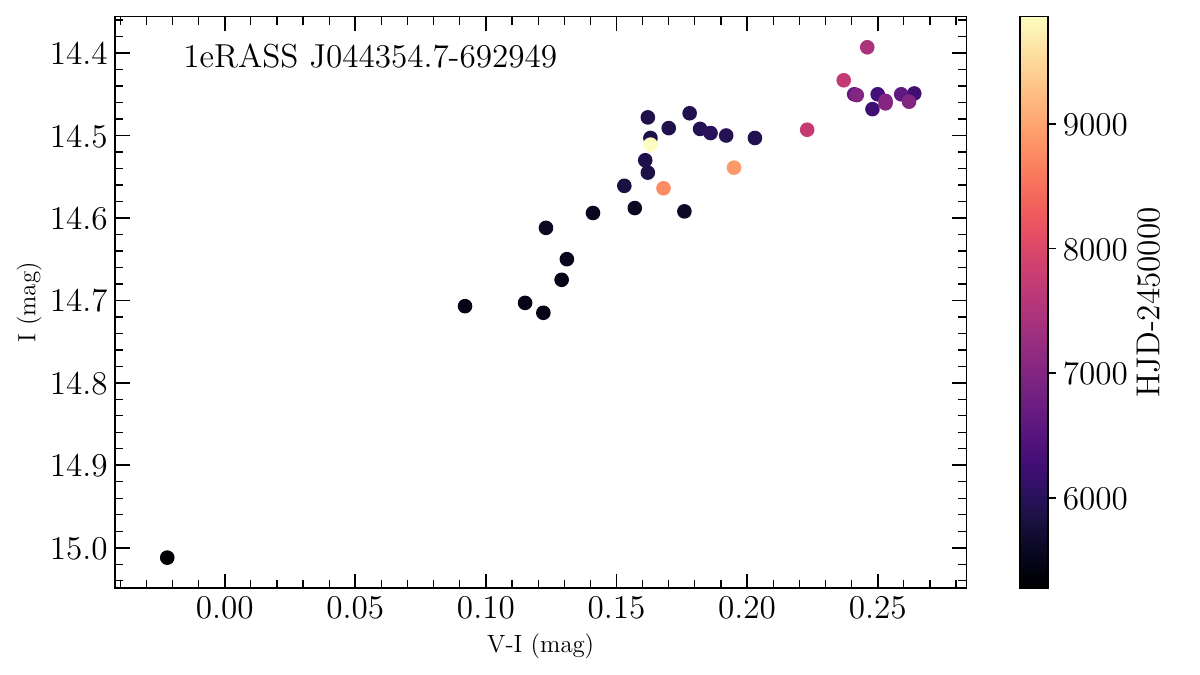}}
                \caption{\ogle photometry of highly variable stars ($>$0.3 mag in I). Left: \ogle I-band (red) and V-band (blue) light curves. Right: Colour (V$-$I) - magnitude (I) diagrams.}
                \label{fig:ogle_IVlc}
        \end{figure*}
\end{document}